\documentclass[11pt,a4paper,oneside]{scrartcl}

\usepackage[T1]{fontenc}
\usepackage[utf8]{inputenc}
\usepackage{libertinus}
\usepackage{amsmath,amsthm,mathtools}
\usepackage[libertinus,slantedGreek,varbb]{newtxmath}
\usepackage[scr=rsfs,frak=euler]{mathalpha}
\DeclareOldFontCommand{\rm}{\normalfont\rmfamily}{\mathrm}
\usepackage{microtype}
\usepackage[a4paper,margin=2.65cm]{geometry}
\usepackage{enumitem}
\usepackage{needspace}
\usepackage{placeins}
\usepackage{xcolor}
\usepackage{tikz}
\usetikzlibrary{arrows.meta,calc,patterns.meta}

\definecolor{phys}{HTML}{0072B2}
\definecolor{aux}{HTML}{D55E00}
\colorlet{notecol}{black!62}
\tikzset{ 
 lppl diagram/.style={x=1cm,y=1cm,font=\small,
   >={Latex[length=4pt,width=3pt]},line width=.5pt,
   line cap=round,line join=round},
 lppl box/.style={draw=black!55,fill=black!3,rounded corners=3pt,
   align=center,inner sep=5pt},
 lppl arrow/.style={->,line width=.6pt},
 fig note/.style={font=\footnotesize,text=notecol},
 fig title/.style={font=\small},
 dim line/.style={{Latex[length=3pt,width=2.4pt]}-{Latex[length=3pt,width=2.4pt]},
   black!75,line width=.45pt},
 excluded/.style={fill=black!4,postaction={pattern={Lines[angle=45,
   distance=4.5pt,line width=.35pt]},pattern color=black!22}},
 observable/.style={fill=phys!16,draw=phys!75!black,line width=.6pt},
 lppl comparison region/.style={fill=phys!4,draw=black!65,line width=.6pt},
 lppl bulk term/.style={phys!45,line width=.9pt},
 lppl boundary term/.style={phys!80!black,line width=1.1pt},
 lppl defect term/.style={aux,line width=1.1pt},
 lppl exterior term/.style={black!35,line width=.9pt},
 lppl term site/.style={circle,inner sep=0pt,minimum size=3.2pt},
 lppl evolution/.style={lppl box,text width=4.8cm},
 lppl restricted evolution/.style={lppl evolution,draw=phys!75!black,
   fill=phys!8,line width=.7pt},
 lppl comparison arrow/.style={{Latex[length=3.5pt,width=3pt]}-{Latex[length=3.5pt,width=3pt]},line width=.55pt},
 lppl comparison label/.style={font=\footnotesize,text=phys!80!black,align=left},
}
\usepackage{hyperref}

\hypersetup{
 hidelinks,
 bookmarksnumbered=true,
 pdftitle={Local determination of low-energy states in gapped fermion systems with weak long-range interactions},
 pdfauthor={Stefan Teufel}
}

\setkomafont{author}{\sffamily}
\setkomafont{date}{\sffamily}
\setkomafont{sectionentry}{\normalfont\normalsize\sffamily}
\DeclareTOCStyleEntry[beforeskip=0pt,linefill=\TOCLineLeaderFill]{tocline}{section}
\DeclareTOCStyleEntry[entryformat=\normalfont\sffamily,pagenumberformat=\normalfont\sffamily]{tocline}{subsection}
\setlist[itemize]{topsep=4pt,itemsep=2pt,parsep=1pt}
\setlist[enumerate]{topsep=4pt,itemsep=2pt,parsep=1pt}
\allowdisplaybreaks
\newtheorem{theorem}{Theorem}[section]
\newtheorem{proposition}[theorem]{Proposition}
\newtheorem{lemma}[theorem]{Lemma}
\newtheorem{corollary}[theorem]{Corollary}
\theoremstyle{definition}
\newtheorem{definition}[theorem]{Definition}
\newtheorem{hypothesis}[theorem]{Hypothesis}
\theoremstyle{remark}
\newtheorem{remark}[theorem]{Remark}

\newcommand{\statementqedsymbol}{\ensuremath{\diamond}}
\let\proofqedsymbol\qedsymbol
\newcommand{\pushStatementQED}{\let\qedsymbol\statementqedsymbol
  \pushQED{\qed}}
\AddToHook{env/definition/begin}{\pushStatementQED}
\AddToHook{env/definition/end}{\popQED}
\AddToHook{env/remark/begin}{\pushStatementQED}
\AddToHook{env/remark/end}{\popQED}
\AddToHook{env/hypothesis/begin}{\pushStatementQED}
\AddToHook{env/hypothesis/end}{\popQED}
\AddToHook{env/proof/begin}{\let\qedsymbol\proofqedsymbol}

\newcommand{\A}{\mathcal A}
\newcommand{\Hsp}{\mathfrak h}
\newcommand{\C}{\mathbb C}
\newcommand{\R}{\mathbb R}
\newcommand{\Z}{\mathbb Z}
\newcommand{\1}{\mathbf 1}
\newcommand{\I}{\mathrm{i}}
\newcommand{\E}{\mathrm{e}}
\newcommand{\dd}{\mathrm d}
\newcommand{\supp}{\operatorname{supp}}
\newcommand{\diam}{\operatorname{diam}}
\newcommand{\dist}{\operatorname{dist}}
\newcommand{\spec}{\operatorname{spec}}
\newcommand{\LG}{\mathrm{LG}}
\newcommand{\ball}{\mathrm B}
\newcommand{\at}{\mathrm{pv}}
\newcommand{\st}{\mathrm{st}}
\newcommand{\sgnh}{\mathsf S}
\newcommand{\vac}{\mathrm{vac}}
\newcommand{\loc}{\mathrm{loc}}
\newcommand{\pol}{\mathrm{pol}}
\newcommand{\nc}{\mathrm{nc}}
\newcommand{\lambdaLED}{\lambda_{\mathrm{LED}}}
\newcommand{\lambdaWC}{\lambda_{\mathrm{wc}}}
\newcommand{\JLED}{\mathcal J_{\mathrm{LED}}}
\newcommand{\JWC}{\mathcal J_{\mathrm{wc}}}

\title{Local determination of low-energy states\\ in gapped fermion systems\\ with weak long-range interactions}
\author{Stefan Teufel\thanks{Fachbereich Mathematik, Universit\"at
T\"ubingen, Auf der Morgenstelle 10, 72076 T\"ubingen, Germany.}}
\date{December 2, 2026}

\begin{document}

\maketitle

\begin{abstract}
We consider bulk-gapped fermion systems whose infinite-volume bulk is obtained by adding sufficiently weak even many-body interactions to a free Hamiltonian with spectral gap \(g>0\), allowing polynomial spatial decay of both parts.
We prove \emph{quantitative local determination of low-energy states}: whenever another interaction agrees with the bulk interaction inside a finite or infinite region \(U\), with bulk terms crossing its boundary unchanged or removed, all its states with excitation energies in \([0,3g/4]\) have local expectations close to those of the bulk ground state deep inside \(U\).
The error decays with the distance to the complement of \(U\), independently of the interaction terms supported outside \(U\) and of the chosen low-energy state.

Quantitative local determination has strong consequences.
(i) The bulk interaction has a unique algebraic ground state.
(ii) A strong “local perturbations perturb locally” (LPPL) principle holds: arbitrary distant defects change local ground-state expectations only by a spatially decaying error.
(iii) In finite samples, every low-energy state remains locally indistinguishable from the bulk ground state far from the boundary, up to a spatially decaying error uniform in time and in the choice of even boundary terms.
(iv) Correlations from deep inside \(U\) to outside \(U\) are small, with quantitative bounds in terms of spatial separation and distance to \(U^c\).

Our proof uses the quasi-adiabatic spectral flow to transport the non-interacting ground state to the interacting bulk ground state.
For the systems considered here, a companion paper proves that the free spectral gap persists under sufficiently weak many-body interactions and establishes the required locality properties of the spectral flow.
After adjoining an auxiliary free system, we transform the combined ground state to a product vacuum and introduce a new condition, called local excitation detectability (LED).
Together with locality estimates for the dynamics, LED yields quantitative local determination of low-energy states.
We prove that LED holds at weak coupling, but the abstract result is nonperturbative and applies whenever the transport and detectability hypotheses hold.
\end{abstract}

\clearpage
\tableofcontents
\clearpage

\section{Introduction}
\label{sec:introduction}

Introducing a boundary or a strong defect into a bulk-gapped material like a topological insulator can create excitations at arbitrarily small energies \cite{HasanKane10}.
We prove that, for fermion systems obtained by adding sufficiently weak many-body interactions to a gapped free system, these excitations have only a spatially decaying influence on the interior.
Every state in a fixed window above the modified ground energy has local expectations close to those of the unmodified bulk ground state deep inside any region \(U\) where the terms supported in \(U\) are unchanged and each bulk term crossing its boundary is unchanged or removed.
The bounds are independent of the strength, extent, and geometry of the exterior changes and remain valid throughout the evolution of every state in the window.
To our knowledge, even qualitative local agreement with the bulk ground state is new in this generality: the exterior perturbation is arbitrary and the low-energy window need not be separated from the rest of the spectrum. 
We call this result quantitative local determination of low-energy states. 
It gives a rigorous basis for the familiar picture of an insulator whose edge can support low-energy excitations while the physics deep in the bulk remains essentially unchanged at all times.

We work on a countable metric space \((\Gamma,d)\) of polynomial volume growth, with \(r\in\mathbb N\) fermionic modes per site.
We write \(X\Subset\Gamma\) for a finite subset of \(\Gamma\).
Let \(\A\) be the quasi-local algebra and, for each \(X\Subset\Gamma\), let \(\A_X\) be its local subalgebra:
\[
 \A:=\operatorname{CAR}(\ell^2(\Gamma;\C^r)),
 \qquad
 \A_X:=\operatorname{CAR}(\ell^2(X;\C^r)).
\]
The free system is specified by a self-adjoint Majorana coefficient operator \(h\) on \(\ell^2(\Gamma;\C^{2r})\), with \(\overline h=-h\), spatial decay, and \(\spec(h)\cap(-g,g)=\varnothing\) for some \(g>0\).
Its interaction \(\Phi_h\), defined in Subsection~\ref{sec:quadratic-reference}, has a unique pure quasifree gapped ground state \(\omega_0\).
Given a spatially decaying self-adjoint even many-body interaction \(\mathcal V\) and \(\lambda\in\R\), set
\[
 \Phi_\lambda:=\Phi_h+\lambda\mathcal V.
\]
The interaction \(\Phi_\lambda\) defines a derivation \(\delta_{\Phi_\lambda}\), given on every local observable \(B\) by the norm-convergent sum
\[
 \delta_{\Phi_\lambda}(B)
 :=\sum_{Z\Subset\Gamma}[\Phi_\lambda(Z),B].
\]
In \cite{StabilityBdG26}, we construct, for sufficiently small \(|\lambda|\) a family of parity-preserving automorphisms \(\beta_\lambda\) of \(\A\), with \(\beta_0=\operatorname{id}\), such that
\[
 \omega_\lambda:=\omega_0\circ\beta_\lambda
\]
is a gapped ground state of \(\Phi_\lambda\).
We refer to this relation as \emph{ground-state transport} and to \(\beta_\lambda\) as the \emph{spectral flow}.
Our comparison results assume this transport and uniform \emph{local excitation detectability} (LED), which converts bounds on filtered annihilation operators into bounds on local excitation densities.
Hypotheses~\ref{hyp:background-transport} and~\ref{hyp:uniform-local-excitation-detectability} state the transport and LED assumptions, including the locality bounds.
In Proposition~\ref{prop:unconditional-weak-coupling} we show that they always hold for sufficiently small coupling by combining the transport result from \cite{StabilityBdG26} with a proof of the LED criterion given here.
The abstract comparison results hold on any coupling set \(\mathcal J\) where these assumptions are satisfied, with no further smallness requirement on \(|\lambda|\).

Theorem~\ref{thm:positive-energy-locality} formulates the main comparison in an arbitrary positive-energy representation.
Fix \(\lambda\in\mathcal J\), let \(U\subset\Gamma\) and let \(\Xi\in\mathfrak J_1\) be a modified interaction whose terms supported entirely in \(U\) agree with those of \(\Phi_\lambda\).
Bulk terms crossing the boundary of \(U\) may or may not be removed, and terms supported entirely in \(U^c\) may be changed arbitrarily, as specified in \eqref{eq:comparison-interaction-criterion}.
Let \(\pi\) be a unital representation of \(\A\) in which a nonnegative self-adjoint operator \(L\) implements the dynamics of \(\Xi\), and set
\[
 P:=\1_{[0,3g/4]}(L).
\]
For the support exponent \(q\le1\) fixed in Section~\ref{sec:main-results}, there are a constant \(C>0\) and a common decay profile
\[
 G(D)=
 \begin{cases}
 \E^{-mD^\theta},&\text{stretched-exponential decay of \(h\) and \(\mathcal V\)},\\
 (1+D)^{-\nu},&\text{polynomial decay of \(h\) and \(\mathcal V\)},
 \end{cases}
\]
such that, for every nonempty finite \(Y\Subset\Gamma\) and every \(A\in\A_Y\),
\[
 \|P(\pi(A)-\omega_\lambda(A)\1)P\|
 \le C(1+|Y|)^q\|A\|G(\dist(Y,U^c)).
\]
Thus every state supported in the possibly infinite-dimensional low-energy subspace \(\operatorname{Ran}P\) has approximately the same local measurement statistics as the bulk ground state deep inside \(U\), with an error that decays with the distance to \(U^c\).
The constant \(C\) and the decay profile \(G\) are independent of the region \(U\), the modified interaction \(\Xi\), and the chosen positive-energy representation \((\pi,L)\).
The estimate holds with the same constants under every matrix amplification, so it also controls local observations in the presence of an arbitrary finite-dimensional ancillary system.

We obtain uniqueness of the algebraic ground state of \(\Phi_\lambda\) directly from Theorem~\ref{thm:positive-energy-locality}, applied with \(U=\Gamma\) and \(\Xi=\Phi_\lambda\) in the GNS representation of any algebraic ground state (Corollary~\ref{cor:ground-state-uniqueness}).
Stability of the bulk gap alone does not imply this uniqueness: distinct ground states of the same interaction may each have a positive GNS gap, as illustrated in Subsection~\ref{sec:hypotheses-discussion}.
To our knowledge, this paper and the independent argument in \cite{ThermalLPPL26} provide the first general proofs of uniqueness among all algebraic ground states for weak many-body perturbations of spatially decaying, gapped free-fermion systems.

For polynomially localized \(h\) and \(\mathcal V\), with weighted decay exponents \(p_h,p_{\mathcal V}>d_{\rm vol}\), the weak-coupling results give the full decay power \(\nu=\min\{p_h,p_{\mathcal V}\}\).
For two-site interactions, our assumptions allow every pointwise decay power \(\varpi>2d_{\rm vol}\).
On \(\Z^d\) with \(d_{\rm vol}=d\), this range reaches the threshold for algebraic Lieb--Robinson light cones for general two-site dynamics \cite{FossFeigEtAl15,TranEtAl21,TranEtAlStateTransfer21}.
For these polynomial bounds, we impose an exponential support-size weight on \(\mathcal V\), automatic for bounded support size, and a cell-partition condition on \(\Gamma\), satisfied by uniformly discrete sets in Euclidean spaces \(\R^d\).
For stretched-exponentially localized \(h\) and \(\mathcal V\) of order \(0<\theta_{\rm in}\le1\), we obtain every fixed \(0<\theta<\theta_{\rm in}\).

For a defect \(\Psi\in\mathfrak J_1\) supported in \(S\), the interaction \(\Xi=\Phi_\lambda+\Psi\) satisfies the comparison criterion with \(U=S^c\).
We apply the theorem in the GNS representation of each ground state of \(\Phi_\lambda+\Psi\) to obtain locality throughout its low-energy window (Corollary~\ref{thm:low-energy-locality}), including strong ground-state LPPL (\emph{local perturbations perturb locally}) at zero energy.
The bound is independent of the defect's strength, extent, geometry, and interaction range, and of the chosen ground state and the spectrum of its GNS Hamiltonian.
This gives a quantitative form of the \emph{nearsightedness principle} of Kohn~\cite{Kohn96} and Prodan and Kohn~\cite{ProdanKohn05}, extending its scope from changes of external potential to arbitrary distant interaction changes and their effect on the full local state.

For long-range interactions, changing the couplings across an interface may also change terms that reach into the observation region \(U\).
In Theorem~\ref{thm:quasilocal-tails}, we bound their effect on local expectations by a contribution with profile \(G\) and an additional contribution controlled by the strength and spatial decay of the changed terms that reach into \(U\).
Corollary~\ref{cor:uniformly-weak-defects} also bounds changes of local expectations by the uniform local strength of a perturbation, even when it extends throughout the system.

For a finite sample, take \(L\) to be its Hamiltonian minus its ground energy.
Corollary~\ref{cor:finite-sample-bulk} compares local expectations in every state in the window \([0,3g/4]\) with those of the infinite-volume bulk ground state, uniformly over arbitrary even boundary operators.
Estimate~\eqref{eq:infinite-region-low-energy} gives the corresponding result in possibly infinite regions.
Since \(P\) commutes with \(\E^{-\I tL}\), the same expectation bound holds uniformly in real time for every normal state supported in the window.
This permits nontrivial low-energy boundary dynamics in a topological insulator while interior expectations remain close to their bulk ground-state values uniformly for all times.
The same conclusion holds under time-dependent driving that preserves this energy window, as explained in Subsection~\ref{sec:finite-samples}.

We also determine local expectations from energy minimality imposed only inside \(U\).
For \(U\subset\Gamma\), let \(\LG_{\Phi_\lambda}(U)\) be the set of states \(\rho\) on \(\A\) satisfying the local ground state inequality, i.e.\ for every finite \(X\Subset U\) and every \(B\in\A_X\),
\[
 \operatorname{Re}\rho\bigl(B^*\delta_{\Phi_\lambda}(B)\bigr)\ge0.
\]
In particular, no unitary supported on a finite subset of \(U\) can lower the energy in \(\rho\).
Theorem~\ref{thm:bulk-uniqueness} gives, with the same \(C,q,G\) as in the low-energy theorem,
\[
 \sup_{\rho\in\LG_{\Phi_\lambda}(U)}
 |\rho(A)-\omega_\lambda(A)|
 \le C(1+|Y|)^q\|A\|G\bigl(\dist(Y,U^c)\bigr)
 \qquad(A\in\A_Y).
\]
The bound is uniform in \(U\), its geometry, and the behavior of \(\rho\) outside \(U\).
For finite-range interactions and self-adjoint \(A\), we bound \(\omega_\lambda(A)\) from below and above by optimizing over density matrices in a finite neighborhood of its support, subject to the local ground-state inequalities.
Theorem~\ref{thm:bulk-uniqueness} gives an explicit rate at which these bounds converge as the neighborhood grows (Subsection~\ref{sec:local-ground-state-results}).

We also control correlations when only one observable lies deep inside \(U\).
For \(A\in\A_X\) and \(B\in\A_Y\), Theorem~\ref{thm:spatial-correlations} bounds their covariance by a term decaying with \(\dist(X,Y)\) and a remainder decaying with \(\dist(X,U^c)\), even when \(Y\) lies outside \(U\).
The bound applies both to the low-energy states described above and to states in \(\LG_{\Phi_\lambda}(U)\).
Corollary~\ref{cor:low-energy-spectral-weight} gives further bounds on low-energy spectral weight and imaginary-time correlations.
Comparing local states also gives trace-distance and entropy estimates, as well as mutual-information estimates for even comparison states.

To prove these results, we adjoin an auxiliary free system and express the combined interacting ground state as a product vacuum by composing a quasi-local Bogoliubov transformation of the doubled free system \cite{Araki70,Hastings19} with the quasi-adiabatic spectral flow \(\beta_\lambda\) \cite{Hastings04,HastingsWen05,BMNS12}.
In the companion paper \cite{StabilityBdG26}, we construct these transformations and bound the error in approximating their images of local observables by observables supported in enlarged neighborhoods.
Using the bulk propagation estimates of \cite{FermionicLR}, we compare the bulk and modified evolutions of the annihilation operators of this vacuum, expressed in the original coordinates.
Agreement of the interaction terms inside \(U\) makes these evolutions close for times controlled by the distance to \(U^c\), and integration against a filter function that decays rapidly in time converts this comparison into spatially decaying bounds on the differences of filtered operators.
We choose the filter to select only energy transfers that lower the energy by more than \(3g/4\). 
Since the Hamiltonian of the stacked system is nonnegative, the operators filtered with the modified dynamics annihilate every vector with energy in \([0,3g/4]\).
Our comparison then shows that the operators at \(x\) filtered with the bulk dynamics almost annihilate every unit vector in the low-energy window, with an error that decays with the distance from \(x\) to \(U^c\).

We then use our local excitation detectability condition (LED) to show that low-energy states have few excitations deep inside \(U\). 
In product-vacuum coordinates, LED bounds the expected excitation number at \(x\) by a weighted sum of the expected excitation numbers at sites \(z\), up to an error that is small wherever the bulk-filtered operators almost annihilate the low-energy vector. 
Since the weights sum to less than one uniformly in \(x\), we can iterate this inequality as in Dobrushin’s comparison argument for classical Gibbs measures \cite[Corollary~2.6]{RebeschiniVanHandel14}. 
The resulting bound on the excitation densities shows that local expectations deep inside \(U\) are close to those of the product vacuum. 
The locality of the change of coordinates then gives the desired agreement with the bulk ground state.
For states satisfying only a local ground-state inequality, we construct a positive operator on a finite local cyclic subspace, as in \cite[Appendix]{Yarotsky05}, and use its dynamics in the same filtering argument.
Subsection~\ref{sec:proof-strategy-roadmap} gives a detailed roadmap to the proof.

For states satisfying the ground-state inequality in the interior of a finite volume, Yarotsky proved an analogous estimate for weak finite-range perturbations of gapped product spin Hamiltonians \cite[Theorem~2]{Yarotsky05}.
Using his estimate, Henheik, Teufel, and Wessel showed that an arbitrarily strong defect changes local ground-state expectations by an amount that decays exponentially with the distance to the defect \cite{HTW22}.
Along gap-preserving paths, spectral flow also yields LPPL for spin systems \cite{BMNS12,DeRoeckSchuetz15} and long-range fermionic systems \cite{TeufelWessel25}.
For weakly interacting fermions, Hastings and De Roeck--Salmhofer established spectral-gap stability uniformly in finite volume \cite{Hastings19,DRS19}.
Giuliani, Mastropietro, and Porta used convergent fermionic expansions to prove rapid decay of truncated equilibrium correlations in weakly interacting two-dimensional periodic band insulators \cite{GMP17}.
Subsection~\ref{sec:related-results} gives a more detailed comparison.

The paper is structured as follows.
In Section~\ref{sec:setting} we introduce the setting and hypotheses and in Section~\ref{sec:main-results} we state the main results.
Sections~\ref{sec:odd-locality}--\ref{sec:filters} establish locality estimates for the dynamics, the transformation to product-vacuum coordinates, and the energy-filtered annihilation operators.
In Section~\ref{sec:reconstruction-bulk}, we show how small local excitation densities force local expectations to be close to those of the bulk ground state.
In Section~\ref{sec:local-determination-proofs}, we prove Theorem~\ref{thm:positive-energy-locality} for low-energy states and Theorem~\ref{thm:bulk-uniqueness} for states satisfying the local ground-state inequality, and derive the finite-sample consequences.
In Section~\ref{sec:detectability} we prove LED at weak coupling.
In Section~\ref{sec:quasilocal-tails} we treat defects with tails, and in Section~\ref{sec:correlation-proofs} we prove the correlation bounds.
In Appendix~\ref{app:first-moment-dynamics} we construct the infinite-volume dynamics for interactions with finite first support moment and show that the ground-state inequality can be tested on local observables.
In Appendix~\ref{app:odd-locality-proofs} we bound the difference between the free and interacting evolutions of linear combinations of Majorana operators.

\section{Setting}
\label{sec:setting}

\subsection{Algebras, interactions, and ground states}
\label{sec:ground-state-conditions}

Let \((\Gamma,d)\) be a countable metric space, the space of sites for the fermions.
For \(x\in\Gamma\) and \(R\ge0\), define \(\ball_R(x):=\{y\in\Gamma\,|\,d(x,y)\le R\}\).
We assume uniform polynomial growth of such balls: there are \(d_{\rm vol}\in[0,\infty)\) and \(C_\Gamma\ge1\) such that, for every \(x\in\Gamma\) and \(R\ge0\),
\begin{equation}
 |\ball_R(x)|\le C_\Gamma(1+R)^{d_{\rm vol}}.
 \label{eq:growth}
\end{equation}
For \(x\in\Gamma\), \(X,Y\subset\Gamma\), and \(R\ge0\), define
\begin{align*}
 \dist(x,X)&:=\inf_{y\in X}d(x,y),
 &\dist(X,Y)&:=\inf_{x\in X,\,y\in Y}d(x,y),\\
 X^{(R)}&:=\{y\in\Gamma\,|\,\dist(y,X)\le R\},
 &\diam X&:=\sup_{x,y\in X}d(x,y).
\end{align*}
We set \(\dist(X,\varnothing)=\infty\) and interpret \((1+\infty)^{-a}=\E^{-\mu\infty^\theta}=0\) for \(a,\mu,\theta>0\).

Fix \(r\in\mathbb N\), the number of fermionic modes per site, and set
\begin{equation*}
 \Hsp:=\ell^2(\Gamma;\C^r),
 \qquad \A:=\operatorname{CAR}(\Hsp).
\end{equation*}
For every \(X\subset\Gamma\), define
\begin{equation*}
 \Hsp_X:=\ell^2(X;\C^r),
 \qquad
 \A_X:=\operatorname{CAR}(\Hsp_X)\subset\A.
\end{equation*}
Set \(\A_\varnothing=\C\1\) and \(\A_{\loc}:=\bigcup_{X\Subset\Gamma}\A_X\).
Let \(a:\mathfrak h\to\mathcal A\) be the canonical antilinear annihilation map.
Fermion parity is the automorphism \(\Theta:\mathcal A\to\mathcal A\) determined by \(\Theta(a(\psi))=-a(\psi)\) for every \(\psi\in\mathfrak h\).
Write \(\A_X^+:=\{A\in\A_X\,|\,\Theta(A)=A\}\) for the even subalgebra of \(\A_X\).

An interaction is a map \(\Phi:Z\mapsto\Phi(Z)=\Phi(Z)^*\in\A_Z^+\) on the finite subsets of \(\Gamma\), with \(\Phi(\varnothing)=0\).
For every interaction \(\Phi\), define
\begin{equation}
 \|\Phi\|_1
 :=\sup_{x\in\Gamma}\sum_{\substack{Z\Subset\Gamma\\x\in Z}}
 |Z|\,\|\Phi(Z)\|,
 \qquad
 \mathfrak J_1:=\{\Phi\,|\,\|\Phi\|_1<\infty\}.
 \label{eq:first-moment-interaction-norm}
\end{equation}
We call \(\|\Phi\|_1\) the \emph{first support moment}.
A path \(s\mapsto\Xi_s\) is called \emph{componentwise norm-continuous} if \(s\mapsto\Xi_s(X)\) is norm-continuous for every \(X\Subset\Gamma\).

For \(\Phi\in\mathfrak J_1\), define the associated derivation \(\delta_\Phi\) on \(\A_{\loc}\) by
\begin{equation}
 \delta_\Phi(A):=\sum_{Z\Subset\Gamma}[\Phi(Z),A].
 \label{eq:energy-derivation-definition}
\end{equation}
The series converges absolutely in norm, and our convention gives \(\delta_\Phi(A^*)=-\delta_\Phi(A)^*\).
For \(\Phi\in\mathfrak J_1\), \(\Lambda\Subset\Gamma\), and \(t\in\R\), set
\begin{equation*}
 H_\Lambda^\Phi:=\sum_{Z\subset\Lambda}\Phi(Z),
 \qquad
 \delta_{\Phi,\Lambda}:=[H_\Lambda^\Phi,\,\cdot\,],
 \qquad
 \tau_t^{\Phi,\Lambda}:=\operatorname{Ad}(\E^{\I tH_\Lambda^\Phi}).
\end{equation*}
The following proposition shows that \(\A_{\loc}\) is a core, which allows the dynamical ground-state condition to be tested on local operators.

\begin{proposition}[Construction of the infinite-volume dynamics]
\label{prop:first-moment-dynamics}
Let \(\Gamma\) be countable, let \(\A=\operatorname{CAR}(\ell^2(\Gamma;\C^r))\), and let \(\Xi\in\mathfrak J_1\).
There exists a strongly continuous parity-preserving dynamics \((\tau_t^\Xi)_{t\in\R}\subset\operatorname{Aut}(\A)\) such that, for every increasing  exhaustion \(\Lambda_n\nearrow\Gamma\) by finite sets \(\Lambda_n\) and every \(A\in\A\), \(\tau_t^{\Xi,\Lambda_n}(A)\to\tau_t^\Xi(A)\) in norm, uniformly for \(t\) in compact intervals.
The local derivation \(\delta_\Xi\) is closable, its closure \(\overline\delta_\Xi\) has \(\A_{\loc}\) as a core, and the infinitesimal generator of \(\tau^\Xi\) is \(\I\overline\delta_\Xi\).
\end{proposition}

The proof in Appendix~\ref{app:first-moment-dynamics} applies the commutator estimate of \cite[Lemma~3.1]{FermionicLR}, based on \cite[Lemma~3.3 and proof of Theorem~3.1]{NSY18}, and the resolvent construction of \cite[Lemmas~4.4--4.7 and Theorem~4.8]{BruPedra17} under the first support-moment assumption alone.

\paragraph{Ground states and local ground-state conditions.}

For \(\Phi\in\mathfrak J_1\), a state \(\rho\) on \(\A\) is called an \emph{algebraic ground state} of \(\tau^\Phi\) if, for every \(A\in\operatorname{Dom}(\overline\delta_\Phi)\),
\begin{equation}
 \operatorname{Re}\rho\!\left(A^*\overline\delta_\Phi(A)\right)\geq0.
 \label{eq:algebraic-ground-state}
\end{equation}
The stationarity argument in Lemma~\ref{lem:local-ground-characterization} makes \(\rho(A^*\overline\delta_\Phi(A))\) real, so \eqref{eq:algebraic-ground-state} is the standard ground-state condition for the generator \(\I\overline\delta_\Phi\) \cite[Definition~5.3.18 and Proposition~5.3.19]{BR97}.
For a derivation \(\delta\) whose domain contains \(\A_{\loc}\), and for \(U\subset\Gamma\), we define
\begin{equation}
 \LG_\delta(U):=\left\{\rho\in \mathfrak S(\A)\,\middle|\,
 \operatorname{Re}\rho(B^*\delta(B))\ge0
 \text{ for all }X\Subset U,
 \ B\in\A_X\right\},
 \label{eq:local-ground-cone}
\end{equation}
where \(\mathfrak S(\A)\) denotes the state space of \(\A\).
We abbreviate \(\LG_\Phi(U):=\LG_{\delta_\Phi}(U)\) and call its elements \emph{local ground states of \(\Phi\) in \(U\)}.
In particular, if \(\rho\in\LG_\Phi(U)\), no local unitary supported in \(U\) can lower the energy: for a unitary \(W\in\A_X\), with \(X\Subset U\), the energy change is
\begin{equation}
 \lim_{\Lambda\nearrow\Gamma}
 \bigl[\rho(W^*H_\Lambda^\Phi W)-\rho(H_\Lambda^\Phi)\bigr]
  =\sum_{Z\Subset\Gamma}
   \rho\!\left(W^*\Phi(Z)W-\Phi(Z)\right) =\rho\!\left(W^*\delta_\Phi(W)\right)\ge0.
\label{eq:local-unitary-energy-change}
\end{equation}
The series converges absolutely because \(\Phi\in\mathfrak J_1\) and the evenness of \(\Phi(Z)\) makes every term with \(Z\cap X=\varnothing\) vanish.

The following lemma shows that it suffices to test the algebraic ground-state condition \eqref{eq:algebraic-ground-state} on local operators, thereby identifying \(\LG_\Phi(\Gamma)\) with the set of algebraic ground states.

\begin{lemma}[Local characterization of ground states]
\label{lem:local-ground-characterization}
Let \(\Phi\in\mathfrak J_1\), and let \(\rho\in \mathfrak S(\A)\).
Then \(\rho\) is an algebraic ground state of \(\tau^\Phi\) if and only if, for every \(B\in\A_{\loc}\),
\begin{equation}
 \operatorname{Re}\rho(B^*\delta_\Phi(B))\geq0.
 \label{eq:local-characterization-ground-state}
\end{equation}
\end{lemma}

The proof is given in Appendix~\ref{app:first-moment-dynamics}.
Every algebraic ground state \(\omega\) of \(\tau^\Phi\), with \(\Phi\in\mathfrak J_1\), is \(\tau^\Phi\)-invariant \cite[Proposition~5.3.19]{BR97}.
Its GNS representation \((\pi_\omega,\mathcal H_\omega,\Omega_\omega)\) carries the canonical implementing Hamiltonian \(L_\omega\ge0\), characterized by
\[
 L_\omega\Omega_\omega=0,\qquad
 \pi_\omega(\tau_t^\Phi(A))
 =\E^{\I tL_\omega}\pi_\omega(A)\E^{-\I tL_\omega}
 \quad(t\in\R,\ A\in\A).
\]
We call \(L_\omega\) the \emph{GNS Hamiltonian} of \(\omega\).
For every \(A\in\A_{\loc}\), the ground-state inequality takes the form
\[
 \bigl\langle\pi_\omega(A)\Omega_\omega,
 L_\omega\pi_\omega(A)\Omega_\omega\bigr\rangle
 =\operatorname{Re}\omega(A^*\delta_\Phi(A))\ge0;
\]
see also \cite[Subsection~7.2]{NSY22} for these standard GNS identities.

For every \(\Phi\in\mathfrak J_1\), the set \(\LG_\Phi(\Gamma)\) is nonempty.
Indeed, extend finite-volume ground states to \(\A\) and take a weak-* cluster point along an increasing finite exhaustion.
For each local \(B\), norm convergence \(\delta_{\Phi,\Lambda}(B)\to\delta_\Phi(B)\) passes the finite-volume ground-state inequality to the limit, and the lemma applies.
For every \(U\subset\Gamma\), the inclusion \(\LG_\Phi(\Gamma)\subset\LG_\Phi(U)\) therefore implies that \(\LG_\Phi(U)\) is nonempty.

\subsection{The bulk interaction and its locality}
\label{sec:quadratic-reference}
\label{sec:locality-assumptions}

We use Majorana coordinates to treat number-conserving band Hamiltonians and quadratic Hamiltonians with pairing in one notation.
We follow the conventions of \cite[Section~2.2]{StabilityBdG26}.
For \(x\in\Gamma\), the canonical isometry \(\iota_x:\C^r\to\Hsp\) is defined, for \(v\in\C^r\) and \(y\in\Gamma\), by \((\iota_xv)(y):=\1_{\{x\}}(y)v\).
For \(x\in\Gamma\) and \(1\le j\le r\), set \(a_{x,j}:=a(\iota_xe_j)\) and define
\begin{equation}
 \gamma_{x,2j-1}:=a_{x,j}+a_{x,j}^*,\qquad
 \gamma_{x,2j}:=\I(a_{x,j}^*-a_{x,j}).
 \label{eq:physical-Majoranas}
\end{equation}
For labels \(u=(x,j)\) with \(1\le j\le2r\), these self-adjoint generators satisfy \(\{\gamma_u,\gamma_v\}=2\delta_{uv}\1\).
Their coefficient space is
\[
 \Hsp_{\mathrm M}:=\ell^2(\Gamma;\C^{2r}).
\]
For a bounded operator \(k\) on \(\ell^2(\Gamma;\C^m)\), with \(m\in\mathbb N\), let \(k_{xy}\in\mathcal B(\C^m)\) be its block from the fiber at \(y\) to that at \(x\).
For \(\mu\ge0\) and \(0<\theta\le1\), define the weighted Schur norm
\begin{equation}
 \|k\|_{\mu,\theta}^{\rm Sch}
 :=\max\left\{
 \sup_{x\in\Gamma}\sum_{y\in\Gamma}
 \E^{\mu d(x,y)^\theta}\|k_{xy}\|,
 \sup_{y\in\Gamma}\sum_{x\in\Gamma}
 \E^{\mu d(x,y)^\theta}\|k_{xy}\|
 \right\}.
 \label{eq:one-particle-Schur}
\end{equation}
For \(p\ge0\), its polynomial counterpart is
\begin{equation}
 \|k\|_p^{\mathrm{Sch,pol}}
 :=\max\left\{
 \sup_{x\in\Gamma}\sum_{y\in\Gamma}
 (1+d(x,y))^p\|k_{xy}\|,
 \sup_{y\in\Gamma}\sum_{x\in\Gamma}
 (1+d(x,y))^p\|k_{xy}\|
 \right\}.
 \label{eq:one-particle-polynomial-Schur}
\end{equation}
Throughout the paper, let \(h\in\mathcal B(\Hsp_{\mathrm M})\) satisfy
\begin{equation}
 h=h^*,\qquad \overline h=-h.
 \label{eq:Majorana-structure}
\end{equation}
Here \(\overline h\) denotes the entrywise complex conjugate of \(h\) in the Majorana coefficient basis.
Thus \(h\) is purely imaginary and \(h^{\mathsf T}=-h\).
These structural conditions ensure that the second-quantization formula \eqref{eq:quadratic-Majorana-interaction} below defines an even self-adjoint quadratic interaction \(\Phi_h\).
Fix \(g>0\), and assume the one-particle gap
\begin{equation}
 \spec(h)\cap(-g,g)=\varnothing
 \label{eq:one-particle-gap}
\end{equation}
and one of the following two locality bounds.
In the stretched-exponential case, fix \(0<\theta_{\rm in}\le1\) and assume, for some \(\mu_h>0\),
\begin{equation}
 C_h:=\|h\|_{\mu_h,\theta_{\rm in}}^{\rm Sch}<\infty.
 \label{eq:h-locality}
\end{equation}
In the polynomial case, we assume \(d_{\rm vol}>0\) and, for some \(p_h>d_{\rm vol}\),
\begin{equation}
 C_h:=\|h\|_{p_h}^{\mathrm{Sch,pol}}<\infty.
 \label{eq:h-polynomial-locality}
\end{equation}

For \(u=(x,j)\), write \(x_u:=x\) and let \(h_{uv}\) denote the scalar matrix entries of \(h\).
Define the canonical quadratic interaction by summing over ordered pairs \((u,v)\) of Majorana labels: for \(X\Subset\Gamma\) with \(1\le|X|\le2\), set
\begin{equation}
 \Phi_h(X):=\frac14
 \sum_{\substack{(u,v)\\\{x_u,x_v\}=X}}h_{uv}\gamma_u\gamma_v,
 \label{eq:quadratic-Majorana-interaction}
\end{equation}
and \(\Phi_h(X):=0\) otherwise.
Since \(\overline h_{uv}=-h_{uv}=h_{vu}\), each component is self-adjoint and even.
Summing the matrix entries of the blocks \(h_{xy}\) gives a constant \(C_r>0\), depending only on \(r\), such that in either locality regime
\[
 \|\Phi_h\|_1\le C_rC_h.
\]
With the normalization in \eqref{eq:quadratic-Majorana-interaction}, the derivation and free dynamics are given by \cite[Section~2.2]{StabilityBdG26}:
\begin{equation}
 \delta_{\Phi_h}(\gamma_u)=-\sum_vh_{uv}\gamma_v,\qquad
 \tau_t^{\Phi_h}(\gamma_u)=\sum_v(\E^{-\I th})_{uv}\gamma_v.
 \label{eq:Majorana-free-dynamics}
\end{equation}
Both series converge absolutely in norm by the Schur bound on \(h\), and the convergence of the second is uniform for \(t\) in compact intervals.
Write \(h_{\nc}=h_{\nc}^*\) and \(\Delta^{\mathsf T}=-\Delta\) for the number-conserving and pairing operators on \(\Hsp\), respectively, with the chemical potential included in \(h_{\nc}\).
An onsite unitary identifies \(h\) with their full Bogoliubov--de Gennes (BdG) operator and preserves its spectrum and block Schur norms \cite[Section~2.2]{StabilityBdG26}.
The gap assumption therefore concerns the full BdG operator, and the pairing terms need not be small.
We use the quasifree CAR formalism of \cite{Araki70}. Hastings \cite[Equation~(1)]{Hastings19} uses the same Majorana relations but writes the quadratic Hamiltonian without the factor \(1/4\) in \eqref{eq:quadratic-Majorana-interaction}, so his coefficient matrix corresponds to \(h/4\).

Set
\begin{equation}
 \sgnh:=\operatorname{sgn}(h),\qquad |h|:=(h^2)^{1/2}.
 \label{eq:free-sign}
\end{equation}
Then \(\overline \sgnh=-\sgnh\), whereas \(|h|\) is real symmetric.
Let \(\omega_0\) be the even quasifree state determined by
\begin{equation}
 \omega_0(\gamma_u\gamma_v)=(\1+\sgnh)_{uv}.
 \label{eq:free-Majorana-state}
\end{equation}
Its odd correlations vanish, and its even correlations are given by Wick's rule.
Since \(\sgnh=\sgnh^*\), \(\sgnh^2=\1\), and \(\overline \sgnh=-\sgnh\), the matrix \((\1+\sgnh)/2\) is a basis projection: it is an orthogonal projection whose conjugate is its complement.
The quasifree construction therefore gives a pure state \cite{Araki70}.
It is a gapped ground state of \(\Phi_h\): for every \(A\in\A_{\loc}\),
\begin{equation}
 \operatorname{Re}\omega_0(A^*\delta_{\Phi_h}(A))
 \ge g\bigl(\omega_0(A^*A)-|\omega_0(A)|^2\bigr).
 \label{eq:free-ground-gap}
\end{equation}
The ground-state property and the gap follow from \cite[Proposition~4.3]{StabilityBdG26} by undoing the transformation to product-vacuum coordinates and testing the resulting gap inequality on \(A\widehat\otimes\1\).
Since \(0\notin\spec(h)\), \(\omega_0\) is the unique algebraic ground state of \(\Phi_h\) by \cite[Theorem~2]{Araki70}.
The product-vacuum representation is described in Section~\ref{sec:atomicization}.
In the number-conserving case \(\Delta=0\), \(\omega_0\) is the Fermi sea with one-particle density \(\1_{(-\infty,0)}(h_{\nc})\).

Under stretched-exponential locality, weighted functional calculus gives a rate \(\mu_1\in(0,\mu_h)\), depending only on \(C_\Gamma,d_{\rm vol},r,C_h/g,\mu_h,\theta_{\rm in}\), such that
\begin{equation}
 \|\sgnh\|_{\mu_1,\theta_{\rm in}}^{\rm Sch}
 +g^{-1}\bigl\||h|\bigr\|_{\mu_1,\theta_{\rm in}}^{\rm Sch}
 \le C(C_\Gamma,d_{\rm vol},r,C_h/g,\mu_h,\theta_{\rm in}).
 \label{eq:free-functional-locality}
\end{equation}
Under polynomial locality it gives the bound
\begin{equation}
 \|\sgnh\|_{p_h}^{\mathrm{Sch,pol}}
 +g^{-1}\bigl\||h|\bigr\|_{p_h}^{\mathrm{Sch,pol}}
 \le C(p_h,d_{\rm vol},C_\Gamma,r,C_h/g).
 \label{eq:free-polynomial-functional-locality}
\end{equation}
These estimates follow from \cite[Lemmas~A.7--A.8]{StabilityBdG26}.

\paragraph{Many-body interactions and locality.}

Let \(\mathcal V\) be an interaction.
We consider the family
\begin{equation}
 \R\ni \lambda\mapsto\Phi_\lambda:=\Phi_h+\lambda\mathcal V.
 \label{eq:background-path}
\end{equation}
We treat stretched-exponential decay of both \(h\) and \(\mathcal V\), including exponential decay when \(\theta_{\rm in}=1\), and polynomial decay of both with possibly different exponents.
For \(\mu\ge0\), \(0<\theta\le1\), and \(k\ge0\), define
\begin{equation}
 \|\Phi\|_{\mu,\theta;k}
 :=\sup_{x\in\Gamma}
 \sum_{\substack{X\Subset\Gamma\\x\in X}}
 |X|^k\E^{\mu\diam(X)^\theta}\|\Phi(X)\|.
 \label{eq:exp-interaction-norm}
\end{equation}
For \(p,k\ge0\), define the polynomial interaction norm by
\begin{equation}
 \|\Phi\|^{\pol}_{p;k}
 :=\sup_{x\in\Gamma}
 \sum_{\substack{X\Subset\Gamma\\x\in X}}
 |X|^k(1+\diam X)^p\|\Phi(X)\|.
 \label{eq:polynomial-interaction-norm}
\end{equation}
For \(p\ge0\) and \(\kappa>0\), define the norm with exponential support-size weight by
\begin{equation}
 \|\Phi\|^{\#}_{p;\kappa}
 :=\sup_{x\in\Gamma}
 \sum_{\substack{X\Subset\Gamma\\x\in X}}
 \E^{\kappa|X|}(1+\diam X)^p\|\Phi(X)\|.
 \label{eq:card-norm}
\end{equation}
In particular, \(\|\Phi\|_1=\|\Phi\|_{0,1;1}=\|\Phi\|^{\pol}_{0;1}\).

In the polynomial regime we impose the following cell-partition condition:  there are constants \(C_{\rm cell}\ge1\) and \(D_{\rm cell}\in\mathbb N_0\) such that, for every \(\ell\ge1\), there is a partition into nonempty finite sets satisfying
\begin{equation}
 \Gamma=\bigsqcup_{b\in\mathcal B_\ell}Q_b^{(\ell)},
 \qquad
 \sup_{b\in\mathcal B_\ell}\diam Q_b^{(\ell)}
       \le C_{\rm cell}\ell.
 \label{eq:cell-partitions}
\end{equation}
The graph joining distinct cells at distance at most \(\ell\) is required to have degree at most \(D_{\rm cell}\).
This is \cite[Hypothesis~3.2]{StabilityBdG26}, equivalently \cite[Hypothesis~2.1]{FermionicLR}.
Euclidean point sets with the standing volume bound satisfy it by \cite[Remark~2.2]{FermionicLR}, even when their embedding dimension differs from \(d_{\rm vol}\).
Let \(\kappa_0=\kappa_0(D_{\rm cell})>0\) be the support-size weight sufficient for \cite[Theorem~2.6]{FermionicLR}.
We assume one of the following:
\begin{enumerate}[label=(\Alph*),ref=\Alph*,leftmargin=*]
\item\label{ass:exponential-locality} \emph{Stretched-exponential locality.}
Assume \eqref{eq:h-locality} and, for some \(\mu_{\mathcal V}>0\), \(\|\mathcal V\|_{\mu_{\mathcal V},\theta_{\rm in};1}<\infty\).
\item\label{ass:polynomial-locality} \emph{Polynomial locality.}
The bound \eqref{eq:h-polynomial-locality} holds, the metric space satisfies \eqref{eq:cell-partitions}, and, for some \(p_{\mathcal V}>d_{\rm vol}\),
\begin{equation}
 \|\mathcal V\|^{\#}_{p_{\mathcal V};\kappa_0}<\infty.
 \label{eq:polynomial-interaction-locality}
\end{equation}
\end{enumerate}
These assumptions imply the following locality bounds for \(\Phi_s\) on every bounded coupling interval \(I\subset\R\).
For each \(k\ge0\), summing the matrix entries of the blocks \(h_{xy}\) gives a constant \(C_{r,k}>0\), depending only on \(r\) and \(k\), such that in the stretched-exponential regime
\[
 \|\Phi_h\|_{\mu_h,\theta_{\rm in};k}\le C_{r,k}C_h.
\]
Since \(\Phi_s=\Phi_h+s\mathcal V\) and \(I\) is bounded,
\begin{equation}
 \sup_{s\in I}
 \|\Phi_s\|_{\min\{\mu_h,\mu_{\mathcal V}\},\theta_{\rm in};1}<\infty.
 \label{eq:background-exponential-locality}
\end{equation}
In the polynomial regime, define the physical-interaction exponent
\begin{equation}
 p_\Phi:=\min\{p_h,p_{\mathcal V}\}>d_{\rm vol}.
 \label{eq:physical-polynomial-exponent}
\end{equation}
For every \(k\ge0\), the quadratic terms satisfy \(\|\Phi_h\|^{\pol}_{p_h;k}\le C_{r,k}C_h\) and \(\|\Phi_h\|^{\#}_{p_h;\kappa_0} \le C_r\E^{2\kappa_0}C_h\).
Combining these estimates with \eqref{eq:polynomial-interaction-locality} gives
\begin{equation}
 \sup_{s\in I}\|\Phi_s\|^{\#}_{p_\Phi;\kappa_0}<\infty.
 \label{eq:background-polynomial-locality}
\end{equation}
Thus \(\Phi_s\in\mathfrak J_1\) for every \(s\in\R\) in both regimes, and Proposition~\ref{prop:first-moment-dynamics} gives the corresponding dynamics.
We call \(\tau^{\Phi_s}\) the \emph{physical dynamics}; at zero coupling it is \(\tau^{\Phi_h}\).

\subsection{Ground-state transport}
\label{sec:ground-state-transport}

We introduce the first of our two main hypotheses: the free ground state \(\omega_0\) can be transported to a ground state \(\omega_\lambda\) of \(\Phi_\lambda\) by a quasi-local family of automorphisms with controlled locality properties.
The hypothesis holds at small \(|\lambda|\) in both locality regimes, by \cite[Theorems~3.1 and~3.3]{StabilityBdG26}.
We nevertheless formulate transport as a hypothesis because it may hold on larger coupling intervals than those established by the weak-coupling results.
Our abstract results require this hypothesis and local excitation detectability, introduced in the next subsection, but no separate smallness assumption on~\(|\lambda|\).

\begin{hypothesis}[Quasi-local ground-state transport]
\label{hyp:background-transport}
There exist \(\lambda_0>0\) and a family \((\eta_s)_{s\in[-\lambda_0,\lambda_0]}\) of interactions satisfying the following conditions.
\begin{enumerate}[label=(\alph*)]
\item \emph{Parameter continuity.}
The path \(s\mapsto\eta_s\) is componentwise norm-continuous.

\item \emph{Uniform locality.}
One of the following two alternatives holds.
\begin{enumerate}[label=(\roman*)]
\item \emph{Stretched-exponential locality.}
Condition~(\ref{ass:exponential-locality}) holds and, for a fixed \(0<\theta<\theta_{\rm in}\) and some \(\mu_\eta>0\),
\begin{equation}
 \sup_{|s|\le\lambda_0}
 \|\eta_s\|_{\mu_\eta,\theta;1}<\infty.
 \label{eq:stretched-flow}
\end{equation}
\item \emph{Polynomial locality.}
Condition~(\ref{ass:polynomial-locality}) holds and, for some \(p_\eta>d_{\rm vol}\),
\begin{equation}
 \sup_{|s|\leq\lambda_0}
 \|\eta_s\|^{\pol}_{p_\eta;1}<\infty.
 \label{eq:finite-polynomial-flow}
\end{equation}
\end{enumerate}
\end{enumerate}
Let \((\beta_{s,t})\) be the cocycle generated by \((\eta_s)\), as constructed in Lemma~\ref{lem:truncated-cocycle}, and write \(\beta_\lambda:=\beta_{0,\lambda}\).
For \(|\lambda|\le\lambda_0\) the state
\begin{equation}
 \omega_\lambda:=\omega_0\circ\beta_{0,\lambda}
 \label{eq:state-transport}
\end{equation}
satisfies
\begin{equation}
 \omega_\lambda\in\LG_{\Phi_\lambda}(\Gamma).
 \label{eq:transported-ground-state}
\end{equation}
\end{hypothesis}
The hypothesis requires \(\omega_\lambda\) to be a ground state but does not assume that its GNS Hamiltonian has a positive spectral gap.
Whenever Hypothesis~\ref{hyp:background-transport} is assumed, we fix one such \(\lambda_0\) and generator family~\((\eta_s)\).

\begin{definition}[Model data]
\label{def:model-data}
The \emph{model data} consist of the numerical parameters
\begin{equation*}
 C_\Gamma,d_{\rm vol},r,g,\lambda_0,
\end{equation*}
a prescribed finite bound \(\overline C_h\ge C_h\), and the following parameters and bounds for the chosen locality alternative:
\begin{enumerate}[label=(\roman*),leftmargin=*]
\item In the stretched-exponential case, \(\mu_h,\mu_{\mathcal V},\theta_{\rm in},\theta,\mu_\eta\), together with prescribed finite bounds \(\overline C_{\mathcal V},\overline C_\eta\) satisfying
\[
 \|\mathcal V\|_{\mu_{\mathcal V},\theta_{\rm in};1}
 \le\overline C_{\mathcal V},
 \qquad
 \sup_{|s|\le\lambda_0}\|\eta_s\|_{\mu_\eta,\theta;1}
 \le\overline C_\eta.
\]
\item In the polynomial case, \(C_{\rm cell},D_{\rm cell},p_h,p_{\mathcal V},p_\eta\), together with prescribed finite bounds \(\overline C_{\mathcal V},\overline C_\eta\) satisfying
\[
 \|\mathcal V\|^{\#}_{p_{\mathcal V};\kappa_0}
 \le\overline C_{\mathcal V},
 \qquad
 \sup_{|s|\le\lambda_0}\|\eta_s\|^{\pol}_{p_\eta;1}
 \le\overline C_\eta.
\]
\end{enumerate}
A constant depending only on the model data can be chosen uniformly over all metric spaces, coefficient operators, interactions, and transport paths satisfying these bounds.

A family has \emph{uniformly controlled model data} if one common choice of these parameters and bounds applies to all its models.
\end{definition}

\subsection{Local excitation detectability and weak coupling}
\label{sec:local-excitation-detectability}
\label{sec:weak-coupling}

The second hypothesis compares local excitation numbers with annihilation operators filtered under the interacting dynamics.
We give the definitions here.
Sections~\ref{sec:atomicization} and~\ref{sec:filters} construct the transformation to product-vacuum coordinates and establish locality and normalization of the filtered annihilation operators.

Let \(\Hsp_{\rm aux}\) be a labeled copy of \(\Hsp\), and set \(\A^{\rm aux}:=\operatorname{CAR}(\Hsp_{\rm aux})\) and \(\A^\st:=\operatorname{CAR}(\Hsp\oplus\Hsp_{\rm aux})\cong\A\widehat\otimes\A^{\rm aux}\), with the graded tensor product.
The label \(\st\) denotes the stacked system.
On the auxiliary factor, let \(h_{\rm aux}\) be the copy of \(-h\), let \(\Phi_{h_{\rm aux}}\) be its quadratic interaction, and denote its pure even quasifree ground state by \(\omega_{\rm aux}\).
For \(Z\Subset\Gamma\), define the stacked interaction by
\begin{equation}
 \Phi_\lambda^\st(Z):=\Phi_\lambda(Z)\widehat\otimes\1
                +\1\widehat\otimes\Phi_{h_{\rm aux}}(Z).
 \label{eq:stacked-interaction}
\end{equation}
Write \(\gamma_{x,j,1}\) and \(\gamma_{x,j,2}\) for the physical and auxiliary Majorana generators.
For \(x\in\Gamma\) and \(j\in\mathscr I:=\{1,\ldots,2r\}\), set
\[
 c_{x,j}:=\frac{\gamma_{x,j,1}+\I\gamma_{x,j,2}}2,
 \qquad n_x:=\sum_{j\in\mathscr I}c_{x,j}^*c_{x,j}.
\]
Let \(\omega_\at\) be the product vacuum of these modes, characterized by \(\omega_\at(c_{x,j}^*c_{x,j})=0\) for every \(x,j\).
Identify the physical and auxiliary Majorana coefficient spaces by their site and mode labels, and set
\begin{equation}
 O_\st:=\frac1{\sqrt2}
 \begin{pmatrix}\1&\I \sgnh\\ \I \sgnh&\1\end{pmatrix}.
 \label{eq:O-st}
\end{equation}
For coefficient labels \(u,v\in\Gamma\times\{1,\ldots,2r\}\) and \(\alpha\in\{1,2\}\), define the Bogoliubov automorphism \(\sigma_\st\) by
\begin{equation}
 \sigma_\st(\gamma_{u,\alpha})
 :=\sum_{\substack{v\\\alpha'\in\{1,2\}}}
 (O_\st)_{(u,\alpha),(v,\alpha')}\gamma_{v,\alpha'}.
 \label{eq:sigma-st}
\end{equation}
Lemma~\ref{lem:atomicization} shows that \(\sigma_\st\) maps the stacked free state to \(\omega_\at\).
We lift the spectral flow to the first factor by \(\beta_{0,\lambda}^{(1)}:=\beta_{0,\lambda}\widehat\otimes\operatorname{id}_{\A^{\rm aux}}\), and set
\begin{equation}
 \sigma_\lambda:=\sigma_\st\circ\beta_{0,\lambda}^{(1)}.
 \label{eq:transported-coordinates}
\end{equation}
Since \(\beta_{0,0}=\operatorname{id}_{\A}\), one has \(\sigma_0=\sigma_\st\).
We call the descriptions before and after \(\sigma_\lambda\) the \emph{original coordinates} and \emph{product-vacuum coordinates} on the same stacked algebra \(\A^\st\).
The transported map satisfies
\[
 (\omega_\lambda\widehat\otimes\omega_{\rm aux})
      \circ\sigma_\lambda^{-1}=\omega_\at.
\]
Write \(\tau^{\rm aux}:=\tau^{\Phi_{h_{\rm aux}}}\) and \(\tau^{\st,\lambda}:=\tau^{\Phi_\lambda}\widehat\otimes\tau^{\rm aux}\) for the auxiliary and stacked dynamics.
Let \(\delta_\lambda^\st:=\overline{\delta_{\Phi_\lambda^\st}}\) denote the closed derivation associated with the stacked interaction \(\Phi_\lambda^\st\).
The dynamics in product-vacuum coordinates is
\[
 \tau_t^{\at,\lambda}:=\sigma_\lambda\circ\tau_t^{\st,\lambda}
                           \circ\sigma_\lambda^{-1}.
\]
For \(k\in L^1(\R)\), we use the Fourier-transform convention \(\widehat k(\omega):=\int_\R k(t)\E^{\I t\omega}\dd t\).
Set \(s_f:=2\) in the polynomial case and \(s_f:=\theta_{\rm in}/\theta>1\) in the stretched-exponential case.
Fix the filter \(f\) defined in \eqref{eq:Fourier} from \(g,\overline C_h,s_f\).
It satisfies
\[
 \widehat f=1\ \text{on }[-\overline C_h,-g],
 \qquad
 \supp\widehat f\subset[-\overline C_h-g,-3g/4].
\]
The cutoff at \(-3g/4\) determines the low-energy window \([0,3g/4]\) used below.
With this choice of filter, LED also implies a bulk GNS gap of at least \(3g/4\), as shown in Corollary~\ref{cor:ground-state-uniqueness}.

For \(x\in\Gamma\), \(j\in\mathscr I\), and \(|\lambda|\le\lambda_0\), filter \(c_{x,j}\) with the bulk dynamics in product-vacuum coordinates, \(\tau^{\at,\lambda}\), and define the associated excitation observables by
\begin{equation}
 c_{f,x,j}^{(\lambda)}
 :=\int_\R f(t)\tau_t^{\at,\lambda}(c_{x,j})\,\dd t,
 \qquad
 n_{f,x}^{(\lambda)}
 :=\sum_{j\in\mathscr I}(c_{f,x,j}^{(\lambda)})^*c_{f,x,j}^{(\lambda)}.
 \label{eq:bulk-filtered def LED}
\end{equation}
We call \(c_{f,x,j}^{(\lambda)}\) the \emph{bulk-filtered annihilation operators}.
The following hypothesis bounds the onsite excitation number \(n_x\) by \(2n_{f,x}^{(\lambda)}\) plus a weighted sum of the excitation numbers \(n_z\), with a uniform contraction bound on its coefficients.
Proposition~\ref{prop:weak-local-excitation-detectability} shows that this hypothesis holds at sufficiently weak coupling for every transport satisfying Hypothesis~\ref{hyp:background-transport}.

\begin{hypothesis}[Uniform local excitation detectability]
\label{hyp:uniform-local-excitation-detectability}
Consider \(\Phi_\lambda=\Phi_h+\lambda\mathcal V\), with the transport fixed in Hypothesis~\ref{hyp:background-transport}, the onsite excitation numbers \(n_x\), and the filtered excitation numbers \(n_{f,x}^{(\lambda)}\) defined in \eqref{eq:bulk-filtered def LED} using the fixed filter \(f\) of \eqref{eq:Fourier}.
We assume that there exist a nonempty set \(\mathcal J\subset[-\lambda_0,\lambda_0]\) and a family \((\mathscr K_\lambda)_{\lambda\in\mathcal J}\) of nonnegative kernels
\begin{equation*}
 \mathscr K_\lambda:\Gamma\times\Gamma\longrightarrow[0,\infty)
\end{equation*}
satisfying the following two conditions.
\begin{enumerate}[label=(\alph*),leftmargin=*]
\item \emph{Excitation comparison.}
For every \(\lambda\in\mathcal J\) and \(x\in\Gamma\), the following operator inequality holds in~\(\A^\st\):
\begin{equation}
 n_x\le 2n_{f,x}^{(\lambda)}
 +\sum_{z\in\Gamma}\mathscr K_\lambda(x,z)n_z.
 \label{eq:uniform-local-excitation-detectability}
\end{equation}
\item \emph{Uniform contraction.}
The kernels satisfy the bounds corresponding to the locality alternative in Hypothesis~\ref{hyp:background-transport}:
\begin{enumerate}[label=(\roman*),leftmargin=*]
\item In the stretched-exponential case, for some \(\mu_{\rm LED}>0\),
\begin{equation}
 \vartheta_{\mathcal J,\mu_{\rm LED}}
 :=\sup_{\lambda\in\mathcal J}\sup_{x\in\Gamma}
 \sum_{z\in\Gamma}\E^{\mu_{\rm LED}d(x,z)^\theta}
 \mathscr K_\lambda(x,z)<1.
 \label{eq:LED-stretched-contraction}
\end{equation}
\Needspace{6\baselineskip}
\item In the polynomial case,
\begin{equation}
 \vartheta_{\mathcal J}
 :=\sup_{\lambda\in\mathcal J}\sup_{x\in\Gamma}
 \sum_{z\in\Gamma}\mathscr K_\lambda(x,z)<1.
 \label{eq:LED-row-contraction}
\end{equation}
\end{enumerate}
\end{enumerate}
\end{hypothesis}

In the stretched-exponential case, set \(\vartheta_{\mathcal J}:=\sup_{\lambda\in\mathcal J}\sup_{x\in\Gamma}\sum_{z\in\Gamma}\mathscr K_\lambda(x,z)\); then \(\vartheta_{\mathcal J}\le\vartheta_{\mathcal J,\mu_{\rm LED}}<1\).
The kernel bounds imply norm convergence of the series in \eqref{eq:uniform-local-excitation-detectability}, since \(0\le n_z\le2r\,\1\).

Whenever Hypothesis~\ref{hyp:uniform-local-excitation-detectability} is assumed, we fix such a set \(\mathcal J\) and a corresponding kernel family.
We call \eqref{eq:uniform-local-excitation-detectability} the \emph{detectability inequality}.
The kernel \(\mathscr K_\lambda\) plays a role analogous to that of the influence matrix in Dobrushin's comparison theorem for classical Gibbs measures \cite{Dobrushin70}, as formulated by Rebeschini and van Handel \cite{RebeschiniVanHandel14}.
For \(\lambda\in\mathcal J\) and a state \(\varrho\) on \(\A^\st\), set
\[
 u_x:=\varrho(n_x),\qquad
 b_x:=2\varrho(n_{f,x}^{(\lambda)}).
\]
Taking expectations in \eqref{eq:uniform-local-excitation-detectability} gives
\[
 u_x\le b_x+\sum_{z\in\Gamma}\mathscr K_\lambda(x,z)u_z.
\]
Since \(0\le u_x\le2r\) and the row sums of \(\mathscr K_\lambda\) are uniformly below one, iteration yields
\[
 u_x\le\sum_{n=0}^{\infty}\sum_{z\in\Gamma}
 \mathscr K_\lambda^n(x,z)b_z.
\]
Dobrushin's theorem uses the corresponding series of powers of an influence matrix to compare local expectations through differences of one-site conditional distributions.
Here the quantities controlled by this iteration are excitation densities, and \(b_x\) measures the expectations of the filtered squares.
For stretched-exponential locality, Lemma~\ref{lem:absorption}(i) bounds \(u_x\) by iterating this inequality with a spatial weight.
In the polynomial case, Lemma~\ref{lem:localized-detectability} first applies the conditional expectation \(\mathbb E^\vac_{\ball_R(x)}\) in \eqref{eq:vacuum-slices} to~LED: it removes \(n_z\) for \(z\notin\ball_R(x)\), and \eqref{eq:localized-detectability} bounds the error by \(C_{\rm loc}(1+R)^{-2\nu}\), with \(\nu\) as in \eqref{eq:main-polynomial-range}.
At \(\lambda=0\), the filter leaves the onsite annihilation operators unchanged, so \(n_{f,x}^{(0)}=n_x\) and one may take \(\mathscr K_0=0\), as proved in Lemma~\ref{lem:filter-normalization}.

\paragraph{Transport and LED at weak coupling.}

The following proposition derives both ground-state transport and local excitation detectability, for sufficiently small \(|\lambda|\), from the free spectral gap and the locality of \(h\) and~\(\mathcal V\).
\begin{proposition}[Transport and local excitation detectability at weak coupling]
\label{prop:unconditional-weak-coupling}
Let \(h\) be a Majorana coefficient operator satisfying \eqref{eq:Majorana-structure} and \eqref{eq:one-particle-gap}.
For each interaction \(\mathcal V\), the following assertions hold.
\begin{enumerate}[label=(\roman*),leftmargin=*]
\item Assume condition~(\ref{ass:exponential-locality}).
Decrease the rate \(\mu_{\mathcal V}>0\), if necessary, so that
\begin{equation}
 \|\mathcal V\|_{\mu_{\mathcal V},\theta_{\rm in};2}<\infty.
 \label{eq:automatic-exponential-input}
\end{equation}
For every fixed \(0<\theta<\theta_{\rm in}\), there exists \(\lambdaWC>0\) such that Hypotheses~\ref{hyp:background-transport} and \ref{hyp:uniform-local-excitation-detectability} hold on \(\JWC:=[-\lambdaWC,\lambdaWC]\) in the stretched-exponential alternative with exponent \(\theta\).
\item Assume condition~(\ref{ass:polynomial-locality}).
There exists \(\lambdaWC>0\) such that both hypotheses hold on \(\JWC:=[-\lambdaWC,\lambdaWC]\) in the polynomial alternative, with \(p_\eta=p_\Phi=\min\{p_h,p_{\mathcal V}\}\).
\end{enumerate}
For families with uniformly controlled geometric and free-system data, fixed locality parameters of \(h\) and \(\mathcal V\), and a common bound \(\overline C_{\mathcal V}\) on the norm of \(\mathcal V\) in the respective case, the coupling threshold \(\lambdaWC\) can be chosen uniformly.
\end{proposition}

\begin{proof}
In case (i), reducing the spatial rate gives \eqref{eq:automatic-exponential-input} from condition~(\ref{ass:exponential-locality}); apply \cite[Lemma~A.2]{StabilityBdG26} in the metric \(d^{\theta_{\rm in}}\).
For the prescribed \(\theta\), apply \cite[Theorem~3.1]{StabilityBdG26} with the common decay order \(\theta_{\rm in}\).
It constructs an even, componentwise norm-continuous generator path satisfying Hypothesis~\ref{hyp:background-transport} on \([-\lambda_0,\lambda_0]\), for some \(\lambda_0>0\).

In case (ii), the norms of \(h\) and \(\mathcal V\) at \(p_\Phi=\min\{p_h,p_{\mathcal V}\}\) are bounded by those at the prescribed exponents.
Apply \cite[Theorem~3.3]{StabilityBdG26} at \(p_\Phi\), with the weight \(\kappa_0\) fixed above.
It constructs the ground states and their transport on \([-\lambda_0,\lambda_0]\), with
\[
 \sup_{|s|\le\lambda_0}\|\eta_s\|^{\pol}_{p_\Phi;1}<\infty.
\]
Thus Hypothesis~\ref{hyp:background-transport} holds with \(p_\eta=p_\Phi\).

In both cases, Proposition~\ref{prop:weak-local-excitation-detectability} gives \(\lambdaLED\in(0,\lambda_0]\) such that LED holds on \([-\lambdaLED,\lambdaLED]\).
Set \(\lambdaWC:=\lambdaLED\); since \(\lambdaLED\le\lambda_0\), the ground-state transport and LED both hold on \(\JWC=[-\lambdaWC,\lambdaWC]\).
For a family as in the statement, the cited theorems from \cite{StabilityBdG26} give a common \(\lambda_0>0\) and common locality bounds on the transport generators \(\eta_s\).
Equation~\eqref{eq:uniform-family-LED-modulus} then gives a common modulus \(\varsigma_{\rm LED}\), so we may choose \(\lambdaLED\in(0,\lambda_0]\) with \(\varsigma_{\rm LED}(\lambdaLED)\le1/2\) independently of the member of the family.
\end{proof}

\section{Main results}
\label{sec:main-results}

We first state our main comparison theorem for entire low-energy subspaces.
Ground-state LPPL then follows by applying the theorem in a ground-state GNS representation.
For finite samples, we use the sample Hamiltonian shifted by its lowest eigenvalue.
We then extend the comparison to states satisfying a ground-state inequality only in a region.

Unless explicitly stated otherwise, let \(h\) satisfy \eqref{eq:Majorana-structure} and \eqref{eq:one-particle-gap}, and assume either locality alternative of Subsection~\ref{sec:locality-assumptions} for \(h\) and \(\mathcal V\).
The comparison results below assume ground-state transport, Hypothesis~\ref{hyp:background-transport}, and uniform local excitation detectability for this transport, Hypothesis~\ref{hyp:uniform-local-excitation-detectability}, on a fixed nonempty set \(\mathcal J\subset[-\lambda_0,\lambda_0]\).
In Proposition~\ref{prop:unconditional-weak-coupling} we showed that these assumptions hold for sufficiently small coupling.

To be able to state the dependence of constants concisely in the results below, we collect the relevant parameters and bounds into two groups: the model data and the LED data.
The \emph{model data}, listed in Definition~\ref{def:model-data}, comprise the geometric bounds, the onsite mode number \(r\), the free gap \(g\), the coupling bound \(\lambda_0\), the spatial decay parameters, and the prescribed norm bounds \(\overline C_h,\overline C_{\mathcal V},\overline C_\eta\) for \(h\), \(\mathcal V\), and the transport generators.
The \emph{LED data on \(\mathcal J\)} consist of:
\begin{itemize}
\item in the stretched-exponential case, the rate \(\mu_{\rm LED}>0\) and a prescribed positive lower bound on \(1-\vartheta_{\mathcal J,\mu_{\rm LED}}\) from \eqref{eq:LED-stretched-contraction};
\item in the polynomial case, a prescribed positive lower bound on \(1-\vartheta_{\mathcal J}\) from \eqref{eq:LED-row-contraction}.
\end{itemize}
Dependence on the \emph{model and LED data} means dependence on both collections.

In the polynomial case, our estimates decay as \((1+D)^{-\nu}\) with distance \(D\) and contain a factor \((1+|Y|)^q\) for observables supported in \(Y\).
To specify these exponents, put \(p_\at:=\min\{p_h,p_\eta\}\) and set
\begin{equation}
 \nu:=\min\{p_h,p_{\mathcal V},p_\eta\},
 \label{eq:main-polynomial-range}
\end{equation}
\begin{equation}
 q:=\max\left\{\frac12+\frac{d_{\rm vol}}{2p_\at},
                         \frac{\nu}{p_\at}\right\}\le1.
 \label{eq:main-support-exponent}
\end{equation}
At the full decay power \(\nu\), the support exponent in \eqref{eq:main-support-exponent} is strictly below one exactly when \(p_{\mathcal V}<p_\at\).
In the weak-coupling setting of Proposition~\ref{prop:unconditional-weak-coupling}(ii), these definitions give \(\nu=p_\Phi\) and \(q=1\).

Under stretched-exponential locality, fix \(1/2<q\le1\).

\subsection{Local determination of low-energy states and LPPL}

The following theorem shows that agreement of an interaction \(\Xi\) with the bulk interaction \(\Phi_\lambda\) inside a region \(U\) already determines the local expectations of its low-energy states deep inside \(U\).
These expectations are close to those of \(\omega_\lambda\), with an error decaying with the distance of the observable's support to \(U^c\).
More precisely, for \(\lambda\in\mathcal J\) and \(U\subset\Gamma\), we require \(\Xi\in\mathfrak J_1\) to satisfy, for every nonempty \(Z\Subset\Gamma\),
\begin{equation}
 \begin{cases}
 \Xi(Z)=\Phi_\lambda(Z),&Z\subset U,\\
 \Xi(Z)\in\{0,\Phi_\lambda(Z)\},
      &Z\cap U\ne\varnothing,\quad Z\cap U^c\ne\varnothing.
 \end{cases}
 \label{eq:comparison-interaction-criterion}
\end{equation}
Terms supported entirely in \(U^c\) are unrestricted apart from \(\Xi\in\mathfrak J_1\).
This includes defects in \(\mathfrak J_1\) confined to \(U^c\) and restrictions to samples with arbitrary even boundary operators or boundary interactions in~\(\mathfrak J_1\).

A \emph{positive-energy representation} of \(\tau^\Xi\) is a pair \((\pi,L)\), where \(\pi:\A\to\mathcal B(\mathcal H)\) is a unital representation and \(L\) is a nonnegative self-adjoint operator satisfying
\[
 \E^{\I tL}\pi(A)\E^{-\I tL}=\pi(\tau_t^\Xi(A))
 \qquad(A\in\A,\ t\in\R).
\]
A standard example is the GNS representation \(\pi_\rho\) of an algebraic ground state \(\rho\) of \(\Xi\), in which the nonnegative GNS Hamiltonian \(L_\rho\) implements \(\tau^\Xi\).
Also note that any self-adjoint implementing Hamiltonian \(\widetilde L\) that is bounded below can be normalized by setting
\[
L:=\widetilde L-\inf\spec(\widetilde L)\,\1.
\]
Then \(\inf\spec(L)=0\), and the low-energy window in the following theorem consists of energies at most \(3g/4\) above \(\inf\spec(\widetilde L)\).

\begin{theorem}[Local determination of low-energy states]
\label{thm:positive-energy-locality}
Assume Hypotheses~\ref{hyp:background-transport} and~\ref{hyp:uniform-local-excitation-detectability} on \(\mathcal J\).
There are constants \(C>0\) and, in the stretched-exponential case, \(m>0\), depending only on the model and LED data and the support exponent \(q\) specified above, such that the following holds for every \(\lambda\in\mathcal J\), every \(U\subset\Gamma\), and every \(\Xi\in\mathfrak J_1\) satisfying \eqref{eq:comparison-interaction-criterion}.
For every positive-energy representation \((\pi,L)\) of \(\tau^\Xi\) on \(\mathcal H\), set
\begin{equation}
 P:=\1_{[0,3g/4]}(L).
 \label{eq:positive-energy-window}
\end{equation}
Then, for every nonempty finite \(Y\Subset\Gamma\) and \(A\in\A_Y\), one has
\begin{equation}
 \|P(\pi(A)-\omega_\lambda(A)\1)P\|
 \le C(1+|Y|)^q\|A\|\,G(\dist(Y,U^c)),
 \label{eq:positive-energy-locality}
\end{equation}
with
\begin{equation}
 G(D):=
 \begin{cases}
  \E^{-mD^\theta},&\text{stretched-exponential locality},\\
  (1+D)^{-\nu},&\text{polynomial locality},
 \end{cases}
 \qquad G(\infty):=0.
 \label{eq:main-comparison-profile}
\end{equation}
The same constants apply to matrices of local observables: for every \(k\in\mathbb N\) and \([A_{ij}]\in M_k(\A_Y)\),
\[
 \left\|\left[P\bigl(\pi(A_{ij})-\omega_\lambda(A_{ij})\1\bigr)P\right]_{i,j=1}^k\right\|
 \le C(1+|Y|)^q\|[A_{ij}]\|G(\dist(Y,U^c)).
\]
The left-hand norm is the operator norm on \(\mathcal H^{\oplus k}\).
We identify \(M_k(\A_Y)\cong\A_Y\otimes M_k\), where \(M_k:=\mathcal B(\C^k)\) is regarded as even.
\end{theorem}

The matrix estimate also controls low-energy states entangled with an arbitrary finite-dimensional ancillary system: their joint state on \(Y\) and the ancilla is close in trace norm to the product of the bulk state on \(Y\) and the ancillary marginal.

The theorem also applies when the bottom of \(\spec(L)\) is not an eigenvalue: it controls the low-energy subspace of the given implementation without requiring a ground-state vector.
The same estimates also hold on every smaller energy window. Indeed, for \(0\le\varepsilon\le3g/4\), the projection \(P_\varepsilon:=\1_{[0,\varepsilon]}(L)\) satisfies \(P_\varepsilon\le P\), so further compression gives
 \eqref{eq:positive-energy-locality} and its matrix extension with \(P_\varepsilon\) in place of \(P\) with the same constants.
In particular, \(P_0\) projects onto the entire kernel of \(L\), without any uniqueness assumption.
Neither finite rank nor spectral isolation of these windows is required.
The estimate applies to every state \(\varphi\) on \(\mathcal B(\mathcal H)\) with \(\varphi(P_\varepsilon)=1\), uniformly in real time since \(P_\varepsilon\) commutes with \(\E^{\I tL}\).

The proof is given in Subsection~\ref{sec:low-energy-locality}.

\begin{corollary}[Uniqueness of the algebraic ground state and a GNS gap]
\label{cor:ground-state-uniqueness}
Assume Hypotheses~\ref{hyp:background-transport} and~\ref{hyp:uniform-local-excitation-detectability} on \(\mathcal J\).
For every \(\lambda\in\mathcal J\), the state \(\omega_\lambda\) is the unique algebraic ground state of \(\Phi_\lambda\):
\[
 \LG_{\Phi_\lambda}(\Gamma)=\{\omega_\lambda\}.
\]
The spectral projection of its GNS Hamiltonian onto the low-energy window is
\[
 \1_{[0,3g/4]}(L_{\omega_\lambda})=|\Omega_{\omega_\lambda}\rangle\langle\Omega_{\omega_\lambda}|.
\]
In particular,
\[
 L_{\omega_\lambda}\ge\frac{3g}{4}
 \bigl(\1-|\Omega_{\omega_\lambda}\rangle
              \langle\Omega_{\omega_\lambda}|\bigr).
\]
\end{corollary}

\begin{proof}
Apply Theorem~\ref{thm:positive-energy-locality} with \(U=\Gamma\) and \(\Xi=\Phi_\lambda\) in the GNS representation \((\pi_\rho,\mathcal H_\rho,\Omega_\rho)\) of any \(\rho\in\LG_{\Phi_\lambda}(\Gamma)\).
The nonnegative GNS Hamiltonian annihilates \(\Omega_\rho\), so \(P\Omega_\rho=\Omega_\rho\) and \(G(\infty)=0\) give \(\rho(A)=\omega_\lambda(A)\) for every local \(A\).
Norm density of the local algebra implies \(\rho=\omega_\lambda\).

For the gap, take \(\rho=\omega_\lambda\) and \(P=\1_{[0,3g/4]}(L_{\omega_\lambda})\).
The same application of the theorem gives, for every local \(A\),
\[
 P\pi_{\omega_\lambda}(A)\Omega_{\omega_\lambda}
 =P\pi_{\omega_\lambda}(A)P\Omega_{\omega_\lambda}
 =\omega_\lambda(A)\Omega_{\omega_\lambda}.
\]
The vectors \(\pi_{\omega_\lambda}(A)\Omega_{\omega_\lambda}\), with \(A\) local, are dense, so \(\operatorname{ran}P=\C\Omega_{\omega_\lambda}\).
The spectral theorem therefore gives \(L_{\omega_\lambda}\ge(3g/4)(\1-P)\).
\end{proof}

Combined with Proposition~\ref{prop:unconditional-weak-coupling}, this gives uniqueness for every sufficiently weak many-body interaction in our class, including long-range interactions and systems without translation invariance.
An independent proof of uniqueness follows from \cite[Theorem~3.1(i)]{ThermalLPPL26}.
Unlike finite-volume uniqueness in gap-stability results such as \cite[Theorem~3]{DRS19}, the conclusion includes ground states not assumed normal in the GNS representation of \(\omega_\lambda\).

\paragraph{Low-energy LPPL.}

We now derive a strong LPPL principle: after adding a defect \(\Psi\), every state in the low-energy window \([0,3g/4]\) remains locally close to the bulk ground state \(\omega_\lambda\) far from the support of \(\Psi\).
To this end, we apply Theorem~\ref{thm:positive-energy-locality} with \(\Xi=\Phi_\lambda+\Psi\) in the GNS representation of any ground state \(\omega_\Psi\) of \(\Phi_\lambda+\Psi\).

\begin{definition}[Defect strength and support]
\label{def:defect}
A defect is an interaction \(\Psi\in\mathfrak J_1\).
Its unweighted local strength and support are, respectively,
\begin{equation}
 J_\Psi:=\sup_{x\in\Gamma}
 \sum_{\substack{Z\Subset\Gamma\\Z\ni x}}
 \|\Psi(Z)\|\leq\|\Psi\|_1,
 \qquad
 \Sigma(\Psi):=\bigcup_{\substack{Z\Subset\Gamma\\\Psi(Z)\ne0}}Z.
 \label{eq:defect-strength-support}
 \qedhere
\end{equation}
\end{definition}

For every \(\lambda\in\R\), the interaction \(\Phi_\lambda+\Psi\) belongs to \(\mathfrak J_1\).
Hence Proposition~\ref{prop:first-moment-dynamics} gives the dynamics \(\tau^{\Phi_\lambda+\Psi}\) and identifies its generator with \(\I(\overline{\delta_{\Phi_\lambda}+\delta_\Psi})\).
For fixed \(\lambda\) and \(\Psi\), we write \(\omega_\Psi\) for an arbitrary ground state of this dynamics.

To define the low-energy sector in infinite volume, fix \(\lambda\in\mathcal J\), a defect \(\Psi\in\mathfrak J_1\), and a ground state \(\omega_\Psi\) of \(\Phi_\lambda+\Psi\).
Let \((\pi_{\omega_\Psi},\mathcal H_{\omega_\Psi},\Omega_{\omega_\Psi})\) be its GNS representation and \(L_{\omega_\Psi}\ge0\) its GNS Hamiltonian, as defined in Subsection~\ref{sec:ground-state-conditions}.
The window \([0,3g/4]\) consists of excitation energies relative to \(\omega_\Psi\); in finite volume, this corresponds to subtracting the modified Hamiltonian's ground energy.
Define the low-energy projection
\begin{equation}
 P_{\omega_\Psi}
 :=\1_{[0,3g/4]}(L_{\omega_\Psi}).
 \label{eq:defect-low-energy-projection}
\end{equation}

\begin{corollary}[Low-energy LPPL under arbitrary defects]
\label{thm:low-energy-locality}
Assume Hypotheses~\ref{hyp:background-transport} and~\ref{hyp:uniform-local-excitation-detectability} on \(\mathcal J\), and use \(C,q,G\) from Theorem~\ref{thm:positive-energy-locality}.
For every \(\lambda\in\mathcal J\), \(S\subset\Gamma\), nonempty finite \(Y\Subset\Gamma\), and \(A\in\A_Y\),
\begin{equation}
 \begin{aligned}
 \sup_{\substack{\Psi\in\mathfrak J_1\\\Sigma(\Psi)\subset S}}
 \ \sup_{\omega_\Psi\in\LG_{\Phi_\lambda+\Psi}(\Gamma)}
 \|P_{\omega_\Psi}
   (\pi_{\omega_\Psi}(A)-\omega_\lambda(A)\1)
   P_{\omega_\Psi}\|\le C(1+|Y|)^q\|A\|\,G(\dist(Y,S)).
 \end{aligned}
 \label{eq:low-energy-locality}
\end{equation}
In particular, strong ground-state LPPL holds with the same constants: for \(A\in\A_Y\),
\begin{equation}
 \sup_{\substack{\Psi\in\mathfrak J_1\\\Sigma(\Psi)\subset S}}
 \ \sup_{\omega_\Psi\in\LG_{\Phi_\lambda+\Psi}(\Gamma)}
 |\omega_\Psi(A)-\omega_\lambda(A)|
 \le C(1+|Y|)^q\|A\|\,G(\dist(Y,S)).
 \label{eq:ground-state-lppl}
\end{equation}
For every \(k\in\mathbb N\) and \([A_{ij}]\in M_k(\A_Y)\), estimate~\eqref{eq:low-energy-locality} holds with \(\|A\|\) replaced by \(\|[A_{ij}]\|\) and the norm inside the suprema replaced by
\[
 \left\|\left[P_{\omega_\Psi}
 \bigl(\pi_{\omega_\Psi}(A_{ij})-\omega_\lambda(A_{ij})\1\bigr)
 P_{\omega_\Psi}\right]_{i,j=1}^k\right\|.
\]
The constants are independent of \(k\).
\end{corollary}

The estimates apply to every state \(\varphi\) on \(\mathcal B(\mathcal H_{\omega_\Psi})\) with \(\varphi(P_{\omega_\Psi})=1\).
Low-energy excitations are therefore localized near \(S\) in the following sense: every state in this energy window agrees with \(\omega_\lambda\) on local observables up to an error decaying with the distance of their supports from \(S\).
No gap above or within this subspace, finite-dimensionality, or uniqueness of the ground state of \(\Phi_\lambda+\Psi\) is assumed.
The same estimates hold with \(P_{\omega_\Psi}\) replaced by \(\1_{[0,\varepsilon]}(L_{\omega_\Psi})\) for every \(0\le\varepsilon\le3g/4\), including the whole zero-energy subspace.
The proof is given in Subsection~\ref{sec:low-energy-locality}.

For a fixed local observable and accuracy, Corollary~\ref{thm:low-energy-locality} gives a distance beyond which a defect changes the expectation of the observable by less than that accuracy.
The same distance works for every defect and every ground state of the modified system.
This extends the nearsightedness of the density response to external potentials in \cite{Kohn96,ProdanKohn05} to the full local state and arbitrary defects in \(\mathfrak J_1\).

\paragraph{Uniform-in-time locality in the low-energy window.}
Fix \(\lambda,S,Y,A,\Psi,\omega_\Psi\) as in Corollary~\ref{thm:low-energy-locality}.
Since \(P_{\omega_\Psi}\) commutes with \(\E^{\I tL_{\omega_\Psi}}\), for every \(t\in\R\),
\[
 P_{\omega_\Psi}\bigl(\pi_{\omega_\Psi}(\tau_t^{\Phi_\lambda+\Psi}(A))
                  -\omega_\lambda(A)\1\bigr)P_{\omega_\Psi}=\E^{\I tL_{\omega_\Psi}}
 P_{\omega_\Psi}\bigl(\pi_{\omega_\Psi}(A)-\omega_\lambda(A)\1\bigr)
 P_{\omega_\Psi}\E^{-\I tL_{\omega_\Psi}}.
\]
Thus every state \(\varphi\) on \(\mathcal B(\mathcal H_{\omega_\Psi})\) with \(\varphi(P_{\omega_\Psi})=1\) satisfies the same local comparison with \(\omega_\lambda\), uniformly for all times under \(\tau^{\Phi_\lambda+\Psi}\).

For a topological insulator with low-energy boundary states, this allows nontrivial boundary dynamics while all interior expectations remain close to their infinite-volume ground-state values, up to the error in Corollary~\ref{thm:low-energy-locality};
see the paragraph “Edges of topological insulators” in Subsection~\ref{sec:finite-samples}, which also discusses time-dependent driving whose evolving states remain in the fixed low-energy window.

\subsection{Local ground-state determination}
\label{sec:local-ground-state-results}

We now strengthen the uniqueness conclusion of Corollary~\ref{cor:ground-state-uniqueness} to quantitative local determination for states satisfying the ground-state inequality only in a region \(U\).
Such states need not admit global positive-energy implementing Hamiltonians.
In Subsection~\ref{sec:local-positive-energy} we construct instead a corresponding local positive-energy operator.

\begin{theorem}[Local determination from the ground-state inequality]
\label{thm:bulk-uniqueness}
Assume Hypotheses~\ref{hyp:background-transport} and~\ref{hyp:uniform-local-excitation-detectability} on \(\mathcal J\).
The constants in Theorem~\ref{thm:positive-energy-locality} can be chosen so that, for every \(\lambda\in\mathcal J\), \(U\subset\Gamma\), nonempty finite \(Y\Subset\Gamma\), and \(A\in\A_Y\),
\begin{equation}
 \sup_{\rho\in\LG_{\Phi_\lambda}(U)}
 |\rho(A)-\omega_\lambda(A)|
 \le C(1+|Y|)^q\|A\|\,G(\dist(Y,U^c)),
 \label{eq:local-ground-state-determination}
\end{equation}
with the same support exponent \(q\) and profile \(G\) from \eqref{eq:main-comparison-profile}.
\end{theorem}
The proof is given in Subsection~\ref{sec:bulk-uniqueness}.

\begin{remark}[Support dependence and spatial decay]
\label{rem:support-decay-tradeoff}
One can reduce the support exponent in \eqref{eq:positive-energy-locality} and \eqref{eq:local-ground-state-determination} at the cost of slower spatial decay.
Both left-hand sides are bounded by \(2\|A\|\), and for every \(0<a<1\) and \(D:=\dist(Y,U^c)\),
\[
 \|A\|\min\{2,C(1+|Y|)^qG(D)\}
 \le 2^{1-a}C^a(1+|Y|)^{aq}\|A\|G(D)^a.
\]
With constant \(2^{1-a}C^a\), the two estimates hold with \((q,\nu)\) replaced by \((aq,a\nu)\) in the polynomial case, and with \((q,m)\) replaced by \((aq,am)\) in the stretched-exponential case, where \(\theta\) is unchanged.
The same argument applies to the matrix estimate in Theorem~\ref{thm:positive-energy-locality}, whose left-hand side is bounded by \(2\|[A_{ij}]\|\).
\end{remark}

\paragraph{Local computation and certification.}
Local energy inequalities can certify approximations to the reduced bulk ground state using only a finite neighborhood of the observation region.
For quantum spin systems, Fawzi, Fawzi, and Scalet \cite[Theorem~2.1]{FawziFawziScalet24} use such inequalities to obtain convergent semidefinite bounds on ground-state expectations.
Theorem~\ref{thm:bulk-uniqueness} gives an explicit convergence rate for these fermionic systems.
Assume its hypotheses, fix \(\lambda\in\mathcal J\), and suppose that \(\Phi_\lambda(Z)=0\) whenever \(\diam Z>a\), for some \(a\ge0\).
For nonempty finite \(Y\subset U\Subset\Gamma\), put \(W:=U^{(a)}\).
Then \(\delta_{\Phi_\lambda}(B)=[H_W^{\Phi_\lambda},B]\) for every \(B\in\A_U\), since every nonzero interaction term meeting \(U\) is contained in \(W\).
Choose a homogeneous complex basis \(B_1,\ldots,B_N\) of the full algebra \(\A_U\), with both even and odd elements.
For a density matrix \(D\) on \(\A_W\), define the Hermitian matrix
\[
 Q_{ij}(D):=\frac12\operatorname{Tr}D
 \left(B_i^*[H_W^{\Phi_\lambda},B_j]
       +[H_W^{\Phi_\lambda},B_i]^*B_j\right).
\]
For \(B=\sum_j c_jB_j\), one has \(c^*Q(D)c=\operatorname{Re}\operatorname{Tr}DB^*[H_W^{\Phi_\lambda},B]\).
Every density matrix \(D\) satisfying \(Q(D)\ge0\) extends to a state in \(\LG_{\Phi_\lambda}(U)\) on \(\A\), because these inequalities involve only operators in \(\A_W\).
The density matrix of \(\omega_\lambda|_{\A_W}\) also satisfies this condition.
For every density matrix \(D\) on \(\A_W\) satisfying \(Q(D)\ge0\), Theorem~\ref{thm:bulk-uniqueness} gives, by taking the supremum over unit-norm observables in \(\A_Y\),
\[
 \|D_Y-\omega_{\lambda,Y}\|_1
 \le C(1+|Y|)^q\,G(\dist(Y,U^c)),
\]
where \(D_Y\) and \(\omega_{\lambda,Y}\) are the density matrices of the restrictions to \(\A_Y\), and \(\|\cdot\|_1\) is the trace norm.
For every self-adjoint \(A\in\A_Y\), minimizing and maximizing \(\operatorname{Tr}(DA)\) under these constraints gives certified lower and upper bounds on \(\omega_\lambda(A)\).
Taking \(U=Y^{(R)}\) and increasing \(R\) therefore determines the local state to arbitrary accuracy from finite-dimensional constraints.

\subsection{Defects with tails and uniformly weak perturbations}

Corollary~\ref{thm:low-energy-locality} applies to defects confined to a prescribed set \(S\).
The next theorem allows additional terms outside \(S\), measured by their local strength.
For a defect \(\Psi\in\mathfrak J_1\) and a nonempty set \(S\subset\Gamma\), define its exterior tail strength at \(z\in\Gamma\) by
\begin{equation}
 t_{\Psi,S}(z)
 :=\1_{S^c}(z)
 \sum_{\substack{Z\Subset\Gamma\\Z\ni z}}\|\Psi(Z)\|.
 \label{eq:exterior-tail-intensity}
\end{equation}
Terms supported entirely in \(S\) do not contribute, whatever their strength.
Crossing terms and terms outside \(S\) are counted at each exterior site they touch.
The definition \eqref{eq:defect-strength-support} of \(J_\Psi\) gives, for every \(z\in\Gamma\),
\begin{equation}
 0\le t_{\Psi,S}(z)\le J_\Psi.
 \label{eq:tail-intensity-bounded}
\end{equation}
For \(M\ge0\), \(\xi>0\), and \(0<\theta<1\), define the decay envelopes
\begin{align}
 J_{\Psi,S}^{M}
 &:=\sup_{z\in S^c}(1+\dist(z,S))^M t_{\Psi,S}(z),
 &&\text{(polynomial tails)},
 \label{eq:polynomial-tail-envelope}\\
 J_{\Psi,S}^{\xi,\theta}
 &:=\sup_{z\in S^c}\E^{\xi\dist(z,S)^\theta}t_{\Psi,S}(z),
 &&\text{(stretched-exponential tails)}.
 \label{eq:stretched-tail-envelope}
\end{align}
Here and below \(\sup\varnothing:=0\).
The following theorem bounds the influence of a defect uniformly in the strength of its terms inside \(S\), with an additional error controlled by the tail envelope.

\begin{theorem}[Locality under defects with tails]
\label{thm:quasilocal-tails}
Assume Hypotheses~\ref{hyp:background-transport} and~\ref{hyp:uniform-local-excitation-detectability} on \(\mathcal J\).
The prefactor \(C\) and, under stretched-exponential bulk locality, the rate \(m\) in Theorem~\ref{thm:positive-energy-locality} can be chosen independently of the tail parameters below, so that the following estimates hold with \(G\) from \eqref{eq:main-comparison-profile}.

For every \(\lambda\in\mathcal J\), defect \(\Psi\in\mathfrak J_1\), nonempty \(S\subset\Gamma\), and positive-energy representation \((\pi,L)\) of \(\tau^{\Phi_\lambda+\Psi}\), set \(P:=\1_{[0,3g/4]}(L)\).
For every nonempty finite \(Y\Subset\Gamma\) and \(A\in\A_Y\), put \(D:=\dist(Y,S)\).
\begin{enumerate}[label=(\roman*),leftmargin=*]
\item \emph{Polynomial tails and bounded exterior strength.}
Fix \(M\ge0\).
Set \(\sigma_0:=0\), and for \(M>0\) define
\[
 \sigma_M:=
 \begin{cases}
 M,&\text{stretched-exponential bulk locality},\\
 \min\{M,\nu\},&\text{polynomial bulk locality}.
 \end{cases}
\]
Set \(\alpha_0:=1\) and, for \(M>0\), \(\alpha_M:=\sigma_M/M\).
There is a constant \(C_M^{\rm tail}>0\), depending only on the model and LED data, \(q\), and \(M\).
Whenever \(J_{\Psi,S}^{M}<\infty\),
\begin{equation}
 \begin{split}
 \|P(\pi(A)-\omega_\lambda(A)\1)P\|
 &\le (1+|Y|)^q\|A\|\Big[C\,G(D)\\
 &\qquad+C_M^{\rm tail}
       \min\!\left\{1,
       \min\{J_{\Psi,S}^{M},(J_{\Psi,S}^{M})^{\alpha_M}\}
                         (1+D)^{-\sigma_M}\right\}\Big].
 \end{split}
 \label{eq:tail-short-polynomial}
\end{equation}

\item \emph{Stretched-exponential tails.}
Under stretched-exponential bulk locality, use the exponent \(\theta\) from Hypothesis~\ref{hyp:background-transport}(b)(i).
For every \(\xi>0\), there are constants \(C_\xi^{\rm tail},m_\xi^{\rm tail}>0\), depending only on the model and LED data, \(q\), and \(\xi\).
Whenever \(J_{\Psi,S}^{\xi,\theta}<\infty\),
\begin{equation}
 \begin{split}
 \|P(\pi(A)-\omega_\lambda(A)\1)P\|
  \le (1+|Y|)^q\|A\|\Big[C\,G(D) +C_\xi^{\rm tail}
 \min\{1,J_{\Psi,S}^{\xi,\theta}\E^{-m_\xi^{\rm tail}D^\theta}\}\Big].
 \end{split}
 \label{eq:tail-short-stretched}
\end{equation}
\end{enumerate}
For every \(k\in\mathbb N\) and \([A_{ij}]\in M_k(\A_Y)\), both estimates hold with the same constants, with \(\|A\|\) replaced by \(\|[A_{ij}]\|\) and their left-hand side replaced by
\[
 \left\|\left[P\bigl(\pi(A_{ij})-\omega_\lambda(A_{ij})\1\bigr)P\right]_{i,j=1}^k\right\|.
\]
\end{theorem}

Taking a ground-state GNS representation and its cyclic vector recovers the estimates for every ground state \(\omega_\Psi\) of \(\Phi_\lambda+\Psi\).

For \(J_{\Psi,S}^{M}\le1\), the tail contribution inside the brackets in \eqref{eq:tail-short-polynomial} reduces to \(C_M^{\rm tail}J_{\Psi,S}^{M}(1+D)^{-\sigma_M}\).
For \(M=0\), the additional error in \eqref{eq:tail-short-polynomial} is \(C_0^{\rm tail}\min\{1,J_{\Psi,S}^{0}\}\), while the bulk term still decays with distance from \(S\).
The terms of \(\Psi\) supported inside \(S\) may therefore be arbitrarily strong, provided that the exterior strength \(J_{\Psi,S}^{0}\) is small.
The proof is given in Section~\ref{sec:quasilocal-tails}.

For a perturbation \(\Psi\) of the whole system, the following corollary gives a bound in terms of the strength \(J_\Psi\) from \eqref{eq:defect-strength-support}.

\begin{corollary}[Local stability under uniformly weak perturbations]
\label{cor:uniformly-weak-defects}
Assume Hypotheses~\ref{hyp:background-transport} and~\ref{hyp:uniform-local-excitation-detectability} on \(\mathcal J\).
There is a constant \(C>0\), depending only on the model and LED data and \(q\), with the following property.
For every \(\lambda\in\mathcal J\), defect \(\Psi\in\mathfrak J_1\), and positive-energy representation \((\pi,L)\) of \(\tau^{\Phi_\lambda+\Psi}\), set \(P:=\1_{[0,3g/4]}(L)\).
Then, for every nonempty finite \(Y\Subset\Gamma\) and \(A\in\A_Y\),
\begin{equation}
 \|P(\pi(A)-\omega_\lambda(A)\1)P\|\le
 \begin{cases}
 \min\{2,C(1+|Y|)^qJ_\Psi\}\|A\|,
       &\text{stretched-exponential case},\\
 \min\{2,C(1+|Y|)^{1/2+d_{\rm vol}/(2p_\at)}J_\Psi\}\|A\|,
       &\text{polynomial case}.
 \end{cases}
 \label{eq:uniformly-weak-defect-comparison}
\end{equation}
For every \(k\in\mathbb N\) and \([A_{ij}]\in M_k(\A_Y)\), the estimate holds with the same constant, with \(\|A\|\) replaced by \(\|[A_{ij}]\|\) and its left-hand side replaced by
\[
 \left\|\left[P\bigl(\pi(A_{ij})-\omega_\lambda(A_{ij})\1\bigr)P\right]_{i,j=1}^k\right\|.
\]
\end{corollary}
In particular, all ground states of \(\Phi_\lambda+\Psi\) are locally close to \(\omega_\lambda\), and hence to one another, with an error linear in \(J_\Psi\).
As \(J_\Psi\to0\), every choice of perturbed ground states converges weak-* to \(\omega_\lambda\).
The proof is given at the end of Section~\ref{sec:quasilocal-tails}.

The ground states of \(\Phi_\lambda+\Psi\) also satisfy an LPPL estimate, up to a remainder linear in \(J_\Psi\).
After adding an arbitrarily strong interaction \(\Upsilon\) confined to \(S\), all low-energy states of \(\Phi_\lambda+\Psi+\Upsilon\) remain locally close to every ground state of \(\Phi_\lambda+\Psi\), with an error decaying with distance from \(S\) plus this remainder.
The uniformly weak interaction \(\Psi\) may have substantially slower spatial decay than \(\mathcal V\).
For two-site \(\Psi\), uniform summability of the local strength suffices; in one dimension, this includes decay \((1+d(x,y))^{-1-\delta}\) for every \(\delta>0\).
Fix \(\lambda\in\mathcal J\), interactions \(\Psi,\Upsilon\in\mathfrak J_1\), and nonempty \(S\subset\Gamma\) with \(\Sigma(\Upsilon)\subset S\).
Choose any ground state \(\omega_\Psi\) of \(\Phi_\lambda+\Psi\) and any ground state \(\omega\) of \(\Phi_\lambda+\Psi+\Upsilon\).
Let \(\pi\) be the GNS representation of \(\omega\), \(L\ge0\) its GNS Hamiltonian, and \(P:=\1_{[0,3g/4]}(L)\).
Since \(J_{\Psi+\Upsilon,S}^{0}\le J_\Psi\), Theorem~\ref{thm:quasilocal-tails}(i), with \(M=0\), and Corollary~\ref{cor:uniformly-weak-defects} give a constant \(C_{\rm tail}>0\), depending only on the model and LED data and \(q\), such that, for every nonempty finite \(Y\Subset\Gamma\) and \(A\in\A_Y\),
\[
 \begin{aligned}
 \|P(\pi(A)-\omega_\Psi(A)\1)P\|
 &\le \|P(\pi(A)-\omega_\lambda(A)\1)P\|
       +|\omega_\Psi(A)-\omega_\lambda(A)|\\
 &\le (1+|Y|)^q\|A\|\Big[CG(\dist(Y,S))
                 +C_{\rm tail}\min\{1,J_\Psi\}\Big].
 \end{aligned}
\]

\subsection{Correlations and low-energy spectral weight}
\label{sec:correlation-results}

For a state \(\rho\) and operators \(A,B\in\A\), define the connected correlation
\[
 \operatorname{Cov}_\rho(A,B):=\rho(AB)-\rho(A)\rho(B).
\]
We first prove clustering of the bulk state \(\omega_\lambda\): its connected correlations decay as the distance between the supports of the observables increases.
We obtain bounds in both locality regimes from the locality of \(\sigma_\lambda\) and the factorization of the product vacuum \(\omega_\at\) on disjoint supports.
Under Hypothesis~\ref{hyp:uniform-local-excitation-detectability}, the GNS gap and Lieb--Robinson bounds give a stronger stretched-exponential estimate through the standard exponential-clustering argument \cite{HastingsKoma06,NachtergaeleSims06} in the metric \(d^{\theta_{\rm in}}\).

For a finite set \(X\Subset\Gamma\), define its metric boundary by
\[
 \partial X:=\{x\in X\,|\,\text{there exists }u\in X^c
                         \text{ with }d(x,u)=\dist(u,X)\}.
\]
\begin{proposition}[Clustering of the bulk ground state]
\label{prop:bulk-clustering}
Assume Hypothesis~\ref{hyp:background-transport}.
\begin{enumerate}[label=(\roman*),leftmargin=*]
\item \emph{Polynomial locality.}
Use \(p_\at=\min\{p_h,p_\eta\}\).
There is \(C>0\), depending only on the model data, such that, for every \(|\lambda|\le\lambda_0\), nonempty finite \(X,Y\Subset\Gamma\), \(A\in\A_X\), and \(B\in\A_Y\),
\begin{equation}
 |\operatorname{Cov}_{\omega_\lambda}(A,B)|
 \le C(1+|X|+|Y|)\|A\|\|B\|
       (1+\dist(X,Y))^{-p_\at}
 \label{eq:bulk-correlation-polynomial}
\end{equation}
and
\begin{equation}
 |\operatorname{Cov}_{\omega_\lambda}(A,B)|
 \le C(1+|\partial X|+|\partial Y|)\|A\|\|B\|
       (1+\dist(X,Y))^{-(p_\at-d_{\rm vol})}.
 \label{eq:bulk-correlation-polynomial-boundary}
\end{equation}
\item \emph{Stretched-exponential locality.}
There are \(C_{\rm tr},\mu_{\rm tr}>0\), depending only on the model data, such that, for every \(|\lambda|\le\lambda_0\), nonempty finite \(X,Y\Subset\Gamma\), \(A\in\A_X\), and \(B\in\A_Y\),
\begin{equation}
 |\operatorname{Cov}_{\omega_\lambda}(A,B)|
 \le C_{\rm tr}\bigl(1+|\partial X|+|\partial Y|\bigr)^{1/2}\|A\|\|B\|
       \E^{-\mu_{\rm tr}\dist(X,Y)^\theta}.
 \label{eq:bulk-correlation-transport-stretched}
\end{equation}
If, in addition, Hypothesis~\ref{hyp:uniform-local-excitation-detectability} holds on \(\mathcal J\), there are \(C,\mu>0\), depending only on the model and LED data on \(\mathcal J\), such that, for every \(\lambda\in\mathcal J\), nonempty finite \(X,Y\Subset\Gamma\), \(A\in\A_X\), and \(B\in\A_Y\),
\begin{equation}
 |\operatorname{Cov}_{\omega_\lambda}(A,B)|
 \le C\bigl(1+\min\{|\partial X|,|\partial Y|\}\bigr)^{1/2}\|A\|\|B\|
       \E^{-\mu\dist(X,Y)^{\theta_{\rm in}}}.
 \label{eq:bulk-correlation-exponential}
\end{equation}
\end{enumerate}
\end{proposition}

Summability of these correlation bounds gives an extensive variance estimate for uniformly localized observables.
Let \(\mathbb F^{\rm tr}_Z:\A\to\A_Z\) denote the tracial conditional expectation described in Section~\ref{sec:odd-locality}.
For \(s>d_{\rm vol}\), \(x\in\Gamma\), and \(A\in\A\), use the metric-ball version of the localization norm in \cite[Definition~2.2]{BTW26}:
\[
 \|A\|_{s,x}:=\|A\|+
     \sup_{R\ge0}(1+R)^s
           \|A-\mathbb F^{\rm tr}_{\ball_R(x)}(A)\|.
\]
For a state \(\rho\) and a self-adjoint \(T\), write \(\operatorname{Var}_\rho(T):=\rho(T^2)-\rho(T)^2\).
Under Hypothesis~\ref{hyp:background-transport}, there is \(C_s>0\), depending only on \(s\) and the model data, such that, for every self-adjoint family \((A_x)_{x\in\Gamma}\subset\A\) with \(J_s:=\sup_x\|A_x\|_{s,x}<\infty\), every finite \(\Lambda\Subset\Gamma\), and every \(|\lambda|\le\lambda_0\),
\begin{equation}
 \operatorname{Var}_{\omega_\lambda}
       \Big(\sum_{x\in\Lambda}A_x\Big)\le C_sJ_s^2|\Lambda|.
 \label{eq:extensive-variance}
\end{equation}
Hence, the variance of the spatial average \(|\Lambda|^{-1}\sum_{x\in\Lambda}A_x\) is bounded by \(C_sJ_s^2/|\Lambda|\).
The proposition and \eqref{eq:extensive-variance} are proved in Subsection~\ref{sec:bulk-clustering-proof}.

Applying local determination (Theorem~\ref{thm:bulk-uniqueness}) to \(AB\), \(A\), and \(B\) also controls the deviation of the covariance of \(A\) and \(B\) from its bulk value when both supports lie deep inside \(U\).
The following theorem controls covariances in positive-energy windows and in local ground states when only one support is far from the modified region or the complement of \(U\).

\begin{theorem}[One-sided clustering]
\label{thm:spatial-correlations}
Assume Hypotheses~\ref{hyp:background-transport} and~\ref{hyp:uniform-local-excitation-detectability} on \(\mathcal J\), and use the common support exponent \(q\) and profile \(G\) from \eqref{eq:main-comparison-profile}.
There is a constant \(C>0\), depending only on the model and LED data and \(q\), such that the following holds.
For \(R\in[0,\infty)\), set
\begin{equation}
 G_{\rm sep}(R):=
 \begin{cases}
 G(R),&\text{stretched-exponential locality},\\
 (1+R)^{-p_\at},&\text{polynomial locality},
 \end{cases}
 \qquad G_{\rm sep}(\infty):=0.
 \label{eq:correlation-separation-profile}
\end{equation}
For every \(\lambda\in\mathcal J\), nonempty finite \(X,Y\Subset\Gamma\), \(A\in\A_X\), and \(B\in\A_Y\), the following assertions hold.
\begin{enumerate}[label=(\roman*),leftmargin=*]
\item \emph{Positive-energy windows.}
Let \(U\subset\Gamma\) and \(\Xi\in\mathfrak J_1\) satisfy \eqref{eq:comparison-interaction-criterion}.
Let \((\pi,L)\) be any positive-energy representation of \(\tau^\Xi\) on \(\mathcal H\), and set \(P:=\1_{[0,3g/4]}(L)\).
For every state \(\varphi\) on \(\mathcal B(\mathcal H)\) with \(\varphi(P)=1\), define \(\rho:=\varphi\circ\pi\).
Then
\begin{equation}
 |\operatorname{Cov}_\rho(A,B)|
 \le C\|A\|\|B\|\bigl[
 (1+|X|+|Y|)G_{\rm sep}(\dist(X,Y)) +(1+|X|)^qG(\dist(X,U^c))\bigr].
 \label{eq:positive-energy-clustering}
\end{equation}
Moreover,
\begin{equation}
 \begin{aligned}
 \bigl\|P\pi(AB)P&-P\pi(A)P\pi(B)P\bigr\| =\bigl\|P\pi(A)(\1-P)\pi(B)P\bigr\|\\
 &\le C\|A\|\|B\|\bigl[
 (1+|X|+|Y|)G_{\rm sep}(\dist(X,Y)) +(1+|X|)^qG(\dist(X,U^c))\bigr].
 \end{aligned}
 \label{eq:operator-one-sided-clustering}
\end{equation}
The operator estimate also holds with the same constant when \(P\) is replaced throughout by \(\1_I(L)\), for any Borel set \(I\subset[0,3g/4]\).

\item \emph{Local ground states.}
For every \(U\subset\Gamma\),
\begin{equation}
 \begin{aligned}
 \sup_{\rho\in\LG_{\Phi_\lambda}(U)}
       |\operatorname{Cov}_\rho(A,B)|
  \le C\|A\|\|B\|\bigl[&
 (1+|X|+|Y|)G_{\rm sep}(\dist(X,Y))\\
 & +(1+|X|)^qG(\dist(X,U^c))\bigr].
 \end{aligned}
 \label{eq:one-sided-clustering}
\end{equation}
\end{enumerate}
\end{theorem}

Note that part (ii) requires only the ground-state inequality inside \(U\); the states need not arise from a positive-energy representation as in part (i).
The proof is given in Subsection~\ref{sec:spatial-correlation-proofs}.

De Roeck and Sch\"utz compare local expectations in impurity eigenstates with bulk ground-state expectations and prove clustering for a single impurity, assuming that the low-energy sector remains uniformly separated from the rest of the spectrum along the perturbation path \cite[Propositions~4.1 and~4.3]{DeRoeckSchuetz15}.

For completeness, we collect two further elementary consequences of local determination: bounds on low-energy spectral weight and imaginary-time correlations, and comparisons of local density matrices, entropies, and mutual information.

\begin{corollary}[Low-energy spectral weight and imaginary time]
\label{cor:low-energy-spectral-weight}
Assume Hypotheses~\ref{hyp:background-transport} and~\ref{hyp:uniform-local-excitation-detectability} on \(\mathcal J\), and use \(C,q,G\) from Theorem~\ref{thm:positive-energy-locality}.
For every \(\lambda\in\mathcal J\), \(U\subset\Gamma\), and \(\Xi\in\mathfrak J_1\) satisfying \eqref{eq:comparison-interaction-criterion}, let \((\pi,L)\) be a positive-energy representation of \(\tau^\Xi\) on \(\mathcal H\) and set \( P:=\1_{[0,3g/4]}(L)\).
For every nonempty finite \(X\Subset\Gamma\) and \(A\in\A_X\), write
\[
 A^\circ:=\pi(A)-\omega_\lambda(A)\1,
 \qquad
 \delta_A:=C(1+|X|)^q\|A\|\,G(\dist(X,U^c)).
\]
For every Borel set \(J\subset[0,3g/4]\), one has the operator inequality
\begin{equation}
 0\le P(A^\circ)^*\1_J(L)A^\circ P\le\delta_A^2P.
 \label{eq:low-energy-spectral-weight}
\end{equation}
For every nonempty finite \(Y\Subset\Gamma\), \(B\in\A_Y\), and \(t\ge0\), with \(B^\circ\) and \(\delta_B\) defined in the same way,
\begin{equation}
 \bigl\|P(A^\circ)^*\E^{-tL}B^\circ P\bigr\|
 \le\|A\|\|B\|\E^{-3gt/4} +\min\{\delta_A,2\|A\|\}\min\{\delta_B,2\|B\|\}.
 \label{eq:mixed-imaginary-time-correlation}
\end{equation}
\end{corollary}

\begin{proof}
Theorem~\ref{thm:positive-energy-locality} and \(\|A^\circ\|\le2\|A\|\) give \(\|PA^\circ P\|\le\min\{\delta_A,2\|A\|\}\), and the analogous bound holds for \(B\).
Since \(\1_J(L)\le P\), one has \(\|\1_J(L)A^\circ P\|\le\delta_A\), which proves \eqref{eq:low-energy-spectral-weight}.
Split the semigroup using \(P+(\1-P)=\1\):
\[
 P(A^\circ)^*\E^{-tL}B^\circ P
 =(PA^\circ P)^*\E^{-tL}(PB^\circ P) +P\pi(A)^*(\1-P)\E^{-tL}(\1-P)\pi(B)P.
\]
The scalar centerings disappear in the second term because \((\1-P)P=0\).
The bounds \(\|\E^{-tL}P\|\le1\) and \(\|\E^{-tL}(\1-P)\|\le\E^{-3gt/4}\) give \eqref{eq:mixed-imaginary-time-correlation}.
\end{proof}

We can also subtract the expectations of \(A\) and \(B\) in the state under consideration, rather than their expectations in \(\omega_\lambda\).
Let \(\varphi\) be a state on \(\mathcal B(\mathcal H)\) with \(\varphi(P)=1\), put \(\rho:=\varphi\circ\pi\), and define \(A_\rho:=\pi(A)-\rho(A)\1\) and \(B_\rho\) analogously.
Since \(\varphi(T)=\varphi(PTP)\) for every \(T\in\mathcal B(\mathcal H)\),
\begin{equation}
 \varphi(A_\rho^*PA_\rho)
 =\varphi\bigl((PA^\circ P)^*(PA^\circ P)\bigr)
       -|\varphi(PA^\circ P)|^2 \le\min\{\delta_A^2,4\|A\|^2\}.
 \label{eq:centered-low-energy-vector}
\end{equation}
Thus \(\varphi(A_\rho^*\1_J(L)A_\rho)\le\delta_A^2\) for every Borel \(J\subset[0,3g/4]\).

For \(t\ge0\), define the imaginary-time correlation by
\begin{equation}
 C^\varphi_{A,B}(t):=\varphi(A_\rho^*\E^{-tL}B_\rho).
 \label{eq:mixed-imaginary-time-definition}
\end{equation}
Splitting the semigroup with \(P+(\1-P)\) as in the proof and applying Cauchy--Schwarz to its \(P\)-part gives
\begin{equation}
 \begin{aligned}
 |C^\varphi_{A,B}(t)|
 &\le\|A\|\|B\|\E^{-3gt/4}
       +\min\{\delta_A,2\|A\|\}\min\{\delta_B,2\|B\|\},\\
 0\le C^\varphi_{A,A}(t)
 &\le\|A\|^2\E^{-3gt/4}+\min\{\delta_A^2,4\|A\|^2\}.
 \end{aligned}
 \label{eq:imaginary-time-correlation}
\end{equation}
At \(t=0\), one has \(C^\varphi_{A,B}(0)=\operatorname{Cov}_\rho(A^*,B)\).
These estimates use the nonnegative Hamiltonian \(L\) with its fixed energy origin; for the vector state of an eigenvector with energy \(e\), replacing \(L\) by \(L-e\1\) multiplies the correlation by \(\E^{et}\).

In particular, let \(\omega_\Psi\) be a ground state of \(\Phi_\lambda+\Psi\), where \(\Psi\in\mathfrak J_1\) and \(\Sigma(\Psi)\subset S\).
Let \((\pi,\mathcal H,\Omega)\) be its GNS representation and \(L\) its GNS Hamiltonian, with \(L\Omega=0\).
Apply the corollary to the positive-energy representation \((\pi,L)\) with \(U=S^c\), and take \(\varphi(T)=\langle\Omega,T\Omega\rangle\).
Let \(\mu_A(J):=\|\1_J(L)\pi(A)\Omega\|^2\) denote the spectral weight of \(\pi(A)\Omega\) in a Borel set \(J\subset[0,\infty)\).
Since \(L\Omega=0\), one has \(\1_J(L)\Omega=0\) for \(J\subset(0,3g/4]\), and \eqref{eq:low-energy-spectral-weight} gives
\[
\mu_A(J)=\|\1_J(L)A^\circ\Omega\|^2\le\delta_A^2.
\]
The spectral weight created by \(A\) in \((0,3g/4]\) therefore decays with \(\dist(X,S)\), even if the defect produces excitations arbitrarily close to zero energy.
In this representation \(C^\varphi_{A,B}(t)=\langle v_A,\E^{-tL}v_B\rangle\), where \(v_A=(\pi(A)-\omega_\Psi(A)\1)\Omega\) and \(v_B=(\pi(B)-\omega_\Psi(B)\1)\Omega\).

Integrating \eqref{eq:mixed-imaginary-time-correlation} against \(\E^{-zt}\), for \(z>0\), also gives
\begin{equation}
 \bigl\|P(A^\circ)^*(L+z)^{-1}B^\circ P\bigr\|
 \le\frac{\|A\|\|B\|}{z+3g/4}+\frac{\min\{\delta_A,2\|A\|\}\min\{\delta_B,2\|B\|\}}{z}.
 \label{eq:elementary-inverse-energy}
\end{equation}
Integrating \eqref{eq:imaginary-time-correlation} gives the same bound for \(|\varphi(A_\rho^*(L+z)^{-1}B_\rho)|\).

We next compare local density matrices, entropies, and mutual information with their bulk counterparts.
The local-determination estimates bound the trace-norm distance between the density matrices, and Audenaert’s continuity bound then controls the entropy differences.
For a density matrix \(\varrho\), write \(S_{\rm vN}(\varrho):=-\operatorname{Tr}(\varrho\log\varrho)\).
We use \(\|\cdot\|_1\) for the trace norm, identifying states on finite-dimensional matrix algebras with their density matrices.
For \(u\in[0,1]\), set \(h_2(u):=-u\log u-(1-u)\log(1-u)\), with \(0\log0:=0\).
For an even state \(\rho\) and disjoint nonempty finite \(X,Y\Subset\Gamma\), define
\[
 I_\rho(X{:}Y):=S_{\rm vN}(\rho_X)+S_{\rm vN}(\rho_Y)
                         -S_{\rm vN}(\rho_{X\cup Y}),
\]
where \(\rho_Z\) is the density matrix of the restriction to \(\A_Z\).

\begin{corollary}[Local trace distance, entropy, and mutual information]
\label{cor:local-entropy}
Assume Hypotheses~\ref{hyp:background-transport} and~\ref{hyp:uniform-local-excitation-detectability} on \(\mathcal J\), and use the common constants \(C,q\) and profile \(G\) from Theorems~\ref{thm:positive-energy-locality} and~\ref{thm:bulk-uniqueness}.
For every \(\lambda\in\mathcal J\) and \(U\subset\Gamma\), let \(\rho\) be a state on \(\A\) of either of the following types:
\begin{enumerate}[label=(\roman*),leftmargin=*]
\item Low-energy state: \(\rho=\varphi\circ\pi\), where \(\Xi\in\mathfrak J_1\) satisfies \eqref{eq:comparison-interaction-criterion}, \((\pi,L)\) is a positive-energy representation of \(\tau^\Xi\) on \(\mathcal H\), and \(\varphi\) is a state on \(\mathcal B(\mathcal H)\) with \(\varphi(P)=1\) for \(P:=\1_{[0,3g/4]}(L)\).
\item Local ground state: \(\rho\in\LG_{\Phi_\lambda}(U)\).
\end{enumerate}
For every nonempty finite \(Z\Subset\Gamma\), let \(\omega_{\lambda,Z}\) be the density matrix of \(\omega_\lambda|_{\A_Z}\).
Then
\begin{equation}
 \varepsilon_Z:=\tfrac12\|\rho_Z-\omega_{\lambda,Z}\|_1
 \le \tfrac C2(1+|Z|)^qG(\dist(Z,U^c)),
 \label{eq:local-trace-distance}
\end{equation}
and
\begin{equation}
 |S_{\rm vN}(\rho_Z)-S_{\rm vN}(\omega_{\lambda,Z})|
 \le r|Z|\,\varepsilon_Z\log2+h_2(\varepsilon_Z).
 \label{eq:local-entropy-comparison}
\end{equation}
If \(\rho\) is even and \(X,Y\Subset\Gamma\) are disjoint and nonempty, then
\begin{equation}
 \begin{aligned}
 |I_\rho(X{:}Y)-I_{\omega_\lambda}(X{:}Y)|
 &\le r\log2\bigl(|X|\varepsilon_X+|Y|\varepsilon_Y
                  +|X\cup Y|\varepsilon_{X\cup Y}\bigr)\\
 &\quad+h_2(\varepsilon_X)+h_2(\varepsilon_Y)
                  +h_2(\varepsilon_{X\cup Y}).
 \end{aligned}
 \label{eq:local-mutual-information-comparison}
\end{equation}
\end{corollary}

\begin{proof}
In case (i), Cauchy--Schwarz and \(\varphi(P)=1\) give \(\varphi(T)=\varphi(PTP)\) for all \(T\in\mathcal B(\mathcal H)\).
Applying this identity to \(T=\pi(A)-\omega_\lambda(A)\1\) gives
\[
 |\rho(A)-\omega_\lambda(A)|
 \le\|P(\pi(A)-\omega_\lambda(A)\1)P\|
 \le C(1+|Z|)^q\|A\|G(\dist(Z,U^c)),
 \qquad A\in\A_Z,
\]
by \eqref{eq:positive-energy-locality}.
In case (ii), estimate~\eqref{eq:local-ground-state-determination} gives the same bound on \(|\rho(A)-\omega_\lambda(A)|\).
Taking the supremum over \(A\in\A_Z\) with \(\|A\|\le1\) proves \eqref{eq:local-trace-distance} in both cases.
Since \(\A_Z\) acts on a Hilbert space of dimension \(2^{r|Z|}\), Audenaert's inequality~\cite[Theorem~1]{Audenaert07} gives \eqref{eq:local-entropy-comparison}.
Applying \eqref{eq:local-entropy-comparison} to \(X\), \(Y\), and \(X\cup Y\) and using the triangle inequality proves \eqref{eq:local-mutual-information-comparison}.
\end{proof}

\subsection{Finite samples and systems with boundaries}
\label{sec:finite-samples}

One usually expects that cutting a sample out of an insulator changes its ground-state expectations only near the boundary, even if low-energy boundary states appear.
For gapped non-interacting fermions this picture can be justified by one-particle methods.
For general interacting quantum systems, however, it is a heuristic that does not follow from a bulk gap alone.
Boundary changes can instead select a different bulk phase: in a zero-field ferromagnetic Ising model, removing a plus boundary condition also allows the minus ground state.
With one fermionic mode per site, the same example is obtained by replacing each Ising spin by \(\1-2a_{x,1}^*a_{x,1}\); this phase coexistence occurs outside the weak-coupling regime considered here.
By contrast, Corollary~\ref{cor:finite-sample-bulk} establishes boundary insensitivity for the weakly interacting fermion systems considered here, uniformly over arbitrary even boundary operators and all states in a fixed window above the sample ground state energy.

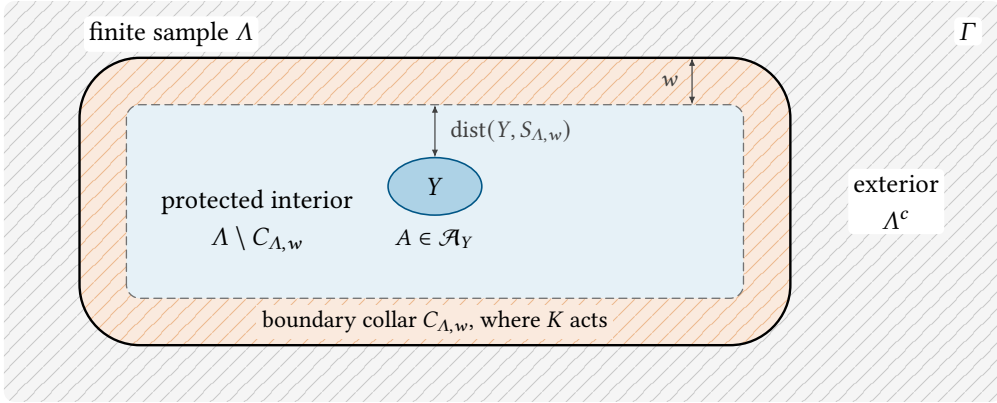
\begin{figure}[htbp]
\centering
\begin{tikzpicture}[lppl diagram]
 \def\w{.62}
 \path[excluded,rounded corners=4pt] (0,0) rectangle (13.2,5.3);
 \path[fill=aux!13,postaction={pattern={Lines[angle=45,distance=4.5pt,
   line width=.35pt]},pattern color=aux!45},rounded corners=.8cm]
   (1,.75) rectangle (10.4,4.55);
 \filldraw[fill=phys!10,draw=black!55,dash pattern=on 3pt off 2pt,
   rounded corners=.18cm] (1+\w,.75+\w) rectangle (10.4-\w,4.55-\w);
 \draw[line width=.9pt,rounded corners=.8cm] (1,.75) rectangle (10.4,4.55);
 \path[observable,fill=phys!32] (5.7,2.85) ellipse[x radius=.62,y radius=.38];
 \node at (5.7,2.85) {$Y$};
 \node[font=\footnotesize,anchor=north] at (5.7,2.43) {$A\in\A_Y$};
 \draw[dim line] (5.7,3.23) -- (5.7,4.55-\w)
   node[midway,right=2pt,font=\footnotesize] {$\dist(Y,S_{\Lambda,w})$};
 \draw[dim line] (9.1,4.55-\w) -- (9.1,4.55)
   node[midway,left=1pt,font=\footnotesize] {$w$};
 \node[align=center] at (3.35,2.4)
   {protected interior\\[2pt]$\Lambda\setminus C_{\Lambda,w}$};
 \node[fill=aux!13,inner sep=2pt,font=\footnotesize] at (5.7,1.06)
   {boundary collar $C_{\Lambda,w}$, where $K$ acts};
 \node[fill=white,inner sep=2pt,rounded corners=2pt,align=center]
   at (11.8,2.65) {exterior\\[2pt]$\Lambda^c$};
 \node[fill=white,inner sep=2pt,rounded corners=2pt,anchor=south west]
   at (1.05,4.62) {finite sample $\Lambda$};
 \node[fill=white,inner sep=2pt,rounded corners=2pt] at (12.75,4.9) {$\Gamma$};
\end{tikzpicture}
\caption{Schematic geometry of a finite sample.
The exterior \(\Lambda^c\) and the inner boundary collar \(C_{\Lambda,w}\) form the excluded region \(S_{\Lambda,w}\), shown hatched; the boundary operator \(K\) acts in the collar.
Corollary~\ref{cor:finite-sample-bulk} compares all states in the spectral window \eqref{eq:sample-low-energy-projection} with \(\omega_\lambda\) on observables supported in \(Y\), uniformly over even boundary operators in the collar.}
\label{fig:finite-sample-geometry}
\end{figure}

For a nonempty proper subset \(\Lambda\subset\Gamma\) and a width \(w\ge0\), define the inner boundary collar and the excluded region by
\begin{equation}
 C_{\Lambda,w}:=\{x\in\Lambda\,|\,\dist(x,\Lambda^c)\le w\},
 \qquad
 S_{\Lambda,w}:=\Lambda^c\cup C_{\Lambda,w}.
 \label{eq:sample-boundary-collar}
\end{equation}
We call \(\Lambda\setminus C_{\Lambda,w}\) the \emph{protected interior}; see Figure~\ref{fig:finite-sample-geometry}.
For finite \(\Lambda\), \(\lambda\in\mathcal J\), and a self-adjoint even boundary operator \(K\in\A_{C_{\Lambda,w}}^+\), set
\begin{equation}
 H_{\Lambda,K}^{(\lambda)}:=H_\Lambda^{\Phi_\lambda}+K
 =\sum_{Z\subset\Lambda}\Phi_\lambda(Z)+K,
 \qquad
 E_{\Lambda,K}^{(\lambda)}:=\min\spec H_{\Lambda,K}^{(\lambda)}.
 \label{eq:sample-Hamiltonian-ground-projection}
\end{equation}
The boundary operator \(K\) represents arbitrary even boundary conditions for the restricted Hamiltonian; its norm and its interaction range within \(C_{\Lambda,w}\) are unrestricted.
The Hamiltonian acts in the finite Fock representation of \(\A_\Lambda\) on
\[
 \mathcal F_\Lambda:=\bigoplus_{k=0}^{r|\Lambda|}
                            \bigwedge^k\Hsp_\Lambda.
\]
Define the low-energy projection
\begin{equation}
 P_{\Lambda,K}^{(\lambda)}
 :=\1_{[E_{\Lambda,K}^{(\lambda)},\,
              E_{\Lambda,K}^{(\lambda)}+3g/4]}
          (H_{\Lambda,K}^{(\lambda)}).
 \label{eq:sample-low-energy-projection}
\end{equation}
A state \(\rho\in \mathfrak S(\A_\Lambda)\) is supported in this spectral subspace exactly when \(\rho(P_{\Lambda,K}^{(\lambda)})=1\).

\begin{corollary}[Low-energy states of finite samples]
\label{cor:finite-sample-bulk}
Assume Hypotheses~\ref{hyp:background-transport} and~\ref{hyp:uniform-local-excitation-detectability} on \(\mathcal J\), and use \(C,q,G\) from Theorem~\ref{thm:positive-energy-locality}.
For every \(\lambda\in\mathcal J\), nonempty finite proper subset \(\Lambda\subset\Gamma\), \(w\ge0\), nonempty \(Y\subset\Lambda\), and \(A\in\A_Y\),
\begin{equation}
 \sup_{\substack{K\in\A_{C_{\Lambda,w}}^+\\K=K^*}}
 \left\|P_{\Lambda,K}^{(\lambda)}
       \bigl(A-\omega_\lambda(A)\1\bigr)
       P_{\Lambda,K}^{(\lambda)}\right\|
 \le C(1+|Y|)^q\|A\|\,G(\dist(Y,S_{\Lambda,w})).
 \label{eq:finite-sample-bulk}
\end{equation}
Thus, for every such \(K\) and every state \(\rho\in\mathfrak S(\A_\Lambda)\) with \(\rho(P_{\Lambda,K}^{(\lambda)})=1\), the same right-hand side bounds \(|\rho(A)-\omega_\lambda(A)|\).
For every \(k\in\mathbb N\), the estimate also holds with the same constant for every \(A\in\A_Y\otimes M_k\), with the left hand side replaced by
\[
 \sup_{\substack{K\in\A_{C_{\Lambda,w}}^+\\K=K^*}}
 \left\|(P_{\Lambda,K}^{(\lambda)}\otimes\1)
 \bigl[A-\1\otimes(\omega_\lambda\otimes\operatorname{id}_{M_k})(A)\bigr]
 (P_{\Lambda,K}^{(\lambda)}\otimes\1)\right\|.
\]
\end{corollary}

The corollary follows directly from Theorem~\ref{thm:positive-energy-locality}, applied to the cut interaction with \(U=\Lambda\setminus C_{\Lambda,w}\) and with the sample ground energy \(E_{\Lambda,K}^{(\lambda)}\) subtracted from \(H_{\Lambda,K}^{(\lambda)}\) (see Subsection~\ref{sec:low-energy-locality}).

\begin{corollary}[Finite-sample correlations and spectral estimates]
\label{cor:finite-sample-correlations}
Assume Hypotheses~\ref{hyp:background-transport} and~\ref{hyp:uniform-local-excitation-detectability} on \(\mathcal J\), and use the sample notation in \eqref{eq:sample-boundary-collar}--\eqref{eq:sample-low-energy-projection}.
Parts (i) and (ii) use the constants of Theorem~\ref{thm:spatial-correlations} and Corollary~\ref{cor:low-energy-spectral-weight}, respectively.
With these choices, the following conclusions hold for every \(\lambda\in\mathcal J\), nonempty finite proper subset \(\Lambda\subset\Gamma\), \(w\ge0\), self-adjoint even \(K\in\A_{C_{\Lambda,w}}\), nonempty \(X,Y\subset\Lambda\), and \(A\in\A_X\), \(B\in\A_Y\).
In the cited estimates, take \(U=\Lambda\setminus C_{\Lambda,w}=S_{\Lambda,w}^c\).
\begin{enumerate}[label=(\roman*),leftmargin=*]
\item For every sample state \(\rho\) with \(\rho(P_{\Lambda,K}^{(\lambda)})=1\), the covariance bound \eqref{eq:positive-energy-clustering} holds.
The operator bound \eqref{eq:operator-one-sided-clustering} holds with \(\pi(A)=A\), \(\pi(B)=B\), and \(P=P_{\Lambda,K}^{(\lambda)}\).
Both conclusions also hold with \(P\) replaced by
\[
 P_{\Lambda,K}^{(\lambda)}(I)
 :=\1_{E_{\Lambda,K}^{(\lambda)}+I}
          (H_{\Lambda,K}^{(\lambda)}),
 \qquad I\subset[0,3g/4]\ \text{a Borel set},
\]
and with \(\rho(P_{\Lambda,K}^{(\lambda)}(I))=1\) in the covariance bound.
\item Put \(L:=H_{\Lambda,K}^{(\lambda)}-E_{\Lambda,K}^{(\lambda)}\1\).
Estimates \eqref{eq:low-energy-spectral-weight}, \eqref{eq:mixed-imaginary-time-correlation}, and \eqref{eq:elementary-inverse-energy} hold with \(\pi(A)=A\) and \(P=P_{\Lambda,K}^{(\lambda)}\).
For every sample state \(\rho\) with \(\rho(P_{\Lambda,K}^{(\lambda)})=1\), estimates \eqref{eq:centered-low-energy-vector} and \eqref{eq:imaginary-time-correlation} hold with \(\varphi=\rho\), \(A_\rho=A-\rho(A)\1\), and \(B_\rho=B-\rho(B)\1\).
The right-hand side of \eqref{eq:elementary-inverse-energy} also bounds \(|\rho(A_\rho^*(L+z)^{-1}B_\rho)|\) for every \(z>0\).
\end{enumerate}
\end{corollary}

Both parts follow by applying the cited estimates in the positive-energy representation used for Corollary~\ref{cor:finite-sample-bulk} (see Section~\ref{sec:correlation-proofs}).

\paragraph{Local indistinguishability.}
Corollary~\ref{cor:finite-sample-bulk} shows that, in every state supported in the range of \(P_{\Lambda,K}^{(\lambda)}\), the expectation of an observable supported deep inside the sample is close to its bulk ground-state expectation.

These estimates relate to several notions of local indistinguishability.
For Gibbs states, local indistinguishability in \cite[Definition~3]{CMTW25} and \emph{locality of temperature} in \cite{KlieschEtAl14} concern the change of local expectations when the surrounding system is restricted.
Corollary~\ref{cor:finite-sample-bulk} gives a zero-temperature counterpart, uniform over boundary Hamiltonians and all states in the prescribed energy windows.

Bravyi, Hastings, and Michalakis \cite[Section~2.2]{BHM10} introduce two conditions for proving gap stability of frustration-free Hamiltonians with commuting local terms.
TQO-1 requires \(P_0AP_0=c_AP_0\) for local observables \(A\), where \(P_0\) projects onto the ground space of the full finite system.
This expresses local indistinguishability of its ground states and is automatic when the ground state is unique.
TQO-2 connects this global ground space to the ground space of a Hamiltonian restricted to a neighborhood of an observation region: the partial traces of the two projections onto that region must have the same kernel.
Together, these conditions support their proof of spectral-gap stability under weak local perturbations.
These exact conditions need not hold for the finite-sample Hamiltonians considered here.

Closer to our finite-sample estimates is local topological quantum order (LTQO) in \cite[Definition~4]{MichalakisZwolak13}, which allows a decaying error in scalar compression on the ground space of a surrounding region.
The scalar there is the expectation in the normalized ground-space projection.
Choose \(\Lambda\) as a neighborhood of \(Y\), set \(K=0\), and restrict the compression in Corollary~\ref{cor:finite-sample-bulk} to the ground space.
Replacing \(\omega_\lambda(A)\) by the expectation in the normalized ground-space projection then gives this form of estimate, with at most twice the original bound by the triangle inequality.
Our bound contains the factor \((1+|Y|)^q\), whereas the decay function in that definition depends only on the distance between the observable's support and the boundary of the surrounding region.
Thus we obtain an LTQO-type estimate with explicit support dependence, rather than the support-uniform condition of that definition.
Jackson, Nachtergaele, and Young prove an LTQO-type estimate with a support-dependent prefactor for the AKLT models on the hexagonal and Lieb lattices \cite[Corollary~1.2]{JNY26}.

\paragraph{Thermodynamic limits.}
Along every increasing finite exhaustion of \(\Gamma\) with uniformly bounded collar widths, every sequence of states in the sample low-energy windows has the thermodynamic limit \(\omega_\lambda\).
The limit is independent of the exhaustion, the boundary Hamiltonians, and the chosen low-energy states.
More generally, fix \(\lambda\in\mathcal J\) and let \(\Lambda_n\) be finite samples with collars of widths \(w_n\).
Assume that, for every \(\varnothing\ne Y\Subset\Gamma\),
\[
 \dist(Y,S_{\Lambda_n,w_n})\longrightarrow\infty.
\]
For each \(n\), choose a self-adjoint even boundary operator \(K_n\in\A_{C_{\Lambda_n,w_n}}\) and a state \(\rho_n\in \mathfrak S(\A_{\Lambda_n})\) satisfying \(\rho_n(P_{\Lambda_n,K_n}^{(\lambda)})=1\).
Then, for every local observable \(A\),
\[
 \rho_n(A)\longrightarrow\omega_\lambda(A).
\]
This follows from Corollary~\ref{cor:finite-sample-bulk}; the expectations are defined once the support of \(A\) lies in \(\Lambda_n\).
Their state extensions to \(\A\) converge weak-* to \(\omega_\lambda\) by norm density of the local algebra.
The boundary Hamiltonians need not converge or have uniformly bounded norms.

\paragraph{Edges of topological insulators.}
The finite-sample estimate is particularly useful for topological insulators, whose boundaries can support excitations below the bulk gap.
In the Hofstadter model, for example, edge states can traverse a bulk gap containing the chemical potential \cite{Hofstadter76,TKNN82,Hatsugai93}.
After adding sufficiently weak many-body interactions satisfying either locality assumption, Corollary~\ref{cor:finite-sample-bulk} shows that all states in the sample's low-energy window still have local expectations close to those of \(\omega_\lambda\) far from the boundary.
We now use Theorem~\ref{thm:positive-energy-locality} to obtain the corresponding estimate for an infinite sample with an edge.

Under the hypotheses of Theorem~\ref{thm:positive-energy-locality}, fix \(\lambda\in\mathcal J\), \(w\ge0\), and a nonempty proper region \(\Lambda\subset\Gamma\), such as a half-space.
Use \(C_{\Lambda,w}\) and \(S_{\Lambda,w}\) from \eqref{eq:sample-boundary-collar}, and let \(\Upsilon\in\mathfrak J_1\) satisfy \(\Sigma(\Upsilon)\subset C_{\Lambda,w}\).
The interaction on \(\Lambda\) is \(\Phi_\lambda^\Lambda+\Upsilon\), where \(\Phi_\lambda^\Lambda(Z):=\1_{\{Z\subset\Lambda\}}\Phi_\lambda(Z)\) for \(Z\Subset\Gamma\).
For any algebraic ground state \(\omega_\Lambda\) of this interaction on \(\operatorname{CAR}(\ell^2(\Lambda;\C^r))\), write \((\pi_\Lambda,\mathcal H_\Lambda,\Omega_\Lambda)\) for its GNS representation and \(L_\Lambda\) for its GNS Hamiltonian, with \(L_\Lambda\Omega_\Lambda=0\).
Set \(P_\Lambda:=\1_{[0,3g/4]}(L_\Lambda)\).
Theorem~\ref{thm:positive-energy-locality} gives, with the same \(C,q,G\), for every nonempty finite \(Y\Subset\Lambda\) and \(A\in\A_Y\),
\begin{equation}
 \left\|P_\Lambda\bigl(\pi_\Lambda(A)-\omega_\lambda(A)\1\bigr)
                 P_\Lambda\right\|
 \le C(1+|Y|)^q\|A\|\,G(\dist(Y,S_{\Lambda,w})).
 \label{eq:infinite-region-low-energy}
\end{equation}
To apply the theorem, we extend \((\pi_\Lambda,L_\Lambda)\) to \(\A\) with trivial dynamics on \(\Lambda^c\), as detailed at the end of Subsection~\ref{sec:low-energy-locality}.

Estimate~\eqref{eq:infinite-region-low-energy} controls interior expectations uniformly in time for every family \((\rho_t)_{t\in\R}\) of states on \(\mathcal B(\mathcal H_\Lambda)\) satisfying \(\rho_t(P_\Lambda)=1\) for all \(t\).
This includes evolution generated by \(L_\Lambda\) and additional boundary driving whenever the full evolving state remains in this fixed low-energy window.

These conclusions complement rigorous results on transport along the boundary.
Antinucci, Mastropietro, and Porta prove quantization of edge conductance for weakly interacting Hall systems with single-mode chiral edge currents \cite{AntinucciMastropietroPorta18}, and Mastropietro and Porta extend this analysis to multiple channels \cite{MastropietroPorta22}.
For noninteracting systems, Porta and Singh derive edge response from time-dependent quantum dynamics under weak, slowly varying perturbations, in the zero-temperature and large-scale limits \cite[Theorem~5.3]{PortaSingh26}.
These works determine edge transport coefficients under hypotheses on the edge spectrum and geometry, whereas \eqref{eq:infinite-region-low-energy} controls all interior observables throughout the low-energy subspace, uniformly in the boundary interactions and without assumptions on the dispersion or multiplicity of boundary modes.

\paragraph{Free fermions with pairing.}
The Kitaev chain \cite{Kitaev01} is a quadratic model on \(\Gamma=\Z\) with one complex fermion mode per site, nearest-neighbor hopping, and creation and annihilation of neighboring fermion pairs.
For real nonzero hopping and pairing amplitudes \(t_0,\Delta_0\) and chemical potential \(\mu\in\R\) satisfying \(|\mu|<2|t_0|\), its positive one-particle energy is
\[
 E(k):=\sqrt{(2t_0\cos k+\mu)^2+4\Delta_0^2\sin^2k},
 \qquad \min_{k\in[-\pi,\pi]}E(k)>0.
\]
The infinite chain therefore has a unique pure quasifree gapped ground state \(\omega_0\).
For \(\mathcal V\) satisfying either locality assumption and sufficiently small \(|\lambda|\), Proposition~\ref{prop:unconditional-weak-coupling} and Corollary~\ref{cor:ground-state-uniqueness} give a unique gapped ground state \(\omega_\lambda\) of \(\Phi_h+\lambda\mathcal V\).

Cutting out a finite chain can nevertheless create excitations below the bulk gap.
For a long unperturbed open chain \(\Lambda=\{1,\ldots,L\}\), the two Majorana modes localized near its ends combine into one fermionic mode.
Its empty and occupied states give the two lowest many-body states, of opposite fermionic parities, called the \emph{boundary doublet}.
Their energy difference is exponentially small in the chain length \(L\), while the gap to the remaining spectrum stays bounded away from zero as \(L\to\infty\)~\cite{Kitaev01}.
Lapa and Levin explain how their stability theorem for the long-range Ising chain applies to the open Kitaev wire near \(t_0=\Delta_0>0\), \(\mu=0\)~\cite{LapaLevin23}.
For the sufficiently weak summable density--density perturbations they consider, the gap above the doublet stays positive uniformly in \(L\), and the splitting is at most \(CL\E^{-cL}\), with \(C,c>0\) independent of \(L\).

Corollary~\ref{cor:finite-sample-bulk} shows that these changes in the low-energy spectrum have only a small effect on local expectations in the interior.
For every ground state \(\rho\) of the finite-chain Hamiltonian \(H_{\Lambda,K}^{(\lambda)}\) and every \(A\in\A_Y\), the difference \(|\rho(A)-\omega_\lambda(A)|\) decays stretched exponentially with the distance of \(Y\) from the end collar under stretched-exponential locality, and polynomially under polynomial locality.
The same bound holds for every state supported in the spectral window \([E_{\Lambda,K}^{(\lambda)},E_{\Lambda,K}^{(\lambda)}+3g/4]\), including coherent superpositions and mixtures of all levels in that window.
The constants are independent of the arbitrary even end terms \(K\), the number of these levels, and their energies relative to \(E_{\Lambda,K}^{(\lambda)}\).

\paragraph{Interfaces with long-range couplings.}
For a nonempty proper region \(\Lambda\subset\Gamma\) and \(w\ge0\), set \(S:=S_{\Lambda,w}\).
An interface with a second material in \(\Lambda^c\) changes the interaction on \(\Gamma\) to \(\Phi_\lambda+\Psi\) for a defect \(\Psi\).
If \(\Psi\in\mathfrak J_1\) and \(\Sigma(\Psi)\subset S\), so that the interaction changes are confined to the collar and the exterior, then Corollary~\ref{thm:low-energy-locality} applies.
Changed long-range couplings can instead meet both \(S\) and \(S^c\), and therefore need not satisfy this support condition.

In the polynomial case, suppose that \(\Psi\in\mathfrak J_1\), \(\Psi(Z)=0\) whenever \(Z\subset S^c\), and the changed crossing terms obey
\[
 C_\Psi:=\sup_{z\in S^c}
 \sum_{\substack{Z\ni z\\Z\cap S\ne\varnothing}}
 (1+\diam Z)^{p_\Phi}\|\Psi(Z)\|<\infty.
\]
For \(z\in S^c\), each term in this sum has \(\diam Z\ge\dist(z,S)\), giving
\[
 t_{\Psi,S}(z)
 \le (1+\dist(z,S))^{-p_\Phi}
 \sum_{\substack{Z\ni z\\Z\cap S\ne\varnothing}}
 (1+\diam Z)^{p_\Phi}\|\Psi(Z)\|
 \le C_\Psi(1+\dist(z,S))^{-p_\Phi}.
\]
Thus \(J_{\Psi,S}^{p_\Phi}\le C_\Psi\).
Theorem~\ref{thm:quasilocal-tails}(i) applies with \(M=p_\Phi\).
Since \(\nu\le p_\Phi\), both the bulk and tail terms have decay power \(\nu\), with the tail prefactor depending on \(C_\Psi\).
Under stretched-exponential bulk locality, the corresponding crossing-term bound with weight \(\E^{a(\diam Z)^{\theta_{\rm in}}}\), for some \(a>0\), gives \(t_{\Psi,S}(z)\le C_\Psi\E^{-a\dist(z,S)^{\theta_{\rm in}}}\).
Because \(\theta<\theta_{\rm in}\), this implies \(J_{\Psi,S}^{\xi,\theta}<\infty\) for every \(\xi>0\), as required by Theorem~\ref{thm:quasilocal-tails}(ii).

\subsection{Discussion of the hypotheses}
\label{sec:hypotheses-discussion}

Under the transport and LED assumptions, Hypotheses~\ref{hyp:background-transport} and~\ref{hyp:uniform-local-excitation-detectability}, Corollary~\ref{cor:ground-state-uniqueness} gives uniqueness and a GNS gap.
We give two examples showing that neither transport together with a GNS gap nor uniqueness alone implies LED for the chosen filtered family.
We also translate the polynomial locality assumptions into pointwise decay conditions on the quadratic coefficients and two-body interaction terms.

\paragraph{Transport and a GNS gap do not imply uniqueness.}
A GNS gap controls excitations in that state's representation, but does not exclude a second ground state in a disjoint representation.
At a first-order transition, for example, the state obtained by quasi-local spectral flow may remain a ground state at the coexistence parameter, while a second, macroscopically distinct state also satisfies the ground-state inequality.
Spontaneous breaking of a discrete internal or translation symmetry can likewise produce several ground states, each gapped in its own GNS representation.
Symmetry covariance of the transport can keep the distinguished state symmetric without excluding other, symmetry-broken ground states.
Rigorous coexistence examples arise from quantum Pirogov--Sinai theory \cite{YarotskiiPS06,BorgsKotecky00,DattaEtAl99}.

The one-dimensional Ising lattice gas shows that exact ground-state transport and a uniform GNS gap do not imply uniqueness.
On \(\Gamma=\Z\), with one fermionic mode per site, put \(n_x^{\rm occ}:=a_x^*a_x\) and define
\[
 \Xi_\lambda(\{x\}):=n_x^{\rm occ},\qquad
 \Xi_\lambda(\{x,x+1\}):=-\lambda n_x^{\rm occ}n_{x+1}^{\rm occ},
 \qquad -1\le\lambda\le1,
\]
with all other terms zero.
At \(\lambda=0\), this is the free onsite interaction with one-particle Hamiltonian \(h_{\nc}=\1\), up to an irrelevant scalar.
At \(\lambda=1\), its formal Hamiltonian becomes
\begin{equation}
 \mathsf H_1
 =\frac12\sum_x(n_x^{\rm occ}-n_{x+1}^{\rm occ})^2
 =\frac14\sum_x(\1-s_xs_{x+1}),
 \qquad s_x:=\1-2n_x^{\rm occ}.
 \label{eq:coexistence-example}
\end{equation}
This is the zero-field ferromagnetic Ising Hamiltonian, with empty and occupied sites corresponding to spins \(+1\) and \(-1\).

Let \(\omega^{(0)}\) and \(\omega^{(1)}\) be the completely empty and completely filled product states.
In the GNS representation of \(\omega^{(0)}\), the vectors \(\psi_S\) with precisely the sites in the finite set \(S\subset\Z\) occupied form an orthonormal basis.
If \(S\ne\varnothing\) consists of \(c(S)\) maximal occupied intervals, it contains \(|S|-c(S)\) occupied nearest-neighbor pairs.
The implementing Hamiltonian \(L_\lambda^{(0)}\), normalized to annihilate the empty vector, therefore satisfies
\[
 L_\lambda^{(0)}\psi_S
 =\bigl[c(S)+(1-\lambda)(|S|-c(S))\bigr]\psi_S.
\]
Every nonempty configuration has energy at least one, and a single occupied site has energy exactly one.
Thus \(\omega^{(0)}\) is a ground state with GNS gap one throughout \([-1,1]\).

At \(\lambda=1\), the particle--hole transformation \(a_x\mapsto a_x^*\) preserves the dynamics and exchanges \(\omega^{(0)}\) and \(\omega^{(1)}\).
Hence the filled state is also a ground state with GNS gap one.
The constant family \(\omega_\lambda=\omega^{(0)}\) therefore admits identity transport and a uniform GNS gap, although uniqueness fails at \(\lambda=1\).
By Corollary~\ref{cor:ground-state-uniqueness}, LED cannot hold there for this transported state.
The next example shows that LED may also fail when the ground state remains unique.

\paragraph{Uniqueness does not imply local excitation detectability.}
Even when the Fock vacuum is the unique ground state and is separated from all excited states by a positive gap, removing one fermion from an excited state can raise its energy.
On \(\Gamma=\{x\}\), take two fermionic modes and define \(n_i:=a_i^*a_i\) for \(i=1,2\).
Choose \(\mu>0\) and \(\mu<J<5\mu/4\).
For \(0\le\lambda\le1\), set
\[
 H_\lambda:=\mu(n_1+n_2)-\lambda J n_1n_2.
\]
The occupation vectors \(\lvert00\rangle,\lvert10\rangle,\lvert01\rangle,\lvert11\rangle\) have energies
\[
 E_{00}=0,\qquad E_{10}=E_{01}=\mu,\qquad
 E_{11}=2\mu-\lambda J>3\mu/4.
\]
The free gap is \(g=\mu\), and the empty state remains the unique ground state, with gap at least \(2\mu-J>3g/4\), throughout \(0\le\lambda\le1\).
Thus \(\omega_\lambda=\omega_0\) is transported by the identity map, and \(\sigma_\lambda=\sigma_{\rm st}\).

Now suppose \(\lambda J>\mu\).
Removing one fermion from the doubly occupied state raises its energy to \(\mu\), whereas removing both fermions lowers it to zero.
The fixed filter \(f\) of \eqref{eq:Fourier} satisfies \(\widehat f(-\mu)=1\) and \(\widehat f(\lambda J-\mu)=0\), so, for \(i=1,2\),
\[
 \int_\R f(t)\E^{\I tH_\lambda}a_i\E^{-\I tH_\lambda}\,\dd t
 =a_i(\1-n_{3-i}).
\]
These filtered operators annihilate the excited pair as well as the vacuum.

This implies that LED cannot hold for the chosen filtered family.
To show this using the stacked excitation operators \(n_x\) and \(n_{f,x}^{(\lambda)}\), adjoin the auxiliary free system, whose ground state has both auxiliary modes filled.
The transformation \(\sigma_\st\) of Section~\ref{sec:atomicization} and the filtering identity above give
\begin{align*}
 \sigma_\st^{-1}(n_x)
 &=n_1+n_2+\sum_{i=1}^2(\1-a_{{\rm aux},i}^*a_{{\rm aux},i}),\\
 \sigma_\st^{-1}\!\left(n_{f,x}^{(\lambda)}\right)
 &=n_1(\1-n_2)+n_2(\1-n_1)
   +\sum_{i=1}^2(\1-a_{{\rm aux},i}^*a_{{\rm aux},i}).
\end{align*}
Let \(\rho\) be the state in the original coordinates with both physical and both auxiliary modes occupied.
Applying \(\rho\circ\sigma_\st^{-1}\) to the detectability inequality \eqref{eq:uniform-local-excitation-detectability} would give
\[
 2\le2\mathscr K_\lambda(x,x)<2.
\]
Thus the ground state remains unique and uniformly gapped, but the chosen filtered annihilation operators do not detect this excited pair.
This example suggests enlarging the family of detecting operators to include products of annihilation operators.
Here \(a_1a_2\) lowers the excited pair to the vacuum, and its filtered version acts nontrivially on that pair.
Adding the corresponding positive squares to the right-hand side of LED could therefore extend its applicability beyond the present choice of single-fermion annihilation operators.

\paragraph{Decay of two-body interactions.}
For interactions of bounded support size, the exponential support-size weight imposes no additional summability requirement: if, for some \(m_0\in\mathbb N\), \(\mathcal V(X)=0\) whenever \(|X|>m_0\), then, for every \(\kappa>0\), the factor \(\E^{\kappa|X|}\) is bounded by \(\E^{\kappa m_0}\) on every nonzero interaction term.
In particular, suppose only onsite and two-site terms of \(\mathcal V\) are present, the onsite terms are uniformly bounded, and there are \(C>0\) and \(\varpi>2d_{\rm vol}\) such that, for all distinct \(x,y\in\Gamma\),
\[
 \|\mathcal V(\{x,y\})\|\le C(1+d(x,y))^{-\varpi}.
\]
For every \(0\le p<\varpi-d_{\rm vol}\) and \(\kappa>0\), polynomial volume growth therefore gives
\[
 \|\mathcal V\|^{\#}_{p;\kappa}<\infty.
\]
Suppose also that the quadratic coefficient blocks satisfy, for some \(C_h'>0\), \(\varpi_h>2d_{\rm vol}\), and all \(x,y\in\Gamma\),
\[
 \|h_{xy}\|\le C_h'(1+d(x,y))^{-\varpi_h}.
\]
Then \(\|h\|^{\mathrm{Sch,pol}}_{p_h}<\infty\) for \(p_h<\varpi_h-d_{\rm vol}\).
Under the cell-partition condition of Proposition~\ref{prop:unconditional-weak-coupling}(ii), every fixed decay exponent \(\nu\) satisfying
\[
 0<\nu<\min\{\varpi_h,\varpi\}-d_{\rm vol}
\]
therefore gives local determination with exponent \(\nu\) at sufficiently small coupling.
To obtain this estimate, choose \(p_h=p_{\mathcal V}>\max\{d_{\rm vol},\nu\}\) below \(\min\{\varpi_h,\varpi\}-d_{\rm vol}\), and apply the weak-coupling result at the smaller power \(\nu\).
Here \(\varpi_h,\varpi\) measure pointwise decay, whereas \(p_h,p_{\mathcal V}\) are weighted summability exponents.

\subsection{Proof strategy}
\label{sec:proof-strategy-roadmap}

Here we explain the proof of Theorem~\ref{thm:positive-energy-locality} and its extension to local ground-state determination in Theorem~\ref{thm:bulk-uniqueness}.
We give enough detail for readers to follow the full argument without having to work through every technical estimate in the later sections.

Fix \(\lambda,U,\Xi,\pi,L\) satisfying the hypotheses of Theorem~\ref{thm:positive-energy-locality}, and set \(P:=\1_{[0,3g/4]}(L)\).
For a unit vector \(\psi\in P\mathcal H\), write \(\rho(A):=\langle\psi,\pi(A)\psi\rangle\).
Our first goal is to prove, uniformly over these vectors and \(\lambda,U,\Xi,\pi,L\),
\[
 |\rho(A)-\omega_\lambda(A)|
 \le C(1+|Y|)^q\|A\|G(\dist(Y,U^c)),
 \qquad A\in\A_Y.
\]
The first example in Subsection~\ref{sec:hypotheses-discussion} shows why transport and a bulk GNS gap alone do not suffice to obtain such a bound.
We use local excitation detectability to obtain this bound.
We first explain in finite dimensions how filtered operators determine a state, then apply this argument locally in Steps~1--4.
Step~5 adapts the same argument to states satisfying only a local ground-state inequality, and Step~6 explains why detectability holds at weak coupling.

\paragraph{Filtered squares and local state comparison.}
Let \(\mathcal H_0\) be a finite-dimensional Hilbert space of dimension at least two, and let \(B_1,\ldots,B_N\in\mathcal B(\mathcal H_0)\) satisfy \(\bigcap_{k=1}^N\ker B_k=\C\Omega\) for a unit vector \(\Omega\).
In the application, the \(B_k\) are local annihilation operators and \(\Omega\) is the local Fock vacuum.
If \(\kappa_{\min}>0\) is the smallest nonzero eigenvalue of \(\sum_kB_k^*B_k\), then
\begin{equation}
 \1-|\Omega\rangle\langle\Omega|
 \le\kappa_{\min}^{-1}\sum_{k=1}^N B_k^*B_k.
 \label{eq:abstract-roadmap-finite-reconstruction}
\end{equation}
For every state \(\varrho\) on \(\mathcal B(\mathcal H_0)\), the pure-state trace-distance estimate from Lemma~\ref{lem:exact-atomic-reconstruction} therefore gives
\begin{equation}
 \|\varrho-\omega_\Omega\|_1^2
 \le4\kappa_{\min}^{-1}\sum_{k=1}^N\varrho(B_k^*B_k),
 \qquad \omega_\Omega(D):=\langle\Omega,D\Omega\rangle.
 \label{eq:abstract-roadmap-state-control}
\end{equation}
Thus small excitation numbers force a state to be close to the vacuum state \(\omega_\Omega\).

To obtain small excitation numbers, we compare two ways of filtering the annihilation operators.
Filtering with the nonnegative Hamiltonian whose low-energy states we study gives operators that annihilate those states, whereas filtering with the bulk dynamics gives the operators in the detectability inequality \eqref{eq:uniform-local-excitation-detectability}.
For the finite-dimensional model, let \(L_0\ge0\), put \(P_*:=\1_{[0,3g/4]}(L_0)\), and take the filter \(f\) of \eqref{eq:Fourier}, which satisfies \(\widehat f(\omega)=0\) for \(\omega\ge-3g/4\).
Filtering \(B_k\) with the evolution generated by \(L_0\) yields
\[
 C_{f,k}:=\int_\R f(t)\,\E^{\I tL_0}B_k\E^{-\I tL_0}\,\dd t.
\]
Writing \(P_E:=\1_{\{E\}}(L_0)\) for the spectral projections of \(L_0\), we obtain
\[
C_{f,k}
=\sum_{E,E'\in\sigma(L_0)}
\widehat f(E-E')\,P_E B_kP_{E'}.
\]
If \(0\le E'\le3g/4\), then \(E\ge0\) implies \(E-E'\ge-3g/4\), so the corresponding coefficient vanishes.
Hence
\begin{equation}
 C_{f,k}P_*=0.
 \label{eq:abstract-roadmap-square-control}
\end{equation}
Lemma~\ref{lem:low-energy-annihilation} proves the same spectral statement for general nonnegative self-adjoint operators.
No gap or uniqueness assumption at zero, and no gap above the window, is used.

Let \(H_{\rm bulk}\ge0\) be the bulk Hamiltonian in this finite-dimensional model, with \(H_{\rm bulk}\Omega=0\), and filter the same operators \(B_k\) with its evolution:
\[
 F_{f,k}:=\int_\R f(t)\,\E^{\I tH_{\rm bulk}}B_k\E^{-\I tH_{\rm bulk}}\,\dd t.
\]
Suppose these operators satisfy the detectability inequality
\[
 \sum_k B_k^*B_k
 \le2\sum_k F_{f,k}^*F_{f,k}
       +\vartheta\sum_k B_k^*B_k,
 \qquad 0\le\vartheta<1.
\]
For every unit vector \(\psi\in P_*\mathcal H_0\), let \(\omega_\psi(D):=\langle\psi,D\psi\rangle\).
Combining this inequality with \eqref{eq:abstract-roadmap-state-control} and \eqref{eq:abstract-roadmap-square-control} gives
\[
 \|\omega_\psi-\omega_\Omega\|_1^2
 \le\frac{8}{\kappa_{\min}(1-\vartheta)}
       \sum_k\|(F_{f,k}-C_{f,k})\psi\|^2.
\]
The remaining task is to compare the two families of filtered annihilation operators, whose definitions in terms of filtering reduce this to comparing the two dynamics.
In the spatial problem, \(\Xi\) and \(\Phi_\lambda\) agree inside \(U\).
Comparing both evolutions with the bulk dynamics restricted to a region inside \(U\), locality estimates show that their action on an operator supported far from \(U^c\) remains close for a time interval that grows with this distance.
The decay of \(f\) controls the contribution from later times, so the comparison bounds \(\|(F_{f,k}-C_{f,k})\psi\|\), the quantities appearing in the preceding estimate.

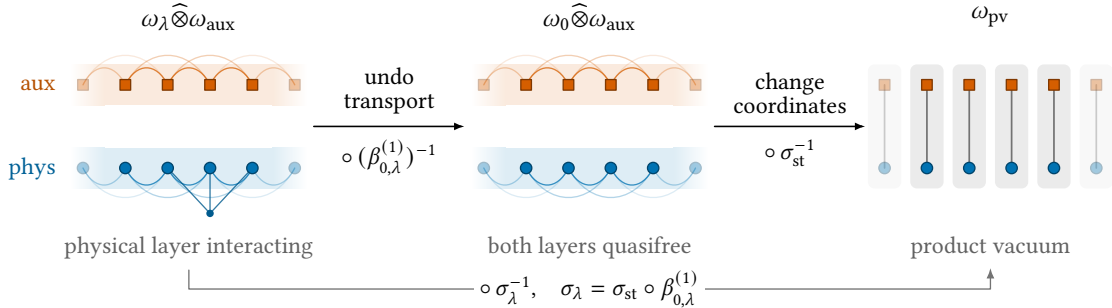
\begin{figure}[htbp]
\centering
\begin{tikzpicture}[lppl diagram]
 \def\dx{.56}\def\yP{0}\def\yA{1.1}\def\hb{.27}
 \newcommand\layerbands[1]{%
  \foreach \y/\c in {\yP/phys,\yA/aux}{
   \shade[left color=white,right color=\c!13]
     ({#1-1.3*\dx},\y-\hb) rectangle ({#1-.5*\dx+.01},\y+\hb);
   \fill[\c!13] ({#1-.5*\dx},\y-\hb) rectangle ({#1+3.5*\dx},\y+\hb);
   \shade[left color=\c!13,right color=white]
     ({#1+3.5*\dx-.01},\y-\hb) rectangle ({#1+4.3*\dx},\y+\hb);}
  \foreach \i/\op in {-1/.3,0/.75,1/.75,2/.75,3/.3}{
   \draw[phys,opacity=\op] ({#1+\i*\dx},\yP) .. controls +(.15,-.3) and +(-.15,-.3) .. ({#1+(\i+1)*\dx},\yP);
   \draw[aux,opacity=\op]  ({#1+\i*\dx},\yA) .. controls +(.15,.3)  and +(-.15,.3)  .. ({#1+(\i+1)*\dx},\yA);}
  \foreach \i/\op in {-1/.14,0/.3,1/.3,2/.14}{
   \draw[phys,opacity=\op] ({#1+\i*\dx},\yP) .. controls +(.28,-.52) and +(-.28,-.52) .. ({#1+(\i+2)*\dx},\yP);
   \draw[aux,opacity=\op]  ({#1+\i*\dx},\yA) .. controls +(.28,.52)  and +(-.28,.52)  .. ({#1+(\i+2)*\dx},\yA);}}
 \newcommand\layersites[1]{%
  \foreach \i/\op in {-1/.35,0/1,1/1,2/1,3/1,4/.35}{
   \begin{scope}[opacity=\op]
    \filldraw[fill=phys,draw=phys!60!black] ({#1+\i*\dx},\yP) circle[radius=.075];
    \filldraw[fill=aux,draw=aux!60!black]
      ({#1+\i*\dx-.07},\yA-.07) rectangle ({#1+\i*\dx+.07},\yA+.07);
   \end{scope}}}
 \layerbands{0}
 \coordinate (v) at (2*\dx,\yP-.6);
 \foreach \k in {1,2,3}
  \draw[phys!80!black,line width=.5pt,opacity=.8] (v) -- (\k*\dx,\yP);
 \fill[phys!80!black] (v) circle[radius=1.4pt];
 \layersites{0}
 \node[fig title] at (1.5*\dx,\yA+.9) {$\omega_\lambda\widehat\otimes\omega_{\rm aux}$};
 \node[fig note] at (1.5*\dx,\yP-1.05) {physical layer interacting};
 \node[anchor=east,text=aux!85!black,font=\footnotesize] at (-1.4*\dx,\yA) {aux};
 \node[anchor=east,text=phys!85!black,font=\footnotesize] at (-1.4*\dx,\yP) {phys};
 \def\xB{5.3}
 \layerbands{\xB}
 \layersites{\xB}
 \node[fig title] at (\xB+1.5*\dx,\yA+.9) {$\omega_0\widehat\otimes\omega_{\rm aux}$};
 \node[fig note] at (\xB+1.5*\dx,\yP-1.05) {both layers quasifree};
 \def\xC{10.6}
 \foreach \i/\op in {-1/.35,0/1,1/1,2/1,3/1,4/.35}{
  \begin{scope}[opacity=\op,shift={(\xC+\i*\dx,0)}]
   \fill[black!8,rounded corners=2.5pt] (-.22,\yP-\hb) rectangle (.22,\yA+\hb);
   \draw[black!55,line width=.65pt] (0,\yP) -- (0,\yA);
   \filldraw[fill=phys,draw=phys!60!black] (0,\yP) circle[radius=.075];
   \filldraw[fill=aux,draw=aux!60!black] (-.07,\yA-.07) rectangle (.07,\yA+.07);
  \end{scope}}
 \node[fig title] at (\xC+1.5*\dx,\yA+.9) {$\omega_\at$};
 \node[fig note] at (\xC+1.5*\dx,\yP-1.05) {product vacuum};
 \draw[lppl arrow] (4.45*\dx,.5*\yA) -- (\xB-1.45*\dx,.5*\yA)
   node[midway,above=1pt,font=\footnotesize,align=center] {undo\\[-2pt]transport}
   node[midway,below=1pt,font=\footnotesize] {$\circ\,(\beta_{0,\lambda}^{(1)})^{-1}$};
 \draw[lppl arrow] (\xB+4.45*\dx,.5*\yA) -- (\xC-1.45*\dx,.5*\yA)
   node[midway,above=1pt,font=\footnotesize,align=center] {change\\[-2pt]coordinates}
   node[midway,below=1pt,font=\footnotesize] {$\circ\,\sigma_\st^{-1}$};
 \draw[black!55,line width=.5pt,-{Latex[length=4pt,width=3pt]}]
   (1.5*\dx,-1.35) -- (1.5*\dx,-1.6)
   -- node[fill=white,inner sep=2pt,font=\footnotesize,text=black]
        {$\circ\,\sigma_\lambda^{-1}$,\quad
         $\sigma_\lambda=\sigma_\st\circ\beta_{0,\lambda}^{(1)}$}
   (\xC+1.5*\dx,-1.6) -- (\xC+1.5*\dx,-1.32);
\end{tikzpicture}
\caption{Construction of product-vacuum coordinates.
Circles represent the physical layer and squares the auxiliary layer at the same sites; shaded regions indicate factors of the state.
Curves and vertical links indicate correlations; the additional vertex represents correlations in \(\omega_\lambda\) that need not satisfy Wick's rule.
On the arrows, \(\circ\,\alpha\) denotes \(\rho\mapsto\rho\circ\alpha\).
Initially the two layers are independent.
Undoing transport returns the physical layer to the quasifree state \(\omega_0\); composition with \(\sigma_\st^{-1}\) then makes the stacked state a product across sites.
The composite transformation \(\sigma_\lambda\) gives the state identity displayed in Step~1.}
\label{fig:atomic-coordinates}
\end{figure}

\paragraph{1. Product-vacuum coordinates for the transported ground state.}
To apply the preceding vacuum comparison locally, we first transform the stacked bulk ground state into a product vacuum.
We adjoin an auxiliary copy of \(\A\) with Majorana coefficient operator \(-h\) and ground state \(\omega_{\rm aux}\).
The Bogoliubov automorphism \(\sigma_\st\) of Lemma~\ref{lem:atomicization} transforms the stacked free ground state \(\omega_0\widehat\otimes\omega_{\rm aux}\) to the product vacuum \(\omega_\at\).
Composing it with the first-factor lift \(\beta_{0,\lambda}^{(1)}=\beta_{0,\lambda}\widehat\otimes\operatorname{id}_{\A^{\rm aux}}\) of the spectral flow gives
\[
 \sigma_\lambda:=\sigma_\st\circ\beta_{0,\lambda}^{(1)},
 \qquad
 (\omega_\lambda\widehat\otimes\omega_{\rm aux})
                  \circ\sigma_\lambda^{-1}=\omega_\at.
\]
Figure~\ref{fig:atomic-coordinates} illustrates this composition.

The transformation \(\sigma_\lambda\) between the original and product-vacuum coordinates is an automorphism of the same stacked algebra \(\mathcal A^\st=\mathcal A\widehat\otimes\mathcal A^{\rm aux}\).
For the comparison state \(\rho\), set
\[
 \rho_\st:=\rho\widehat\otimes\omega_{\rm aux},
 \qquad
 \rho_\at:=\rho_\st\circ\sigma_\lambda^{-1}.
\]
Lemma~\ref{lem:stacked-positive-representation} realizes \(\rho_\st\) as the vector state of \(\chi:=\psi\otimes\Omega_{\rm aux}\) in a graded representation \(\pi_\st\).
The operator
\[
 \mathbb L:=L\otimes\1+\1\otimes L_{\rm aux}\ge0
\]
implements the modified dynamics \(\tau^{\Xi,\st}:=\tau^\Xi\widehat\otimes\tau^{\rm aux}\) on the stacked algebra.
Since \(L_{\rm aux}\Omega_{\rm aux}=0\), the vector \(\chi\) still lies in the window \([0,3g/4]\).
This construction does not require \(\rho\) to be even or stationary.

We compare dynamics in original coordinates, where \eqref{eq:comparison-interaction-criterion} expresses agreement of the interaction terms supported entirely in \(U\).
The automorphism \(\sigma_\lambda\) does not in general preserve local supports.
In product-vacuum coordinates, however, \(\omega_\at\) is the vacuum of the onsite annihilation operators \(c_{x,j}\), so comparison with \(\rho_\at\) reduces to excitation densities.
The quantitative quasi-locality estimates for \(\sigma_\lambda\) in Lemma~\ref{lem:coordinate-locality} connect these two descriptions.

\paragraph{2. From excitation densities to local expectations.}
This step takes place in product-vacuum coordinates.
The onsite annihilation operators \(c_{x,j}\), \(j\in\mathscr I\), determine the local vacuum: the intersection of their kernels in the onsite Fock space is its one-dimensional span.
Recall \(n_x:=\sum_{j\in\mathscr I}c_{x,j}^*c_{x,j}\).
For finite \(Z\Subset\Gamma\), the operator \(N_Z:=\sum_{x\in Z}n_x\) has nonnegative integer spectrum, and \(P_Z^\vac:=\1_{\{0\}}(N_Z)\) is the rank-one local vacuum projection.
Thus \(\1-P_Z^\vac\le N_Z\).
The pure-state trace-distance estimate gives, for \(D\in\A_Z^\st\),
\begin{equation}
 |\rho_\at(D)-\omega_\at(D)|^2
 \le4\|D\|^2\rho_\at(\1-P_Z^\vac)
 \le4\|D\|^2\sum_{x\in Z}\rho_\at(n_x).
 \label{eq:abstract-roadmap-reconstruction}
\end{equation}
This estimate, proved in Lemma~\ref{lem:exact-atomic-reconstruction}, reduces local state comparison to bounds on \(u_x:=\rho_\at(n_x)\) deep inside \(U\).
The corresponding estimate also holds with an ancillary system, with a constant independent of its dimension.

\paragraph{3. Comparing the bulk and modified dynamics.}
We work in original coordinates and write \(\widetilde c_{x,j}^{\,(\lambda)}:=\sigma_\lambda^{-1}(c_{x,j})\).
The bulk-filtered annihilation operators and their positive squares are
\begin{align*}
 \widetilde c_{f,x,j}^{\,(\lambda)}
 :=\int_\R f(t)\tau_t^{\st,\lambda}
             (\widetilde c_{x,j}^{\,(\lambda)})\,\dd t, \qquad
 \widetilde n_{f,x}^{(\lambda)}
  :=\sum_{j\in\mathscr I}
       (\widetilde c_{f,x,j}^{\,(\lambda)})^*
        \widetilde c_{f,x,j}^{\,(\lambda)}.
\end{align*}
Here \(\tau^{\st,\lambda}\) is generated by \(\Phi_\lambda\) on the physical factor and the auxiliary free interaction on the second factor.
The operators \(\pi_\st(\widetilde c_{f,x,j}^{\,(\lambda)})\) play the role of \(F_{f,k}\) in the finite-dimensional example, with \((x,j)\) replacing \(k\).

Filtering the same fields \(\widetilde c_{x,j}^{\,(\lambda)}\) with the modified dynamics \(\tau^{\Xi,\st}\) gives operators
\[
 C_{x,j}^{\Xi}:=\pi_\st\!\left(
   \int_\R f(t)\tau_t^{\Xi,\st}
          (\widetilde c_{x,j}^{\,(\lambda)})\,\dd t\right)
 \quad\text{with}\quad C_{x,j}^{\Xi}\chi=0.
\]
The operators \(C_{x,j}^{\Xi}\) play the role of \(C_{f,k}\), and their annihilation of \(\chi\) is precisely the spectral statement \eqref{eq:abstract-roadmap-square-control}, applied to \(\mathbb L\) and \(\chi\).

Fix \(x\) deep inside \(U\) and a ball \(\Lambda\subset U\) around \(x\), with radius comparable to \(\dist(x,U^c)\).
Choose a local approximant to \(\widetilde c_{x,j}^{\,(\lambda)}\) supported well inside \(\Lambda\), using Lemma~\ref{lem:coordinate-locality}.
Compare its bulk evolution and conjugation by \(\E^{\I t\mathbb L}\) with the same dynamics generated by the bulk Hamiltonian restricted to \(\Lambda\).
All interaction terms inside \(\Lambda\) agree, and evenness ensures that terms supported in \(U^c\) commute with the local approximant and its restricted evolution.
The remaining commutators involve bulk terms crossing the boundary of \(\Lambda\).
Each such physical term is either unchanged or set to zero by the modified interaction, while the auxiliary interaction is unchanged.
Lemma~\ref{lem:local-spectral-comparison} bounds both differences from the restricted evolution by sums of commutator norms involving bulk terms crossing the boundary of \(\Lambda\).

We integrate the dynamical comparison against the filter \(f\), using locality of the bulk dynamics and the decay of \(f\) to control the error.
Since \(C_{x,j}^{\Xi}\chi=0\), the operators filtered with the bulk dynamics almost annihilate \(\chi\) at sites far from \(U^c\).
Lemmas~\ref{lem:common-filtered-occupation} and~\ref{lem:positive-representation-locality} give the quantitative bound
\begin{equation}
 b_x:=2\rho_\at(n_{f,x}^{(\lambda)})
       =2\rho_\st(\widetilde n_{f,x}^{(\lambda)}) =2\sum_{j\in\mathscr I}
       \|\pi_\st(\widetilde c_{f,x,j}^{\,(\lambda)})\chi\|^2
       \le C G_{\rm filt}(\dist(x,U^c)),
 \label{eq:abstract-roadmap-filtered-bound}
\end{equation}
where
\[
 G_{\rm filt}(D):=
 \begin{cases}
  \E^{-\mu_{\rm filt}D^\theta},&\text{stretched-exponential locality},\\
  (1+D)^{-2\nu},&\text{polynomial locality}.
 \end{cases}
\]
Here \(\mu_{\rm filt}>0\), and \(\nu\) is the exponent in \eqref{eq:main-polynomial-range}.
The bounds hold for every unit vector in \(P\mathcal H\).
If \(U=\Gamma\), the modified interaction \(\Xi\) equals the bulk interaction \(\Phi_\lambda\) and \(b=0\) exactly.

\paragraph{4. From LED to local agreement with the bulk ground state.}
Step~3 shows that the fields filtered with the bulk dynamics almost annihilate \(\chi=\psi\otimes\Omega_{\rm aux}\) when the site \(x\) lies deep inside~\(U\).
Next return to product-vacuum coordinates and use LED:
\begin{equation}
 n_x\le2n_{f,x}^{(\lambda)}
          +\sum_{z\in\Gamma}\mathscr K_\lambda(x,z)n_z.
 \label{eq:abstract-roadmap-detectability}
\end{equation}
Taking the expectation in \(\rho_\at\) gives \(u\le b+\mathscr K_\lambda u\), with \(0\le u_x\le2r\).
The strict row contraction in Hypothesis~\ref{hyp:uniform-local-excitation-detectability} implies
\begin{equation}
 u\le(\1-\mathscr K_\lambda)^{-1}b
       =\sum_{n=0}^\infty\mathscr K_\lambda^n b.
 \label{eq:abstract-roadmap-absorption}
\end{equation}
The series converges in \(\ell^\infty(\Gamma)\).
In the stretched-exponential case, the weighted contraction already gives the required bound \(u_x\le C\E^{-\mu_0\dist(x,U^c)^\theta}\), with the same power \(\theta\) and some \(\mu_0>0\) (Lemma~\ref{lem:absorption}).

For polynomial decay, the row-sum bound on \(\mathscr K_\lambda\) alone does not control contributions from distant sites.
We therefore derive a version of LED involving only excitation numbers near \(x\), using locality of the filtered annihilation operators to bound the error.
This gives \(u_x\le C(1+\dist(x,U^c))^{-2\nu}\) uniformly over unit vectors in the low-energy window (Lemma~\ref{lem:localized-detectability}).

We now use the excitation-density bounds to compare \(\rho(A)\) and \(\omega_\lambda(A)\) for \(A\in\A_Y\).
To use the estimate from Step~2, express \(A\) in product-vacuum coordinates by setting \(A_\at:=\sigma_\lambda(A\widehat\otimes\1)\), so that
\[
 \rho(A)-\omega_\lambda(A)=(\rho_\at-\omega_\at)(A_\at).
\]
The transformed observable \(A_\at\) need not have finite support, so we decompose it into local terms supported in increasing neighborhoods of \(Y\).
We apply \eqref{eq:abstract-roadmap-reconstruction} to terms supported deep inside \(U\), using the excitation-density bounds, and control the remaining terms by locality of \(\sigma_\lambda\).
Summing these estimates gives the desired bound on \(|\rho(A)-\omega_\lambda(A)|\) (Lemma~\ref{lem:coordinate-reconstruction}).
Uniformity over the low-energy vectors gives the operator-norm estimate for \(P(\pi(A)-\omega_\lambda(A)\1)P\), and the ancillary version of Step~2 gives its matrix extension (Lemma~\ref{lem:positive-representation-cb}).

\paragraph{5. Local ground-state determination without global stationarity.}
To prove Theorem~\ref{thm:bulk-uniqueness}, fix \(U\subset\Gamma\) and \(\rho\in\LG_{\Phi_\lambda}(U)\).
The information available about \(\rho\) is the local ground-state inequality
\begin{equation}
 \operatorname{Re}\rho\bigl(C^*\delta_{\Phi_\lambda}(C)\bigr)\ge0
 \qquad(C\text{ local and supported in }U).
 \label{eq:abstract-roadmap-local-ground}
\end{equation}
Such a state need not be globally stationary, so its GNS construction does not in general give the positive-energy representation required by Theorem~\ref{thm:positive-energy-locality}.
We again apply Steps~2 and~4.
Set \(\rho_\st:=\rho\widehat\otimes\omega_{\rm aux}\) and \(\rho_\at:=\rho_\st\circ\sigma_\lambda^{-1}\).
To bound \(\rho_\at(n_{f,x}^{(\lambda)})\), we construct a positive operator on a local cyclic subspace.

Lemma~\ref{lem:graded-product} transfers \eqref{eq:abstract-roadmap-local-ground} to \(\rho_\st\) for the stacked bulk derivation \(\delta_\lambda^\st\).
We use this inequality in original coordinates, where supports remain inside \(U\).
For a ball \(\Lambda\subset U\), let \(P_{\rho_\st,\Lambda}\) project onto the finite-dimensional subspace
\[
 \mathcal H_{\rho_\st,\Lambda}
 :=\pi_{\rho_\st}(\A_\Lambda^\st)\Omega_{\rho_\st}
\]
of the GNS representation of \(\rho_\st\), and let \(\pi_{\rho_\st,\Lambda}\) denote the local representation restricted to this subspace.
Lemma~\ref{lem:local-positive-generator} constructs \(L_{\rho_\st,\Lambda}\ge0\) from the quadratic-form identity
\[
 \langle\pi_{\rho_\st}(B)\Omega_{\rho_\st},
 L_{\rho_\st,\Lambda}\pi_{\rho_\st}(B)\Omega_{\rho_\st}\rangle
 =\operatorname{Re}\rho_\st\bigl(B^*\delta_\lambda^\st(B)\bigr)\ge0
 \qquad(B\in\A_\Lambda^\st),
\]
where the inequality follows from \eqref{eq:abstract-roadmap-local-ground} after stacking.
This operator also satisfies
\[
 L_{\rho_\st,\Lambda}\Omega_{\rho_\st}=0
\]
and, for \(B\in\A_\Lambda^\st\),
\begin{equation}
 [L_{\rho_\st,\Lambda},\pi_{\rho_\st,\Lambda}(B)]
 =\left.P_{\rho_\st,\Lambda}\pi_{\rho_\st}(\delta_\lambda^\st(B))
       P_{\rho_\st,\Lambda}\right|_{\mathcal H_{\rho_\st,\Lambda}}.
 \label{eq:roadmap-compressed-commutator}
\end{equation}
A similar positive-Hamiltonian construction from a local ground-state inequality is given by Yarotsky in the appendix of \cite{Yarotsky05}.

Choose a local approximant to \(\widetilde c_{x,j}^{\,(\lambda)}\), supported well inside \(\Lambda\), and filter it with the evolution generated by \(L_{\rho_\st,\Lambda}\).
The resulting operator annihilates \(\Omega_{\rho_\st}\), since \(L_{\rho_\st,\Lambda}\ge0\), \(L_{\rho_\st,\Lambda}\Omega_{\rho_\st}=0\), and the filter has only negative frequencies.
Using \eqref{eq:roadmap-compressed-commutator}, we compare this evolution and the bulk evolution with the same restricted bulk dynamics, as in Step~3.
This gives
\[
 2\rho_\at(n_{f,x}^{(\lambda)})
 \le C G_{\rm filt}(\dist(x,U^c))
\]
(Proposition~\ref{prop:filtered-coercivity}).
Step~4 then converts this bound into local agreement of \(\rho\) with \(\omega_\lambda\), proving Theorem~\ref{thm:bulk-uniqueness}.
We prove these estimates in Subsection~\ref{sec:bulk-uniqueness}.

\paragraph{6. Establishing LED at weak coupling.}
The preceding argument applies whenever transport and LED hold.
It remains to establish LED at weak coupling.
At \(\lambda=0\), the free dynamics acts on the annihilation operators through the matrix \(\E^{-\I t|h|}\), as given in \eqref{eq:atomic-free-dynamics}.
Since \(\operatorname{spec}|h|\subset[g,\overline C_h]\) and \(\widehat f=1\) on \([-\overline C_h,-g]\), filtering gives
\[
 \int_\R f(t)\,\E^{-\I t|h|}\,\dd t
 =\widehat f(-|h|)=\1.
\]
Thus \(c_{f,x,j}^{(0)}=c_{x,j}\) (Lemma~\ref{lem:filter-normalization}).
To prove \eqref{eq:abstract-roadmap-detectability}, consider the corrections
\[
 e_{x,j}^{(\lambda)}:=c_{f,x,j}^{(\lambda)}-c_{x,j}.
\]
Lemma~\ref{lem:small-correction} bounds the changes caused by the spectral flow and the bulk dynamics, giving \(\|e_{x,j}^{(\lambda)}\|\le C_{\rm filt}|\lambda|\).
Using \(c_{x,j}=c_{f,x,j}^{(\lambda)}-e_{x,j}^{(\lambda)}\) gives
\[
 n_x\le2n_{f,x}^{(\lambda)}
        +2\sum_j(e_{x,j}^{(\lambda)})^*e_{x,j}^{(\lambda)}.
\]
Lemmas~\ref{lem:exact-atomic-reconstruction} and~\ref{lem:filter-normalization} give
\[
 \pi_\at(c_{f,x,j}^{(\lambda)})\Omega_\at
 =\pi_\at(c_{x,j})\Omega_\at=0.
\]
Thus the corrections annihilate the target product-vacuum vector \(\Omega_\at\).

Using vacuum-preserving conditional expectations, write each correction as a norm-convergent sum
\[
 e_{x,j}^{(\lambda)}=\sum_{n\ge0}E_{x,j,n}^{(\lambda)},
 \qquad E_{x,j,n}^{(\lambda)}\in\A_{\ball_{R_n}(x)}^\st,
\]
where every shell annihilates the local vacuum (Lemma~\ref{lem:weighted-vacuum-shells}).
For such an operator \(E\) supported in \(X\), one has \(EP_X^\vac=0\), and therefore
\[
 E^*E=(\1-P_X^\vac)E^*E(\1-P_X^\vac)
 \le\|E\|^2(\1-P_X^\vac)\le\|E\|^2N_X.
\]
Weighted Cauchy--Schwarz and the preceding excitation-number bound give
\[
 2\sum_j(e_{x,j}^{(\lambda)})^*e_{x,j}^{(\lambda)}
 \le C\sum_{j,n}(n+1)^2\|E_{x,j,n}^{(\lambda)}\|^2
        \sum_{z\in\ball_{R_n}(x)}n_z.
\]
Collecting the coefficient of each \(n_z\) defines \(\mathscr K_\lambda(x,z)\) (Proposition~\ref{prop:local-excitation-detectability}).
In Section~\ref{sec:detectability}, we use the smallness and spatial decay of these shells to establish LED for sufficiently small \(|\lambda|\), completing the weak-coupling proof.
Sections~\ref{sec:quasilocal-tails} and~\ref{sec:correlation-proofs} extend the comparison argument to defects with tails and derive correlation bounds.

\subsection{Discussion of related literature}
\label{sec:related-results}

We discuss how our results relate to earlier work on ground-state uniqueness, spectral stability, and locality in lattice fermion and quantum spin systems.

\paragraph{Strong LPPL and local ground-state determination.}
For weak finite-range perturbations of sums of onsite spin Hamiltonians with unique, uniformly gapped ground states, Yarotsky proves exponential local determination uniformly in the boundary conditions~\cite{Yarotsky05}.
Henheik, Teufel, and Wessel use this result to obtain strong LPPL, with exponential spatial decay and constants independent of the strength of a bounded defect, and explicitly anticipate an extension to weakly interacting fermions~\cite{HTW22}.
Bachmann, De Roeck, Donvil, and Fraas prove strong LPPL by joining the unperturbed ground state in the interior to a comparison state outside~\cite{BDDF22}.
They assume frustration freeness, local indistinguishability of open-boundary ground states, gaps on open-boundary balls bounded below by an inverse polynomial in their radius, and invertibility, which permits a locally generated transformation to a product state after adding an auxiliary system.
These results compare ground states, whereas Theorem~\ref{thm:positive-energy-locality} controls all states in a fixed low-energy window of the modified system.

\paragraph{LPPL along gapped paths.}
The following results assume that the perturbation can be introduced along a path preserving a spectral gap, either above the ground state or above a chosen low-energy sector.
Bachmann, Michalakis, Nachtergaele, and Sims establish LPPL along gapped paths by approximating the unitary spectral flow with unitaries supported near the perturbation, with an almost-exponentially decaying error \cite[Theorem~3.4]{BMNS12}.
De Roeck and Sch\"utz subsequently obtain exponentially local transformations of low-energy sectors that remain separated from the rest of the spectrum along the path \cite[Theorem~3.2 and Section~4]{DeRoeckSchuetz15}.
Their impurity applications allow small level splittings within this sector.
Bachmann, Du, Fraas, and Wessel use Gaussian filters to prove exponential LPPL along uniformly gapped paths with controlled interaction derivatives \cite[Theorem~13]{BachmannDuFraasWessel25}.
For power-law interactions, Wang and Hazzard prove polynomial locality under gap-preserving perturbations~\cite{WangHazzard23}.
When the unperturbed interaction has commuting long-range terms, Lemm and Wessel obtain improved spatial decay of LPPL estimates, assuming a unique ground state and a uniform gap throughout the perturbation path~\cite[Theorem~13]{LemmWessel25}.
Teufel and Wessel establish automorphic equivalence and polynomial LPPL for long-range interacting fermions, assuming an isolated spectral sector throughout the interpolation \cite[Theorems~9 and~10]{TeufelWessel25}.
Li controls the change of local ground-state expectations under interaction-range truncation by the discarded interaction strength near each site, uniformly in the volume and assuming a gap-preserving interpolation~\cite{Li26}.
In Theorem~\ref{thm:positive-energy-locality}, the window \([0,3g/4]\) need not be separated from the rest of the spectrum, and we require no gap-preserving path from the bulk interaction to the modified interaction \(\Xi\).
The constants are independent of \(\Xi\) and of its positive-energy representation.

\paragraph{Local indistinguishability of finite-volume ground states.}
Fannes, Nachtergaele, and Werner construct finitely correlated states and finite-range parent Hamiltonians \cite{FNW92}.
Their construction includes parent Hamiltonians with a unique infinite-volume frustration-free state, although the ground space of a finite open chain may be degenerate.
Kennedy, Lieb, and Tasaki prove uniqueness of the infinite-volume frustration-free ground state and exponential decay of truncated correlations for the spin-\(3/2\) AKLT antiferromagnet on the hexagonal lattice \cite{KLT88}.
Jackson, Nachtergaele, and Young prove exponential local indistinguishability for certain two-dimensional AKLT models, uniformly over the ground-state vectors along specified finite-volume exhaustions \cite[Theorem~1.1 and Corollary~1.2]{JNY26}.
The latter estimates show that all ground states of these finite-volume AKLT Hamiltonians have approximately the same expectations of observables supported far from the boundary, and imply uniqueness of the infinite-volume frustration-free state.
They require no gap for local indistinguishability, but assume a uniform finite-volume gap for their gap-stability corollary \cite[Corollary~1.3]{JNY26}.
Corollary~\ref{cor:finite-sample-bulk} controls local observables on the entire window \([0,3g/4]\) above the sample ground energy, uniformly over arbitrary even boundary operators.

\paragraph{Spectral stability and transport of a ground state.}
Perturbative constructions of ground states include the expansions of Kirkwood and Thomas for quantum Ising systems \cite{KirkwoodThomas83} and the ground-state and one-quasiparticle expansions of Datta and Kennedy \cite{DattaKennedy02}.
Using local Lie--Schwinger block diagonalization, Fr\"ohlich and Pizzo and their collaborators prove persistence of a unique finite-volume ground state and a volume-uniform spectral gap under weak finite-range perturbations of sums of onsite Hamiltonians with unique, gapped ground states~\cite{FroehlichPizzo20,DelVecchioEtAl22}.
Their results cover arbitrary spatial dimensions and include certain unbounded interactions~\cite{DelVecchioEtAl22,DelVecchioFroehlichPizzo23}.
For weakly interacting fermions, De Roeck and Salmhofer prove persistence of a simple finite-volume ground state and a gap uniform in the volume \cite[Theorem~3]{DRS19}.
Hastings proves stability of the free-fermion gap using a doubling construction and a reduction to frustration-free form \cite{Hastings19}, and Koma subsequently refines this argument \cite{Koma20}.

The quasi-local spectral flow used here goes back to the quasi-adiabatic continuation introduced by Hastings \cite{Hastings04} and developed further by Hastings and Wen \cite{HastingsWen05}.
Bachmann, Michalakis, Nachtergaele, and Sims use this construction to establish automorphic equivalence within gapped phases \cite{BMNS12}.
For finite-range quantum spin systems, Moon and Ogata prove that the quasi-local spectral flow transports the bulk ground state along a path with a uniform bulk gap, provided that ground-state expectations of sufficiently localized observables are differentiable along the path \cite{MoonOgata20}.
Becker, Teufel, and Wesle prove a corresponding transport theorem for differentiable paths of locally unique gapped fermionic ground states when the interactions decay faster than any inverse power \cite{BTW26}.
For the interactions considered here, we prove stability of the gap and construct the ground states \(\omega_\lambda\) and their quasi-local transport \(\beta_{0,\lambda}\) directly in infinite volume in \cite{StabilityBdG26}.
These results show that the transport hypothesis holds under the assumptions of Proposition~\ref{prop:unconditional-weak-coupling}.

\paragraph{Fermionic expansions and equilibrium correlations.}
Giuliani, Mastropietro, and Porta construct thermodynamic and zero-temperature limits of current and density correlations for weakly interacting, translation-invariant two-dimensional band insulators \cite[Proposition~4.1 and Lemma~5.2]{GMP17}.
Assuming a gap of the free Hamiltonian, they prove analyticity in the coupling and rapid decay of truncated correlations, uniformly in volume and inverse temperature.
If corresponding correlation bounds hold uniformly along a weak local perturbation path, the Gibbs differentiation formula also gives spatially decaying changes of local equilibrium expectations.
Theorem~\ref{thm:positive-energy-locality} instead controls every state in the window \([0,3g/4]\) of the modified interaction \(\Xi\), with constants independent of its terms supported outside \(U\).

\paragraph{Ground-state uniqueness in specific models.}
Uniqueness among translation-invariant or frustration-free ground states is weaker than uniqueness among all algebraic ground states.
For a class of quantum spin systems in sufficiently strong fields, Matsui proves uniqueness of the translation-invariant ground state \cite{Matsui90}.
For the infinite XY chain, Araki and Matsui use its quadratic-fermion description to determine the full ground-state set, including regimes with a unique ground state and regimes with symmetry breaking \cite{ArakiMatsui85}.
The spin-one AKLT chain has a unique translation-invariant ground state \cite{AKLT88}, and Matsui proves that every algebraic ground state is translation invariant, establishing algebraic uniqueness \cite{Matsui97}.
For certain finite-range parent Hamiltonians of matrix-product states, Ogata likewise extends uniqueness from frustration-free states to all algebraic ground states \cite[Lemma~4.22 and Corollary~5.6]{Ogata17}.
The toric code on the plane illustrates why frustration-free uniqueness does not imply algebraic uniqueness: its frustration-free ground state is unique, but it also has algebraic ground states that are not frustration-free \cite{ChaNaaijkensNachtergaele18}.

\paragraph{Local ground-state inequalities and positive local energy operators.}
In Theorem~\ref{thm:bulk-uniqueness}, we extend the comparison to states satisfying the ground-state inequality only in \(U\), following Yarotsky's use of this inequality to derive and iterate excitation estimates \cite[Sections~2--3]{Yarotsky05}.
Yarotsky uses cluster expansions to construct a nonunitary similarity transformation relating the perturbed ground state to the unperturbed product state.
Here the quasi-local \(*\)-automorphism \(\beta_{0,\lambda}\) transports the free ground state to \(\omega_\lambda\), and \(\sigma_\st\) reduces the stacked free ground state to a product vacuum.

To apply energy filtering to \(\rho\in\LG_{\Phi_\lambda}(U)\), we need a positive operator that describes its local energy variations.
Averaging the interaction over exterior states does not generally provide such an operator: Fannes and Werner show that even the restriction of a global ground state need not be a mixture of ground states of the resulting Hamiltonians~\cite{FannesWerner95}.
Instead, for finite \(\Lambda\subset U\), we construct \(L_{\rho,\Lambda}\) on \(\pi_\rho(\A_\Lambda)\Omega_\rho\) in the GNS representation of \(\rho\), directly from
\[
 \langle\pi_\rho(B)\Omega_\rho,
 L_{\rho,\Lambda}\pi_\rho(B)\Omega_\rho\rangle
 =\rho(B^*\delta_{\Phi_\lambda}(B))\ge0,
 \qquad B\in\A_\Lambda.
\]
Its commutators are compressions of the bulk derivation (Lemma~\ref{lem:local-positive-generator}).
This identity allows us to compare its evolution with the restricted bulk dynamics and hence compare the corresponding filtered annihilation operators.

\paragraph{Detecting operators and sums of squares.}
The name local excitation detectability refers to control of excitation numbers by filtered annihilation operators.
The conventional detectability lemma instead bounds the error in approximating the global ground-space projection by an ordered product of local ground-space projections for a frustration-free Hamiltonian \cite{AharonovAradLandauVazirani09}, with a simplified proof in \cite{AnshuAradVidick16}.

Yarotsky's proof already uses vacuum annihilation to bound squared operators by local excitation observables and derives a contractive inequality for their expectations \cite[Lemmas~2--3]{Yarotsky05}.
LED formulates a related operator comparison for energy-filtered annihilation operators, which we combine with locality of the dynamics to control entire low-energy subspaces.

Hastings uses perturbative corrections to fermionic annihilation operators to obtain lower bounds on ground-state energies from sums of their squares \cite{HastingsSoS22,HastingsSoS24}.
In this construction, self-adjoint even polynomial remainders annihilating the free vacuum are written as sums of squares plus linear combinations of occupation operators \cite[Lemma~1]{HastingsSoS22}.
At weak coupling, the latter terms can be absorbed into the positive quadratic part.
Hastings also extends this energy-bound construction to certain density--density interactions that need not be perturbatively small \cite{HastingsSoS26}.

We instead use squares of filtered annihilation operators to control local excitation densities and thereby determine the local expectations of low-energy states.

\paragraph{Uniqueness of KMS states.}
Fr\"ohlich and Ueltschi prove KMS uniqueness for quantum-spin systems on \(\mathbb Z^n\) at sufficiently high temperature under a summability condition with an exponential support-size weight \cite[Theorem~2.1]{FroehlichUeltschi15}.
Their proof derives a contractive linear equation for local expectations directly from the KMS condition.
Under a spatially weighted smallness condition, they also prove decay of connected correlations by a cluster expansion \cite[Theorem~3.2]{FroehlichUeltschi15}.

For one-dimensional spin systems, Araki and Kishimoto prove KMS uniqueness at every positive temperature under a uniform bound on the interaction across the ends of finite intervals \cite{Araki75,Kishimoto76}.
Uniformly bounded finite-range interactions satisfy this condition.
Such positive-temperature uniqueness does not imply uniqueness of algebraic ground states.

For weakly interacting fermionic models with a gapped free Hamiltonian, we prove KMS uniqueness and strong LPPL at every positive temperature on one temperature-independent weak-coupling interval in \cite[Theorems~3.1(i) and~3.7]{ThermalLPPL26}.
Theorem~3.1(i), Proposition~3.2 and Corollary~3.6 of that paper give ground-state uniqueness, a GNS gap and strong LPPL for ground states.
For each local observable, Corollary~3.5 bounds the difference between its KMS and ground-state expectations by an error exponentially small in the inverse temperature.
Our zero-temperature results go further: we determine local expectations in every state in the energy window \([0,3g/4]\), including coherent superpositions.
Under the modified dynamics, these states remain locally indistinguishable from the bulk ground state deep inside any region \(U\) where the interaction agrees with the gapped bulk interaction, with a spatially decaying error uniform for all times.

\section{Locality estimates for CAR dynamics}
\label{sec:odd-locality}

In this section, \(\A\) denotes a CAR algebra over \(\Gamma\) with a fixed finite number of modes per site. 
Later it will be the physical algebra or a fixed finite stack.
We collect the conditional expectations, shell decompositions, and locality estimates used to transport and filter the odd annihilation operators.
We use the tracial conditional expectations of Araki and Moriya \cite[Theorem~4.7]{AM03} and the fermionic propagation estimates of \cite{FermionicLR}.
Nachtergaele, Sims, and Young discuss the corresponding conditional expectations for quantum spin systems in \cite{NSY19}.
Appendix~\ref{app:odd-locality-proofs} proves the comparison with free dynamics used to show LED in the stretched-exponential regime.

In the stretched-exponential case, we apply some exponential estimates in the metric
\begin{equation}
 d'(x,y):=d(x,y)^{\theta_{\rm in}},\qquad
 \dist_{d'}(X,Y)=\dist_d(X,Y)^{\theta_{\rm in}}.
 \label{eq:input-metric-reduction}
\end{equation}
Subadditivity of \(u\mapsto u^{\theta_{\rm in}}\) proves the triangle inequality for \(d'\).
Its balls satisfy, for \(R\ge0\),
\[
 |\ball_R^{d'}(x)|=|\ball_{R^{1/\theta_{\rm in}}}^{d}(x)|
 \le C_\Gamma(1+R)^{d_{\rm vol}/\theta_{\rm in}}.
\]
With a superscript \(d'\) on norms computed in this metric, for \(\mu\ge0\) and \(k\ge0\) one has
\[
 \|h\|_{\mu,1}^{\mathrm{Sch},d'}=\|h\|_{\mu,\theta_{\rm in}}^{\rm Sch},
 \qquad
 \|\Phi\|_{\mu,1;k}^{d'}=\|\Phi\|_{\mu,\theta_{\rm in};k}.
\]
We use this reduction for the support-moment estimate in Proposition~\ref{prop:unconditional-weak-coupling}, the free-dynamics comparison in Appendix~\ref{app:odd-locality-proofs}, and the clustering argument in Proposition~\ref{prop:bulk-clustering}.
In each application, the volume-growth exponent becomes \(d_{\rm vol}/\theta_{\rm in}\), and a neighborhood of radius \(R^{\theta_{\rm in}}\) in \(d'\) is the neighborhood of radius \(R\) in \(d\).
All locality statements and shell decompositions in this paper use the original metric \(d\).

\subsection{Conditional expectations and shell decompositions}

We first define the conditional expectations used to decompose quasi-local observables.
Let \(\operatorname{tr}\) denote the unique normalized tracial state on \(\A\).
For \(\Lambda\Subset\Gamma\), the tracial CAR conditional expectation
\begin{equation}
 \mathbb F^{\rm tr}_\Lambda:\A\longrightarrow\A_\Lambda
 \label{eq:tracial-expectation-definition}
\end{equation}
is characterized, for every \(A\in\A\) and \(B\in\A_\Lambda\), by
\begin{equation*}
 \operatorname{tr}(AB)
 =\operatorname{tr}\!\left(\mathbb F^{\rm tr}_\Lambda(A)B\right).
\end{equation*}
It is unital, positive, contractive, an \(\A_\Lambda\)-bimodule map, fixes \(\A_\Lambda\), and commutes with parity.
For every \(\Lambda_1,\Lambda_2\Subset\Gamma\),
\begin{equation}
 \mathbb F^{\rm tr}_{\Lambda_1}
 \mathbb F^{\rm tr}_{\Lambda_2}
 =\mathbb F^{\rm tr}_{\Lambda_1\cap\Lambda_2}.
 \label{eq:tracial-expectation-composition}
\end{equation}
For the characterization and properties of \(\mathbb F^{\rm tr}_\Lambda\), see \cite[Theorem~4.7]{AM03} and \cite[Proposition~2.1]{BTW26}.
For every nonempty \(X\Subset\Gamma\) and every \(A\in\A\),
\begin{equation}
 \lim_{R\to\infty}\mathbb F^{\rm tr}_{X^{(R)}}(A)=A
 \quad\text{in norm}.
 \label{eq:tracial-expectation-convergence}
\end{equation}

For \(A\in\A\) and a linear subspace \(\mathcal B\subset\A\), write \(\dist(A,\mathcal B):=\inf_{B\in\mathcal B}\|A-B\|\).
For every nonempty \(X\Subset\Gamma\), \(D\in\A\), and \(R\ge0\), contractivity gives
\begin{equation}
 \|D-\mathbb F^{\rm tr}_{X^{(R)}}D\|
 \le2\dist(D,\A_{X^{(R)}}).
 \label{eq:general-tracial-shell-bounds}
\end{equation}
Indeed, \(\mathbb F^{\rm tr}_{X^{(R)}}\) fixes every \(B\in\A_{X^{(R)}}\), so the left-hand side is at most \(2\|D-B\|\); take the infimum over \(B\).
We use the dyadic radii and shells of \cite[Subsections~6.2--6.3]{StabilityBdG26}:
\begin{equation}
 \begin{aligned}
 R_0&:=0,\qquad R_n:=2^{n-1}\quad(n\ge1),\\
 \Delta_{X,0}^{\rm tr}
 &:=\mathbb F^{\rm tr}_X\,,\qquad
 \Delta_{X,n}^{\rm tr}
 :=\mathbb F^{\rm tr}_{X^{(R_n)}}
   -\mathbb F^{\rm tr}_{X^{(R_{n-1})}}\quad(n\ge1).
 \end{aligned}
 \label{eq:dyadic-tracial-shells}
\end{equation}
Telescoping gives \(D=\sum_{n\ge0}\Delta_{X,n}^{\rm tr}D\) in norm, and contractivity gives
\[
 \|\Delta_{X,0}^{\rm tr}D\|\le\|D\|,
 \qquad
 \|\Delta_{X,n}^{\rm tr}D\|
 \le2\dist(D,\A_{X^{(R_{n-1})}})\quad(n\ge1).
\]
For the second bound, both conditional expectations fix every element of \(\A_{X^{(R_{n-1})}}\), so the same subtraction argument applies.
Suppose that, for some \(a,C>0\) and every \(R\ge0\), \(\dist(D,\A_{X^{(R)}})\le C(1+R)^{-a}\).
Then, for every \(n\ge0\) and \(0\le M<a\),
\begin{equation}
 \|\Delta_{X,n}^{\rm tr}D\|
 \le C'(1+R_n)^{-a},\qquad
 \sum_{n\ge0}(1+R_n)^M
 \|\Delta_{X,n}^{\rm tr}D\|<\infty.
 \label{eq:dyadic-locality-to-shells}
\end{equation}
Here \(C'\) can be chosen in terms of \(a\), \(C\), and \(\|D\|\); \(\|D\|\) controls the zeroth shell.
The bounds follow from the preceding shell estimates and a geometric series.
For every \(R\ge1\), choose \(n_{\max}\) with \(R_{n_{\max}}\le R<R_{n_{\max}+1}\).
Then \(R/2<R_{n_{\max}}\le R\), so estimates at dyadic radii extend to arbitrary radii with the same polynomial decay exponent.

\subsection{Locality and comparison of dynamics}

For stretched-exponential physical dynamics, we use the weighted commutator estimate of \cite[Lemma~3.1]{FermionicLR} with weight \(\E^{\mu R^{\theta_{\rm in}}}\).
The following lemma is proved in \cite[Corollary~2.8]{FermionicLR}.

\begin{lemma}
\label{lem:all-parity-polynomial}
Assume \eqref{eq:cell-partitions}.
Fix \(p>d_{\rm vol}\) and \(0\le J<\infty\), and set
\begin{equation}
 m_{p,0}:=\left(1+\frac{d_{\rm vol}+1}{p-d_{\rm vol}}\right)
                 (p+d_{\rm vol}).
 \label{eq:polynomial-time-exponent}
\end{equation}
There is a constant \(C>0\), depending only on \(p,J,C_\Gamma,d_{\rm vol},C_{\rm cell},D_{\rm cell}\), with the following property.
For every interval \(I\subset\R\) and every componentwise norm-continuous path of interactions \((\Xi_u)_{u\in I}\) satisfying \(\sup_{u\in I}\|\Xi_u\|^{\#}_{p;\kappa_0}\le J\), the associated evolution \(\tau_{s,t}^{\Xi}\) satisfies, for every \(s,t\in I\), nonempty \(X\Subset\Gamma\), \(A\in\A_X\), and \(R\ge0\),
\begin{equation}
 \|(\operatorname{id}-\mathbb F^{\rm tr}_{X^{(R)}})\tau_{s,t}^{\Xi}(A)\|
 \le C|X|\|A\|(1+|t-s|)^{m_{p,0}}(1+R)^{-p}.
 \label{eq:all-parity-polynomial-LR}
\end{equation}
The estimate holds for either parity, on a fixed finite stack, and uniformly under finite restrictions or deletion of Hamiltonian terms.
\end{lemma}

The restriction and deletion clauses in the propagation estimates keep the same constants because the interaction-norm bounds do not increase.
For decompositions into physical and auxiliary interactions, we keep the two commutator sums separate, as in Lemmas~\ref{lem:exponential-crossing-comparison} and~\ref{lem:polynomial-crossing-comparison}.

The next lemma gives locality estimates for the ground-state transport \(\beta_{s,t}\) of Hypothesis~\ref{hyp:background-transport}, using \cite[Theorem~2.9 and Remark~2.10]{FermionicLR} with polynomial or stretched-exponential distance weights.
The locality constant is controlled by the integrated first support moment of the generator on the coupling interval.
\begin{lemma}
\label{lem:truncated-cocycle}
Let \(I\subset\R\) be a compact interval and let \((\eta_s)_{s\in I}\) be a family of interactions such that, for every \(Z\Subset\Gamma\), the map \(I\ni s\mapsto\eta_s(Z)\in\A_Z\) is norm-continuous.
Assume either
\[
 \sup_{s\in I}\|\eta_s\|^{\pol}_{p;1}<\infty
 \quad\text{for some }p>0,
\]
or
\[
 \sup_{s\in I}\|\eta_s\|_{\mu,\theta;1}<\infty
 \quad\text{for some }\mu>0,\quad 0<\theta<1.
\]
Then \((\eta_s)_{s\in I}\) generates a unique strongly continuous cocycle \((\beta_{s,t})_{s,t\in I}\) of parity-preserving automorphisms of \(\A\).
For every \(s\in I\), every \(t\) in the interior of \(I\), and every \(A\in\A_{\loc}\), it satisfies in norm
\begin{equation*}
 \tfrac{\dd}{\dd t}\beta_{s,t}(A)
 =\I\beta_{s,t}\!\left(\delta_{\eta_t}(A)\right).
\end{equation*}

Let \(w(R)=(1+R)^p\) or \(w(R)=\E^{\mu R^\theta}\), and write \(\|\eta_u\|_{w;1}\) for the corresponding first support moment above.
For \(s,t\in I\), set \(M_w(s,t):=\int_{\min(s,t)}^{\max(s,t)}\|\eta_u\|_{w;1}\,\dd u\).
If \(\varnothing\ne X\subseteq\Lambda\Subset\Gamma\) and \(A\in\A_X\), the evolution \(\beta_{s,t}^\Lambda\) generated by the terms of \(\eta_u\) contained in \(\Lambda\) satisfies
\begin{equation}
 \|\beta_{s,t}(A)-\beta_{s,t}^\Lambda(A)\|
 \le \|A\|\bigl(\E^{2M_w(s,t)}-1\bigr)
       \sum_{x\in X}\frac{1}{w(\dist(x,\Lambda^c))}.
 \label{eq:sitewise-transport-restriction}
\end{equation}

For every \(T\in[0,\infty)\), there are \(C_T<\infty\) and a function \(G_\eta:[0,\infty]\to[0,\infty)\) such that, for nonempty \(X\Subset\Gamma\), \(R\ge0\), \(s,t\in I\) with \(|t-s|\le T\), and \(A\in\A_X\),
\begin{equation}
 \|\beta_{s,t}(A)
       -\mathbb F^{\rm tr}_{X^{(R)}}\beta_{s,t}(A)\|
 \le C_T(1+|X|)\|A\|G_\eta(R).
 \label{eq:truncated-cocycle}
\end{equation}
In the polynomial case one may take
\begin{equation}
 G_\eta(R):=(1+R)^{-p},\qquad
 C_T:=2\left(\exp\!\Big(2T\sup_{u\in I}
            \|\eta_u\|^{\pol}_{p;1}\Big)-1\right).
 \label{eq:finite-polynomial-cocycle-profile}
\end{equation}
In the stretched-exponential case one may take
\[
 G_\eta(R):=\E^{-\mu R^\theta},\qquad
 C_T:=2\left(\exp\!\Big(2T\sup_{u\in I}
            \|\eta_u\|_{\mu,\theta;1}\Big)-1\right).
\]

There is also a constant \(C_{\rm flow}<\infty\), depending only on the geometry and the uniform generator bounds, such that, for every nonempty \(X\Subset\Gamma\), every countable set \(\mathscr R\subset[0,\infty)\), and every expansion
\[
 A=\sum_{R\in\mathscr R}D_R,\qquad
 D_R\in\A_{X^{(R)}},\qquad
 \sum_{R\in\mathscr R}|X^{(R)}|\|D_R\|<\infty,
\]
the following estimate holds for every \(s,t\in I\):
\begin{equation}
 \|(\beta_{s,t}-\operatorname{id})(A)\|
 \le C_{\rm flow}|t-s|
 \sum_{R\in\mathscr R}|X^{(R)}|\|D_R\|.
 \label{eq:flow-local-continuity}
\end{equation}
The assertions also hold on a fixed finite stack of CAR algebras.
\end{lemma}

\begin{proof}
Theorem~2.9 and Remark~2.10 of \cite{FermionicLR} give the cocycle and its differential equation.
To obtain \eqref{eq:sitewise-transport-restriction}, keep the initial site \(x\in X\) fixed in the finite-volume comparison in the proof of that theorem.
Each term in the estimate connects \(x\) to a site \(y\in\Lambda\) by the restricted dynamics and then to a site \(z\in\Lambda^c\) in an interaction term containing \(y\).
Both weights are nondecreasing and satisfy \(w(a+b)\le w(a)w(b)\), so \(w(\dist(x,\Lambda^c))\le w(d(x,y))w(d(y,z))\).
For \(s\le t\), the weighted row bounds for the restricted evolution from \(x\) to \(y\) and the interaction from \(y\) to \(z\) bound the contribution from \(x\) by
\[
 \frac{\|A\|}{w(\dist(x,\Lambda^c))}
 \int_s^t 2\|\eta_u\|_{w;1}\E^{2M_w(u,t)}\,\dd u
 =\frac{\|A\|(\E^{2M_w(s,t)}-1)}{w(\dist(x,\Lambda^c))}.
\]
Now sum over \(x\in X\) and pass to infinite volume as in the cited theorem.
Reversing the path gives the same estimate for \(t<s\).
For \(\Lambda=X^{(R)}\), the tracial conditional expectation fixes \(\beta_{s,t}^\Lambda(A)\), so the left side of \eqref{eq:truncated-cocycle} is at most twice that of \eqref{eq:sitewise-transport-restriction}.
Now \(\dist(x,\Lambda^c)\ge R\) for \(x\in X\), and \(M_w(s,t)\le T\sup_{u\in I}\|\eta_u\|_{w;1}\) when \(|t-s|\le T\).

For \(D\in\A_Z\), evenness of the interaction terms gives
\[
 \|\delta_{\eta_u}(D)\|
 \le2|Z|\|D\|\sup_x\sum_{W\ni x}\|\eta_u(W)\|
 \le C_{\rm flow}|Z|\|D\|.
\]
Integrating the differential equation and using isometry proves \(\|(\beta_{s,t}-\operatorname{id})(D)\| \le C_{\rm flow}|t-s|\,|Z|\|D\|\).
Apply this bound to each \(D_R\), \(R\in\mathscr R\), and sum the absolutely norm-convergent series to obtain \eqref{eq:flow-local-continuity}.
\end{proof}

\paragraph{Comparison with free dynamics.}

Establishing local excitation detectability at small \(|\lambda|\) requires a bound on the difference between the interacting and free evolutions of an annihilation operator.
In the polynomial regime, we compare these evolutions using \cite[Proposition~5.3(iii)]{FermionicLR}.
In the stretched-exponential regime, the next lemma gives this comparison for linear Majorana fields.
Both bounds grow polynomially in time, so they can be integrated against the fixed filter \(f\) from \eqref{eq:Fourier} in the proof of Lemma~\ref{lem:small-correction}.

\begin{lemma}
\label{lem:dynamics-comparison}
Fix \(\mu_h>0\), \(0<\theta_{\rm in}\le1\), finite bounds \(\overline C_h,J\ge0\), and \(0<\mu\le\mu_h\).
There is a constant \(C_\mu<\infty\), depending only on \(\mu,\theta_{\rm in},r,\overline C_h,J,C_\Gamma,d_{\rm vol}\), with the following property.
Let \(h\) satisfy \eqref{eq:Majorana-structure} and \eqref{eq:h-locality} with \(C_h\le\overline C_h\), and let \(\mathcal V\in\mathfrak J_1\) satisfy \(\|\mathcal V\|_1\le J\).
For every \(x\in\Gamma\) and complex coefficient family \((v_u)_{u\in\Gamma\times\{1,\ldots,2r\}}\) such that
\[
 \sum_u \E^{\mu d(x,x_u)^{\theta_{\rm in}}}|v_u|<\infty,
\]
define \(A_v:=\sum_u v_u\gamma_u\), where the series converges in norm.
Then, for every \(t,\lambda\in\R\),
\begin{equation}
 \|\tau_t^{\Phi_\lambda}(A_v)-\tau_t^{\Phi_h}(A_v)\|
 \le C_\mu|\lambda|\,|t|(1+|t|)^{d_{\rm vol}/(2\theta_{\rm in})}
             \sum_u\E^{\mu d(x,x_u)^{\theta_{\rm in}}}|v_u|.
 \label{eq:dynamics-comparison}
\end{equation}
\end{lemma}

The proof is given in Appendix~\ref{app:odd-locality-proofs}.

\section{Product-vacuum coordinates and positive-energy representations}
\label{sec:atomicization}

Araki's self-dual CAR formalism relates pure quasifree states to Fock vacua through Bogoliubov transformations \cite[Sections~2 and~4]{Araki70}.
Our original Majorana algebra already has this self-dual form.
To obtain a transformation with controlled spatial decay, we adjoin independent auxiliary fermions with coefficient operator \(-h\) and use Hastings's explicit transformation of the doubled system \(h\oplus(-h)\) \cite[Section~III, equations~(8)--(15)]{Hastings19}.
We use the conventions and quantitative locality estimates of \cite[Section~4]{StabilityBdG26}.
We then prove a local state comparison estimate, establish locality of \(\sigma_\lambda\) and its inverse, and construct the stacked positive-energy representation.

\subsection{Product vacuum and local state comparison}

For a parity-homogeneous element \(A\), define \(|A|\in\{0,1\}\) by \(\Theta(A)=(-1)^{|A|}A\).
For homogeneous \(A,B\), their graded commutator is \([A,B]_{\rm gr}:=AB-(-1)^{|A||B|}BA\).
Our graded tensor-product conventions are
\begin{align*}
 (A_1\widehat\otimes B_1)(A_2\widehat\otimes B_2)
 &:=(-1)^{|B_1||A_2|}A_1A_2\widehat\otimes B_1B_2,\\
 (A\widehat\otimes B)^*
 &:=(-1)^{|A||B|}A^*\widehat\otimes B^*.
\end{align*}
If at least one of two states \(\rho_1,\rho_2\) is even, their graded product state exists by \cite[Theorem~11.2]{AM03} and satisfies
\[
 (\rho_1\widehat\otimes\rho_2)(A\widehat\otimes B)
 :=\rho_1(A)\rho_2(B).
\]
Recall the auxiliary copy \(\Hsp_{\rm aux}\) and the stacked algebra \(\A^\st\) introduced before Hypothesis~\ref{hyp:uniform-local-excitation-detectability}.
Thus
\[
 \A^{\rm aux}=\operatorname{CAR}(\Hsp_{\rm aux}),\qquad
 \A^\st=\operatorname{CAR}(\Hsp\oplus\Hsp_{\rm aux})
 \cong\A\widehat\otimes\A^{\rm aux}.
\]
The corresponding Majorana coefficient space is \(\mathfrak h_{\rm M}\oplus\mathfrak h_{{\rm M},{\rm aux}}\).
Let \(J_{\rm aux}:\mathfrak h_{\rm M}\to \mathfrak h_{{\rm M},{\rm aux}}\) be the real, site-preserving copy unitary; then
\[
 h_{\rm aux}=-J_{\rm aux}hJ_{\rm aux}^*.
\]
The auxiliary state \(\omega_{\rm aux}\) is the pure even quasifree ground state defined by \eqref{eq:free-Majorana-state} for \(h_{\rm aux}\).
Under the copy identification its Majorana covariance is \(\1-\sgnh\).
Let \(\gamma_{{\rm aux},u}\) be the labeled copies of the generators in \eqref{eq:physical-Majoranas}, and write \(\gamma_{u,1}:=\gamma_u\widehat\otimes\1\) and \(\gamma_{u,2}:=\1\widehat\otimes\gamma_{{\rm aux},u}\).
These two families together satisfy the Majorana CAR.

For \(X\subset\Gamma\), define
\[
 \A_X^{\rm aux}:=\operatorname{CAR}(\ell^2(X;\mathbb C^r)_{\rm aux}),
 \qquad
 \A_X^\st:=\A_X\widehat\otimes\A_X^{\rm aux}.
\]
Set
\[
 \A_\loc^{\rm aux}:=\bigcup_{X\Subset\Gamma}\A_X^{\rm aux},
 \qquad
 \A_\loc^\st:=\bigcup_{X\Subset\Gamma}\A_X^\st.
\]
Auxiliary parity \(\Theta_{\rm aux}\) changes the sign of every auxiliary Majorana generator, and total parity is \(\Theta_\st:=\Theta\widehat\otimes\Theta_{\rm aux}\).
For \(U\subset\Gamma\) and a derivation \(\delta\) whose domain contains \(\A_\loc^\st\), define
\[
 \LG_\delta^\st(U):=
 \left\{\rho\in \mathfrak S(\A^\st)\,\middle|\,
 \operatorname{Re}\rho(B^*\delta(B))\ge0
 \text{ for every }X\Subset U,\ B\in\A_X^\st\right\}.
\]
The identification of the Majorana coefficient spaces used in \eqref{eq:O-st} is \(J_{\rm aux}^*\).
The following lemma gives the algebraic and locality properties of \(\sigma_\st\) defined in \eqref{eq:sigma-st}.

\begin{lemma}
\label{lem:atomicization}
Let \(h\) satisfy the Majorana conditions \eqref{eq:Majorana-structure}, the spectral-gap assumption \eqref{eq:one-particle-gap}, and either \eqref{eq:h-locality} or \eqref{eq:h-polynomial-locality}.
Then \(O_\st\) is real orthogonal.
With
\[
 h_\at:=\begin{pmatrix}0&\I|h|\\-\I|h|&0\end{pmatrix},
\]
it satisfies
\begin{equation}
 O_\st^{\mathsf T}
 \begin{pmatrix}h&0\\0&-h\end{pmatrix}O_\st
 =h_\at,
 \qquad
 O_\st^{\mathsf T}
 \left(\1+\begin{pmatrix}\sgnh&0\\0&-\sgnh\end{pmatrix}\right)O_\st
 =\1+\begin{pmatrix}0&\I\1\\-\I\1&0\end{pmatrix}.
 \label{eq:atomicization-identities}
\end{equation}
Under \eqref{eq:h-locality}, the blocks of \(O_\st\) and \(O_\st^{\mathsf T}\) have finite Schur norm with a weight \(\E^{\mu d(x,y)^{\theta_{\rm in}}}\), for some \(\mu>0\).
There are constants \(C_\sigma<\infty\) and \(a_\sigma>0\) such that, for every nonempty \(X\Subset\Gamma\), \(A\in\A_X^\st\), and \(R\ge0\),
\begin{equation}
 \dist\!\left(\sigma_\st^{\pm1}(A),\A_{X^{(R)}}^\st\right)
 \le C_\sigma |X|\,\|A\|\E^{-a_\sigma R^{\theta_{\rm in}}}.
 \label{eq:sigma-exponential-locality}
\end{equation}
Under \eqref{eq:h-polynomial-locality}, there is a constant \(C_\sigma<\infty\) such that, for every such \(X,A,R\),
\begin{align}
 \|O_\st\|_{p_h}^{\mathrm{Sch,pol}}
 +\|O_\st^{\mathsf T}\|_{p_h}^{\mathrm{Sch,pol}}\le C_\sigma,
 \qquad
 \dist\!\left(\sigma_\st^{\pm1}(A),\A_{X^{(R)}}^\st\right)
 \le C_\sigma|X|\|A\|(1+R)^{-p_h}.
 \label{eq:sigma-polynomial-locality}
\end{align}
\end{lemma}

\begin{proof}
The algebraic identities and the Bogoliubov construction follow from \cite[Lemma~4.1 and Section~4.2]{StabilityBdG26}.
The locality estimates follow from \cite[Proposition~4.2]{StabilityBdG26}.
\end{proof}

\paragraph{Onsite annihilation operators and local state comparison.}

Recall the product vacuum introduced before Hypothesis~\ref{hyp:uniform-local-excitation-detectability}:
\begin{equation}
 \omega_\at=(\omega_0\widehat\otimes\omega_{\rm aux})
 \circ\sigma_\st^{-1}.
 \label{eq:atomic-state}
\end{equation}
Equation~\eqref{eq:atomicization-identities} gives its Majorana covariance.
For \(x\in\Gamma\) and \(j\in\mathscr I=\{1,\ldots,2r\}\), its annihilation operators are
\begin{equation}
 c_{x,j}=\frac{\gamma_{x,j,1}+\I\gamma_{x,j,2}}2.
 \label{eq:atomic-annihilators}
\end{equation}
Their CAR are \(\{c_{x,j},c_{y,k}\}=0\) and \(\{c_{x,j},c_{y,k}^*\}=\delta_{xy}\delta_{jk}\1\).
Conversely, \(\gamma_{x,j,1}=c_{x,j}+c_{x,j}^*\) and \(\gamma_{x,j,2}=-\I(c_{x,j}-c_{x,j}^*)\); hence the \(c_{x,j}\) generate the full stacked algebra.

The transformed covariance implies \(\omega_\at(c_{x,j}^*c_{y,k})=0\).
The quasifree pairing rule therefore identifies \(\omega_\at\) as the product vacuum of these \(2r\) complex modes per site.
Equation~\eqref{eq:atomicization-identities} shows that each onsite layer has the tracial marginal, although the joint onsite state is pure.
For \(x\in\Gamma\), recall the onsite excitation-number operator \(n_x\); for \(X\Subset\Gamma\), let \(N_X\) be its sum over \(X\):
\begin{equation}
 n_x=\sum_{j\in\mathscr I}c_{x,j}^*c_{x,j},\qquad
 N_X:=\sum_{x\in X}n_x.
 \label{eq:atomic-number}
\end{equation}
In particular, \(0\le n_x\le2r\,\1\).
The projection onto the vacuum of the \(c\)-modes in \(X\) is
\begin{equation}
 P_X^\vac:=\prod_{x\in X}\prod_{j\in\mathscr I}
 (\1-c_{x,j}^*c_{x,j})=\1_{\{0\}}(N_X).
 \label{eq:local-vacuum-projection}
\end{equation}
The equality follows because the occupation projections commute.
Let \((\pi_\at,\mathcal H_\at,\Omega_\at)\) be the GNS triple of \(\omega_\at\), with
\begin{equation}
 \omega_\at(A)=
 \langle\Omega_\at,\pi_\at(A)\Omega_\at\rangle,\qquad
 \overline{\pi_\at(\A^\st)\Omega_\at}=\mathcal H_\at.
 \label{eq:atomic-GNS}
\end{equation}
The word vacuum below refers to this state of the \(c\)-modes.

\begin{lemma}
\label{lem:exact-atomic-reconstruction}
For every \(x\in\Gamma\) and \(j\in\mathscr I\),
\begin{equation}
 \pi_\at(c_{x,j})\Omega_\at=0.
 \label{eq:atomic-fields-annihilate}
\end{equation}
For \(X\Subset\Gamma\), let \(\Omega_{\at,X}\) be the local vacuum vector representing \(\omega_\at|_{\A_X^\st}\) in the local Fock representation.
Then
\begin{equation}
 \ker N_X
 =\bigcap_{\substack{x\in X\\j\in\mathscr I}}\ker c_{x,j}
 =\mathbb C\Omega_{\at,X}.
 \label{eq:atomic-common-kernel}
\end{equation}
Thus \(P_X^\vac=|\Omega_{\at,X}\rangle\langle\Omega_{\at,X}|\).
More generally, let \(\mathfrak B\) be a unital \(C^*\)-algebra, let \(P\in\mathfrak B\) be a projection, and let \(\varrho\) be a state with \(\varrho(P)>0\).
Define \(\varrho^P(T):=\varrho(PTP)/\varrho(P)\) for \(T\in\mathfrak B\).
Their distance in the dual norm on \(\mathfrak B^*\) satisfies
\begin{equation}
 \|\varrho-\varrho^P\|\le2\sqrt{\varrho(\1-P)}.
 \label{eq:vacuum-projection-state-distance}
\end{equation}
For every state \(\varrho\) on \(\A^\st\) and \(A\in\A_X^\st\),
\begin{equation}
 |\varrho(A)-\omega_\at(A)|^2
 \le4\|A\|^2 \sum_{x\in X}\varrho(n_x) \,.
 \label{eq:product-reconstruction}
\end{equation}
\end{lemma}

\begin{proof}
The GNS representation of the product vacuum is the Fock representation \cite[Proposition~4.3]{StabilityBdG26}.
It identifies the \(c_{x,j}\) with the canonical annihilation operators and \(N_X\) with their number operator, giving \eqref{eq:atomic-fields-annihilate} and \eqref{eq:atomic-common-kernel}.
The normalized projection estimate \eqref{eq:vacuum-projection-state-distance} is the gentle measurement bound \cite[Lemma~9.4.1]{Wilde17}.
For a state on a general \(C^*\)-algebra, apply its pure-state form to the GNS vector \(\Omega_\varrho\) and the normalized vector \(\pi_\varrho(P)\Omega_\varrho/\sqrt{\varrho(P)}\), and restrict the resulting vector states to \(\pi_\varrho(\mathfrak B)\).

For \(P=P_X^\vac\) and \(A\in\A_X^\st\), the identity \(PAP=\omega_\at(A)P\) gives \(\varrho^P(A)=\omega_\at(A)\) whenever \(\varrho(P)>0\).
Hence
\begin{equation}
 |\varrho(A)-\omega_\at(A)|
 \le2\|A\|\varrho(\1-P_X^\vac)^{1/2}.
 \label{eq:product-reconstruction-block-bound}
\end{equation}
If \(\varrho(P_X^\vac)=0\), this follows from the norm bound for the difference of two states.
Finally, \(\1-P_X^\vac=\1_{[1,\infty)}(N_X)\le N_X\) gives \eqref{eq:product-reconstruction}.
\end{proof}

\subsection{Transported coordinates and their locality}

Recall that the auxiliary interaction \(\Phi_{h_{\rm aux}}\) has terms given by the labeled copies of \(-\Phi_h(Z)\).
The free-ground-state property \eqref{eq:free-ground-gap} applies to \(h_{\rm aux}\).
In particular, for every \(B\in\A_\loc^{\rm aux}\),
\begin{equation}
 \operatorname{Re}\omega_{\rm aux}
 \!\left(B^*\delta_{\Phi_{h_{\rm aux}}}(B)\right)\ge0.
 \label{eq:auxiliary-ground-state-local}
\end{equation}
For \(|\lambda|\le\lambda_0\) and \(Z\Subset\Gamma\), recall the stacked interaction from \eqref{eq:stacked-interaction}:
\[
 \Phi_\lambda^\st(Z)
 =\Phi_\lambda(Z)\widehat\otimes\1
   +\1\widehat\otimes\Phi_{h_{\rm aux}}(Z).
\]
In the stretched-exponential regime, there exists \(\mu>0\) such that
\begin{equation}
 \sup_{|\lambda|\le\lambda_0}
 \|\Phi_\lambda^\st\|_{\mu,\theta_{\rm in};1}<\infty,
 \label{eq:stacked-interaction-locality}
\end{equation}
where the norm is the expression in \eqref{eq:exp-interaction-norm} with \(\A_Z\) replaced by \(\A_Z^\st\).
In the polynomial regime, the Schur bound at \(p_h\) for both quadratic layers and \eqref{eq:polynomial-interaction-locality} give instead
\begin{equation}
 \sup_{|\lambda|\le\lambda_0}
 \|\Phi_\lambda^\st\|^{\#}_{p_\Phi;\kappa_0}<\infty.
 \label{eq:stacked-interaction-polynomial-locality}
\end{equation}
The same bound holds for each layer separately: lifting preserves the interaction norm, and the auxiliary quadratic terms have at most two-site support.
In either case \(\sup_{|\lambda|\le\lambda_0}\|\Phi_\lambda^\st\|_1<\infty\).
For \(|\lambda|\le\lambda_0\) and \(t\in\R\), recall the stacked dynamics
\begin{equation*}
 \tau_t^{\st,\lambda}
 =\tau_t^{\Phi_\lambda}\widehat\otimes\tau_t^{\rm aux}.
\end{equation*}
On every parity-homogeneous local simple tensor, the product dynamics has generator \(\I\delta_{\Phi_\lambda^\st}\), since
\begin{equation*}
 \delta_{\Phi_\lambda^\st}(A\widehat\otimes B)
 =\delta_{\Phi_\lambda}(A)\widehat\otimes B
   +A\widehat\otimes\delta_{\Phi_{h_{\rm aux}}}(B),
\end{equation*}
and this formula extends linearly to \(\A_\loc^\st\).
Proposition~\ref{prop:first-moment-dynamics}, applied to the stacked algebra, makes \(\A_\loc^\st\) a core for the generator of \(\tau^{\Phi_\lambda^\st}\), so the two dynamics agree.
Define the lifted generator path and its cocycle by
\begin{equation*}
 \eta_s^{(1)}(Z):=\eta_s(Z)\widehat\otimes\1,
 \qquad
 \beta_{s,t}^{(1)}:=\beta_{s,t}\widehat\otimes
 \operatorname{id}_{\A^{\rm aux}}.
\end{equation*}
The first-layer lift preserves the locality norm of \(\eta_s\), so \((\eta_s^{(1)})\) satisfies the hypotheses of Lemma~\ref{lem:truncated-cocycle} on the stacked algebra.
For \(|\lambda|\le\lambda_0\), recall the transported coordinate map from~\eqref{eq:transported-coordinates}:
\[
 \sigma_\lambda=\sigma_\st\circ\beta_{0,\lambda}^{(1)}.
\]
Both \(\beta_{0,\lambda}^{(1)}\) and \(\sigma_\st\) commute with \(\Theta_\st\) and so do \(\sigma_\lambda\) and \(\sigma_\lambda^{-1}\).
For every \(|\lambda|\le\lambda_0\), the definitions \eqref{eq:transported-coordinates} and \eqref{eq:atomic-state}, together with the transport identity \eqref{eq:state-transport}, give
\begin{equation}
 \begin{aligned}
 (\omega_\lambda\widehat\otimes\omega_{\rm aux})\circ\sigma_\lambda^{-1}
  =\bigl((\omega_\lambda\circ\beta_{0,\lambda}^{-1})
   \widehat\otimes\omega_{\rm aux}\bigr)\circ\sigma_\st^{-1} =(\omega_0\widehat\otimes\omega_{\rm aux})\circ\sigma_\st^{-1}
 =\omega_\at.
 \end{aligned}
 \label{eq:exact-atomic-state}
\end{equation}

Figure~\ref{fig:atomic-coordinates} illustrates the two changes of coordinates.

We next compute the free dynamics in product-vacuum coordinates and identify the frequency interval on which the Fourier transform of the filter must equal one to leave the annihilation operators \(c_{x,j}\) unchanged.
Let \(\Phi_{h_\at}\) be the canonical quadratic interaction associated with \(h_\at\) from Lemma~\ref{lem:atomicization}.
Equation~\eqref{eq:atomicization-identities} identifies its dynamics with the conjugated dynamics at zero coupling:
\begin{equation*}
 \tau_t^{\Phi_{h_\at}}=\sigma_\st\circ\tau_t^{\st,0}\circ\sigma_\st^{-1}=\tau_t^{\at,0}.
\end{equation*}
The free evolution formula in \cite[proof of Proposition~4.3]{StabilityBdG26}, together with its norm derivative at \(t=0\), gives
\begin{equation}
 \delta_{\Phi_{h_\at}}(c_u)=-\sum_v|h|_{uv}c_v,
 \qquad
 \tau_t^{\at,0}(c_u)=\sum_v(\E^{-\I t|h|})_{uv}c_v .
 \label{eq:atomic-free-dynamics}
\end{equation}
For \(\varphi\in L^1(\R)\), multiplying the evolution formula in \eqref{eq:atomic-free-dynamics} by \(\varphi(t)\) and integrating over time gives
\begin{equation}
 \int_{\R}\varphi(t)\tau_t^{\at,0}(c_u)\,\dd t
 =\sum_v\bigl(\widehat\varphi(-|h|)\bigr)_{uv}c_v.
 \label{eq:atomic-filter-functional-calculus}
\end{equation}
The CAR norm identity \(\|\sum_v z_vc_v\|=\|z\|_{\ell^2}\) and the spectral calculus for \(|h|\) justify the series and the time integral in \eqref{eq:atomic-filter-functional-calculus}.
Since \(\spec(|h|)\subset[g,\|h\|]\), this time integral vanishes whenever \(\widehat\varphi\) vanishes on \([-\|h\|,-g]\).
This gives the exact free filter normalization in Section~\ref{sec:filters}.
We use local approximants of \(\sigma_\lambda^{-1}(c_{x,j})\) to compare filtered annihilation operators inside a prescribed region.
Local approximants of \(\sigma_\lambda(A\widehat\otimes\1)\) then allow us to deduce bounds on local expectations from excitation-density estimates.
The next lemma bounds the approximation errors for \(\sigma_\lambda\) and \(\sigma_\lambda^{-1}\).

\begin{lemma}
\label{lem:coordinate-locality}
Assume Hypothesis~\ref{hyp:background-transport}.
In the polynomial case, use
\begin{equation}
 p_\at=\min\{p_h,p_\eta\}>d_{\rm vol}.
 \label{eq:coordinate-polynomial-exponent}
\end{equation}
There exist constants \(C_{\rm map}<\infty\) and, in the stretched-exponential case, \(\mu_\at>0\), depending only on the model data, such that, with \(G_\at:[0,\infty]\to[0,\infty)\) defined by
\begin{equation}
 G_\at(R):=
 \begin{cases}
 (1+R)^{-p_\at},&\text{polynomial case},\\
 \E^{-\mu_\at R^\theta},&\text{stretched-exponential case},
 \end{cases}
 \label{eq:atomic-coordinate-profile}
\end{equation}
one has, for every \(|\lambda|\leq\lambda_0\), nonempty \(X\Subset\Gamma\), \(R\ge0\), \(A\in\A_X^\st\), and either choice of sign,
\begin{equation}
 \dist\!\left(
   \sigma_\lambda^{\pm1}(A),\A_{X^{(R)}}^\st\right)
 \leq C_{\rm map}(1+|X|)
       \|A\|G_\at(R).
\label{eq:gamma-locality-explicit}
\end{equation}
The constants can be chosen so that the following estimate also holds uniformly in the matrix dimension.
For every \(k\in\mathbb N\), \(\lambda,X,R\) as above, \(A\in\A_X^\st\otimes M_k\), and either choice of sign,
\begin{equation}
 \left\|\bigl((\operatorname{id}-\mathbb F^{\rm tr}_{X^{(R)}})\otimes\operatorname{id}_{M_k}\bigr)
       (\sigma_\lambda^{\pm1}\otimes\operatorname{id}_{M_k})(A)\right\|
 \le C_{\rm map}(1+|X|)\|A\|G_\at(R).
 \label{eq:coordinate-matrix-locality}
\end{equation}
\end{lemma}

\begin{proof}
Use the concatenated path from \cite[proof of Lemma~6.13]{StabilityBdG26}, formed from the quadratic path generating \(\sigma_\st\) and the path generating \(\beta_{0,\lambda}^{(1)}\).
The quadratic path has Majorana coefficient operator \(-\frac\pi4\bigl(\begin{smallmatrix}0&\sgnh\\\sgnh&0\end{smallmatrix}\bigr)\), whose locality follows from \eqref{eq:free-functional-locality} and \eqref{eq:free-polynomial-functional-locality}.
Together with Hypothesis~\ref{hyp:background-transport}, these bounds give a uniform first support moment for the concatenated path with weight \((1+R)^{p_\at}\) or \(\E^{\mu R^\theta}\) for some \(\mu>0\), respectively.
Apply Lemma~\ref{lem:truncated-cocycle} to this path in both parameter directions to obtain \eqref{eq:gamma-locality-explicit}.

For \eqref{eq:coordinate-matrix-locality}, place the trivially graded algebra \(M_k\) at one site \(x_0\in X\) and tensor the generating interaction terms with its identity.
Their norms are unchanged, and the enlarged conditional expectation is \(\mathbb F^{\rm tr}_{X^{(R)}}\otimes\operatorname{id}_{M_k}\).
By \cite[Remark~6.6]{FermionicLR}, Theorem~2.9 and Remark~2.10 of that paper apply with constants independent of \(k\).
Using the weights \((1+R)^{p_\at}\) and \(\E^{\mu R^\theta}\), respectively, and applying the estimates in both parameter directions proves \eqref{eq:coordinate-matrix-locality}.
\end{proof}

\subsection{Stacking positive-energy representations}

We stack an arbitrary positive-energy representation with the auxiliary ground-state representation. No parity invariance of its vector states is needed.
\begin{lemma}
\label{lem:stacked-positive-representation}
Let \(\Xi\in\mathfrak J_1\), let \(\pi:\A\to\mathcal B(\mathcal H)\) be a unital representation, and let a nonnegative self-adjoint operator \(L\) implement \(\tau^\Xi\) in \(\pi\).
Write \((\pi_{\rm aux},\mathcal H_{\rm aux},\Omega_{\rm aux})\) for the GNS representation of the even auxiliary ground state, with canonical implementing Hamiltonian \(L_{\rm aux}\ge0\).
Its parity operator \(\mathcal P_{\rm aux}\) is the self-adjoint unitary determined, for \(C\in\A^{\rm aux}\), by
\[
 \mathcal P_{\rm aux}\pi_{\rm aux}(C)\Omega_{\rm aux}
 :=\pi_{\rm aux}(\Theta_{\rm aux}(C))\Omega_{\rm aux}.
\]
For homogeneous \(A\in\A\) and \(C\in\A^{\rm aux}\), the formula
\begin{equation}
 \pi_\st(A\widehat\otimes C)
 :=\pi(A)\otimes\mathcal P_{\rm aux}^{|A|}\pi_{\rm aux}(C)
 \label{eq:low-energy-stacked-representation}
\end{equation}
extends to a unital representation of \(\A^\st\).
The nonnegative self-adjoint operator
\[
 \mathbb L:=L\otimes\1+\1\otimes L_{\rm aux},
\]
defined by the joint spectral calculus, implements \(\tau^{\Xi,\st}:=\tau^\Xi\widehat\otimes\tau^{\rm aux}\).
This is the dynamics of the stacked interaction
\(\Xi^\st(Z):=\Xi(Z)\widehat\otimes\1+\1\widehat\otimes\Phi_{h_{\rm aux}}(Z)\).
For every unit vector \(\psi\in\mathcal H\), the vector \(\psi\otimes\Omega_{\rm aux}\) represents \(\rho_\psi\widehat\otimes\omega_{\rm aux}\), where \(\rho_\psi(A):=\langle\psi,\pi(A)\psi\rangle\).
For every \(\varepsilon\ge0\),
\[
 \1_{[0,\varepsilon]}(\mathbb L)(\psi\otimes\Omega_{\rm aux})
 =(\1_{[0,\varepsilon]}(L)\psi)\otimes\Omega_{\rm aux}.
\]
If \((\pi,\mathcal H,\Omega)\) is the GNS representation of a ground state \(\omega\) of \(\Xi\) and \(L\Omega=0\), then \(\Omega\otimes\Omega_{\rm aux}\) is cyclic for \(\pi_\st\), and \(\mathbb L\) is the canonical implementing Hamiltonian of the product state.
\end{lemma}

\begin{proof}
Evenness of \(\omega_{\rm aux}\) gives the parity unitary \(\mathcal P_{\rm aux}\), which fixes \(\Omega_{\rm aux}\), implements \(\Theta_{\rm aux}\), and commutes with \(\E^{\I tL_{\rm aux}}\).
The operators \(\pi(\gamma_u)\otimes\mathcal P_{\rm aux}\) and \(\1\otimes\pi_{\rm aux}(\gamma_{{\rm aux},v})\) satisfy the Majorana CAR, so they define the representation \eqref{eq:low-energy-stacked-representation}.
Conjugation by \(\E^{\I tL}\otimes\E^{\I tL_{\rm aux}}\) implements the product dynamics, whose local generator is \(\I\delta_{\Xi^\st}\); Proposition~\ref{prop:first-moment-dynamics} identifies it with \(\tau^{\Xi^\st}\).
The product-state identity follows from \(\mathcal P_{\rm aux}\Omega_{\rm aux}=\Omega_{\rm aux}\), and the spectral-projection identity follows from \(L_{\rm aux}\Omega_{\rm aux}=0\) by the joint spectral calculus.
In the GNS case, applying homogeneous simple tensors to \(\Omega\otimes\Omega_{\rm aux}\) gives, up to signs, the dense set of vectors \(\pi(A)\Omega\otimes\pi_{\rm aux}(C)\Omega_{\rm aux}\).
Thus the product vector is cyclic, and \(\mathbb L\), which annihilates it, is the canonical implementing Hamiltonian.
\end{proof}

\section{Filtered annihilation operators and their locality}
\label{sec:filters}

We construct the bulk-filtered annihilation operators \(c_{f,x,j}^{(\lambda)}\) and prove their normalization and spatial locality.
Filtering is time convolution with a function whose Fourier transform selects negative energy transfers.

\subsection{Construction and normalization of the filtered operators}

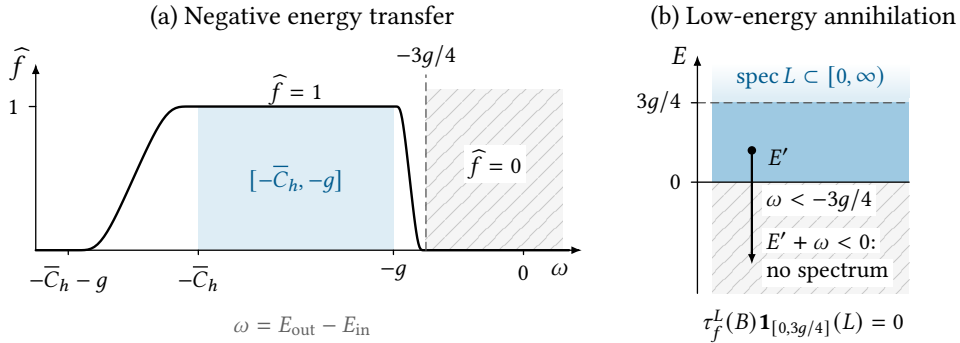
\begin{figure}[htbp]
\centering
\begin{tikzpicture}[lppl diagram]
 \begin{scope}[x=1.72cm,y=1.9cm]
  \node[fig title] at (-1.7,1.62) {(a) Negative energy transfer};
  \path[excluded] (-.75,0) rectangle (.3,1.12);
  \fill[phys!13] (-2.5,0) rectangle (-1,1);
  \node[font=\footnotesize,text=phys!80!black] at (-1.75,.5) {$[-\overline C_h,-g]$};
  \draw[->] (-3.75,0) -- (.45,0) node[below left=1pt] {$\omega$};
  \draw[->] (-3.75,0) -- (-3.75,1.3) node[left] {$\widehat f$};
  \draw (-3.75,1) -- ++(-.05,0) node[left,font=\footnotesize] {$1$};
  \draw[line width=.9pt] plot coordinates {(-3.75,0.0) (-3.5,0.0) (-3.475,0.0) (-3.45,0.0) (-3.425,0.0) (-3.4,0.0001) (-3.375,0.0011) (-3.35,0.0041) (-3.325,0.011) (-3.3,0.023) (-3.275,0.0409) (-3.25,0.065) (-3.225,0.0947) (-3.2,0.1296) (-3.175,0.1686) (-3.15,0.211) (-3.125,0.256) (-3.1,0.3029) (-3.075,0.3512) (-3.05,0.4003) (-3.025,0.45) (-3.0,0.5) (-2.975,0.55) (-2.95,0.5997) (-2.925,0.6488) (-2.9,0.6971) (-2.875,0.744) (-2.85,0.789) (-2.825,0.8314) (-2.8,0.8704) (-2.775,0.9053) (-2.75,0.935) (-2.725,0.9591) (-2.7,0.977) (-2.675,0.989) (-2.65,0.9959) (-2.625,0.9989) (-2.6,0.9999) (-2.575,1.0) (-2.55,1.0) (-2.525,1.0) (-2.5,1.0) (-1.0,1.0) (-0.9917,1.0) (-0.9833,1.0) (-0.975,0.9999) (-0.9667,0.9982) (-0.9583,0.9918) (-0.95,0.977) (-0.9417,0.9517) (-0.9333,0.9158) (-0.925,0.8704) (-0.9167,0.8176) (-0.9083,0.7592) (-0.9,0.6971) (-0.8917,0.6325) (-0.8833,0.5666) (-0.875,0.5) (-0.8667,0.4334) (-0.8583,0.3675) (-0.85,0.3029) (-0.8417,0.2408) (-0.8333,0.1824) (-0.825,0.1296) (-0.8167,0.0842) (-0.8083,0.0483) (-0.8,0.023) (-0.7917,0.0082) (-0.7833,0.0018) (-0.775,0.0001) (-0.7667,0.0) (-0.7583,0.0) (-0.75,0.0) (0.35,0.0)};
  \draw[black!65,dash pattern=on 3pt off 2pt] (-.75,0) -- (-.75,1.22)
    node[above=-1pt,font=\footnotesize,text=black] {$-3g/4$};
  \node[font=\footnotesize] at (-1.75,1.12) {$\widehat f=1$};
  \node[font=\footnotesize,fill=white,inner sep=1.5pt,align=center] at (-.22,.6)
    {$\widehat f=0$};
  \foreach \x/\t in {-3.5/{-\overline C_h-g},-2.5/{-\overline C_h},-1/{-g},0/{0}}
   \draw (\x,0) -- ++(0,-.03) node[below,font=\footnotesize] {$\t$};
  \node[fig note] at (-1.7,-.526) {$\omega=E_{\rm out}-E_{\rm in}$};
 \end{scope}
 \begin{scope}[shift={(2.3,.9)},x=1cm,y=1.4cm]
  \node[fig title] at (1.4,1.557) {(b) Low-energy annihilation};
  \path[excluded] (.2,-1.05) rectangle (2.8,0);
  \shade[bottom color=phys!13,top color=white] (.2,.8) rectangle (2.8,1.12);
  \fill[phys!13] (.2,0) rectangle (2.8,.8);
  \fill[phys!38] (.2,0) rectangle (2.8,.75);
  \draw[black!65,dash pattern=on 3pt off 2pt] (0,.75) -- (2.8,.75);
  \draw[->] (0,-1.05) -- (0,1.2) node[left] {$E$};
  \foreach \y/\t in {0/0,.75/{3g/4}}
   \draw (0,\y) -- ++(-.08,0) node[left,font=\footnotesize] {$\t$};
  \draw (0,0) -- (2.8,0);
  \node[font=\footnotesize,text=phys!80!black] at (1.5,.99) {$\spec L\subset[0,\infty)$};
  \draw[lppl arrow,line width=.85pt] (.72,.3) -- (.72,-.78);
  \fill (.72,.3) circle[radius=1.6pt];
  \node[font=\footnotesize,anchor=west] at (.8,.24) {$E'$};
  \node[font=\footnotesize,anchor=west,fill=white,inner sep=1.5pt] at (.85,-.17)
    {$\omega<-3g/4$};
  \node[font=\footnotesize,anchor=west,fill=white,inner sep=1.5pt] at (.85,-.55)
    {$E'+\omega<0$:};
  \node[font=\footnotesize,anchor=west,fill=white,inner sep=1.5pt] at (.85,-.86)
    {no spectrum};
  \node[font=\footnotesize] at (1.4,-1.357)
    {$\tau_f^L(B)\1_{[0,3g/4]}(L)=0$};
 \end{scope}
\end{tikzpicture}
\caption{Two consequences of the choice of filter.
In (a), \(\widehat f\) is shown for \(s_f=2\) and \(\overline C_h=5g/2\): it equals one on the shaded interval \([-\overline C_h,-g]\), which contains the free annihilators' transfer spectrum \(\spec(-|h|)\), and vanishes for \(\omega\ge-3g/4\).
This gives \(c_{f,x,j}^{(0)}=c_{x,j}\).
In (b), let \(L\ge0\) be an implementing Hamiltonian and let \(B\) be bounded.
Every transfer selected by the filter lowers the energy by more than \(3g/4\), so from an initial energy \(E'\in[0,3g/4]\) it would reach a negative final energy, where \(L\) has no spectrum.
Thus the filtered operator annihilates this spectral subspace, as proved in Lemma~\ref{lem:low-energy-annihilation}; no gap in the nonnegative spectrum is used.}
\label{fig:energy-filter}
\end{figure}

For \(s>1\), define the compactly supported Gevrey class
\begin{equation*}
 G_c^s(\R):=\left\{\varphi\in C_c^\infty(\R)\,\middle|\,
 \begin{array}{l}
 \text{there exist }C_\varphi,R_\varphi>0\text{ such that}\\[-1mm]
 \displaystyle
 \sup_{\omega\in\R}|\varphi^{(n)}(\omega)|
 \le C_\varphi R_\varphi^n(n!)^s\text{ for every }n\in\mathbb N_0
 \end{array}\right\}.
\end{equation*}
Recall the Gevrey order fixed before Hypothesis~\ref{hyp:uniform-local-excitation-detectability}: \(s_f=2\) in the polynomial case and \(s_f=\theta_{\rm in}/\theta>1\) in the stretched-exponential case.
For definiteness, we specify the filter explicitly, so that all constants associated with its choice are controlled by the model data.
For \(s>1\), define the standard Gevrey step
\begin{equation*}
 \chi_s(t):=
 \begin{cases}
  0,&t\leq0,\\
  \displaystyle
  \frac{\exp(-t^{-1/(s-1)})}
  {\exp(-t^{-1/(s-1)})+
   \exp(-(1-t)^{-1/(s-1)})},&0<t<1,\\[2mm]
  1,&t\geq1,
 \end{cases}
\end{equation*}
and define the real-valued function \(\widehat f\in G_c^{s_f}(\R)\) by
\begin{equation*}
 \widehat f(\omega):=
 \chi_{s_f}\!\left(
  \frac{\omega+\overline C_h+g}{g}\right)
 \chi_{s_f}\!\left(
  \frac{-\omega-3g/4}{g/4}\right).
\end{equation*}
It satisfies
\begin{equation}
 \widehat f(\omega)=1\quad\text{for }\omega\in[-\overline C_h,-g],
 \qquad
 \supp\widehat f\subset[-\overline C_h-g,-3g/4].
 \label{eq:filter-frequency}
\end{equation}
This is the standard compactly supported Gevrey cutoff construction, see \cite[Example~1.4.9 and Proposition~1.4.6]{Rodino93}.
Since \(\|h\|\leq C_h\le\overline C_h\), the function \(\widehat f\) equals one on \(\spec(-|h|)\).
The filter's norms and decay constants are controlled by \(g,\overline C_h\) and, in the stretched-exponential case, \(\theta_{\rm in},\theta\), all of which belong to the model data.
For a uniformly controlled family, the common choices of \(g,\overline C_h\), and \(s_f\) give one filter and the window \([0,3g/4]\) for every model.
Figure~\ref{fig:energy-filter} illustrates the normalization of \(\widehat f\) and the spectral exclusion used in the low-energy argument.

Throughout the paper, let \(f\) be the inverse Fourier transform of \(\widehat f\):
\begin{equation}
 f(t):=\frac1{2\pi}\int_\R
 \E^{-\I t\omega}\widehat f(\omega)\dd\omega.
 \label{eq:Fourier}
\end{equation}
The Gevrey Fourier-decay estimate gives \(|f(t)|\le C\exp(-c|t|^{1/s_f})\); see \cite[Chapter~1]{Rodino93} and \cite[Section~5.2]{StabilityBdG26}.
Since \(s_f=\theta_{\rm in}/\theta\) in the stretched-exponential case, there are constants \(C_{\rm time},c_{\rm time}>0\) such that
\begin{equation}
|f(t)|\le C_{\rm time}\E^{-c_{\rm time}|t|^{\theta/\theta_{\rm in}}}.
 \label{eq:filter-time}
\end{equation}

For a strongly continuous automorphism group \(\tau=(\tau_t)_{t\in\R}\) on a \(C^*\)-algebra \(\mathcal B\) and \(k\in L^1(\R)\), define the integrated action \(\tau_k:\mathcal B\to\mathcal B\) by setting, for every \(A\in\mathcal B\),
\begin{equation}
 \tau_k(A):=\int_\R k(t)\tau_t(A)\dd t.
 \label{eq:integrated-action}
\end{equation}
This Bochner integral converges in norm.
When the dynamics is \(\tau^{\st,\lambda}\) or the dynamics in product-vacuum coordinates \(\tau^{\at,\lambda}\) defined before Hypothesis~\ref{hyp:uniform-local-excitation-detectability}, we write the integrated action as \(\tau_k^{\st,\lambda}\) or \(\tau_k^{\at,\lambda}\), respectively.
For \(|\lambda|\le\lambda_0\), \(x\in\Gamma\), and \(j\in\mathscr I\), pull \(c_{x,j}\) back to the original coordinates and filter it with the stacked bulk dynamics \(\tau^{\st,\lambda}\):
\begin{equation}
 \widetilde c_{x,j}^{\,(\lambda)}
 :=\sigma_\lambda^{-1}(c_{x,j}),
 \qquad
 \widetilde c_{f,x,j}^{\,(\lambda)}
 :=\tau_f^{\st,\lambda}(\widetilde c_{x,j}^{\,(\lambda)}),
 \label{eq:pulled-filter}
\end{equation}
and express the bulk-filtered field in product-vacuum coordinates by
\begin{equation}
 c_{f,x,j}^{(\lambda)}
 =\sigma_\lambda(\widetilde c_{f,x,j}^{\,(\lambda)}).
 \label{eq:coordinate-filter}
\end{equation}
Thus \(c_{f,x,j}^{(\lambda)}\) is obtained by passing to the original coordinates, filtering with \(\tau^{\st,\lambda}\), and returning to product-vacuum coordinates:
\begin{equation*}
 c_{x,j}
 \xmapsto{\ \sigma_\lambda^{-1}\ }
 \widetilde c_{x,j}^{\,(\lambda)}
 \xmapsto{\ \tau_f^{\st,\lambda}\ }
 \widetilde c_{f,x,j}^{\,(\lambda)}
 \xmapsto{\ \sigma_\lambda\ }
 c_{f,x,j}^{(\lambda)}.
\end{equation*}
For \(|\lambda|\le\lambda_0\) and \(x\in\Gamma\), recall the filtered excitation observable
\begin{equation}
 n_{f,x}^{(\lambda)}
 =\sum_{j\in\mathscr I}
 (c_{f,x,j}^{(\lambda)})^*c_{f,x,j}^{(\lambda)}.
 \label{eq:filtered-density}
\end{equation}
Equivalently, for every \(j\in\mathscr I\),
\begin{equation}
 c_{f,x,j}^{(\lambda)}
 =\tau_f^{\at,\lambda}(c_{x,j}),
 \qquad
 \tau_t^{\at,\lambda}
 =\sigma_\lambda\circ\tau_t^{\st,\lambda}\circ\sigma_\lambda^{-1}.
 \label{eq:intertwined-filter}
\end{equation}
Formula~\eqref{eq:pulled-filter} expresses filtering through the dynamics generated by the stacked bulk interaction \(\Phi_\lambda^\st\).
We can therefore compare this dynamics with its finite-region restrictions and with the modified dynamics by estimating the interaction terms in which they differ.
We filter and localize the operators in the original coordinates on \(\A^\st\), then apply \(\sigma_\lambda\).

\begin{lemma}
\label{lem:filter-normalization}
Under Hypothesis~\ref{hyp:background-transport}, for every \(|\lambda|\le\lambda_0\), \(x\in\Gamma\), and \(j\in\mathscr I\),
\begin{equation}
 c_{f,x,j}^{(0)}=c_{x,j},
 \qquad
 \pi_\at(c_{f,x,j}^{(\lambda)})\Omega_\at=0.
 \label{eq:filter-normalization}
\end{equation}
In particular,
\begin{equation}
 n_{f,x}^{(0)}=n_x.
 \label{eq:filtered-density-normalization}
\end{equation}
Here \((\pi_\at,\mathcal H_\at,\Omega_\at)\) is the GNS triple fixed in \eqref{eq:atomic-GNS}.
\end{lemma}

\begin{proof}
Fix \(x\in\Gamma\) and \(j\in\mathscr I\).
The definitions and the intertwining relation give
\begin{align*}
 c_{f,x,j}^{(0)}
 =\sigma_0\!\left(
   \tau_f^{\st,0}(\sigma_0^{-1}(c_{x,j}))\right) =\tau_f^{\at,0}(c_{x,j}).
\end{align*}
The plateau condition \eqref{eq:filter-frequency} gives \(\widehat f(-|h|)=\1_{\mathfrak h_{\rm M}}\).
Substituting \(\varphi=f\) into \eqref{eq:atomic-filter-functional-calculus} therefore yields \(c_{f,x,j}^{(0)}=c_{x,j}\) for every \(x\in\Gamma\) and \(j\in\mathscr I\).
Summing the corresponding positive squares over \(j\in\mathscr I\) proves \eqref{eq:filtered-density-normalization}.

For the second identity, ground states are invariant under their dynamics.
Thus \(\omega_\lambda\widehat\otimes\omega_{\rm aux}\) is invariant under \(\tau^{\st,\lambda}\).
The state identification \eqref{eq:exact-atomic-state} implies that \(\omega_\at\) is invariant under \(\tau^{\at,\lambda}\).
Applying Cauchy--Schwarz in the state seminorm to \(c_{f,x,j}^{(\lambda)}=\tau_f^{\at,\lambda}(c_{x,j})\) gives
\begin{align*}
 \bigl\|\pi_\at(c_{f,x,j}^{(\lambda)})\Omega_\at\bigr\|^2
 &=\omega_\at\bigl((c_{f,x,j}^{(\lambda)})^*c_{f,x,j}^{(\lambda)}\bigr)\\
 &\le \|f\|_{L^1(\R)}\int_{\R}|f(t)|
       \omega_\at\bigl(\tau_t^{\at,\lambda}(c_{x,j}^*c_{x,j})\bigr)\,\dd t\\
 &=\|f\|_{L^1(\R)}^2\omega_\at(c_{x,j}^*c_{x,j})=0,
\end{align*}
where the last equality is \eqref{eq:atomic-fields-annihilate}.
\end{proof}

\begin{lemma}
\label{lem:low-energy-annihilation}
Let \(L\ge0\) be a self-adjoint operator on a Hilbert space \(\mathcal H\).
Fix \(E\ge0\) and \(k\in L^1(\R)\) with \(\widehat k(\omega)=0\) for every \(\omega\ge-E\).
For \(B\in\mathcal B(\mathcal H)\), define the bounded operator
\[
 \tau_k^L(B):=\int_\R k(t)\E^{\I tL}B\E^{-\I tL}\dd t
\]
as a strong operator integral.
For every \(0\le\varepsilon\le E\),
\begin{equation}
 \tau_k^L(B)\1_{[0,\varepsilon]}(L)=0.
 \label{eq:low-energy-annihilation}
\end{equation}
\end{lemma}

\begin{proof}
The strong integral exists and has norm at most \(\|k\|_{L^1(\R)}\|B\|\).
Fix \(0\le\varepsilon\le E\) and \(N>0\).
For each \(n\in\mathbb N\), partition \([0,N]\) and \([0,\varepsilon]\) into finitely many nonempty Borel sets \(I_i\) and \(J_j\) of diameter at most \(1/n\), and choose \(a_i\in I_i\) and \(b_j\in J_j\).
Since \(a_i-b_j\ge-\varepsilon\ge-E\),
\[
 \begin{aligned}
 &\int_\R k(t)
    \left(\sum_i\E^{\I ta_i}\1_{I_i}(L)\right)B
    \left(\sum_j\E^{-\I tb_j}\1_{J_j}(L)\right)\dd t\\
 &\qquad=\sum_{i,j}\widehat k(a_i-b_j)\1_{I_i}(L)B\1_{J_j}(L)=0.
 \end{aligned}
\]
For fixed \(t\), the two finite sums approximate \(\E^{\I tL}\1_{[0,N]}(L)\) and \(\E^{-\I tL}\1_{[0,\varepsilon]}(L)\), respectively, with norm error at most \(|t|/n\) by the spectral theorem.
Both sums have norm at most one, so dominated convergence against \(|k(t)|\) gives \(\1_{[0,N]}(L)\tau_k^L(B)\1_{[0,\varepsilon]}(L)=0\).
Letting \(N\to\infty\) proves \eqref{eq:low-energy-annihilation}.
\end{proof}

\subsection{Restricted dynamics and bounds on filtered operators}
\label{sec:restricted-filter-locality}

The following estimate compares the infinite-volume dynamics with the dynamics restricted to a finite region.
It will also control the comparison with the state-dependent positive operator constructed in Lemma~\ref{lem:local-positive-generator}.
For an interaction \(\Phi\in\mathfrak J_1\), a finite region \(\Lambda\), \(A\in\A_\Lambda\), and \(k\in L^1(\R)\), define the integrated boundary error by
\begin{equation}
 \mathfrak b_k^\Phi(A;\Lambda)
 :=\int_{\R}|k(t)|\min\Bigl\{2\|A\|,\,
 \int_0^{|t|}\max_{\epsilon=\pm1}
       \sum_{\substack{Z\cap\Lambda\ne\varnothing\\
                       Z\cap\Lambda^c\ne\varnothing}}
 \|[\Phi(Z),\tau_{\epsilon s}^{\Phi,\Lambda}(A)]\|
       \,\dd s\Bigr\}\,\dd t .
 \label{eq:integrated-boundary-error}
\end{equation}
The first entry of the minimum is the norm bound for the difference of two isometric evolutions.
For the stacked bulk dynamics, write \(\mathfrak b_k^{\st,\lambda}(A;\Lambda)\) for the integral in \eqref{eq:integrated-boundary-error} with each crossing norm replaced by
\begin{equation}
 \|[\Phi_\lambda(Z)\widehat\otimes\1,
                 \tau_{\epsilon s}^{\Phi_\lambda^\st,\Lambda}(A)]\|
 +\|[\1\widehat\otimes\Phi_{h_{\rm aux}}(Z),
                 \tau_{\epsilon s}^{\Phi_\lambda^\st,\Lambda}(A)]\|.
 \label{eq:stacked-crossing-norms}
\end{equation}

\begin{lemma}
\label{lem:local-spectral-comparison}
Let \(\mathcal B\) be a CAR algebra with a fixed finite number of modes per site, or a fixed finite stack, and let \(\Phi\in\mathfrak J_1\).
Fix a nonempty finite \(B\Subset\Gamma\), a unital representation \(\pi:\mathcal B\to\mathcal B(\mathcal H)\), and an orthogonal projection \(Q\) reducing \(\pi(\mathcal B_B)\).
Write \(V:Q\mathcal H\to\mathcal H\) for the inclusion and \(\pi_B(D):=V^*\pi(D)V\) for \(D\in\mathcal B_B\).
Let \(L\) be self-adjoint on \(Q\mathcal H\).
Assume that every \(\pi_B(D)\), \(D\in\mathcal B_B\), preserves \(\operatorname{Dom}(L)\), that \([L,\pi_B(D)]\) extends to a bounded operator, and that the discrepancy
\[
 \mathcal R_B(D):=[L,\pi_B(D)]-
                    \pi_B(\delta_{\Phi,B}(D))
\]
satisfies
\begin{equation}
 \|\mathcal R_B(D)\|
 \le\sum_{Z\text{ crossing }B}\|[\Phi(Z),D]\|
 \qquad(D\in\mathcal B_B).
 \label{eq:local-comparison-discrepancy}
\end{equation}
For \(k\in L^1(\R)\), define the strong operator integral
\[
 T_k^L(A):=\int_\R k(t)\E^{\I tL}\pi_B(A)
                              \E^{-\I tL}\,\dd t,
 \qquad A\in\mathcal B_B.
\]
Then
\begin{equation}
 \|\pi(\tau_k^\Phi(A))V-VT_k^L(A)\|
 \le2\mathfrak b_k^\Phi(A;B).
 \label{eq:local-filter-comparison}
\end{equation}
If, in addition, \(L\ge0\) and \(\widehat k=0\) on \([-E,\infty)\) for some \(E\ge0\), then, for every \(0\le\varepsilon\le E\),
\begin{equation}
 \bigl\|\pi(\tau_k^\Phi(A))V\1_{[0,\varepsilon]}(L)\bigr\|
 \le2\mathfrak b_k^\Phi(A;B).
 \label{eq:local-spectral-comparison}
\end{equation}
If \(\Phi=\sum_{i=1}^N\Phi_i\) is a finite sum of interactions, estimates \eqref{eq:local-filter-comparison} and \eqref{eq:local-spectral-comparison} also hold when each norm \(\|[\Phi(Z),D]\|\) in \eqref{eq:local-comparison-discrepancy} is replaced by
\[
 \sum_{i=1}^N\|[\Phi_i(Z),D]\|
\]
and each commutator norm in the definition \eqref{eq:integrated-boundary-error} of \(\mathfrak b_k^\Phi(A;B)\) is replaced by
\[
 \sum_{i=1}^N\|[\Phi_i(Z),\tau_{\epsilon s}^{\Phi,B}(A)]\|.
\]
The restricted dynamics in this integral is still generated by the total interaction \(\Phi\).

More generally, let \(\Psi\in\mathfrak J_1\) and replace \eqref{eq:local-comparison-discrepancy} by
\begin{equation}
 \|\mathcal R_B(D)\|
 \le\sum_{Z\text{ crossing }B}\|[\Phi(Z),D]\|
       +\|\delta_\Psi(D)\|
 \qquad(D\in\mathcal B_B).
 \label{eq:local-comparison-defect-discrepancy}
\end{equation}
Then the right-hand sides of \eqref{eq:local-filter-comparison} and \eqref{eq:local-spectral-comparison} acquire the additional term
\begin{equation}
 \int_0^\infty w_k(s)\max_{\epsilon=\pm1}
       \|\delta_\Psi(\tau_{\epsilon s}^{\Phi,B}(A))\|\,\dd s,
 \qquad
 w_k(s):=\int_{|t|\ge s}|k(t)|\,\dd t,
 \label{eq:local-comparison-defect-error}
\end{equation}
whenever this integral is finite.
The replacements just specified in \eqref{eq:local-comparison-discrepancy} and \eqref{eq:integrated-boundary-error} also apply to \eqref{eq:local-comparison-defect-discrepancy} and the resulting bounds, without changing the additional integral \eqref{eq:local-comparison-defect-error}.
\end{lemma}

\begin{proof}
Put \(D_s:=\tau_s^{\Phi,B}(A)\).
The map \(s\mapsto D_s\) is continuously differentiable in the finite-dimensional algebra \(\mathcal B_B\).
The bounded commutator extensions define a bounded linear map on this algebra.
The domain assumptions therefore justify the product rule on \(\operatorname{Dom}(L)\); integration and density give, for \(t\ge0\), the strong operator identity
\begin{equation}
 \E^{\I tL}\pi_B(A)\E^{-\I tL}
       -\pi_B(\tau_t^{\Phi,B}(A))
 =\I\int_0^t\E^{\I(t-s)L}\mathcal R_B(D_s)
                         \E^{-\I(t-s)L}\,\dd s.
 \label{eq:local-positive-Duhamel}
\end{equation}
This argument requires no operator-norm differentiability of conjugation by an unbounded \(L\).
The integral is strongly continuous on each vector, and its norm is bounded by the integral of the norms.
The corresponding algebraic identity is
\[
 \tau_t^\Phi(A)-\tau_t^{\Phi,B}(A)
 =\I\int_0^t\tau_{t-s}^\Phi
       \bigl((\delta_\Phi-\delta_{\Phi,B})(D_s)\bigr)\,\dd s.
\]
Proposition~\ref{prop:first-moment-dynamics} justifies this product rule, since \(D_s\in\mathcal B_B\).
Only bulk terms crossing \(B\) contribute to its integrand.
By \eqref{eq:local-comparison-discrepancy}, both norm differences are bounded by
\[
 h_A(t):=\min\left\{2\|A\|,
       \int_0^{|t|}\max_{\epsilon=\pm1}
       \sum_{Z\text{ crossing }B}
          \|[\Phi(Z),\tau_{\epsilon s}^{\Phi,B}(A)]\|
                                    \,\dd s\right\}.
\]
Negative times follow by reversing the integrals; the first bound in the minimum follows from contractivity and unitary conjugation.
Since \(Q\) reduces \(\pi(\mathcal B_B)\), one has
\(\pi(\tau_t^{\Phi,B}(A))V=V\pi_B(\tau_t^{\Phi,B}(A))\).
Insert this intermediate operator and integrate against \(k\).
As \(\int_\R |k(t)|h_A(t)\,\dd t=\mathfrak b_k^\Phi(A;B)\), this proves \eqref{eq:local-filter-comparison}.
Under the additional spectral assumptions, Lemma~\ref{lem:low-energy-annihilation} gives \(T_k^L(A)\1_{[0,\varepsilon]}(L)=0\), including \(\varepsilon=0\).
Multiplying \eqref{eq:local-filter-comparison} by this projection proves \eqref{eq:local-spectral-comparison}.

Under \eqref{eq:local-comparison-defect-discrepancy}, the algebraic comparison with the bulk dynamics is unchanged.
In \eqref{eq:local-positive-Duhamel}, the additional contribution has norm at most
\[
 \int_0^{|t|}\max_{\epsilon=\pm1}
       \|\delta_\Psi(\tau_{\epsilon s}^{\Phi,B}(A))\|\,\dd s.
\]
The bound \(2\|A\|\) still holds for the difference on the left of \eqref{eq:local-positive-Duhamel}.
The inequality \(\min\{M,u+v\}\le\min\{M,u\}+v\) therefore bounds the norm difference in \eqref{eq:local-positive-Duhamel} by \(h_A(t)\) plus the displayed integral.
Integration against \(|k(t)|\), followed by Tonelli's theorem, gives \eqref{eq:local-comparison-defect-error}.
The spectral annihilation argument is unchanged.
For \(\Phi=\sum_{i=1}^N\Phi_i\), apply the triangle inequality to the commutators with \(\Phi(Z)\) in the two Duhamel comparisons and use the assumed discrepancy bound with \(\sum_{i=1}^N\|[\Phi_i(Z),D]\|\).
This proves \eqref{eq:local-filter-comparison} and \eqref{eq:local-spectral-comparison} with the specified sums over interaction components.
\end{proof}

We apply the lemma both to the stacked implementing Hamiltonian \(\mathbb L\) and to the positive local energy operator \(L_{\rho_\st,\Lambda}\).
For an interaction \(\Xi\) satisfying \eqref{eq:comparison-interaction-criterion}, we use the stacked representation and Hamiltonian \(\mathbb L\) of Lemma~\ref{lem:stacked-positive-representation}, with \(Q=\1\), \(\Phi=\Phi_\lambda^\st\), and \(B\subset U\).
The discrepancy \(\mathcal R_B(D)\) consists of commutators with the physical bulk terms that cross the boundary of \(B\) and are present in \(\Xi\), together with the auxiliary terms crossing that boundary.
We estimate the two sums separately as in \eqref{eq:stacked-crossing-norms}.
For a state satisfying only a local ground-state inequality, we use \(L_{\rho_\st,\Lambda}\) on the local cyclic subspace constructed in Lemma~\ref{lem:local-positive-generator}, for which the commutator identity \eqref{eq:local-positive-commutator} gives the bound \eqref{eq:local-comparison-discrepancy}.
No global invariance of the latter state is required.

\begin{lemma}
\label{lem:exponential-crossing-comparison}
Fix \(\mu,J>0\) and \(0<\theta_{\rm in}\le1\).
There exist \(C,v,a>0\), depending only on \(\mu,\theta_{\rm in},J,C_\Gamma,d_{\rm vol}\), such that, for every interaction \(\Phi\) with \(\|\Phi\|_{\mu,\theta_{\rm in};1}\le J\), every nonempty \(X\subset\Lambda\Subset\Gamma\), \(A\in\A_X\), and \(t\in\R\),
\begin{align}
  \sum_{\substack{Z\cap\Lambda\ne\varnothing\\
                  Z\cap\Lambda^c\ne\varnothing}}
 \|[\Phi(Z),\tau_t^{\Phi,\Lambda}(A)]\|
 &\le C|X|\|A\|\E^{v|t|-a\dist(X,\Lambda^c)^{\theta_{\rm in}}},
 \label{eq:exponential-crossing-sum}\\
 \|\tau_t^\Phi(A)-\tau_t^{\Phi,\Lambda}(A)\|
 &\le C|X|\|A\|\E^{v|t|-a\dist(X,\Lambda^c)^{\theta_{\rm in}}}.
 \label{eq:exponential-restriction-comparison}
\end{align}
Fix \(0<\theta_k\le1\) and \(C_k,c_k>0\).
There are \(C,m>0\), depending only on \(\mu,\theta_{\rm in},J,C_\Gamma,d_{\rm vol}\) and \(\theta_k,C_k,c_k\), such that, for every interaction \(\Phi\) with \(\|\Phi\|_{\mu,\theta_{\rm in};1}\le J\), every filter \(k\) satisfying \(|k(t)|\le C_k\E^{-c_k|t|^{\theta_k}}\), every nonempty \(X\subset\Lambda\Subset\Gamma\), and \(A\in\A_X\),
\begin{equation}
 \mathfrak b_k^\Phi(A;\Lambda)
 \le C|X|\|A\|\E^{-m\dist(X,\Lambda^c)^{\theta_{\rm in}\theta_k}}.
 \label{eq:integrated-exponential-boundary}
\end{equation}
The fixed filter \(f\) from \eqref{eq:Fourier} satisfies this hypothesis with \(\theta_k=\theta/\theta_{\rm in}\), by \eqref{eq:filter-time}.
If \(\Phi=\Phi_1+\Phi_2\) with \(\|\Phi_1\|_{\mu,\theta_{\rm in};1}+\|\Phi_2\|_{\mu,\theta_{\rm in};1}\le J\), estimates \eqref{eq:exponential-crossing-sum} and \eqref{eq:integrated-exponential-boundary} hold with the same constants when we replace each crossing commutator norm \(\|[\Phi(Z),\tau_u^{\Phi,\Lambda}(A)]\|\) by
\[
 \sum_{i=1}^2\|[\Phi_i(Z),\tau_u^{\Phi,\Lambda}(A)]\|,
\]
with \(u=t\) in \eqref{eq:exponential-crossing-sum} and \(u=\epsilon s\) in the integral \eqref{eq:integrated-boundary-error}.
\end{lemma}

\begin{proof}
Put \(D:=\dist(X,\Lambda^c)\) and \(J_\Phi(x,y):=\sum_{Z\ni x,y}\|\Phi(Z)\|\).
The interaction bound gives \(\sup_x\sum_y\E^{\mu d(x,y)^{\theta_{\rm in}}}J_\Phi(x,y)\le J\).
Apply the matrix commutator estimate of \cite[Lemma~3.1]{FermionicLR} to each crossing term and sum over a site of that term outside \(\Lambda\).
The weighted row estimate in the same lemma gives
\[
 \sum_{Z\text{ crossing }\Lambda}
 \|[\Phi(Z),\tau_t^{\Phi,\Lambda}(A)]\|
 \le2\|A\|\sum_{x\in X,y\notin\Lambda}
                 (\E^{2|t|J_\Phi}J_\Phi)(x,y) \le2J|X|\|A\|\E^{2J|t|-\mu D^{\theta_{\rm in}}}.
\]
The summands are nonnegative and the rows are summable, so the crossing sum converges absolutely.
Each interaction term is even, and hence this argument applies to either parity of \(A\).
Duhamel's formula for the restricted and full dynamics, followed by integration of this bound from \(0\) to \(|t|\), proves \eqref{eq:exponential-restriction-comparison}.

For \(1\le D<\infty\), split \eqref{eq:integrated-boundary-error} at \(|t|=c_0D^{\theta_{\rm in}}\), with \(c_0>0\) small enough that the exponential growth in time is absorbed by half the spatial exponent.
Using the crossing estimate on this interval and the norm cap outside it gives
\[
 \mathfrak b_k^\Phi(A;\Lambda)
 \le C|X|\|A\|\E^{-aD^{\theta_{\rm in}}/2}
       +2\|A\|\int_{|t|>c_0D^{\theta_{\rm in}}}|k(t)|\,\dd t
 \le C'|X|\|A\|\E^{-mD^{\theta_{\rm in}\theta_k}}.
\]
This is the time integration used in \cite[Lemma~6.8]{StabilityBdG26}, applied in the metric \(d^{\theta_{\rm in}}\).
For \(D<1\), use \(2\|k\|_{L^1(\R)}\|A\|\); for \(\Lambda^c=\varnothing\), the crossing sum is empty.
For \(\Phi=\Phi_1+\Phi_2\), apply the commutator estimate separately to \(\Phi_1(Z)\) and \(\Phi_2(Z)\), keeping the same evolution \(\tau_t^{\Phi,\Lambda}\).
In the preceding crossing estimate, this replaces
\[
 \E^{2|t|J_\Phi}J_\Phi
 \quad\text{by}\quad
 \E^{2|t|J_\Phi}(J_{\Phi_1}+J_{\Phi_2}),
 \qquad
 J_{\Phi_i}(x,y)=\sum_{Z\ni x,y}\|\Phi_i(Z)\|.
\]
The weighted row sums of both \(J_\Phi\) and \(J_{\Phi_1}+J_{\Phi_2}\) are bounded by \(J\), so the same calculation proves the estimates for the sum of the two commutator norms.
\end{proof}

For the polynomial boundary estimates we use \cite[Proposition~5.3(ii)]{FermionicLR}.

\begin{lemma}
\label{lem:polynomial-crossing-comparison}
Assume \eqref{eq:cell-partitions}.
Fix \(p>d_{\rm vol}\) and \(J>0\), and use \(m_{p,0}\) from \eqref{eq:polynomial-time-exponent}.
There is a constant \(C>0\), depending only on \(p,J,C_\Gamma,d_{\rm vol},C_{\rm cell},D_{\rm cell}\), such that, for every interaction \(\Phi\) with \(\|\Phi\|^{\#}_{p;\kappa_0}\le J\), every nonempty \(X\subset\Lambda\Subset\Gamma\), \(A\in\A_X\), and \(t\in\R\),
\begin{align}
 &\sum_{\substack{Z\cap\Lambda\ne\varnothing\\
                  Z\cap\Lambda^c\ne\varnothing}}
 \|[\Phi(Z),\tau_t^{\Phi,\Lambda}(A)]\|
 \le C|X|\|A\|(1+|t|)^{m_{p,0}}
       (1+\dist(X,\Lambda^c))^{-p}.                 \label{eq:polynomial-crossing-sum}
\end{align}
For every such \(\Phi,X,\Lambda,A\) and every \(k\in L^1(\R)\) with
\[
 I_p(k):=\int_\R |k(t)|\,|t|(1+|t|)^{m_{p,0}}\,\dd t<\infty,
\]
one has, with the same constant \(C\),
\begin{equation}
 \mathfrak b_k^\Phi(A;\Lambda)
 \le C I_p(k)|X|\|A\|(1+\dist(X,\Lambda^c))^{-p}.
 \label{eq:integrated-polynomial-boundary}
\end{equation}
The same constant applies if terms are deleted from the restricted Hamiltonian or omitted from the crossing sum.
If \(\Phi=\Phi_1+\Phi_2\) with \(\|\Phi_1\|^{\#}_{p;\kappa_0}+\|\Phi_2\|^{\#}_{p;\kappa_0}\le J\), estimates \eqref{eq:polynomial-crossing-sum} and \eqref{eq:integrated-polynomial-boundary} hold with the same constants when we replace each crossing commutator norm \(\|[\Phi(Z),\tau_u^{\Phi,\Lambda}(A)]\|\) by
\[
 \sum_{i=1}^2\|[\Phi_i(Z),\tau_u^{\Phi,\Lambda}(A)]\|,
\]
with \(u=t\) in \eqref{eq:polynomial-crossing-sum} and \(u=\epsilon s\) in the integral \eqref{eq:integrated-boundary-error}.
\end{lemma}

\begin{proof}
If \(\Lambda^c=\varnothing\), both estimates hold because the crossing sum is empty.
Suppose \(\Lambda^c\ne\varnothing\), and set \(D:=\dist(X,\Lambda^c)\).
Apply \cite[Proposition~5.3(ii)]{FermionicLR} with Hamiltonian interaction \(\Phi\), source interaction \(\Psi=\Phi\), and \(a=p\) to obtain \eqref{eq:polynomial-crossing-sum}.
For every \(s\ge0\), the same bound controls the maximum over the two times \(s\) and \(-s\) in \eqref{eq:integrated-boundary-error}.
Substituting \eqref{eq:polynomial-crossing-sum} into \eqref{eq:integrated-boundary-error} gives
\[
 \begin{aligned}
 \mathfrak b_k^\Phi(A;\Lambda)
 &\le C|X|\|A\|(1+D)^{-p}
       \int_\R |k(t)|\int_0^{|t|}(1+s)^{m_{p,0}}\,\dd s\,\dd t
 \\
 &\le C I_p(k)|X|\|A\|(1+D)^{-p}.
 \end{aligned}
\]
For the fixed filter \(f\) from \eqref{eq:Fourier}, \(I_p(f)<\infty\) because \(\widehat f\in C_c^\infty(\R)\).
For \(\Phi=\Phi_1+\Phi_2\), apply the proposition to the sum of commutators with \(\Psi_1=\Phi_1\) and \(\Psi_2=\Phi_2\), using the same restricted dynamics \(\tau^{\Phi,\Lambda}\) for \(A\).
The sum \(\|\Phi_1\|^{\#}_{p;\kappa_0}+\|\Phi_2\|^{\#}_{p;\kappa_0}\) is bounded by \(J\), so the same constant applies.
\end{proof}

For \(s\ge0\), set
\begin{equation}
 w_f(s):=\int_{|t|\ge s}|f(t)|\,\dd t.
 \label{eq:tail-time-weight}
\end{equation}
All polynomial moments of this weight are finite.
In the stretched-exponential regime it also satisfies \(w_f(s)\le C\E^{-c s^{\theta/\theta_{\rm in}}}\), with a possibly smaller \(c>0\).
For every measurable \(q:[0,\infty)\to[0,\infty]\), Tonelli's theorem gives
\[
 \int_\R |f(t)|\int_0^{|t|}q(s)\,\dd s\,\dd t
 =\int_0^\infty w_f(s)q(s)\,\dd s.
\]

\begin{lemma}
\label{lem:filtered-norm-locality}
Assume Hypothesis~\ref{hyp:background-transport}.
In the polynomial case, \eqref{eq:main-polynomial-range} relates the decay exponent \(\nu\) to the propagation exponent \(p_\Phi\) of the bulk dynamics and the locality exponent \(p_\at\) of \(\sigma_\lambda\) by
\begin{equation}
 \nu=\min\{p_\Phi,p_\at\}>d_{\rm vol}.
 \label{eq:polynomial-filter-exponent}
\end{equation}
There are constants \(C,m>0\) in the stretched-exponential case and \(C<\infty\) in the polynomial case, depending only on the model data, with the following properties.
For every \(|\lambda|\le\lambda_0\), \(x\in\Gamma\), \(j\in\mathscr I\), and dyadic index \(n_{\max}\ge0\), define
\[
 A_{n_{\max}}:=\mathbb F^{\rm tr}_{\ball_{R_{n_{\max}}}(x)}\,
                   \widetilde c_{x,j}^{\,(\lambda)}.
\]
For every finite region \(B\supset\ball_{R_{n_{\max}}}(x)\), and also for \(B=\Gamma\), set \(\tau_t^B:=\tau_t^{\Phi_\lambda^\st,B}\).
For either weight \(k(s):=|f(s)|+|f(-s)|\) or \(k(s):=w_f(s)\), and every \(R_{\rm in}\ge0\),
\begin{equation}
 \int_0^\infty k(s)\max_{\epsilon=\pm1}
 \|(\operatorname{id}-\mathbb F^{\rm tr}_{\ball_{R_{\rm in}}(x)})\tau_{\epsilon s}^B(A_{n_{\max}})\|
 \,\dd s
 \le
 \begin{cases}
 C\E^{-m R_{\rm in}^\theta},&\text{stretched-exponential case},\\
 C(1+R_{\rm in})^{-\nu},&\text{polynomial case}.
 \end{cases}
 \label{eq:tail-integrated-locality}
\end{equation}
For \(R\ge16\) with \(R_{n_{\max}}\le R/4\), the separate-layer crossing integral defined by \eqref{eq:integrated-boundary-error} with the replacement \eqref{eq:stacked-crossing-norms} also satisfies
\begin{equation}
 \mathfrak b_f^{\st,\lambda}(A_{n_{\max}};\ball_R(x))
 \le
 \begin{cases}
 C\E^{-m R^\theta},&\text{stretched-exponential case},\\
 C(1+R)^{-\nu},&\text{polynomial case}.
 \end{cases}
 \label{eq:seed-integrated-boundary}
\end{equation}
The fields \(\widetilde c_{f,x,j}^{\,(\lambda)}\) satisfy, for every \(R\ge0\),
\begin{align}
 \|(\operatorname{id}-\mathbb F^{\rm tr}_{\ball_R(x)})\widetilde c_{f,x,j}^{\,(\lambda)}\|
 &\le C\E^{-mR^\theta}
 &&\text{in the stretched-exponential case},
 \label{eq:exponential-filter-norm-tail}\\
 \|(\operatorname{id}-\mathbb F^{\rm tr}_{\ball_R(x)})\widetilde c_{f,x,j}^{\,(\lambda)}\|
 &\le C(1+R)^{-\nu}
 &&\text{in the polynomial case}.
 \label{eq:polynomial-odd-filter-norm-tail}
\end{align}
In the polynomial case, for every dyadic index \(n\ge0\), the exact tracial shells satisfy
\begin{align}
 \|\Delta_{\{x\},n}^{\rm tr}
                 \widetilde c_{f,x,j}^{\,(\lambda)}\|
 &\le C(1+R_n)^{-\nu}.
 \label{eq:polynomial-filter-exact-shell}
\end{align}
The bulk-filtered fields in product-vacuum coordinates satisfy, for every \(R\ge0\),
\begin{equation}
 \| (\operatorname{id}-\mathbb F^{\rm tr}_{\ball_R(x)})
                         c_{f,x,j}^{(\lambda)}\|
 \le
 \begin{cases}
 C\E^{-mR^\theta},&\text{stretched-exponential case},\\
 C(1+R)^{-\nu},&\text{polynomial case}.
 \end{cases}
 \label{eq:coordinate-filter-norm-tail}
\end{equation}
\end{lemma}

\begin{proof}
Decompose \(\widetilde c_{x,j}^{\,(\lambda)}\) into the tracial shells \(D_n:=\Delta_{\{x\},n}^{\rm tr}\widetilde c_{x,j}^{\,(\lambda)}\).
Lemma~\ref{lem:coordinate-locality} gives \(\|D_n\|\le C(1+R_n)^{-p_\at}\) in the polynomial case and \(\|D_n\|\le C\E^{-cR_n^\theta}\) in the stretched-exponential case.
Thus \(A_{n_{\max}}=\sum_{n=0}^{n_{\max}}D_n\), \(\|A_{n_{\max}}\|\le1\), and
\begin{equation}
 \sum_{n\ge0}|\ball_{R_n}(x)|\,\|D_n\|\le C.
 \label{eq:tail-seed-moment}
\end{equation}

First suppose that \(R_{\rm in}\ge4\).
For a shell with \(R_n\le R_{\rm in}/4\), its support has an \(R_{\rm in}/2\)-neighborhood inside \(\ball_{R_{\rm in}}(x)\).
In the polynomial case, apply Lemma~\ref{lem:all-parity-polynomial} to this shell and integrate against \(k\).
Both choices of \(k\) have a finite moment of order \(m_{p_\Phi,0}\).
Bounding the omitted shells by their norms gives
\[
 \begin{aligned}
 \int_0^\infty k(s)&\max_{\epsilon=\pm1}
 \|(\operatorname{id}-\mathbb F^{\rm tr}_{\ball_{R_{\rm in}}(x)})
       \tau_{\epsilon s}^B(A_{n_{\max}})\|\,\dd s\\
 &\quad\le C(1+R_{\rm in})^{-p_\Phi}
       \sum_{R_n\le R_{\rm in}/4}|\ball_{R_n}(x)|\,\|D_n\|
       +C\sum_{R_n>R_{\rm in}/4}\|D_n\|\\
 &\quad\le C'(1+R_{\rm in})^{-\min\{p_\Phi,p_\at\}}.
 \end{aligned}
\]
In the stretched-exponential case, compare the evolution of each included shell with the dynamics restricted to \(B\cap\ball_{R_{\rm in}}(x)\), using Lemma~\ref{lem:exponential-crossing-comparison} on the metric subspace \(B\).
Sum with \eqref{eq:tail-seed-moment} and split the time integral at \(c_0R_{\rm in}^{\theta_{\rm in}}\), as in the proof of that lemma.
The omitted shells contribute at most \(C\E^{-cR_{\rm in}^\theta}\), and both weights satisfy \(\int_T^\infty k(s)\,\dd s\le C\E^{-cT^{\theta/\theta_{\rm in}}}\), giving \eqref{eq:tail-integrated-locality}.
For \(R_{\rm in}<4\), the estimate follows from \(\|A_{n_{\max}}\|\le1\).
These bounds are uniform in \(n_{\max}\) and in the restriction \(B\).

For \eqref{eq:seed-integrated-boundary}, every shell of \(A_{n_{\max}}\) is supported in \(\ball_{R/4}(x)\), at distance at least \(3R/4\) from \(\ball_R(x)^c\).
Apply the separate-component bounds in Lemmas~\ref{lem:exponential-crossing-comparison} and~\ref{lem:polynomial-crossing-comparison} to the physical and auxiliary crossing terms for each shell, and sum with \eqref{eq:tail-seed-moment}.
In the polynomial case, integration in \eqref{eq:integrated-boundary-error} gives
\[
 \mathfrak b_f^{\st,\lambda}(A_{n_{\max}};\ball_R(x))
 \le C I_{p_\Phi}(f)(1+R)^{-p_\Phi},
\]
which implies the required power \(\nu\).
In the stretched-exponential case, the same time split as in \eqref{eq:integrated-exponential-boundary}, with the cap \(2\|A_{n_{\max}}\|\le2\), gives order \(\theta\).

Take \(B=\Gamma\) and let \(n_{\max}\to\infty\) in \eqref{eq:tail-integrated-locality}, using \(k(s)=|f(s)|+|f(-s)|\).
Norm convergence and dominated convergence give \eqref{eq:exponential-filter-norm-tail} and \eqref{eq:polynomial-odd-filter-norm-tail}, and \eqref{eq:dyadic-locality-to-shells} gives \eqref{eq:polynomial-filter-exact-shell}.
Finally apply \cite[Lemma~6.10]{StabilityBdG26} to the tracial shells of \(\widetilde c_{f,x,j}^{\,(\lambda)}\) and the map \(\sigma_\lambda\).
The shells have polynomial order \(\nu\), the map has order \(p_\at\), and their support-weighted norm sum is finite because \(\nu>d_{\rm vol}\).
The composition therefore has order \(\min\{\nu,p_\at\}=\nu\); in the stretched-exponential case, the lemma preserves the order \(\theta\).
Since \(c_{f,x,j}^{(\lambda)}=\sigma_\lambda(\widetilde c_{f,x,j}^{\,(\lambda)})\), estimate~\eqref{eq:general-tracial-shell-bounds} converts this local approximation into \eqref{eq:coordinate-filter-norm-tail}.
\end{proof}

\paragraph{Action on low-energy subspaces.}

We now apply Lemma~\ref{lem:local-spectral-comparison} to the quasi-local fields \(\widetilde c_{x,j}^{\,(\lambda)}\).
Estimate~\eqref{eq:common-filtered-occupation} bounds the squared norm of the action of \(\widetilde c_{f,x,j}^{\,(\lambda)}\) on the low-energy subspace by a term decaying with the comparison radius \(R\) and the square of the defect integral \eqref{eq:filtered-occupation-defect-error}.
In Sections~\ref{sec:local-determination-proofs} and~\ref{sec:quasilocal-tails}, we choose the positive operator \(L\) for each application and estimate the defect term.
For an interaction \(\Psi\in\mathfrak J_1\) on the physical algebra, define its first-layer lift by \(\Psi^{(1)}(Z):=\Psi(Z)\widehat\otimes\1\), and denote its derivation by \(\delta_\Psi^{(1)}\).

\begin{lemma}
\label{lem:common-filtered-occupation}
Assume Hypothesis~\ref{hyp:background-transport} and, in the polynomial case, use \(\nu\) from \eqref{eq:polynomial-filter-exponent}.
There are constants \(C>0\) and, in the stretched-exponential case, \(\mu>0\), depending only on the model data, with the following property.
For every \(|\lambda|\le\lambda_0\), \(x\in\Gamma\), \(j\in\mathscr I\), and \(R\ge16\), set \(B:=\ball_R(x)\), choose \(n_{\max}\) maximal with \(R_{n_{\max}}\le R/4\), and define
\[
 A_R:=\mathbb F^{\rm tr}_{\ball_{R_{n_{\max}}}(x)}
                  \widetilde c_{x,j}^{\,(\lambda)}.
\]
Let \(\pi,Q,L,V\) satisfy the representation and domain assumptions of Lemma~\ref{lem:local-spectral-comparison} on \(\mathcal B=\A^\st\), with \(L\ge0\).
Let \(\Psi\in\mathfrak J_1\) be a physical interaction, and assume \eqref{eq:local-comparison-defect-discrepancy} with bulk interaction \(\Phi_\lambda^\st\), additional interaction \(\Psi^{(1)}\), and the separate-layer crossing sums of \eqref{eq:stacked-crossing-norms}.
Define
\begin{equation}
 e_R:=
 \int_0^\infty w_f(s)\max_{\epsilon=\pm1}
 \bigl\|\delta_\Psi^{(1)}
       (\tau_{\epsilon s}^{\Phi_\lambda^\st,B}(A_R))\bigr\|\,\dd s.
 \label{eq:filtered-occupation-defect-error}
\end{equation}
Then
\begin{equation}
 \left\|\pi(\widetilde c_{f,x,j}^{\,(\lambda)})
                   V\1_{[0,3g/4]}(L)\right\|^2
 \le
 \begin{cases}
 C[\E^{-\mu R^\theta}+e_R^2],
       &\text{stretched-exponential case},\\
 C[(1+R)^{-2\nu}+e_R^2],
       &\text{polynomial case}.
 \end{cases}
 \label{eq:common-filtered-occupation}
\end{equation}
The constants are independent of \(\pi,Q,L,R,\Psi\).
Taking \(\Psi=0\) gives \eqref{eq:common-filtered-occupation} under the discrepancy bound \eqref{eq:local-comparison-discrepancy}, with \(e_R=0\).
\end{lemma}

\begin{proof}
Put \(P:=\1_{[0,3g/4]}(L)\).
The fixed filter from \eqref{eq:Fourier} satisfies \(\widehat f=0\) on \([-3g/4,\infty)\).
Lemma~\ref{lem:local-spectral-comparison}, including \eqref{eq:local-comparison-defect-error}, therefore gives
\[
 \begin{aligned}
 \|\pi(\widetilde c_{f,x,j}^{\,(\lambda)})VP\|
 &\le \|f\|_{L^1(\R)}
       \|\widetilde c_{x,j}^{\,(\lambda)}-A_R\|
       +\|\pi(\tau_f^{\st,\lambda}(A_R))VP\|\\
 &\le \|f\|_{L^1(\R)}
       \|\widetilde c_{x,j}^{\,(\lambda)}-A_R\|
       +2\mathfrak b_f^{\st,\lambda}(A_R;B)+e_R.
 \end{aligned}
\]
The restricted dynamics preserves \(\A_B^\st\), where \(\|\delta_\Psi^{(1)}(D)\|\le2|B|J_\Psi\|D\|\).
Since \(w_f\) is integrable, \(e_R\) is finite.
Since \(R/8<R_{n_{\max}}\le R/4\), Lemma~\ref{lem:coordinate-locality} bounds the truncation error by \(C\E^{-cR^\theta}\), respectively \(C(1+R)^{-p_\at}\).
Estimate~\eqref{eq:seed-integrated-boundary} bounds the boundary term by \(C\E^{-cR^\theta}\), respectively \(C(1+R)^{-\nu}\).
In the polynomial case \(p_\at\ge \nu\).
Squaring and using \((u+v)^2\le2u^2+2v^2\) proves \eqref{eq:common-filtered-occupation}.
\end{proof}

\section{Local state estimates from excitation densities}
\label{sec:reconstruction-bulk}

For a state \(\varrho\) on \(\A^\st\), we use LED to turn bounds on the filtered expectations \(\varrho(n_{f,x}^{(\lambda)})\) into bounds on the excitation densities \(\varrho(n_x)\), and then compare local expectations with those of the product vacuum.
For a unital completely positive map \(\mathcal Q\), we also prove operator-valued versions for \(\mathcal Q(n_{f,x}^{(\lambda)})\) and \(\mathcal Q(n_x)\), which apply to compressions in a representation.
In the polynomial case, we use the localized form of LED to preserve the full decay power.
The same estimates will be used for low-energy subspaces, local ground states, and defects with tails.

\subsection{Spatial decay from a kernel inequality}

The following lemma transfers spatial decay from \(b\) to a bounded nonnegative function \(u\) satisfying \(u\le b+Ku\).
For stretched-exponential decay, it assumes that the corresponding weighted row sums of \(K\) are uniformly below one.
For polynomial decay, it assumes \(\sup_x\sum_zK(x,z)<1\) and a finite polynomial row moment.
We use this polynomial conclusion for defects with polynomial tails under stretched-exponential bulk locality in Section~\ref{sec:quasilocal-tails}.
For a nonnegative kernel \(K\) on \(\Gamma\times\Gamma\), \(\mu\ge0\), \(0<\theta<1\), and \(M>0\), write
\begin{equation}
 \begin{aligned}
 \vartheta(K)&:=\sup_{x\in\Gamma}\sum_{z\in\Gamma}K(x,z),\\
 \vartheta_\mu(K)&:=\sup_{x\in\Gamma}\sum_{z\in\Gamma}\E^{\mu d(x,z)^\theta}K(x,z),\\
 \mathfrak M_M(K)&:=\sup_{x\in\Gamma}\sum_{z\in\Gamma}(1+d(x,z))^M K(x,z).
 \end{aligned}
 \label{eq:single-kernel-bounds}
\end{equation}

\begin{lemma}
\label{lem:absorption}
Let \(U\subset\Gamma\), let \(u,b:\Gamma\to[0,\infty)\), and let \(K:\Gamma\times\Gamma\to[0,\infty)\).
For \(v:\Gamma\to[0,\infty)\), define
\begin{equation*}
 (Kv)(x):=\sum_{z\in\Gamma}K(x,z)v(z),
\end{equation*}
whenever the sum is finite.
Assume that \(u\) is bounded and that, for every \(x\in\Gamma\),
\begin{equation}
 u_x\le b_x+(Ku)(x).
 \label{eq:abstract-convolution}
\end{equation}
\begin{enumerate}[label=(\roman*)]
\item Fix \(0<\theta<1\).
Suppose that there are \(C_b,\mu_b,\mu_K>0\) such that, for every \(x\in\Gamma\),
\begin{equation}
  b_x\le C_b\E^{-\mu_b\dist(x,U^c)^\theta}\qquad\mbox{and}\qquad\vartheta_{\mu_K}(K)<1.
 \label{eq:abstract-weighted-contraction}
\end{equation}
Define
\begin{equation*}
 m:=\min\{\mu_b,\mu_K\}>0.
\end{equation*}
Then, for every \(x\in\Gamma\),
\begin{equation}
 u_x\le \frac{C_b}{1-\vartheta_{\mu_K}(K)}
 \E^{-m\dist(x,U^c)^\theta}.
 \label{eq:absorb-stretched}
\end{equation}
\item Let \(M>0\) and assume \(\vartheta(K)<1\) and \(\mathfrak M_M(K)<\infty\).
If there is \(C_b>0\) such that \(b_x\le C_b(1+\dist(x,U^c))^{-M}\) for every \(x\in\Gamma\), then, for every \(x\in\Gamma\),
\begin{equation}
 u_x\le C'(1+\dist(x,U^c))^{-M},
 \label{eq:absorb-finite-polynomial}
\end{equation}
where \(C'\) depends only on \(M,C_b,\vartheta(K),\mathfrak M_M(K)\).
\end{enumerate}
\end{lemma}

\begin{proof}
If \(U=\Gamma\), then \(b=0\) and \(\|u\|_\infty\le\vartheta(K)\|u\|_\infty\), so \(u=0\) in both cases.
Assume \(U\ne\Gamma\).
In (i), set \(w_T(x):=\min\{\E^{m\dist(x,U^c)^\theta},T\}\) for \(T\ge1\).
Subadditivity gives \(w_T(x)\le\E^{m d(x,z)^\theta}w_T(z)\); multiplying \eqref{eq:abstract-convolution} by this bounded weight and taking the supremum yields
\[
 \|w_Tu\|_\infty
 \le C_b+\vartheta_{\mu_K}(K)\|w_Tu\|_\infty.
\]
Rearranging and letting \(T\to\infty\) proves \eqref{eq:absorb-stretched}.

For (ii), the inequality
\[
 (1+r/L)^M-1\le C_M L^{-\min\{M,1\}}(1+r)^M
 \qquad(r\ge0,\ L\ge1)
\]
allows us to choose \(L\) so that
\[
 \mathfrak M_{M,L}(K):=\sup_x\sum_z(1+d(x,z)/L)^M K(x,z)
 \le\vartheta(K)+C_M L^{-\min\{M,1\}}\mathfrak M_M(K)<1.
\]
Now use \(w_{L,T}(x):=\min\{(1+\dist(x,U^c)/L)^M,T\}\).
The triangle inequality gives \(w_{L,T}(x)\le(1+d(x,z)/L)^Mw_{L,T}(z)\), so the preceding supremum argument gives
\[
 \|w_{L,T}u\|_\infty
 \le C_b+\mathfrak M_{M,L}(K)\|w_{L,T}u\|_\infty.
\]
Rearrange, let \(T\to\infty\), and use \((1+D/L)^{-M}\le L^M(1+D)^{-M}\).
This proves \eqref{eq:absorb-finite-polynomial} with \(C'=C_bL^M/(1-\mathfrak M_{M,L}(K))\).
\end{proof}

For polynomial bulk locality, we obtain a local form of LED by taking the product-vacuum expectation outside a finite region.
This removes the excitation numbers outside that region, and locality of the filtered annihilation operators controls the resulting error in Lemma~\ref{lem:localized-detectability}.
For finite \(X\subset\Gamma\), use the graded factorization \(\A^\st\cong\A_X^\st\widehat\otimes\A_{X^c}^\st\).
For homogeneous \(B\in\A_X^\st\) and \(C\in\A_{X^c}^\st\), define
\begin{equation}
 \mathbb E_X^\vac(B\widehat\otimes C)
 :=\omega_\at(C)B.
 \label{eq:vacuum-slices}
\end{equation}
By \cite[Lemma~6.6]{StabilityBdG26}, this formula extends to a unital completely positive contraction onto \(\A_X^\st\).
The map fixes \(\A_X^\st\), is an \(\A_X^\st\)-bimodule map, preserves parity and \(\omega_\at\), and satisfies \(\mathbb E_X^\vac\mathbb E_Y^\vac=\mathbb E_X^\vac\) for \(X\subset Y\Subset\Gamma\).
For \(X=\varnothing\), it is \(\mathbb E_\varnothing^\vac(A)=\omega_\at(A)\1\).

\begin{lemma}
\label{lem:localized-detectability}
Assume polynomial locality and Hypothesis~\ref{hyp:background-transport}, with \(\nu\) as in \eqref{eq:main-polynomial-range}.
Fix \(0\le\kappa<1\) and suppose that \eqref{eq:uniform-local-excitation-detectability} holds on \(\mathcal J\), with
\[
 \sup_{\lambda\in\mathcal J}\sup_x\sum_z\mathscr K_\lambda(x,z)
 \le\kappa.
\]
There is a constant \(C_{\rm loc}\), depending only on the model data, such that, for every \(\lambda\in\mathcal J\), \(x\in\Gamma\), and \(R\ge0\),
\begin{equation}
 n_x\le4n_{f,x}^{(\lambda)}
       +\sum_{z\in\ball_R(x)}\mathscr K_\lambda(x,z)n_z
       +C_{\rm loc}(1+R)^{-2\nu}\1.
 \label{eq:localized-detectability}
\end{equation}

Choose \(\epsilon\in(0,1)\), depending only on \(\kappa,\nu\), such that \(\kappa(1-\epsilon)^{-2\nu}<1\), and fix \(C_b\ge0\).
There is \(C>0\), depending only on \(C_b,\kappa\) and the model data, with the following property.
Let \(\lambda\in\mathcal J\), let \(S\subset\Gamma\) be nonempty, and let \(\mathcal Q:\A^\st\to\mathfrak B\) be a unital positive map to a unital \(C^*\)-algebra.
Write \(D_x:=\dist(x,S)\), and define the localized kernel by
\begin{equation}
 \mathscr K_{\lambda,S}^{\rm loc}(x,z)
 :=\mathscr K_\lambda(x,z)
       \1_{\{d(x,z)\le\epsilon(1+D_x)\}}.
 \label{eq:localized-LED-kernel}
\end{equation}
Suppose that a bounded nonnegative function \(b^{\rm tail}\) satisfies, for every \(x\in\Gamma\),
\[
 2\mathcal Q(n_{f,x}^{(\lambda)})
 \le\bigl[C_b(1+\dist(x,S))^{-2\nu}+b_x^{\rm tail}\bigr]\1_{\mathfrak B}.
\]
Then, with the inverse defined by its norm-convergent Neumann series on \(\ell^\infty(\Gamma)\),
\begin{equation}
 0\le\mathcal Q(n_x)
 \le\bigl[C(1+\dist(x,S))^{-2\nu}
       +2((\1-\mathscr K_{\lambda,S}^{\rm loc})^{-1}b^{\rm tail})_x\bigr]
       \1_{\mathfrak B}.
 \label{eq:localized-density-with-tail}
\end{equation}
The constant \(C\) is independent of \(S,\lambda,\mathfrak B,\mathcal Q\), and \(b^{\rm tail}\).
For every \(0\le s\le2\nu\), the function \(w_s(x):=(1+D_x)^{-s}\) satisfies
\begin{equation}
 ((\1-\mathscr K_{\lambda,S}^{\rm loc})^{-1}w_s)_x
 \le\frac{(1+D_x)^{-s}}{1-\kappa(1-\epsilon)^{-2\nu}}.
 \label{eq:localized-kernel-profile}
\end{equation}
\end{lemma}

\begin{proof}
Fix \(\lambda,x,R\), and put \(B:=\ball_R(x)\).
The product-vacuum conditional expectation \eqref{eq:vacuum-slices} gives
\[
 \mathbb E_B^\vac(n_z)=
 \begin{cases}
 n_z,&z\in B,\\
 0,&z\notin B.
 \end{cases}
\]
Applying it to LED therefore gives
\[
 n_x\le2\mathbb E_B^\vac(n_{f,x}^{(\lambda)})
          +\sum_{z\in B}\mathscr K_\lambda(x,z)n_z.
\]
The original series converges in norm, so continuity justifies applying the expectation term by term.

For \(j\in\mathscr I\), write
\[
 a_j:=c_{f,x,j}^{(\lambda)},\qquad
 a_{j,R}:=\mathbb E_B^\vac(a_j),\qquad
 r_{j,R}:=a_j-a_{j,R}.
\]
The locality estimate \eqref{eq:coordinate-filter-norm-tail} gives
\[
 \|r_{j,R}\|\le2\dist(a_j,\A_B^\st)
                \le C(1+R)^{-\nu}.
\]
Since \(\mathbb E_B^\vac(r_{j,R})=0\), the bimodule property eliminates the mixed terms and yields
\[
 \mathbb E_B^\vac(a_j^*a_j)
 =a_{j,R}^*a_{j,R}
       +\mathbb E_B^\vac(r_{j,R}^*r_{j,R})\le2a_j^*a_j+3\|r_{j,R}\|^2\1.
\]
Here we used \(a_{j,R}=a_j-r_{j,R}\), the inequality \((A-B)^*(A-B)\le2A^*A+2B^*B\), and contractivity of the conditional expectation.
Summing over \(j\) proves \eqref{eq:localized-detectability}.

To prove \eqref{eq:localized-density-with-tail}, set \(u_x:=\|\mathcal Q(n_x)\|\le2r\) and apply \(\mathcal Q\) to \eqref{eq:localized-detectability} with \(R=\epsilon(1+D_x)\).
Since \(1+R\ge\epsilon(1+D_x)\), this gives
\[
 u_x\le (2C_b+C_{\rm loc}\epsilon^{-2\nu})(1+D_x)^{-2\nu}
          +2b_x^{\rm tail}
          +(\mathscr K_{\lambda,S}^{\rm loc}u)_x.
\]
The row sums of \(\mathscr K_{\lambda,S}^{\rm loc}\) are at most \(\kappa<1\), so the Neumann series for \((\1-\mathscr K_{\lambda,S}^{\rm loc})^{-1}\) converges in operator norm on \(\ell^\infty(\Gamma)\).
Iterating the inequality and using \(\|(\mathscr K_{\lambda,S}^{\rm loc})^n u\|_\infty\le\kappa^n\|u\|_\infty\to0\) yields
\begin{equation}
 u\le(2C_b+C_{\rm loc}\epsilon^{-2\nu})
             (\1-\mathscr K_{\lambda,S}^{\rm loc})^{-1}w_{2\nu}
       +2(\1-\mathscr K_{\lambda,S}^{\rm loc})^{-1}b^{\rm tail}.
 \label{eq:localized-density-inverse}
\end{equation}
For \(z\in\ball_{\epsilon(1+D_x)}(x)\), the triangle inequality gives
\[
 1+D_z\ge(1-\epsilon)(1+D_x).
\]
Thus, for \(0\le s\le2\nu\),
\[
 (\mathscr K_{\lambda,S}^{\rm loc}w_s)_x
 \le\kappa(1-\epsilon)^{-s}w_s(x).
\]
Since \(\mathscr K_{\lambda,S}^{\rm loc}(x,z)\ge0\), applying this kernel preserves pointwise inequalities, and iteration gives
\[
 ((\1-\mathscr K_{\lambda,S}^{\rm loc})^{-1}w_s)_x
 \le\sum_{n\ge0}\bigl[\kappa(1-\epsilon)^{-s}\bigr]^n w_s(x)
 \le\frac{w_s(x)}{1-\kappa(1-\epsilon)^{-2\nu}}.
\]
This proves \eqref{eq:localized-kernel-profile}.
Substituting \eqref{eq:localized-kernel-profile} with \(s=2\nu\) into \eqref{eq:localized-density-inverse}, and using \(0\le\mathcal Q(n_x)\le u_x\1_{\mathfrak B}\), proves \eqref{eq:localized-density-with-tail}.
\end{proof}

\subsection{Local state comparison in product-vacuum coordinates}

Lemma~\ref{lem:exact-atomic-reconstruction} bounds the difference of local expectations by the excitation number in the observable's support.
The following lemma extends this comparison to \(\sigma(A)\), assuming that for \(A\in\A_Y^\st\), the operator \(\sigma(A)\) can be approximated in \(\A_{Y^{(R)}}^\st\) with an error decaying polynomially or stretched-exponentially in \(R\).

\begin{lemma}
\label{lem:coordinate-reconstruction}
Let \(\omega_{\rm p}\) be the graded product of even pure onsite states on \(\A^\st\), with rank-one support projections \(P_x\in\A_{\{x\}}^\st\).
Let \(Q_x\in\A_{\{x\}}^\st\) be positive operators satisfying \(\1-P_x\le Q_x\).
Suppose that an automorphism \(\sigma\) satisfies, for every nonempty finite \(Y\Subset\Gamma\), \(A\in\A_Y^\st\), and \(R\ge0\),
\[
 \dist(\sigma(A),\A_{Y^{(R)}}^\st)
 \le C_\sigma |Y|\|A\|G_\sigma(R).
\]
In the polynomial case assume \(G_\sigma(R)=(1+R)^{-p}\), with \(p>d_{\rm vol}\), and set \(\alpha_{\rm rec}:=d_{\rm vol}/(2p)\).
In the stretched-exponential case assume \(G_\sigma(R)=\E^{-cR^\theta}\), with \(c>0\) and \(0<\theta<1\), and fix \(0<\alpha_{\rm rec}\le1/2\).
There is a constant \(C\), depending only on these parameters and the geometry, with the following property.
Let \(S\subset\Gamma\) be nonempty, let \(\varrho\) be a state on \(\A^\st\), and let \(H:[0,\infty)\to[0,\infty)\) be bounded and nonincreasing, with \(\varrho(Q_x)\le H(\dist(x,S))\) for every \(x\in\Gamma\).
For every nonempty finite \(Y\Subset\Gamma\) and \(A\in\A_Y^\st\), set \(D:=\dist(Y,S)\).
If \(D\ge4\), then
\begin{equation}
 |(\varrho-\omega_{\rm p})(\sigma(A))|
 \le C\|A\|\left[
 (1+|Y|)^{1/2+\alpha_{\rm rec}}H(D/2)^{1/2}
 +\min\{1,(1+|Y|)G_\sigma(D/4)\}\right].
 \label{eq:coordinate-reconstruction}
\end{equation}
Estimate~\eqref{eq:coordinate-reconstruction} holds on a single CAR algebra or a fixed finite stack.

Suppose also that, for every \(k\in\mathbb N\), nonempty finite \(Y\Subset\Gamma\), \(A\in\A_Y^\st\otimes M_k\), and \(R\ge0\),
\begin{equation}
 \dist\!\left((\sigma\otimes\operatorname{id}_{M_k})(A),
              \A_{Y^{(R)}}^\st\otimes M_k\right)
 \le C_\sigma|Y|\|A\|G_\sigma(R),
 \label{eq:reconstruction-amplified-locality}
\end{equation}
with the same constant \(C_\sigma\), independent of \(k\).
For every such \(k\), \(S,H\) as above, and state \(\varrho\) on \(\A^\st\otimes M_k\) satisfying \(\varrho(Q_x\otimes\1)\le H(\dist(x,S))\), define \(\varrho_{\rm anc}(T):=\varrho(\1\otimes T)\).
For every nonempty finite \(Y\Subset\Gamma\) with \(D\ge4\) and \(A\in\A_Y^\st\otimes M_k\),
\begin{equation}
 \begin{split}
 &\left|(\varrho-\omega_{\rm p}\otimes\varrho_{\rm anc})
                 ((\sigma\otimes\operatorname{id}_{M_k})(A))\right|\\
 &\qquad\le C\|A\|\left[
 (1+|Y|)^{1/2+\alpha_{\rm rec}}H(D/2)^{1/2}
 +\min\{1,(1+|Y|)G_\sigma(D/4)\}\right],
 \end{split}
 \label{eq:ancillary-coordinate-reconstruction}
\end{equation}
with \(C\) independent of \(k\).
Alternatively, if a state \(\varrho\) on \(\A^\st\otimes M_k\) satisfies \(\varrho(Q_x\otimes\1)\le v_x\) for a bounded \(v:\Gamma\to[0,\infty)\), then, for every nonempty finite \(Y\Subset\Gamma\) and \(A\in\A_Y^\st\otimes M_k\),
\begin{equation}
\left|(\varrho-\omega_{\rm p}\otimes\varrho_{\rm anc})
                  ((\sigma\otimes\operatorname{id}_{M_k})(A))\right|\le\min\left\{2,
 C(1+|Y|)^{1/2+\alpha_{\rm rec}}
             \|v\|_{\ell^\infty(\Gamma)}^{1/2}\right\}\|A\|.
 \label{eq:uniform-ancillary-coordinate-reconstruction}
\end{equation}
The scalar version is the case \(k=1\).
\end{lemma}

\begin{proof}
For finite \(X\Subset\Gamma\), set \(P_X:=\prod_{x\in X}P_x\).
These even projections commute, and for every \(C\in\A_X^\st\),
\[
 \1-P_X\le\sum_{x\in X}Q_x,\qquad
 P_XCP_X=\omega_{\rm p}(C)P_X.
\]
When \(\varrho(P_X)>0\), the normalized compressed state therefore agrees with \(\omega_{\rm p}\) on \(\A_X^\st\).
Estimate~\eqref{eq:vacuum-projection-state-distance} gives
\begin{equation}
 |\varrho(C)-\omega_{\rm p}(C)|
 \le2\|C\|\sqrt{\varrho(\1-P_X)}
 \le2\|C\|\left(\sum_{x\in X}\varrho(Q_x)\right)^{1/2}.
 \label{eq:general-product-reconstruction}
\end{equation}
For \(\varrho(P_X)=0\), the bound by \(2\|C\|\) gives the same inequality.

Use the dyadic tracial shells \(A_n:=\Delta_{Y,n}^{\rm tr}\sigma(A)\).
They are supported in \(Y^{(R_n)}\) and sum to \(\sigma(A)\) in norm.
Locality and contractivity give
\[
 \|A_n\|\le C\|A\|
 \begin{cases}
 \min\{1,(1+|Y|)(1+R_n)^{-p}\},&\text{polynomial case},\\
 \min\{1,(1+|Y|)\E^{-c'R_n^\theta}\},&\text{stretched-exponential case},
 \end{cases}
\]
where \(c'>0\) accounts for the ratio of consecutive dyadic radii.
We claim that
\begin{equation}
 \sum_{n\ge0}(1+R_n)^{d_{\rm vol}/2}\|A_n\|
 \le C (1+|Y|)^{\alpha_{\rm rec}}\|A\|.
 \label{eq:coordinate-weighted-shells}
\end{equation}
In the polynomial case with \(d_{\rm vol}>0\), use the norm bound below the scale \(1+R_n=(1+|Y|)^{1/p}\) and the locality bound above it.
The two contributions, after removing \(C\|A\|\), satisfy
\[
 \begin{aligned}
 \sum_{1+R_n\le (1+|Y|)^{1/p}}(1+R_n)^{d_{\rm vol}/2}
 &\le C (1+|Y|)^{d_{\rm vol}/(2p)},\\
 (1+|Y|)\sum_{1+R_n>(1+|Y|)^{1/p}}(1+R_n)^{d_{\rm vol}/2-p}
 &\le C (1+|Y|)^{d_{\rm vol}/(2p)}.
 \end{aligned}
\]
Both are geometric sums over dyadic radii.
If \(d_{\rm vol}=0\), volume growth implies \(|\Gamma|\le C_\Gamma\).
Thus \(1+|Y|\le1+C_\Gamma\), and the inequality \(\min\{1,u\}\le u^{1/2}\) gives a uniformly bounded sum, proving \eqref{eq:coordinate-weighted-shells} with \(\alpha_{\rm rec}=0\).
In the stretched-exponential case, use \(\min\{1,u\}\le u^{\alpha_{\rm rec}}\).
The remaining series \(\sum_n(1+R_n)^{d_{\rm vol}/2}\E^{-\alpha_{\rm rec} c'R_n^\theta}\) converges.

Let \(n_{\max}\) be the largest index with \(R_{n_{\max}}\le D/2\), so that \(D/4<R_{n_{\max}}\le D/2\).
For \(n\le n_{\max}\), every site of \(Y^{(R_n)}\) has distance at least \(D/2\) from \(S\).
Since \(|Y^{(R_n)}|\le C_\Gamma|Y|(1+R_n)^{d_{\rm vol}}\), \eqref{eq:general-product-reconstruction} and \eqref{eq:coordinate-weighted-shells} imply
\[
 \sum_{n=0}^{n_{\max}}|(\varrho-\omega_{\rm p})(A_n)|
 \le C (1+|Y|)^{1/2+\alpha_{\rm rec}}\|A\|H(D/2)^{1/2}.
\]
The omitted shells sum to \(\sigma(A)-\mathbb F^{\rm tr}_{Y^{(R_{n_{\max}})}}\sigma(A)\).
Their contribution is at most \(4\|A\|\), by contractivity, and at most \(C (1+|Y|)\|A\|G_\sigma(D/4)\), by locality.
Taking the smaller of these bounds proves \eqref{eq:coordinate-reconstruction}.

To prove \eqref{eq:ancillary-coordinate-reconstruction}, apply \eqref{eq:vacuum-projection-state-distance} in \(\A^\st\otimes M_k\), with the projection \(P_X\otimes\1\).
Suppose first that \(\varrho(P_X\otimes\1)>0\), and denote the normalized compressed state by \(\varrho^{P_X}\).
Because \(P_X\) has rank one on the local algebra, this state's restriction to \(\A_X^\st\otimes M_k\) is \(\omega_{\rm p}|_{\A_X^\st}\otimes\varrho^{P_X}_{\rm anc}\).
We must also compare the compressed ancillary marginal with the original one.
Restriction to the ancillary algebra gives
\[
 \|\varrho^{P_X}_{\rm anc}-\varrho_{\rm anc}\|_1
 \le \|\varrho^{P_X}-\varrho\|
 \le2\sqrt{\varrho((\1-P_X)\otimes\1)}.
\]
The triangle inequality yields
\begin{equation}
 \left\|\varrho|_{\A_X^\st\otimes M_k}
       -\omega_{\rm p}|_{\A_X^\st}\otimes\varrho_{\rm anc}\right\|_1
 \le4\sqrt{\varrho((\1-P_X)\otimes\1)}
 \le4\left(\sum_{x\in X}\varrho(Q_x\otimes\1)\right)^{1/2}.
 \label{eq:vacuum-ancillary-decoupling}
\end{equation}
When \(\varrho(P_X\otimes\1)=0\), the norm bound by two proves the same estimate.
Complete contractivity and amplified locality give \eqref{eq:coordinate-weighted-shells} for the shells of \((\sigma\otimes\operatorname{id}_{M_k})(A)\), uniformly in \(k\).
Apply \eqref{eq:vacuum-ancillary-decoupling} to each shell with \(R_n\le D/2\).
Their sum is bounded by \(C(1+|Y|)^{1/2+\alpha_{\rm rec}}\|A\|H(D/2)^{1/2}\), and the omitted shells by \(C\|A\|\min\{1,(1+|Y|)G_\sigma(D/4)\}\).
This proves \eqref{eq:ancillary-coordinate-reconstruction}.
To prove \eqref{eq:uniform-ancillary-coordinate-reconstruction}, apply \eqref{eq:vacuum-ancillary-decoupling} to every shell and sum, using volume growth and \eqref{eq:coordinate-weighted-shells}:
\[
 \begin{aligned}
 \left|(\varrho-\omega_{\rm p}\otimes\varrho_{\rm anc})
                  ((\sigma\otimes\operatorname{id}_{M_k})(A))\right| &\le4\sum_{n\ge0}\|A_n\|
                   \left(\sum_{x\in Y^{(R_n)}}v_x\right)^{1/2}\\
 & \le C|Y|^{1/2}\|v\|_{\ell^\infty(\Gamma)}^{1/2}
                \sum_{n\ge0}(1+R_n)^{d_{\rm vol}/2}\|A_n\|\\
 & \le C(1+|Y|)^{1/2+\alpha_{\rm rec}}
                  \|v\|_{\ell^\infty(\Gamma)}^{1/2}\|A\|.
 \end{aligned}
\]
This also proves absolute summability of the shell expectations.
The difference of two states has norm at most two, giving the other bound in \eqref{eq:uniform-ancillary-coordinate-reconstruction}.
\end{proof}

\begin{lemma}
\label{lem:positive-representation-cb}
Let \(\omega_{\rm p},Q_x,\sigma,G_\sigma\), and \(\alpha_{\rm rec}\) satisfy the assumptions of Lemma~\ref{lem:coordinate-reconstruction}, including \eqref{eq:reconstruction-amplified-locality}.
Let \(\mathfrak B\) be a unital \(C^*\)-algebra, let \(\mathcal Q:\A^\st\to\mathfrak B\) be unital and completely positive, and suppose that a bounded \(v:\Gamma\to[0,\infty)\) satisfies
\begin{equation}
 0\le\mathcal Q(Q_x)\le v_x\1_{\mathfrak B}
 \qquad(x\in\Gamma).
 \label{eq:compressed-excitation-density}
\end{equation}
For nonempty finite \(Y\Subset\Gamma\), define a map on \(\A_Y^\st\) by
\[
 \mathcal D_Y(A):=\mathcal Q(\sigma(A))
                   -\omega_{\rm p}(\sigma(A))\1_{\mathfrak B}.
\]
There is a constant \(C>0\), depending only on the parameters and locality bounds in Lemma~\ref{lem:coordinate-reconstruction}, such that the following bounds hold for every \(k\in\mathbb N\) and \(A\in\A_Y^\st\otimes M_k\).
In particular, \(C\) is independent of \(\mathfrak B\), \(\mathcal Q\), and \(k\).
\begin{enumerate}[label=(\roman*),leftmargin=*]
\item One has
\begin{equation}
 \| (\mathcal D_Y\otimes\operatorname{id}_{M_k})(A)\|
 \le \min\{2,C(1+|Y|)^{1/2+\alpha_{\rm rec}}
                  \|v\|_{\ell^\infty(\Gamma)}^{1/2}\}\|A\|.
 \label{eq:uniform-density-compression}
\end{equation}
\item Suppose that \(S\subset\Gamma\) is nonempty and that \(H:[0,\infty)\to[0,\infty)\) is bounded and nonincreasing, with \(v_x\le H(\dist(x,S))\) for every \(x\).
For \(D:=\dist(Y,S)\ge4\),
\begin{equation}
 \begin{split}
 \| (\mathcal D_Y\otimes\operatorname{id}_{M_k})(A)\|
 \le C\|A\|\bigl[&(1+|Y|)^{1/2+\alpha_{\rm rec}}H(D/2)^{1/2}\\
 &+\min\{1,(1+|Y|)G_\sigma(D/4)\}\bigr].
 \end{split}
 \label{eq:positive-representation-cb}
\end{equation}
\end{enumerate}
Both estimates also hold after restriction to a local unital \(C^*\)-subalgebra of \(\A_Y^\st\).
\end{lemma}

\begin{proof}
Fix a state \(\zeta\) on \(\mathfrak B\otimes M_k\), set
\(\varrho:=\zeta\circ(\mathcal Q\otimes\operatorname{id}_{M_k})\), and denote the ancillary marginal of \(\zeta\) by \(\zeta_{\rm anc}\).
Complete positivity and unitality make \(\varrho\) a state with the same ancillary marginal and with \(\varrho(Q_x\otimes\1)\le v_x\).
The definitions give
\begin{equation}
 \begin{split}
 \zeta\bigl((\mathcal D_Y\otimes\operatorname{id}_{M_k})(A)\bigr)
 ={}&(\varrho-\omega_{\rm p}\otimes\zeta_{\rm anc})\\
 &\bigl((\sigma\otimes\operatorname{id}_{M_k})(A)\bigr).
 \end{split}
 \label{eq:tail-amplified-expectation}
\end{equation}
Apply \eqref{eq:uniform-ancillary-coordinate-reconstruction} for (i) and \eqref{eq:ancillary-coordinate-reconstruction} for (ii).
The estimates are uniform over \(\zeta\) and \(k\).
For self-adjoint \(A\), taking the supremum over states \(\zeta\) gives the operator norm.
For arbitrary \(A\), apply the self-adjoint estimate at matrix size \(2k\) to \(\bigl(\begin{smallmatrix}0&A\\A^*&0\end{smallmatrix}\bigr)\).
The norm of this matrix is \(\|A\|\), and its image has norm \(\|(\mathcal D_Y\otimes\operatorname{id}_{M_k})(A)\|\), so no additional constant occurs.
Restriction to a subalgebra does not increase the norms.
\end{proof}

We now apply Lemma~\ref{lem:positive-representation-cb} to compressions of observables in a representation of the original algebra \(\A\).
Assume Hypotheses~\ref{hyp:background-transport} and~\ref{hyp:uniform-local-excitation-detectability} on \(\mathcal J\) and fix \(\lambda\in\mathcal J\).
Take \(\omega_{\rm p}=\omega_\at\), \(Q_x=n_x\), and \(\sigma=\sigma_\lambda\).
Estimate~\eqref{eq:coordinate-matrix-locality} implies \eqref{eq:reconstruction-amplified-locality} for \(\sigma=\sigma_\lambda\) and \(G_\sigma=G_\at\).
Thus \(\alpha_{\rm rec}:=d_{\rm vol}/(2p_\at)\) in the polynomial case, and we take \(\alpha_{\rm rec}:=q-1/2\) in the stretched-exponential case.
The constants in \eqref{eq:uniform-density-compression} and \eqref{eq:positive-representation-cb} therefore depend only on the model and LED data and \(q\).

Let \(\pi:\A\to\mathcal B(\mathcal H)\) be any unital representation, let \(P\) be an orthogonal projection on \(\mathcal H\), put \(\mathcal C:=P\mathcal H\), and let \(V:\mathcal C\to\mathcal H\) be the inclusion.
The case \(P=0\) is trivial, so suppose that \(P\ne0\).
Let \((\pi_{\rm aux},\mathcal H_{\rm aux},\Omega_{\rm aux})\) be the auxiliary GNS representation from Lemma~\ref{lem:stacked-positive-representation}.
The formula \eqref{eq:low-energy-stacked-representation} defines \(\pi_\st\) on \(\mathcal H\otimes\mathcal H_{\rm aux}\) for this arbitrary unital \(\pi\): its construction in that lemma uses only \(\pi\) and the even auxiliary state, not an implementing Hamiltonian.
The following map expresses this compression in product-vacuum coordinates:
\begin{equation}
 W\psi:=V\psi\otimes\Omega_{\rm aux},\qquad
 \mathcal Q(B):=W^*\pi_\st(\sigma_\lambda^{-1}(B))W
 \quad(B\in\A^\st).
 \label{eq:tail-amplification-maps}
\end{equation}
This is a unital completely positive map to \(\mathcal B(\mathcal C)\).
For a unit vector \(\psi\in\mathcal C\), define
\[
 \rho_\psi(A):=\langle V\psi,\pi(A)V\psi\rangle,
 \qquad
 \rho_{\psi,\at}:=(\rho_\psi\widehat\otimes\omega_{\rm aux})
                       \circ\sigma_\lambda^{-1}.
\]
Then \(\langle\psi,\mathcal Q(B)\psi\rangle=\rho_{\psi,\at}(B)\).
The condition \(\rho_{\psi,\at}(n_x)\le v_x\) for every unit vector \(\psi\in\mathcal C\) is equivalent to \eqref{eq:compressed-excitation-density}.
Let \(\iota:\A\to\A^\st\) be the first-layer embedding, \(\iota(A):=A\widehat\otimes\1\).
Composing the lemma's difference map with \(\iota\) gives a map on \(\A_Y\), which we denote by the same symbol:
\[
 \mathcal D_Y(A)=V^*\pi(A)V-\omega_\lambda(A)\1_{\mathcal C}
 \quad(A\in\A_Y),
\]
where the scalar identity follows from \eqref{eq:exact-atomic-state}.
Thus \eqref{eq:uniform-density-compression} and \eqref{eq:positive-representation-cb} apply to \(\mathcal D_Y\) on \(\A_Y\), with \(G_\sigma=G_\at\).
Conjugation by \(V\otimes\1\) identifies \((\mathcal D_Y\otimes\operatorname{id}_{M_k})(A)\) with
\[
 (P\otimes\1)\bigl[(\pi\otimes\operatorname{id}_{M_k})(A)
 -\1\otimes(\omega_\lambda\otimes\operatorname{id}_{M_k})(A)\bigr]
 (P\otimes\1),
\]
with the same norm, because \(VV^*=P\).

\begin{corollary}
\label{cor:filtered-reconstruction}
Assume Hypotheses~\ref{hyp:background-transport} and~\ref{hyp:uniform-local-excitation-detectability} on \(\mathcal J\) and fix \(\lambda\in\mathcal J\).
Let \(\varrho\) be a state on \(\A^\st\) and let \(b:\Gamma\to[0,\infty)\) be bounded, with \(2\varrho(n_{f,x}^{(\lambda)})\le b_x\) for every \(x\in\Gamma\).
Then, for every \(x\in\Gamma\),
\begin{equation}
 \varrho(n_x)\le[(\1-\mathscr K_\lambda)^{-1}b]_x.
 \label{eq:filtered-reconstruction-density}
\end{equation}
More generally, if \(\mathcal Q:\A^\st\to\mathfrak B\) is a unital positive map to a unital \(C^*\)-algebra and
\(2\mathcal Q(n_{f,x}^{(\lambda)})\le b_x\1_{\mathfrak B}\) for every \(x\), then
\begin{equation}
 0\le\mathcal Q(n_x)
 \le[(\1-\mathscr K_\lambda)^{-1}b]_x\1_{\mathfrak B}.
 \label{eq:filtered-reconstruction-operator-density}
\end{equation}
If \(\mathcal Q\) is completely positive, Lemma~\ref{lem:positive-representation-cb} applies with \(v_x=[(\1-\mathscr K_\lambda)^{-1}b]_x\), \(Q_x=n_x\), and \(\sigma=\sigma_\lambda\).

Fix \(C_b>0\) and, in the stretched-exponential case, \(\mu_b>0\).
There are constants \(C>0\) and, in that case, \(\mu_0,m>0\), depending only on these parameters, the model and LED data, and \(q\), such that the following holds uniformly in \(\lambda\), \(\mathfrak B\), and \(\mathcal Q\).
For any \(S\subset\Gamma\), including \(S=\varnothing\), suppose that
\begin{equation}
 b_x\le C_b
 \begin{cases}
 \E^{-\mu_b\dist(x,S)^\theta},&\text{stretched-exponential case},\\
 (1+\dist(x,S))^{-2\nu},&\text{polynomial case},
 \end{cases}
 \qquad x\in\Gamma,
 \label{eq:filtered-density-decay-assumption}
\end{equation}
with both profiles equal to zero at infinity.
Then
\begin{equation}
 0\le\mathcal Q(n_x)\le C
 \begin{cases}
 \E^{-\mu_0\dist(x,S)^\theta},&\text{stretched-exponential case},\\
 (1+\dist(x,S))^{-2\nu},&\text{polynomial case}
 \end{cases}
 \1_{\mathfrak B}.
 \label{eq:filtered-decay-density}
\end{equation}
If \(\mathcal Q\) is completely positive, then, for every nonempty finite \(Y\Subset\Gamma\), \(k\in\mathbb N\), and \(A\in\A_Y\otimes M_k\), with \(D:=\dist(Y,S)\),
\begin{equation}
 \begin{split}
 &\left\|((\mathcal Q\circ\sigma_\lambda\circ\iota)\otimes\operatorname{id}_{M_k})(A)
       -\1_{\mathfrak B}\otimes(\omega_\lambda\otimes\operatorname{id}_{M_k})(A)\right\|\\
 &\qquad\le C(1+|Y|)^q\|A\|
 \begin{cases}
 \E^{-mD^\theta},&\text{stretched-exponential case},\\
 (1+D)^{-\nu},&\text{polynomial case},
 \end{cases}
 \end{split}
 \label{eq:filtered-decay-comparison}
\end{equation}
where \(\iota(A)=A\widehat\otimes\1\).
For \(\mathfrak B=\mathbb C\), \(\mathcal Q=\varrho\), and \(k=1\), estimate~\eqref{eq:filtered-decay-comparison} bounds \(|\varrho(\sigma_\lambda(A\widehat\otimes\1))-\omega_\lambda(A)|\).
\end{corollary}

\begin{proof}
To prove \eqref{eq:filtered-reconstruction-operator-density}, set \(u_x:=\mathcal Q(n_x)\).
Positivity and unitality give \(0\le u_x\le2r\1_{\mathfrak B}\), and applying \(\mathcal Q\) to LED gives
\(u_x\le b_x\1_{\mathfrak B}+\sum_z\mathscr K_\lambda(x,z)u_z\).
The sums converge in norm, since the kernel has finite row sums and the \(u_z\) are uniformly bounded.
Iterating this inequality in the positive order yields
\[
 u_x\le\left[\sum_{n=0}^N\mathscr K_\lambda^n b\right]_x\1_{\mathfrak B}
           +[\mathscr K_\lambda^{N+1}u]_x.
\]
The last term has norm at most \(2r\vartheta_{\mathcal J}^{N+1}\), uniformly in \(x\), so it tends to zero.
This proves \eqref{eq:filtered-reconstruction-operator-density}; taking \(\mathfrak B=\mathbb C\) and \(\mathcal Q=\varrho\) gives \eqref{eq:filtered-reconstruction-density}.
For completely positive \(\mathcal Q\), apply \eqref{eq:uniform-density-compression} with \(v=(\1-\mathscr K_\lambda)^{-1}b\).

For the explicit decay estimates, first suppose that \(S\ne\varnothing\).
In the stretched-exponential case, set \(v:=(\1-\mathscr K_\lambda)^{-1}b\).
This bounded nonnegative function satisfies \(v=b+\mathscr K_\lambda v\).
Apply Lemma~\ref{lem:absorption}(i), with \(U=S^c\) and \(K=\mathscr K_\lambda\), using the weighted contraction \eqref{eq:LED-stretched-contraction}.
It gives \eqref{eq:filtered-decay-density} with \(\mu_0:=\min\{\mu_b,\mu_{\rm LED}\}\).
In the polynomial case, \eqref{eq:localized-density-with-tail} with \(b^{\rm tail}=0\) gives \eqref{eq:filtered-decay-density}.

Suppose that \(\mathcal Q\) is completely positive and \(S\ne\varnothing\).
For \(D\ge4\), apply Lemma~\ref{lem:positive-representation-cb}(ii), with \(H\) equal to the density profile in \eqref{eq:filtered-decay-density}, and restrict to the first layer.
In the polynomial case, taking its square root gives decay power \(\nu\) and support exponent \(1/2+d_{\rm vol}/(2p_\at)\le q\).
Since \(0<\nu\le p_\at\), the term \(\min\{1,(1+|Y|)G_\at(D/4)\}\) in \eqref{eq:positive-representation-cb} can be bounded using
\begin{equation}
 \min\{1,(1+|Y|)(1+D)^{-p_\at}\}
 \le (1+|Y|)^{\nu/p_\at}(1+D)^{-\nu}
 \le (1+|Y|)^q(1+D)^{-\nu}.
 \label{eq:support-tail-interpolation}
\end{equation}
The fixed rescalings \(D/2,D/4\) change only the constants.
In the stretched-exponential case, take \(\alpha_{\rm rec}=q-1/2\).
The term containing \(H(D/2)^{1/2}\) in \eqref{eq:positive-representation-cb} is at most \(C(1+|Y|)^q\|A\|\E^{-\mu_0D^\theta/2^{1+\theta}}\), and the other term satisfies
\[
 \min\{1,(1+|Y|)\E^{-\mu_\at(D/4)^\theta}\}
 \le (1+|Y|)^q\E^{-q\mu_\at D^\theta/4^\theta}.
\]
Choose \(m:=\min\{\mu_0/2^{1+\theta},q\mu_\at/4^\theta\}\).
This proves \eqref{eq:filtered-decay-comparison} for \(D\ge4\); for \(D<4\), use the norm bound \(2\|A\|\) and increase \(C\).
If \(S=\varnothing\), then \(b=0\), and \eqref{eq:filtered-reconstruction-operator-density} gives \(\mathcal Q(n_x)=0\) for every \(x\).
This proves \eqref{eq:filtered-decay-density} in that case, and Lemma~\ref{lem:positive-representation-cb}(i) gives exact equality with the bulk expectation.
\end{proof}

\section{Proofs of local determination}
\label{sec:local-determination-proofs}

We first prove Theorem~\ref{thm:positive-energy-locality} in positive-energy representations, together with its finite-sample consequences.
We then construct positive local energy operators for states satisfying the ground-state inequality only in \(U\) and use them to prove Theorem~\ref{thm:bulk-uniqueness}.

\subsection{Low-energy states in positive-energy representations}
\label{sec:low-energy-locality}

We first prove \eqref{eq:low-energy-filtered-square}, which bounds the expectations of \((c_{f,x,j}^{(\lambda)})^*c_{f,x,j}^{(\lambda)}\) in low-energy states expressed in product-vacuum coordinates.
LED then gives \eqref{eq:low-energy-density} for the expectations of \(n_x\), and \eqref{eq:filtered-decay-comparison} yields the estimate \eqref{eq:positive-energy-locality} for \(P(\pi(A)-\omega_\lambda(A)\1)P\).
The proof works in every positive-energy representation, and uniformly for \(0\le\varepsilon\le3g/4\).
It does not require the modified dynamics to be gapped or its energy window to have finite dimension.
The constants depend only on the model and LED data, because comparison with restricted bulk dynamics involves no uncontrolled terms supported in \(U^c\).

The spectral-window comparison uses a nonnegative Hamiltonian implementing the stacked dynamics and the fact that tensoring with the auxiliary ground vector preserves the energy bound.
We first explain why filtering the annihilation operators with the modified dynamics \(\tau^{\Xi,\st}\) makes them annihilate the stacked low-energy vectors.

Let \(\pi:\A\to\mathcal B(\mathcal H)\) be a unital representation, and let a nonnegative self-adjoint operator \(L\) implement \(\tau^\Xi\) in this representation.
Fix \(0\le\varepsilon\le3g/4\), and let \(\psi\) be a unit vector in \(\1_{[0,\varepsilon]}(L)\mathcal H\).
Use the stacked representation \(\pi_\st\) and Hamiltonian \(\mathbb L\) from Lemma~\ref{lem:stacked-positive-representation}.
The vector \(\psi\otimes\Omega_{\rm aux}\) represents \(\rho_\st:=\rho\widehat\otimes\omega_{\rm aux}\), where \(\rho(A):=\langle\psi,\pi(A)\psi\rangle\), and lies in \(\1_{[0,\varepsilon]}(\mathbb L)(\mathcal H\otimes\mathcal H_{\rm aux})\).
Lemma~\ref{lem:low-energy-annihilation} therefore gives, for every \(x\in\Gamma\) and \(j\in\mathscr I\),
\begin{equation}
 \pi_\st\!\left(
 \tau_f^{\Xi,\st}(\widetilde c_{x,j}^{\,(\lambda)})\right)
 (\psi\otimes\Omega_{\rm aux})=0.
 \label{eq:low-energy-stacked-annihilation}
\end{equation}
The norm integral defining the algebraic filter intertwines with the strong operator integral in the lemma.
Lemma~\ref{lem:common-filtered-occupation} uses this annihilation and comparison with the restricted bulk dynamics to bound the bulk-filtered fields on the same vectors.

The comparison criterion \eqref{eq:comparison-interaction-criterion} includes \(\Xi=\Phi_\lambda+\Psi\) with \(\Sigma(\Psi)\subset S\), taking \(U=S^c\).
It also includes every cut interaction
\begin{equation}
 \Xi(Z):=\1_{\{Z\subset\Lambda\}}\Phi_\lambda(Z)+\Upsilon(Z),
 \qquad \Sigma(\Upsilon)\subset C_{\Lambda,w},\qquad
 U:=\Lambda\setminus C_{\Lambda,w}=S_{\Lambda,w}^c,
 \label{eq:cut-comparison-interaction}
\end{equation}
where \(\varnothing\ne\Lambda\subsetneq\Gamma\), \(w\ge0\), and \(\Upsilon\in\mathfrak J_1\).
Indeed, terms supported in \(U\) agree with the bulk.
Terms crossing its boundary are unchanged if their supports lie in \(\Lambda\) and zero otherwise, while the boundary interaction is confined to \(U^c\).
The first support moment is finite because \(\|\Xi\|_1\le\|\Phi_\lambda\|_1+\|\Upsilon\|_1\).
For a finite sample with boundary operator \(K\), take \(\Upsilon(C_{\Lambda,w})=K\) and zero elsewhere when the collar is nonempty; an empty collar contributes only a scalar, which does not affect the dynamics.

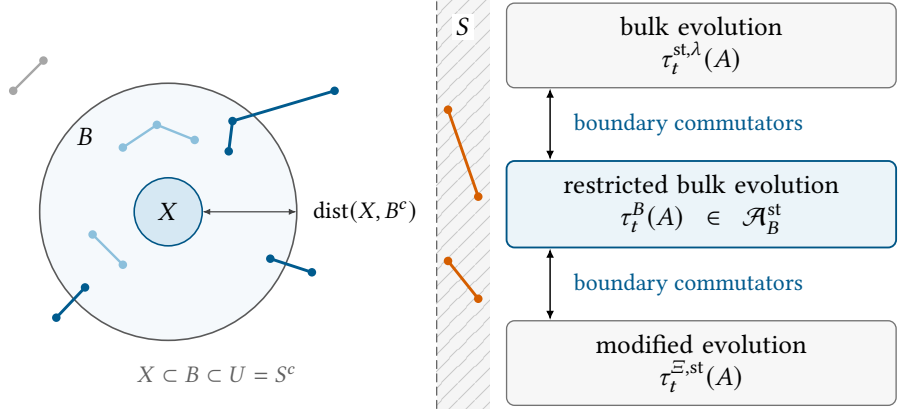
\begin{figure}[htbp]
\centering
\begin{tikzpicture}[lppl diagram]
 \path[excluded] (6.15,.4) rectangle (6.85,5.8);
 \draw[black!60,dash pattern=on 3pt off 2pt] (6.15,.4) -- (6.15,5.8);
 \node[fill=white,inner sep=1.5pt] at (6.5,5.45) {$S$};
 \foreach \a/\b in {{6.3,4.35}/{6.7,3.2},{6.3,2.35}/{6.7,1.85}}{
  \draw[lppl defect term] (\a) -- (\b);
  \node[lppl term site,fill=aux] at (\a) {};
  \node[lppl term site,fill=aux] at (\b) {};}
 \path[lppl comparison region] (2.6,3) circle[radius=1.7];
 \path[observable] (2.6,3) circle[radius=.45];
 \node at (2.6,3) {$X$};
 \node at (1.5,4) {$B$};
 \draw[dim line] (3.05,3) -- (4.3,3);
 \node[font=\footnotesize,anchor=west] at (4.38,3) {$\dist(X,B^c)$};
 \draw[lppl bulk term] (2.0,3.85) -- (2.45,4.15) -- (2.95,3.95);
 \foreach \p in {(2.0,3.85),(2.45,4.15),(2.95,3.95)}
  \node[lppl term site,fill=phys!45] at \p {};
 \foreach \a/\b in {{1.6,2.7}/{2.0,2.3}}{
  \draw[lppl bulk term] (\a) -- (\b);
  \node[lppl term site,fill=phys!45] at (\a) {};
  \node[lppl term site,fill=phys!45] at (\b) {};}
 \draw[lppl boundary term] (3.4,3.8) -- (3.45,4.2) -- (4.8,4.6);
 \foreach \p in {(3.4,3.8),(3.45,4.2),(4.8,4.6)}
  \node[lppl term site,fill=phys!80!black] at \p {};
 \foreach \a/\b in {{1.5,2}/{1.12,1.6},{3.95,2.38}/{4.5,2.2}}{
  \draw[lppl boundary term] (\a) -- (\b);
  \node[lppl term site,fill=phys!80!black] at (\a) {};
  \node[lppl term site,fill=phys!80!black] at (\b) {};}
 \draw[lppl exterior term] (.55,4.6) -- (.95,5);
 \node[lppl term site,fill=black!35] at (.55,4.6) {};
 \node[lppl term site,fill=black!35] at (.95,5) {};
 \node[fig note] at (3.25,.85) {$X\subset B\subset U=S^c$};
 \begin{scope}[shift={(9.65,0)}]
  \node[lppl evolution] (bulk) at (0,5.2)
    {bulk evolution\\$\tau_t^{\st,\lambda}(A)$};
  \node[lppl restricted evolution] (restricted) at (0,3.1)
    {restricted bulk evolution\\$\tau_t^B(A)\in\A_B^\st$};
  \node[lppl evolution] (modified) at (0,1)
    {modified evolution\\$\tau_t^{\Xi,\st}(A)$};
  \foreach \u/\v in {bulk/restricted,restricted/modified}{
   \draw[lppl comparison arrow]
     ($(\u.south)+(-2,0)$) -- ($(\v.north)+(-2,0)$)
     node[midway,right=5pt,lppl comparison label]
       {boundary commutators};}
 \end{scope}
\end{tikzpicture}
\caption{Comparison with a common restricted dynamics.
The stacked bulk dynamics \(\tau_t^{\st,\lambda}=\tau_t^{\Phi_\lambda}\widehat\otimes\tau_t^{\rm aux}\) combines the physical evolution with the auxiliary free evolution.
For a defect confined to \(S=U^c\), the modified dynamics \(\tau_t^{\Xi,\st}\) replaces \(\Phi_\lambda\) by \(\Xi=\Phi_\lambda+\Psi\), where \(\Sigma(\Psi)\subset S\) (orange terms), and leaves the auxiliary evolution unchanged.
For \(A\in\A_X^\st\) and \(X\subset B\subset U\), restricting the stacked bulk interaction to terms supported in \(B\) (light blue) gives \(\tau_t^B:=\tau_t^{\Phi_\lambda^\st,B}\), with \(\tau_t^B(A)\in\A_B^\st\).
The orange defect terms supported in \(S\) and the gray bulk term supported in \(B^c\) are even and therefore commute with \(\tau_t^B(A)\).
Both Duhamel comparisons thus involve only bulk terms crossing the boundary of \(B\), shown in dark blue.
For a finite sample, \(\Xi\) also omits bulk terms not contained in the sample.
The comparison of the modified evolution with \(\tau^B\) uses only the crossing terms present in \(\Xi^\st\), whereas the comparison of the bulk evolution with \(\tau^B\) uses all bulk crossing terms, including those deleted from \(\Xi^\st\).}
\label{fig:restricted-comparison}
\end{figure}

Figure~\ref{fig:restricted-comparison} illustrates the comparison of \(\tau^{\st,\lambda}\) and \(\tau^{\Xi,\st}\) with the common restricted dynamics \(\tau_t^B:=\tau_t^{\Phi_\lambda^\st,B}\), for a finite region \(B\subset U=S^c\) and a defect confined to \(S\).
\FloatBarrier

\begin{lemma}
\label{lem:positive-representation-locality}
Assume Hypotheses~\ref{hyp:background-transport} and~\ref{hyp:uniform-local-excitation-detectability} on \(\mathcal J\).
In the polynomial case, use \(\nu\) from \eqref{eq:main-polynomial-range}.
There are constants \(C>0\) and, in the stretched-exponential case, \(\mu_{\rm filt},m_1>0\), depending only on the model and LED data, with the following property.
For every \(\lambda\in\mathcal J\), \(U\subset\Gamma\), and \(\Xi\in\mathfrak J_1\) satisfying \eqref{eq:comparison-interaction-criterion}, let \(\pi:\A\to\mathcal B(\mathcal H)\) be a unital representation in which a nonnegative self-adjoint \(L\) implements \(\tau^\Xi\).
For every unit vector \(\psi\in\1_{[0,3g/4]}(L)\mathcal H\), define the states \(\rho\) and \(\rho_\at\) by setting, for \(A\in\A\),
\[
 \rho(A):=\langle\psi,\pi(A)\psi\rangle,\qquad
 \rho_\at:=(\rho\widehat\otimes\omega_{\rm aux})
                        \circ\sigma_\lambda^{-1}.
\]
For every \(x\in\Gamma\) and \(j\in\mathscr I\),
\begin{equation}
 \rho_\at\!\left(
 (c_{f,x,j}^{(\lambda)})^*c_{f,x,j}^{(\lambda)}\right)
 \le
 \begin{cases}
 C\E^{-\mu_{\rm filt}\dist(x,U^c)^\theta},&\text{stretched-exponential case},\\
 C(1+\dist(x,U^c))^{-2\nu},&\text{polynomial case},
 \end{cases}
 \label{eq:low-energy-filtered-square}
\end{equation}
and
\begin{equation}
 \rho_\at(n_x)\le
 \begin{cases}
 C\E^{-m_1\dist(x,U^c)^\theta},&\text{stretched-exponential case},\\
 C(1+\dist(x,U^c))^{-2\nu},&\text{polynomial case}.
 \end{cases}
 \label{eq:low-energy-density}
\end{equation}
\end{lemma}

\begin{proof}
Use the stacked representation and nonnegative Hamiltonian \(\mathbb L\) from Lemma~\ref{lem:stacked-positive-representation}.
The vector \(\psi\otimes\Omega_{\rm aux}\) belongs to \(\1_{[0,3g/4]}(\mathbb L)(\mathcal H\otimes\mathcal H_{\rm aux})\).
For any finite \(B\subset U\), the comparison criterion implies \eqref{eq:local-comparison-discrepancy} on the stack, with \(Q=\1\), \(L=\mathbb L\), and the componentwise crossing sums of \eqref{eq:stacked-crossing-norms}.
For local \(D\), the map \(t\mapsto\E^{\I t\mathbb L}\pi_\st(D)\E^{-\I t\mathbb L}\) is norm differentiable at zero, with derivative \(\I\pi_\st(\delta_{\Xi^\st}(D))\).
The product rule applied to its action on \(\E^{\I t\mathbb L}\zeta\), for \(\zeta\in\operatorname{Dom}\mathbb L\), gives
\[
 \pi_\st(D)\zeta\in\operatorname{Dom}\mathbb L,\qquad
 [\mathbb L,\pi_\st(D)]\zeta
 =\pi_\st(\delta_{\Xi^\st}(D))\zeta.
\]
These are the domain and commutator conditions of Lemma~\ref{lem:local-spectral-comparison}.

Fix \(x,j\), and first suppose that \(32\le D:=\dist(x,U^c)<\infty\).
Take \(R:=D/2\), so that \(B=\ball_R(x)\subset U\).
Apply Lemma~\ref{lem:common-filtered-occupation} with \(\pi=\pi_\st\), \(Q=\1\), \(L=\mathbb L\), and \(\Psi=0\).
Since
\[
 \rho_\at\bigl((c_{f,x,j}^{(\lambda)})^*c_{f,x,j}^{(\lambda)}\bigr)
 =\|\pi_\st(\widetilde c_{f,x,j}^{\,(\lambda)})(\psi\otimes\Omega_{\rm aux})\|^2,
\]
the estimate \eqref{eq:common-filtered-occupation} gives \eqref{eq:low-energy-filtered-square}.
For \(D<32\), use \(\|c_{f,x,j}^{(\lambda)}\|\le\|f\|_{L^1(\R)}\).
If \(U=\Gamma\), then \(\Xi=\Phi_\lambda\), and \eqref{eq:low-energy-stacked-annihilation} gives zero.
Set \(b_x:=2\rho_\at(n_{f,x}^{(\lambda)})\).
Since \(|\mathscr I|=2r\), summing \eqref{eq:low-energy-filtered-square} gives the bound on \(b_x\) required by Corollary~\ref{cor:filtered-reconstruction} with \(S=U^c\), \(C_b=4rC\), and, in the stretched-exponential case, \(\mu_b=\mu_{\rm filt}\).
Its density estimate \eqref{eq:filtered-decay-density}, applied to \(\mathcal Q=\rho_\at\), proves \eqref{eq:low-energy-density}.
\end{proof}

\begin{proof}[Proof of Theorem~\ref{thm:positive-energy-locality}]
Fix \(\lambda,U,\Xi,\pi,L\) satisfying the hypotheses of the theorem and put \(P:=\1_{[0,3g/4]}(L)\).
If \(P=0\), the compressed operator is zero.
Otherwise use the map \(\mathcal Q:\A^\st\to\mathcal B(P\mathcal H)\) from \eqref{eq:tail-amplification-maps}, which expresses compression to \(P\mathcal H\) in product-vacuum coordinates.
For every unit vector \(\psi\in P\mathcal H\), one has
\[
 \langle\psi,\mathcal Q(n_{f,x}^{(\lambda)})\psi\rangle
 =\rho_{\psi,\at}(n_{f,x}^{(\lambda)}).
\]
Sum \eqref{eq:low-energy-filtered-square} over \(j\in\mathscr I\) and take the supremum over \(\psi\).
The result is the operator hypothesis \(2\mathcal Q(n_{f,x}^{(\lambda)})\le b_x\1_{P\mathcal H}\) of Corollary~\ref{cor:filtered-reconstruction} with \(S=U^c\), where \(b_x\) is bounded by \(4rC\) times the profile in \eqref{eq:low-energy-filtered-square}.
For the map \(\mathcal Q\) in \eqref{eq:tail-amplification-maps}, the amplified difference in \eqref{eq:filtered-decay-comparison} is precisely the compression by \(P\otimes\1\), as shown after its definition.
This gives the matrix estimate in Theorem~\ref{thm:positive-energy-locality}.
Taking matrix dimension one gives \eqref{eq:positive-energy-locality}.
The constants depend only on the model and LED data and \(q\), and the corollary also covers \(U=\Gamma\) by taking \(S=\varnothing\).
\end{proof}

\begin{proof}[Proof of Corollary~\ref{thm:low-energy-locality}]
Fix \(\lambda\in\mathcal J\), \(S\subset\Gamma\), and nonempty \(Y\Subset\Gamma\).
Let \(\Psi\in\mathfrak J_1\) satisfy \(\Sigma(\Psi)\subset S\), and let \(\omega_\Psi\) be a ground state of \(\Phi_\lambda+\Psi\).
The GNS Hamiltonian \(L_{\omega_\Psi}\) is nonnegative and implements \(\tau^{\Phi_\lambda+\Psi}\), by Subsection~\ref{sec:ground-state-conditions}.
Apply Theorem~\ref{thm:positive-energy-locality} with \(U:=S^c\), \(\Xi:=\Phi_\lambda+\Psi\), \(\pi:=\pi_{\omega_\Psi}\), and \(L:=L_{\omega_\Psi}\).
Theorem~\ref{thm:positive-energy-locality} gives the bounds for matrices of local observables with constants independent of the matrix size \(k\).
Taking \(k=1\) gives the bound for a single local observable.
The constants are independent of \(\Psi\) and \(\omega_\Psi\); taking the two suprema proves \eqref{eq:low-energy-locality} in both decay regimes.
The cyclic vector satisfies \(L_{\omega_\Psi}\Omega_{\omega_\Psi}=0\), and hence \(P_{\omega_\Psi}\Omega_{\omega_\Psi}=\Omega_{\omega_\Psi}\).
Taking its expectation in the scalar compression bound and then the same suprema proves \eqref{eq:ground-state-lppl}.
\end{proof}

\begin{proof}[Proof of Corollary~\ref{cor:finite-sample-bulk}]
Fix \(\lambda,\Lambda,w,K\) as in the corollary, and let \(\Xi\) be the cut interaction in \eqref{eq:cut-comparison-interaction} associated with this sample.
Since \(\Xi\) has only finitely many nonzero terms, its dynamics is inner:
\[
 H_{\Lambda,K}^{(\lambda)}-\sum_Z\Xi(Z)\in\C\1,
 \qquad
 \tau_t^\Xi=\operatorname{Ad}(\E^{\I tH_{\Lambda,K}^{(\lambda)}})
 \quad\text{on }\A.
\]
Choose a faithful unital representation \(\pi\) of \(\A\), and set
\[
 L:=\pi(H_{\Lambda,K}^{(\lambda)}-E_{\Lambda,K}^{(\lambda)}\1)\ge0,
 \qquad
 \1_{[0,3g/4]}(L)=\pi(P_{\Lambda,K}^{(\lambda)}).
\]
The operator \(L\) implements \(\tau^\Xi\), so Theorem~\ref{thm:positive-energy-locality} applies with \(U=\Lambda\setminus C_{\Lambda,w}=S_{\Lambda,w}^c\).
Faithfulness of \(\pi\otimes\operatorname{id}_{M_k}\) identifies the norm of its compressed operator with the sample compression norm, proving the matrix estimates uniformly in \(k\).
Taking \(k=1\) gives \eqref{eq:finite-sample-bulk} in both locality regimes.
For a state \(\rho\) supported in \(P_{\Lambda,K}^{(\lambda)}\),
\[
 |\rho(A)-\omega_\lambda(A)|
 \le\|P_{\Lambda,K}^{(\lambda)}
          (A-\omega_\lambda(A)\1)P_{\Lambda,K}^{(\lambda)}\|,
\]
which gives the expectation estimates.
The constants are independent of \(K\), so the bounds hold after taking the suprema.
\end{proof}

\begin{proof}[Proof of \eqref{eq:infinite-region-low-energy}]
Use the regional ground-state representation \((\pi_\Lambda,\mathcal H_\Lambda,\Omega_\Lambda)\) defined in Subsection~\ref{sec:finite-samples}.
Let \((\pi_{\rm ext},\mathcal H_{\rm ext},\Omega_{\rm ext})\) be the GNS representation of the tracial state on \(\A_{\Lambda^c}\).
The exterior tracial state is even, so its canonical parity unitary \(\mathcal P_{\rm ext}\) fixes \(\Omega_{\rm ext}\).
Repeat the representation construction in the proof of Lemma~\ref{lem:stacked-positive-representation}, using the regional system and the exterior tracial state with zero exterior implementing Hamiltonian.
For homogeneous \(B\in\A_\Lambda\) and \(C\in\A_{\Lambda^c}\), this gives
\[
 \pi^\Gamma(B\widehat\otimes C)
 =\pi_\Lambda(B)\otimes\mathcal P_{\rm ext}^{|B|}\pi_{\rm ext}(C),
 \qquad L^\Gamma=L_\Lambda\otimes\1\ge0.
\]
The operator \(L^\Gamma\) implements \(\tau^{\Phi_\lambda^\Lambda+\Upsilon}\) by Proposition~\ref{prop:first-moment-dynamics}.
The isometry \(V\psi:=\psi\otimes\Omega_{\rm ext}\) satisfies, for every \(A\in\A_\Lambda\),
\[
 \pi^\Gamma(A)V=V\pi_\Lambda(A),\qquad
 \1_{[0,3g/4]}(L^\Gamma)V=VP_\Lambda.
\]
Apply Theorem~\ref{thm:positive-energy-locality} to \(\pi^\Gamma,L^\Gamma\), using the cut interaction \eqref{eq:cut-comparison-interaction} with \(U:=\Lambda\setminus C_{\Lambda,w}=S_{\Lambda,w}^c\).
Compress the estimate from Theorem~\ref{thm:positive-energy-locality} by the restriction of \(V\) to \(P_\Lambda\mathcal H_\Lambda\), and take matrix dimension one to obtain \eqref{eq:infinite-region-low-energy}.
All comparison constants are independent of \(\Upsilon\).
\end{proof}

\subsection{Positive local energy operators}
\label{sec:local-positive-energy}

The states considered in this subsection satisfy a ground-state inequality only inside \(U\), and need not be globally stationary.
We use the filtering and local state-comparison arguments from the proof for low-energy states.
We construct a positive operator \(L_{\rho_\st,\Lambda}\) on the local cyclic subspace with \(L_{\rho_\st,\Lambda}\Omega_{\rho_\st}=0\).
Comparing the evolution generated by this operator with the bulk dynamics gives \eqref{eq:general-local-positive-filter} for every \(A\in\A^\st_\Lambda\).

We first lift the physical local ground-state inequality to the two-layer stack.
Stacking with the even auxiliary ground state preserves that inequality on the same region.

\begin{lemma}
\label{lem:graded-product}
Let \(\delta_1\) be a derivation on \(\A\) satisfying \(\A_\loc\subset\operatorname{Dom}(\delta_1)\) and \(\delta_1\circ\Theta=\Theta\circ\delta_1\), and let \(\delta_2\) be a derivation on \(\A^{\rm aux}\) satisfying \(\A_\loc^{\rm aux}\subset\operatorname{Dom}(\delta_2)\) and \(\delta_2\circ\Theta_{\rm aux}=\Theta_{\rm aux}\circ\delta_2\).
Define the stacked local derivation \(\delta_{1,2}^\st\) on homogeneous local simple tensors by
\begin{equation*}
 \delta_{1,2}^\st(A\widehat\otimes B)
 :=\delta_1(A)\widehat\otimes B
   +A\widehat\otimes\delta_2(B),
\end{equation*}
and extend it linearly to \(\A_\loc^\st\).
Let \(U\subset\Gamma\) and   \(\rho_1\in\LG_{\delta_1}(U)\), and let \(\rho_2\in \mathfrak S(\A^{\rm aux})\) satisfy \(\rho_2\circ\Theta_{\rm aux}=\rho_2\) and, for every \(B\in\A_\loc^{\rm aux}\),
\begin{equation*}
 \operatorname{Re}\rho_2(B^*\delta_2(B))\ge0.
\end{equation*}
Then
\begin{equation}
 \rho_1\widehat\otimes\rho_2
 \in\LG_{\delta_{1,2}^\st}^\st(U).
 \label{eq:graded-product-ground}
\end{equation}
\end{lemma}

\begin{proof}
The graded product exists because \(\rho_2\) is even.
Fix \(X\Subset U\).
On \(\A_X\) for \(i=1\) and on \(\A_X^{\rm aux}\) for \(i=2\), define the Hermitian forms
\begin{align*}
 g_i(A,B)&:=\rho_i(A^*B),\\
 q_i(A,B)&:=\tfrac12\bigl[
 \rho_i(A^*\delta_i(B))+\rho_i(\delta_i(A)^*B)\bigr].
\end{align*}
Positivity of \(\rho_i\) gives \(g_i(A,A)\ge0\), while the local ground-state inequality gives \(q_i(A,A)=\operatorname{Re}\rho_i(A^*\delta_i(A))\ge0\).
Tensor products of positive semidefinite forms are positive semidefinite, so the form
\[
 Q:=q_1\otimes g_2+g_1\otimes q_2
\]
is positive semidefinite on the ordinary algebraic tensor product \(\A_X\otimes\A_X^{\rm aux}\).

Define a linear map \(T:\A_X^\st\to\A_X\otimes\A_X^{\rm aux}\) by
\[
 T(A\widehat\otimes B):=\Theta^{|B|}(A)\otimes B
\]
for homogeneous \(B\), and extend linearly.
For every \(C\in\A_X^\st\), the graded multiplication rules give
\[
 \operatorname{Re}(\rho_1\widehat\otimes\rho_2)
 \bigl(C^*\delta_{1,2}^\st(C)\bigr)
 =Q(TC,TC)\ge0.
\]
To check the equality, expand \(C=\sum_k A_k\widehat\otimes B_k\) into homogeneous factors.
Terms with \(|B_k|\ne|B_\ell|\) vanish on both sides because \(\rho_2\) is even and \(\delta_2\) preserves parity.
For \(|B_k|=|B_\ell|=b\), the graded sign is \((-1)^{b(|A_k|+|A_\ell|)}\), which is the sign produced by applying \(\Theta^b\) to both physical factors, since \(\delta_1\) preserves parity.
This proves \eqref{eq:graded-product-ground}.
\end{proof}

For \(|\lambda|\le\lambda_0\), \(U\subset\Gamma\), and \(\rho\in\LG_{\Phi_\lambda}(U)\), define the stacked state \(\rho_\st:=\rho\widehat\otimes\omega_{\rm aux}\).
According to Lemma~\ref{lem:graded-product}, it satisfies the stacked local ground-state inequality on the same region \(U\):
\begin{equation}
 \rho_\st\in\LG_{\delta_\lambda^\st}^\st(U).
 \label{eq:stacked-local-ground}
\end{equation}

\paragraph{Construction of the positive local energy operator.}

A state in \(\LG_\Phi(U)\) need not be invariant under the infinite-volume dynamics.
It therefore need not have a GNS Hamiltonian implementing that dynamics.
Yarotsky gives a finite-dimensional positive-Hamiltonian realization of a state satisfying the local ground-state inequality \cite[Appendix]{Yarotsky05}.
The following lemma constructs such a positive operator on a finite-dimensional local cyclic subspace and identifies its compressed commutators.
The construction applies to any fixed finite number of fermionic modes per site, so it covers both the physical algebra \(\A\) and the stacked algebra \( \A^\st\).

\begin{lemma}
\label{lem:local-positive-generator}
For \(r_{\mathcal B}\in\mathbb N\), let
\[
 \mathcal B:=\operatorname{CAR}(\ell^2(\Gamma;\C^{r_{\mathcal B}})),
 \qquad
 \mathcal B_X:=\operatorname{CAR}(\ell^2(X;\C^{r_{\mathcal B}}))
 \quad(X\subset\Gamma).
\]
Let \(\Phi\) be an interaction on \(\mathcal B\) with \(\|\Phi\|_1<\infty\), let \(U\subset\Gamma\), and let \(\rho\in\LG_\Phi(U)\).
Denote its GNS representation by \((\pi_\rho,\mathcal H_\rho,\Omega_\rho)\).
For a nonempty finite \(\Lambda\subset U\), define
\[
 \mathcal H_{\rho,\Lambda}:=
       \pi_\rho(\mathcal B_\Lambda)\Omega_\rho,
 \qquad
 P_{\rho,\Lambda}:=
       \text{the orthogonal projection onto }\mathcal H_{\rho,\Lambda}.
\]
The subspace \( \mathcal H_{\rho,\Lambda}\) is finite dimensional and reduces \(\pi_\rho(\mathcal B_\Lambda)\).
There is a unique operator \(L_{\rho,\Lambda}\) on \(\mathcal H_{\rho,\Lambda}\) such that, for all \(A,B\in\mathcal B_\Lambda\),
\begin{equation}
 \langle\pi_\rho(A)\Omega_\rho,
          L_{\rho,\Lambda}\pi_\rho(B)\Omega_\rho\rangle
 =\rho(A^*\delta_\Phi(B)).
 \label{eq:local-positive-form}
\end{equation}
Equivalently, for every \(B\in\mathcal B_\Lambda\),
\[
 L_{\rho,\Lambda}\pi_\rho(B)\Omega_\rho
 =P_{\rho,\Lambda}\pi_\rho(\delta_\Phi(B))\Omega_\rho.
\]
It is self-adjoint and satisfies
\begin{equation}
 L_{\rho,\Lambda}\ge0,\qquad
 L_{\rho,\Lambda}\Omega_\rho=0.
 \label{eq:local-positive-energy}
\end{equation}
For every \(A\in\mathcal B_\Lambda\), its commutator is
\begin{equation}
[L_{\rho,\Lambda},\pi_\rho(A)]
=P_{\rho,\Lambda}\pi_\rho(\delta_\Phi(A))P_{\rho,\Lambda}
\quad\text{on }\mathcal H_{\rho,\Lambda}.
 \label{eq:local-positive-commutator}
\end{equation}
\end{lemma}

\begin{proof}
The argument with \(\1+zB\) in the proof of Lemma~\ref{lem:local-ground-characterization}, now restricted to \(\mathcal B_\Lambda\), gives \(\rho(\delta_\Phi(B))=0\) there.
Applying this identity to \(A^*B\) shows that \(\mathfrak q(A,B):=\rho(A^*\delta_\Phi(B))\) is Hermitian, and the local ground-state inequality makes it nonnegative.
If \(\pi_\rho(B)\Omega_\rho=0\), then
\[
 \mathfrak q(B,A)
 =\langle\pi_\rho(B)\Omega_\rho,
          \pi_\rho(\delta_\Phi(A))\Omega_\rho\rangle=0,
\]
and Hermiticity also gives \(\mathfrak q(A,B)=0\).
Thus the form descends to the finite-dimensional space \(\mathcal H_{\rho,\Lambda}\), where it defines the unique positive operator in \eqref{eq:local-positive-form}.
The spanning vectors \(\pi_\rho(B)\Omega_\rho\) give its equivalent action formula, and \(\delta_\Phi(\1)=0\) gives \(L_{\rho,\Lambda}\Omega_\rho=0\).

The space \(\mathcal H_{\rho,\Lambda}\) is invariant under \(\pi_\rho(\mathcal B_\Lambda)\) and its adjoints, so it is reducing.
For \(A,B,C\in\mathcal B_\Lambda\), the derivation rule gives
\[
 \langle\pi_\rho(C)\Omega_\rho,
 [L_{\rho,\Lambda},\pi_\rho(A)]\pi_\rho(B)\Omega_\rho\rangle
 =\rho(C^*\delta_\Phi(AB))-\rho(C^*A\delta_\Phi(B))=\rho(C^*\delta_\Phi(A)B).
\]
Since these vectors span \(\mathcal H_{\rho,\Lambda}\), this proves \eqref{eq:local-positive-commutator}.
\end{proof}

\begin{corollary}
\label{lem:local-positive-filter}
Let \(\mathcal B\) and its local algebras \(\mathcal B_X\) be as in Lemma~\ref{lem:local-positive-generator}.
For every interaction \(\Phi\) with \(\|\Phi\|_1<\infty\), \(k\in L^1(\R)\) with \(\widehat k=0\) on \([0,\infty)\), \(U\subset\Gamma\), \(\rho\in\LG_\Phi(U)\), nonempty finite \(\Lambda\subset U\), and \(A\in\mathcal B_\Lambda\),
\begin{equation}
 \rho\!\left(\tau_k^\Phi(A)^*\tau_k^\Phi(A)\right)^{1/2}
 \le2\mathfrak b_k^\Phi(A;\Lambda).
 \label{eq:general-local-positive-filter}
\end{equation}
\end{corollary}

For \(A\in\mathcal B_X\), Lemmas~\ref{lem:exponential-crossing-comparison} and~\ref{lem:polynomial-crossing-comparison} bound \(\mathfrak b_k^\Phi(A;\Lambda)\) by a function decaying with \(\dist(X,\Lambda^c)\), stretched-exponentially or polynomially under their respective assumptions on \(\Phi\) and \(k\).

\begin{figure}[!htbp]
\centering
\begin{tikzpicture}[lppl diagram]
 \draw[black!45,dash pattern=on 3pt off 2pt,rounded corners=3pt]
   (0,.4) rectangle (6.5,5.8);
 \node at (.35,5.45) {$U$};
 \path[lppl comparison region]
   (2.6,3) circle[radius=1.7];
 \path[observable] (2.6,3) circle[radius=.45];
 \node at (2.6,3) {$X$};
 \node at (1.5,4) {$\Lambda$};
 \draw[dim line] (3.05,3) -- (4.3,3);
 \node[font=\footnotesize,anchor=west] at (4.38,3)
   {$\dist(X,\Lambda^c)$};
 \draw[lppl bulk term] (2.0,3.85) -- (2.45,4.15) -- (2.95,3.95);
 \foreach \p in {(2.0,3.85),(2.45,4.15),(2.95,3.95)}
  \node[lppl term site,fill=phys!45] at \p {};
 \foreach \a/\b in {{1.6,2.7}/{2.0,2.3}}{
  \draw[lppl bulk term] (\a) -- (\b);
  \node[lppl term site,fill=phys!45] at (\a) {};
  \node[lppl term site,fill=phys!45] at (\b) {};}
 \draw[lppl boundary term] (3.4,3.8) -- (3.45,4.2) -- (4.8,4.6);
 \foreach \p in {(3.4,3.8),(3.45,4.2),(4.8,4.6)}
  \node[lppl term site,fill=phys!80!black] at \p {};
 \foreach \a/\b in {{1.5,2}/{1.12,1.6},{3.95,2.38}/{4.5,2.2}}{
  \draw[lppl boundary term] (\a) -- (\b);
  \node[lppl term site,fill=phys!80!black] at (\a) {};
  \node[lppl term site,fill=phys!80!black] at (\b) {};}
 \draw[lppl exterior term] (.55,4.6) -- (.95,5);
 \node[lppl term site,fill=black!35] at (.55,4.6) {};
 \node[lppl term site,fill=black!35] at (.95,5) {};
 \node[fig note] at (3.25,.85) {$\rho\in\LG_\Phi(U)$};
 \begin{scope}[shift={(9.65,0)}]
  \node[lppl evolution] (bulk) at (0,5.2)
    {bulk evolution\\$\pi_\rho(\tau_t^\Phi(A))$};
  \node[lppl restricted evolution]
    (restricted) at (0,3.1)
    {restricted bulk evolution\\$\pi_\rho(\tau_t^{\Phi,\Lambda}(A))$};
  \node[lppl evolution] (local) at (0,1)
    {$L_{\rho,\Lambda}$-evolution\\$\E^{\I tL_{\rho,\Lambda}}\pi_{\rho,\Lambda}(A)\E^{-\I tL_{\rho,\Lambda}}$};
  \draw[lppl comparison arrow]
    ($(bulk.south)+(-2,0)$) -- ($(restricted.north)+(-2,0)$)
    node[midway,right=5pt,lppl comparison label]
      {boundary commutator\\[2pt]$\pi_\rho([\Phi(Z),D])$};
  \draw[lppl comparison arrow]
    ($(restricted.south)+(-2,0)$) -- ($(local.north)+(-2,0)$)
    node[midway,right=5pt,lppl comparison label]
      {compressed boundary commutator\\[2pt]
       $\textcolor{black!65}{P_{\rho,\Lambda}}\,\pi_\rho([\Phi(Z),D])\,\textcolor{black!65}{P_{\rho,\Lambda}}$};
 \end{scope}
\end{tikzpicture}
\caption{For a local ground state \(\rho\in\LG_\Phi(U)\) and \(A\in\mathcal B_X\), with \(X\subset\Lambda\subset U\) and \(\Lambda\) finite, we compare the bulk evolution \(\tau^\Phi\) with conjugation by \(\E^{\I tL_{\rho,\Lambda}}\) in the GNS representation of \(\rho\).
The positive local energy operator \(L_{\rho,\Lambda}\) is constructed from the local ground-state inequality on \(\mathcal H_{\rho,\Lambda}=\pi_\rho(\mathcal B_\Lambda)\Omega_\rho\) (Lemma~\ref{lem:local-positive-generator}).
Here \(\pi_{\rho,\Lambda}(A):=\pi_\rho(A)|_{\mathcal H_{\rho,\Lambda}}\).
Both evolutions are compared with the restricted bulk evolution \(\tau^{\Phi,\Lambda}\), generated by the terms supported in \(\Lambda\) (light blue).
The dark blue terms cross the boundary of \(\Lambda\) and contribute to the comparisons, while the gray term supported outside \(\Lambda\) does not.
The labels at the arrows show the boundary commutators in the two Duhamel comparisons, with \(D=\tau_s^{\Phi,\Lambda}(A)\) and \(Z\) crossing the boundary of \(\Lambda\).
The lower comparison takes place on \(\mathcal H_{\rho,\Lambda}\), and the compression follows from \eqref{eq:local-positive-commutator}.}
\label{fig:local-positive-comparison}
\end{figure}
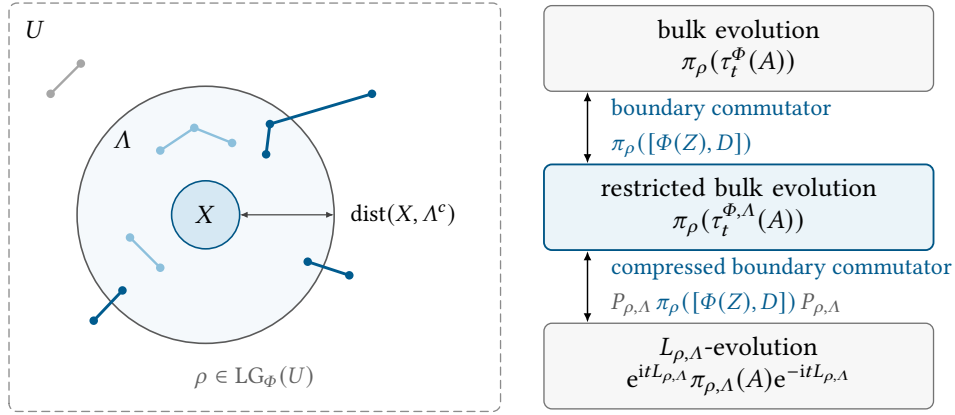

Figure~\ref{fig:local-positive-comparison} illustrates the two comparisons used in the proof.

\begin{proof}
Use \(Q=P_{\rho,\Lambda}\), \(L=L_{\rho,\Lambda}\), and \(\pi=\pi_\rho\) from Lemma~\ref{lem:local-positive-generator}, with \(B=\Lambda\).
The local cyclic subspace is finite dimensional and reduces \(\pi_\rho(\mathcal B_\Lambda)\), so the domain conditions of Lemma~\ref{lem:local-spectral-comparison} are automatic.
Equation~\eqref{eq:local-positive-commutator} gives
\[
 \mathcal R_\Lambda(D)
 =\left.P_{\rho,\Lambda}\pi_\rho
       \bigl((\delta_\Phi-\delta_{\Phi,\Lambda})(D)\bigr)
                      P_{\rho,\Lambda}\right|_{\mathcal H_{\rho,\Lambda}}.
\]
Only terms crossing \(\Lambda\) contribute, and compression is contractive; hence \eqref{eq:local-comparison-discrepancy} holds.
Since \(L_{\rho,\Lambda}\Omega_\rho=0\), apply \eqref{eq:local-spectral-comparison} with \(E=\varepsilon=0\) to \(\Omega_\rho\).
This gives \eqref{eq:general-local-positive-filter}.
\end{proof}

\subsection{Local determination from the ground-state inequality}
\label{sec:bulk-uniqueness}
\label{sec:filtered-occupation}

We apply \eqref{eq:general-local-positive-filter} to local approximations of \(\widetilde c_{x,j}^{\,(\lambda)}\) to bound the expectations of the filtered squares in \(\rho_\st\).

\begin{proposition}
\label{prop:filtered-coercivity}
Assume Hypothesis~\ref{hyp:background-transport}.
In the polynomial case, use \(\nu\) from \eqref{eq:main-polynomial-range}.
The model data determine a constant \(C_{\rm occ}<\infty\) in either locality regime and a rate \(\mu_{\rm filt}>0\) in the stretched-exponential case.
For every \(|\lambda|\le\lambda_0\), \(U\subset\Gamma\), \(\rho\in\LG_{\Phi_\lambda}(U)\), \(x\in\Gamma\), and \(j\in\mathscr I\), form \(\rho_\st\) as in \eqref{eq:stacked-local-ground}, and define
\begin{equation}
 \rho_\at:=\rho_\st\circ\sigma_\lambda^{-1}.
 \label{eq:coordinate-state}
\end{equation}
If \(U=\Gamma\), then
\begin{equation}
 \rho_\at\left(
 (c_{f,x,j}^{(\lambda)})^*c_{f,x,j}^{(\lambda)}\right)=0.
 \label{eq:filtered-occupation-zero}
\end{equation}
If \(U\ne\Gamma\), then
\begin{equation}
 \rho_\at\left(
 (c_{f,x,j}^{(\lambda)})^*c_{f,x,j}^{(\lambda)}\right)
 \le
 \begin{cases}
  C_{\rm occ}(1+\dist(x,U^c))^{-2\nu},
  &\text{polynomial case},\\
  C_{\rm occ}\E^{-\mu_{\rm filt}\dist(x,U^c)^\theta},
  &\text{stretched-exponential case}.
 \end{cases}
 \label{eq:filtered-occupation}
\end{equation}
\end{proposition}

\begin{proof}
By Lemma~\ref{lem:graded-product}, the stacked state \(\rho_\st\) satisfies the local ground-state inequality on \(U\).
Fix \(x,j\) and take \(R\ge16\) with \(\Lambda:=\ball_R(x)\subset U\).
Let \(s\) be the largest dyadic radius with \(s\le R/4\), and set \(A_R:=\mathbb F^{\rm tr}_{\ball_s(x)}\widetilde c_{x,j}^{\,(\lambda)}\).
Apply \eqref{eq:general-local-positive-filter} to \(A_R\) with \(\Phi=\Phi_\lambda^\st\), \(k=f\), and \(\rho=\rho_\st\).
The norm of \(\tau_f^{\st,\lambda}\) is at most \(\|f\|_{L^1(\R)}\), so the triangle inequality gives
\[
 \rho_\at\bigl((c_{f,x,j}^{(\lambda)})^*c_{f,x,j}^{(\lambda)}\bigr)^{1/2}
 =\|\pi_{\rho_\st}(\widetilde c_{f,x,j}^{\,(\lambda)})\Omega_{\rho_\st}\| \le \|f\|_{L^1(\R)}\,\|\widetilde c_{x,j}^{\,(\lambda)}-A_R\|
      +2\mathfrak b_f^{\Phi_\lambda^\st}(A_R;\Lambda).
\]
Since \(s>R/8\), estimate~\eqref{eq:coordinate-matrix-locality} for \(\sigma_\lambda^{-1}\) with \(k=1\) bounds the first term by \(C\E^{-cR^\theta}\) under stretched-exponential locality and by \(C(1+R)^{-p_\at}\) under polynomial locality.
The triangle inequality in \eqref{eq:stacked-crossing-norms} gives \(\mathfrak b_f^{\Phi_\lambda^\st}(A_R;\Lambda)\le\mathfrak b_f^{\st,\lambda}(A_R;\Lambda)\), and \eqref{eq:seed-integrated-boundary} bounds the latter by \(C\E^{-cR^\theta}\) and \(C(1+R)^{-\nu}\) in the two cases, respectively.
If \(32\le D:=\dist(x,U^c)<\infty\), taking \(R=D/2\) and squaring gives \eqref{eq:filtered-occupation}, since \(\nu\le p_\at\) in the polynomial case.
For \(D<32\), use \(\|c_{f,x,j}^{(\lambda)}\|\le\|f\|_{L^1(\R)}\) and increase \(C_{\rm occ}\).
If \(U=\Gamma\), every \(R\ge16\) is allowed, and letting \(R\to\infty\) gives \eqref{eq:filtered-occupation-zero}.
\end{proof}

\begin{proposition}
\label{prop:density-decay}
Assume Hypotheses~\ref{hyp:background-transport} and~\ref{hyp:uniform-local-excitation-detectability} on \(\mathcal J\).

There is a constant \(C<\infty\), depending only on the model and LED data on \(\mathcal J\), such that, for every \(\lambda\in\mathcal J\), \(U\subset\Gamma\), \(\rho\in\LG_{\Phi_\lambda}(U)\), and \(x\in\Gamma\),
\begin{equation}
 \rho_\at(n_x)
 \le C(1+\dist(x,U^c))^{-2\nu}.
 \label{eq:density-finite-polynomial}
\end{equation}
Here \(\rho_\at\) is the state defined in \eqref{eq:coordinate-state}.

In the stretched-exponential case there exist constants \(C,\mu_{\rm den}>0\), depending only on the model and LED data on \(\mathcal J\), such that, for every \(\lambda\in\mathcal J\), \(U\subset\Gamma\), \(\rho\in\LG_{\Phi_\lambda}(U)\), and \(x\in\Gamma\), the state \(\rho_\at\) satisfies
\begin{equation}
 \rho_\at(n_x)
 \le C\E^{-\mu_{\rm den}\dist(x,U^c)^\theta}
\,. \label{eq:density-stretched}
\end{equation}
\end{proposition}

\begin{proof}
Set \(b_x:=2\rho_\at(n_{f,x}^{(\lambda)})\).
Summing \eqref{eq:filtered-occupation} over \(j\in\mathscr I\), with \(|\mathscr I|=2r\), gives \eqref{eq:filtered-density-decay-assumption} for \(S=U^c\), \(C_b=4rC_{\rm occ}\), and, in the stretched-exponential case, \(\mu_b=\mu_{\rm filt}\).
For \(U=\Gamma\), use \eqref{eq:filtered-occupation-zero}.
Applying \eqref{eq:filtered-decay-density} with \(\mathcal Q=\rho_\at\) proves \eqref{eq:density-finite-polynomial} and \eqref{eq:density-stretched}.
\end{proof}

\begin{proof}[Proof of Theorem~\ref{thm:bulk-uniqueness}]
Fix \(\lambda\in\mathcal J\), \(U\subset\Gamma\), and \(\rho\in\LG_{\Phi_\lambda}(U)\).
Use the stacked state \(\rho_\st\) from \eqref{eq:stacked-local-ground} and the state \(\rho_\at\) in product-vacuum coordinates from \eqref{eq:coordinate-state}.
Summing the estimates of Proposition~\ref{prop:filtered-coercivity} over \(j\in\mathscr I\) bounds \(2\rho_\at(n_{f,x}^{(\lambda)})\) by \(4rC_{\rm occ}\) times the corresponding decay function in \eqref{eq:filtered-occupation}, with zero on the right when \(U=\Gamma\).
We can therefore apply Corollary~\ref{cor:filtered-reconstruction} with \(\mathcal Q=\rho_\at\) and \(S=U^c\).
Apply its scalar comparison \eqref{eq:filtered-decay-comparison} and use, for \(A\in\A_Y\),
\[
 (\rho_\at-\omega_\at)\bigl(\sigma_\lambda(A\widehat\otimes\1)\bigr)
 =\rho(A)-\omega_\lambda(A).
\]
This proves the local-determination estimate in both regimes, with constants uniform in \(\lambda,U,\rho,Y,A\).
Choose the larger prefactor and, in the stretched-exponential case, the smaller rate between this bound and Theorem~\ref{thm:positive-energy-locality}.
Both estimates then hold with the same \(C\) and \(G\).
Taking the supremum over \(\rho\in\LG_{\Phi_\lambda}(U)\) completes the proof.
\end{proof}

\section{Local excitation detectability at weak coupling}
\label{sec:detectability}

In this section we prove Proposition~\ref{prop:weak-local-excitation-detectability}, which establishes LED for sufficiently small \(|\lambda|\).
We first estimate the differences between the bulk-filtered annihilation operators \(c_{f,x,j}^{(\lambda)}\) and the onsite annihilation operators \(c_{x,j}\).
We then prove the detectability inequality \eqref{eq:uniform-local-excitation-detectability} and show that the row sums of \(\mathscr K_\lambda\) are uniformly below one.
We also establish weighted row contraction in the stretched-exponential case and finite polynomial row moments in the polynomial case.

\subsection{Local estimates for the differences of annihilation operators}

\begin{lemma}
\label{lem:small-correction}
Assume Hypothesis~\ref{hyp:background-transport}.
For \(|\lambda|\le\lambda_0\), \(x\in\Gamma\), and \(j\in\mathscr I\), define
\begin{equation}
 e_{x,j}^{(\lambda)}
 :=c_{f,x,j}^{(\lambda)}-c_{x,j}.
 \label{eq:correction-field}
\end{equation}
Then there is a constant \(C_{\rm filt}<\infty\), depending only on the model data, such that, for every \(|\lambda|\le\lambda_0\),
\begin{equation}
 \sup_{\substack{x\in\Gamma\\j\in\mathscr I}}
 \|e_{x,j}^{(\lambda)}\|
 \leq C_{\rm filt}|\lambda|.
 \label{eq:correction-small-norm}
\end{equation}
In the polynomial case, use \(\nu\) from \eqref{eq:main-polynomial-range}.
There are constants \(C>0\) and, in the stretched-exponential case, \(\mu_{\rm sc}>0\), depending only on the model data, such that, for every \(|\lambda|\le\lambda_0\), \(x\in\Gamma\), \(j\in\mathscr I\), and \(R\ge0\), there exists \(e_{x,j}^{(\lambda),[R]}\in\A_{\ball_R(x)}^\st\) satisfying
\begin{equation*}
 \|e_{x,j}^{(\lambda)}-e_{x,j}^{(\lambda),[R]}\|
 \le
 \begin{cases}
 C(1+R)^{-\nu},&\text{in the polynomial case},\\
 C\E^{-\mu_{\rm sc} R^\theta},
 &\text{in the stretched-exponential case}.
 \end{cases}
\end{equation*}
\end{lemma}

\begin{proof}
Fix \(x\in\Gamma\) and \(j\in\mathscr I\), and write \(\widetilde c_\lambda:=\sigma_\lambda^{-1}(c_{x,j})\).
Isometry and the free normalization \(\tau_f^{\st,0}(\widetilde c_0)=\widetilde c_0\) give
\begin{align}
 \|e_{x,j}^{(\lambda)}\|
  =\|(\tau_f^{\st,\lambda}-\operatorname{id})
                         (\widetilde c_\lambda)\| \le(\|f\|_{L^1(\R)}+1)\|\widetilde c_\lambda-\widetilde c_0\|
   +\|(\tau_f^{\st,\lambda}-\tau_f^{\st,0})
                         (\widetilde c_0)\|.
 \label{eq:smallness-split}
\end{align}
To bound the first difference, use the locality of \(\sigma_\st^{-1}\) in Lemma~\ref{lem:atomicization}.
The dyadic shells \(D_n^{(0)}:=\Delta_{\{x\},n}^{\rm tr}\widetilde c_0 \in\A_{\ball_{R_n}(x)}^\st\) satisfy
\[
 \widetilde c_0=\sum_{n\ge0}D_n^{(0)},
 \qquad
 \sum_{n\ge0}|\ball_{R_n}(x)|\,\|D_n^{(0)}\|\le C,
\]
uniformly in \(x,j\).
In the polynomial case, \(\|D_n^{(0)}\|\le C(1+R_n)^{-p_h}\), so the support-weighted sum is bounded by \(C\sum_{n\ge0}(1+R_n)^{d_{\rm vol}-p_h}<\infty\).
The shell bound \(C\E^{-cR_n^{\theta_{\rm in}}}\) gives the same conclusion in the stretched-exponential case.
Since \(\widetilde c_\lambda =(\beta_{0,\lambda}^{(1)})^{-1}(\widetilde c_0)=\beta_{\lambda,0}^{(1)}(\widetilde c_0)\), estimate~\eqref{eq:flow-local-continuity} applied to this expansion gives \(\|\widetilde c_\lambda-\widetilde c_0\|\le C|\lambda|\).

For the second difference, the free interaction on the stack is \(\Phi_0^\st\), and the perturbation has terms \(\mathcal V^{(1)}(Z)=\mathcal V(Z)\widehat\otimes\1\).
In the polynomial regime, their \(\|\cdot\|^{\#}_{p_\Phi;\kappa_0}\)-norms are bounded in terms of the model data: the free terms have at most two-site support and polynomial Schur decay of order \(p_h\), and the first-layer lift preserves the perturbation norm.
The preceding bound on \(\sum_n|\ball_{R_n}(x)|\,\|D_n^{(0)}\|\) allows us to apply \cite[Proposition~5.3(iii)]{FermionicLR} to \(\widetilde c_0=\sum_n D_n^{(0)}\), with Hamiltonian interaction \(\Phi_0^\st\), perturbation \(\mathcal V^{(1)}\), and exponent \(a=p_\Phi\).
For Hamiltonian exponent \(p_\Phi\), the required filter moment in that proposition is
\[
 \int_\R |f(t)|\,|t|
 (1+|t|)^{d_{\rm vol}(p_\Phi+1)/(p_\Phi-d_{\rm vol})}\,\dd t<\infty,
\]
where finiteness follows from \(\widehat f\in C_c^\infty(\R)\).
That proposition gives
\[
 \|(\tau_f^{\st,\lambda}-\tau_f^{\st,0})(\widetilde c_0)\|
 \le C|\lambda|\|\mathcal V\|^{\#}_{0;\kappa_0}
 \sum_{n\ge0}|\ball_{R_n}(x)|\,\|D_n^{(0)}\|
 \le C|\lambda|.
\]

In the stretched-exponential regime, the Bogoliubov formula for \(\sigma_\st^{-1}\) expresses \(\widetilde c_0\) as a linear Majorana field whose coefficients have uniformly bounded \(\ell^1\)-norm with weight \(\E^{\mu d(x,x_u)^{\theta_{\rm in}}}\), for some \(\mu>0\).
Since \(h_{\rm aux}=-h\) on the auxiliary copy, \(h\oplus h_{\rm aux}\) has the same Majorana conditions, gap, and Schur bound as \(h\).
Lemma~\ref{lem:dynamics-comparison}, applied on the stack with Majorana coefficient operator \(h\oplus h_{\rm aux}\), therefore gives
\[
 \|(\tau_f^{\st,\lambda}-\tau_f^{\st,0})(\widetilde c_0)\|
 \le C|\lambda|\int_{\R}|f(t)|\,|t|
             (1+|t|)^{d_{\rm vol}/(2\theta_{\rm in})}\,\dd t
 \le C|\lambda|.
\]
Together with \eqref{eq:smallness-split}, this proves \eqref{eq:correction-small-norm}.

For locality, define, for every \(R\ge0\),
\[
 e_{x,j}^{(\lambda),[R]}
 :=\mathbb F^{\rm tr}_{\ball_R(x)}(c_{f,x,j}^{(\lambda)})-c_{x,j}.
\]
Since \(c_{x,j}\in\A_{\{x\}}^\st\), this approximant belongs to \(\A_{\ball_R(x)}^\st\), and
\[
 \|e_{x,j}^{(\lambda)}-e_{x,j}^{(\lambda),[R]}\|
 =\|(\operatorname{id}-\mathbb F^{\rm tr}_{\ball_R(x)})(c_{f,x,j}^{(\lambda)})\|.
\]
We apply \eqref{eq:coordinate-filter-norm-tail} to \(c_{f,x,j}^{(\lambda)}\) in the preceding equality and set \(\mu_{\rm sc}=m\) in the stretched-exponential case.
\end{proof}

\paragraph{Local decomposition.}

We use conditional expectations preserving \(\omega_\at\) to decompose \(e_{x,j}^{(\lambda)}\) into local terms that annihilate the vacuum.
Each such term \(E\), supported in \(X\), satisfies \(E^*E\le\|E\|^2N_X\).

Use the product-vacuum conditional expectations \(\mathbb E_X^\vac\) from \eqref{eq:vacuum-slices}.
For nonempty \(X\Subset\Gamma\), use the dyadic radii of \eqref{eq:dyadic-tracial-shells} and define
\begin{equation}
 \Delta_{X,0}^\vac:=\mathbb E_X^\vac,\qquad
 \Delta_{X,n}^\vac
 :=\mathbb E_{X^{(R_n)}}^\vac
       -\mathbb E_{X^{(R_{n-1})}}^\vac\quad(n\ge1).
 \label{eq:dyadic-vacuum-shells}
\end{equation}
These are the maps \(\Delta_{X,n}^\vac\) of \cite[Section~6.2]{StabilityBdG26}.
These shell maps preserve parity and have norm at most two.
For every \(x\in\Gamma\) and \(A\in\A^\st\), they give
\[
 A=\sum_{n\ge0}\Delta_{\{x\},n}^\vac(A)
 \quad\text{in norm}.
\]
Estimate~\eqref{eq:dyadic-locality-to-shells} also holds with \(\Delta_{X,n}^\vac\) in place of \(\Delta_{X,n}^{\rm tr}\), since the vacuum conditional expectations are contractions and fix the corresponding local algebras.

\begin{lemma}
\label{lem:weighted-vacuum-shells}
Under the assumptions of Lemma~\ref{lem:small-correction}, let \(e_{x,j}^{(\lambda)}\) and \(C_{\rm filt}\) be as in that lemma.
For \(|\lambda|\le\lambda_0\), \(x\in\Gamma\), \(j\in\mathscr I\), and \(n\ge0\), define
\[
 E_{x,j,n}^{(\lambda)}
 :=\Delta_{\{x\},n}^\vac e_{x,j}^{(\lambda)},
 \qquad
 \varepsilon_{x,j,n}^{(\lambda)}
 :=\|E_{x,j,n}^{(\lambda)}\|.
\]
Each shell annihilates the local vacuum vector and satisfies
\begin{equation}
 (E_{x,j,n}^{(\lambda)})^*E_{x,j,n}^{(\lambda)}
 \le(\varepsilon_{x,j,n}^{(\lambda)})^2N_{\ball_{R_n}(x)}.
 \label{eq:dyadic-vacuum-division}
\end{equation}
In the polynomial case, use \(\nu\) from \eqref{eq:main-polynomial-range}.
In the stretched-exponential case there is a constant \(\mu_{\rm sh}>0\), depending only on the model data.
Define the shell envelope, for \(R\ge0\), by
\[
 F_{\rm sh}(R):=
 \begin{cases}
 (1+R)^{-\nu},&\text{polynomial case},\\
 \E^{-\mu_{\rm sh}R^\theta},&\text{stretched-exponential case}.
 \end{cases}
\]
With a constant \(C\) depending only on the model data, uniformly in all shell indices,
\begin{equation}
 \varepsilon_{x,j,n}^{(\lambda)}
 \le C\min\{C_{\rm filt}|\lambda|,F_{\rm sh}(R_n)\}.
 \label{eq:vacuum-shell-envelope}
\end{equation}
\end{lemma}

\begin{proof}
Lemma~\ref{lem:filter-normalization} and \eqref{eq:atomic-fields-annihilate} give \(\pi_\at(e_{x,j}^{(\lambda)})\Omega_\at=0\).
By \cite[Lemma~6.7]{StabilityBdG26}, each localized operator satisfies \(\mathbb E_X^\vac(e_{x,j}^{(\lambda)})P_X^\vac=0\).
The nested vacuum projections then give \(E_{x,j,n}^{(\lambda)}P^\vac_{\ball_{R_n}(x)}=0\), as in the shell construction of \cite[Section~6.2]{StabilityBdG26}.
It follows that
\[
 (E_{x,j,n}^{(\lambda)})^*E_{x,j,n}^{(\lambda)}
 \le(\varepsilon_{x,j,n}^{(\lambda)})^2
       (\1-P^\vac_{\ball_{R_n}(x)})
 \le(\varepsilon_{x,j,n}^{(\lambda)})^2N_{\ball_{R_n}(x)},
\]
which is \eqref{eq:dyadic-vacuum-division}.

Contractivity and Lemma~\ref{lem:small-correction} give \(\varepsilon_{x,j,n}^{(\lambda)}\le2C_{\rm filt}|\lambda|\).
For \(n\ge1\), both expectations defining the shell fix the approximant \(e_{x,j}^{(\lambda),[R_{n-1}]}\), so
\[
 \varepsilon_{x,j,n}^{(\lambda)}
 \le2\|e_{x,j}^{(\lambda)}-e_{x,j}^{(\lambda),[R_{n-1}]}\|.
\]
The locality bound in Lemma~\ref{lem:small-correction} gives \(CF_{\rm sh}(R_n)\) for \(n\ge2\), using \(R_n=2R_{n-1}\) and \(\mu_{\rm sh}=2^{-\theta}\mu_{\rm sc}\) in the stretched-exponential case.
Increasing \(C\) covers \(n=0,1\) and proves \eqref{eq:vacuum-shell-envelope}.
\end{proof}

\subsection{Construction of the detectability kernel}

\begin{proposition}
\label{prop:local-excitation-detectability}
Use the shells \(E_{x,j,n}^{(\lambda)}\) and their norms \(\varepsilon_{x,j,n}^{(\lambda)}\) from Lemma~\ref{lem:weighted-vacuum-shells}.
For \(|\lambda|\le\lambda_0\) and \(x,z\in\Gamma\), define
\begin{equation}
 \mathscr K_\lambda^{\rm wc}(x,z)
 :=\frac{\pi^2}{3}\sum_{j\in\mathscr I}
 \sum_{n:\,R_n\ge d(x,z)}
       (n+1)^2\bigl(\varepsilon_{x,j,n}^{(\lambda)}\bigr)^2.
 \label{eq:weak-detectability-witness}
\end{equation}
For every \(x\in\Gamma\), the following operator inequality holds:
\begin{equation}
 n_x\le2n_{f,x}^{(\lambda)}
       +\sum_z\mathscr K_\lambda^{\rm wc}(x,z)n_z,
 \label{eq:weak-detectability-inequality}
\end{equation}
where the series converges in norm.
For every nondecreasing weight \(w:[0,\infty)\to[1,\infty)\),
\begin{equation}
 \sum_z w(d(x,z))\mathscr K_\lambda^{\rm wc}(x,z)
 \le C\sum_{j\in\mathscr I}\sum_{n\ge0}
 (n+1)^2(1+R_n)^{d_{\rm vol}}w(R_n)
       \bigl(\varepsilon_{x,j,n}^{(\lambda)}\bigr)^2.
 \label{eq:weak-witness-weighted-row}
\end{equation}
Here \(C\) depends only on the volume-growth bound.
The rows \(z\mapsto\mathscr K_\lambda^{\rm wc}(x,z)\) have finite polynomial moments of every order \(M<2\nu-d_{\rm vol}\) in the polynomial case and finite stretched-exponential moments of every rate \(\mu<2\mu_{\rm sh}\) in the stretched-exponential case, uniformly in~\(\lambda\).
\end{proposition}

\begin{proof}
Fix \(x,j,\lambda\).
For every \(n_{\max}\ge0\), weighted Cauchy--Schwarz on each vector, with weights \((n+1)^{-2}\), using \(\sum_{n\ge0}(n+1)^{-2}=\pi^2/6\) gives
\[
 \left(\sum_{n=0}^{n_{\max}} E_{x,j,n}^{(\lambda)}\right)^*
 \left(\sum_{n=0}^{n_{\max}} E_{x,j,n}^{(\lambda)}\right)
 \le \frac{\pi^2}{6}\sum_{n=0}^{n_{\max}}(n+1)^2
       (E_{x,j,n}^{(\lambda)})^*E_{x,j,n}^{(\lambda)}.
\]
By \eqref{eq:dyadic-vacuum-division}, the right-hand side is at most the corresponding sum of \((\pi^2/6)(n+1)^2(\varepsilon_{x,j,n}^{(\lambda)})^2 N_{\ball_{R_n}(x)}\).
The shell bounds and volume growth imply
\[
 \sum_{n\ge0}(n+1)^2(1+R_n)^{d_{\rm vol}}
       \bigl(\varepsilon_{x,j,n}^{(\lambda)}\bigr)^2<\infty.
\]
The partial sums of \(E_{x,j,n}^{(\lambda)}\) converge in norm to \(e_{x,j}^{(\lambda)}\).
The last bound also gives norm convergence of the series involving \(N_{\ball_{R_n}(x)}\), since \(\|N_{\ball_{R_n}(x)}\|\le2r|\ball_{R_n}(x)|\).
Passing to the limit in the finite-sum inequality therefore yields
\begin{equation}
 (e_{x,j}^{(\lambda)})^*e_{x,j}^{(\lambda)}
 \le \frac{\pi^2}{6}\sum_{n\ge0}(n+1)^2
       \bigl(\varepsilon_{x,j,n}^{(\lambda)}\bigr)^2
       N_{\ball_{R_n}(x)}.
 \label{eq:operator-CS}
\end{equation}
Apply \(c_{x,j}^*c_{x,j}\le 2(c_{f,x,j}^{(\lambda)})^*c_{f,x,j}^{(\lambda)} +2(e_{x,j}^{(\lambda)})^*e_{x,j}^{(\lambda)}\) and sum over \(j\).
Expanding each local excitation number and rearranging the absolutely convergent series gives \eqref{eq:weak-detectability-inequality} with the kernel \eqref{eq:weak-detectability-witness}.

For its weighted rows, use \(w(d(x,z))\le w(R_n)\) on \(\ball_{R_n}(x)\), sum over \(z\), and apply volume growth.
This proves \eqref{eq:weak-witness-weighted-row}.
For polynomial weights the resulting series is bounded by
\[
 C\sum_{n\ge0}(n+1)^2
                 (1+R_n)^{M+d_{\rm vol}-2\nu},
\]
which converges when \(M<2\nu-d_{\rm vol}\), since the radii are dyadic.
For a stretched-exponential weight, the squared shell envelope absorbs the polynomial factors whenever \(\mu<2\mu_{\rm sh}\).
\end{proof}

\begin{proposition}
\label{prop:weak-local-excitation-detectability}
Assume \eqref{eq:Majorana-structure}, \eqref{eq:one-particle-gap}, \eqref{eq:background-path}, and Hypothesis~\ref{hyp:background-transport}.
There is \(\lambdaLED\in(0,\lambda_0]\), depending only on the model data, such that
\[
 \JLED:=[-\lambdaLED,\lambdaLED]
\]
is a parameter set on which Hypothesis~\ref{hyp:uniform-local-excitation-detectability} holds.
We use the kernels \(\mathscr K_\lambda=\mathscr K_\lambda^{\rm wc}\) defined in \eqref{eq:weak-detectability-witness} for every \(|\lambda|\le\lambda_0\).
For a family with uniformly controlled model data in the sense of Definition~\ref{def:model-data}, the comparison kernels may be chosen with a common \(\mu_{\rm LED}>0\) in the stretched-exponential case and with a nondecreasing modulus \(\varsigma_{\rm LED}:[0,\lambda_0]\to[0,\infty)\), independent of the model, such that \(\varsigma_{\rm LED}(t)\to0\) as \(t\searrow0\) and, for every \(|\lambda|\le\lambda_0\),
\begin{equation}
 \begin{cases}
 \displaystyle
 \sup_x\sum_z\mathscr K_\lambda(x,z),
 &\text{in the polynomial case},\\[2mm]
 \displaystyle
 \sup_x\sum_z\E^{\mu_{\rm LED}d(x,z)^\theta}
       \mathscr K_\lambda(x,z),
 &\text{in the stretched-exponential case}
 \end{cases}
 \leq\varsigma_{\rm LED}(|\lambda|).
 \label{eq:uniform-family-LED-modulus}
\end{equation}
\end{proposition}

\begin{proof}
Use the comparison kernels \(\mathscr K_\lambda^{\rm wc}\) defined in \eqref{eq:weak-detectability-witness}.
In the polynomial case, use \(w_0(R):=1\) in the weighted row bound \eqref{eq:weak-witness-weighted-row}.
In the stretched-exponential case choose \(\mu_{\rm LED}:=\mu_{\rm sh}/2\) and use \(w_0(R):=\E^{\mu_{\rm LED}R^\theta}\).

For a uniformly controlled family, the constants in \eqref{eq:vacuum-shell-envelope} and the envelope \(F_{\rm sh}\) can be chosen in common.
For \(0\le t\le\lambda_0\), define
\begin{equation}
 \begin{aligned}
 \varsigma_{\rm LED}(t):=C\sum_{n\ge0}
 (n+1)^2(1+R_n)^{d_{\rm vol}}w_0(R_n)
 \min\{C_{\rm filt}t,F_{\rm sh}(R_n)\}^2.
 \end{aligned}
 \label{eq:LED-common-modulus}
\end{equation}
Substituting \eqref{eq:vacuum-shell-envelope} into \eqref{eq:weak-witness-weighted-row} and using \(|\mathscr I|=2r\), we can choose \(C\) in \eqref{eq:LED-common-modulus} so that, for every \(|\lambda|\le\lambda_0\),
\[
 \sup_{x\in\Gamma}\sum_{z\in\Gamma}
 w_0(d(x,z))\mathscr K_\lambda^{\rm wc}(x,z)
 \le\varsigma_{\rm LED}(|\lambda|).
\]
By the definition of \(w_0\), this is \eqref{eq:uniform-family-LED-modulus}.
The summands defining \(\varsigma_{\rm LED}\) are bounded by
\[
 C(n+1)^2(1+R_n)^{d_{\rm vol}}
             w_0(R_n)F_{\rm sh}(R_n)^2.
\]
In the polynomial case their sum is finite because \(2\nu>d_{\rm vol}\).
In the stretched-exponential case it is finite because \(\mu_{\rm LED}<2\mu_{\rm sh}\).
Each minimum in \eqref{eq:LED-common-modulus} is nondecreasing in \(t\) and tends to zero as \(t\searrow0\).
Dominated convergence therefore gives \(\varsigma_{\rm LED}(t)\to0\), and \(\varsigma_{\rm LED}\) is nondecreasing.

Choose \(\lambdaLED\in(0,\lambda_0]\) with \(\varsigma_{\rm LED}(\lambdaLED)\le1/2\).
Then
\[
 \sup_{|\lambda|\le\lambdaLED}\sup_{x\in\Gamma}
 \sum_{z\in\Gamma}w_0(d(x,z))\mathscr K_\lambda^{\rm wc}(x,z)
 \le\varsigma_{\rm LED}(\lambdaLED)\le\frac12.
\]
With \(\mathscr K_\lambda:=\mathscr K_\lambda^{\rm wc}\), this proves \eqref{eq:LED-row-contraction} in the polynomial case and \eqref{eq:LED-stretched-contraction} in the stretched-exponential case, on \(\JLED\).
Since \(\varsigma_{\rm LED}\) is common to the family, the same \(\lambdaLED\) applies to every member.

Finally, \(e_{x,j}^{(0)}=0\) and \eqref{eq:weak-detectability-witness} give \(\mathscr K_0^{\rm wc}=0\).
\end{proof}

\section{Defects with tails and uniformly weak perturbations}
\label{sec:quasilocal-tails}

We prove Theorem~\ref{thm:quasilocal-tails} by applying Lemma~\ref{lem:common-filtered-occupation} in a positive-energy representation of \(\Phi_\lambda+\Psi\).
The additional interaction contributes \(\delta_\Psi^{(1)}\) to the comparison with restricted bulk dynamics.
We choose a comparison ball \(B\subset S^c\) and bound the integral in \eqref{eq:filtered-occupation-defect-error} by \(C\sup_{z\in B}t_{\Psi,S}(z)\).
The terms \(\Psi(Z)\) with \(Z\subset S\) commute with operators in \(\A_B^\st\) and make no contribution.
Estimate~\eqref{eq:tail-filtered-coercivity} bounds \(\rho_\at((c_{f,x,j}^{(\lambda)})^*c_{f,x,j}^{(\lambda)})\) by a term decaying with \(\dist(x,S)\) and \(CT_{\Psi,S}(x)^2\).
Applying LED and iterating the resulting inequality bounds \(\rho_\at(n_x)\).
We then use \eqref{eq:positive-representation-cb} to estimate \(P(\pi(A)-\omega_\lambda(A)\1)P\), including matrices of observables.
At the end of the section, a uniform bound on the same defect integral proves Corollary~\ref{cor:uniformly-weak-defects}.
Figure~\ref{fig:tail-supports} shows the supports of the defect terms.

\begin{figure}[!htbp]
\KOMAoptions{captions=nooneline}
\setcapwidth{\linewidth}
\centering
\begin{tikzpicture}[lppl diagram]
 \path[excluded] (8.8,.6) rectangle (13.2,4.7);
 \draw[black!60,dash pattern=on 3pt off 2pt] (8.8,.6) -- (8.8,4.7);
 \node[fill=white,inner sep=1.5pt] at (11,4.35) {$S$};
 \node at (.7,4.45) {$S^c$};
 \path[observable] (3.4,2.9) ellipse[x radius=2.4,y radius=1.2];
 \node at (3.4,3.8) {$X$};
 \node[font=\footnotesize] at (3.4,2.05) {$A\in\A_X$};
 \draw[lppl defect term,aux!45] (10.1,3.4) -- (11.2,3.9) -- (12.2,3.1);
 \draw[lppl defect term,aux!45] (5.7,4.45) -- (6.75,4.65);
 \foreach \p in {(10.1,3.4),(11.2,3.9),(12.2,3.1),(5.7,4.45),(6.75,4.65)}
  \node[lppl term site,fill=aux!45] at \p {};
 \draw[lppl defect term] (10.7,2.1) -- (7.1,2.4) -- (5.2,3.05);
 \draw[lppl defect term] (3.05,2.65) -- (.5,1.9);
 \foreach \p in {(10.7,2.1),(7.1,2.4),(.5,1.9)}
  \node[lppl term site,fill=aux] at \p {};
 \foreach \x/\y/\l in {5.2/3.05/z_2,3.05/2.65/z_3}{
  \filldraw[fill=phys!85!black,draw=white,line width=.5pt] (\x,\y) circle[radius=2.3pt];
  \node[font=\footnotesize,anchor=south,inner sep=2pt] at (\x,\y+.16) {$\l$};}
 \node[font=\footnotesize,fill=white,inner sep=1.5pt] at (11.2,2.8) {$Z_1\subset S$};
 \node[font=\footnotesize] at (7.1,2.85) {$Z_2$};
 \node[font=\footnotesize,anchor=west] at (6.9,4.55) {$Z_4\subset S^c$};
 \node[font=\footnotesize,anchor=north] at (1.05,1.65) {$Z_3\subset S^c$};
\end{tikzpicture}
\caption[Defect terms in the exterior commutator estimate.]{Defect terms in the exterior commutator estimate.
Let \(\Psi\in\mathfrak J_1\) and \(A\in\A_X\), with \(X\subset S^c\) finite.
Since the terms \(\Psi(Z)\) are even, those supported in \(S\), such as \(\Psi(Z_1)\), commute with \(A\), as do other terms whose supports miss \(X\), such as \(\Psi(Z_4)\); these are drawn faintly.
Crossing terms \(\Psi(Z_2)\) and exterior terms \(\Psi(Z_3)\) can contribute only when their supports meet \(X\), at sites such as \(z_2\) and \(z_3\).
The exterior tail strength \(t_{\Psi,S}\), defined in \eqref{eq:exterior-tail-intensity}, counts the full norm of each term at every exterior site in its support.
Summing over interaction terms whose supports meet \(X\) gives
\[
 \sum_{Z\Subset\Gamma}\|[\Psi(Z),A]\|
 \le 2\|A\|\sum_{z\in X}t_{\Psi,S}(z).
\]
This bound is independent of the strength of terms supported entirely in \(S\).
Dots represent sites of \(\Gamma\); each connected group depicts the support of one interaction term.}
\label{fig:tail-supports}
\end{figure}
\FloatBarrier

\subsection{Commutator bounds and filtered excitation densities}

We estimate commutators with the first-layer lift \(\Psi^{(1)}(Z)=\Psi(Z)\widehat\otimes\1\) along the restricted bulk dynamics.

\begin{lemma}
\label{lem:tail-integrated-commutator}
Assume Hypothesis~\ref{hyp:background-transport} in either locality regime.
In the polynomial case, use \(\nu\) from \eqref{eq:main-polynomial-range}.
There is a constant \(C\), depending only on the model data, with the following property.
For every \(|\lambda|\le\lambda_0\), \(x\in\Gamma\), \(j\in\mathscr I\), and dyadic index \(n_{\max}\ge0\), define
\[
 A_{n_{\max}}:=\mathbb F^{\rm tr}_{\ball_{R_{n_{\max}}}(x)}
                  \widetilde c_{x,j}^{\,(\lambda)}.
\]
For every defect \(\Psi\in\mathfrak J_1\) and finite \(B\supset\ball_{R_{n_{\max}}}(x)\), let \(\tau_t^B:=\tau_t^{\Phi_\lambda^\st,B}\) be the restricted bulk dynamics.
Then
\begin{equation}
 \int_0^\infty w_f(s)
       \max_{\epsilon=\pm1}
       \|\delta_\Psi^{(1)}(\tau_{\epsilon s}^B(A_{n_{\max}}))\|\,\dd s
 \le CJ_\Psi.
 \label{eq:global-integrated-defect-commutator}
\end{equation}
If, in addition, \(S\subset\Gamma\) is nonempty and \(B\subset S^c\), then
\begin{equation}
 \int_0^\infty w_f(s)
       \max_{\epsilon=\pm1}
       \|\delta_\Psi^{(1)}(\tau_{\epsilon s}^B(A_{n_{\max}}))\|\,\dd s
 \le C\sup_{z\in B}t_{\Psi,S}(z).
 \label{eq:tail-integrated-commutator}
\end{equation}
\end{lemma}

\begin{proof}
Fix \(\lambda,x,j,n_{\max},\Psi\) and a finite \(B\supset\ball_{R_{n_{\max}}}(x)\) as in the statement.
For \(t\in\R\) and \(n\ge0\), define
\[
 D_n(t):=\Delta_{\{x\},n}^{\rm tr}\tau_t^B(A_{n_{\max}}).
\]
Since \(\tau_t^B(A_{n_{\max}})\in\A_B^\st\), the composition rule \eqref{eq:tracial-expectation-composition} gives, for every \(R\ge0\),
\[
 \mathbb F^{\rm tr}_{\ball_R(x)}\tau_t^B(A_{n_{\max}})
 =\mathbb F^{\rm tr}_{\ball_R(x)}\mathbb F^{\rm tr}_B\tau_t^B(A_{n_{\max}})
 =\mathbb F^{\rm tr}_{B\cap\ball_R(x)}\tau_t^B(A_{n_{\max}}).
\]
Thus \(D_n(t)\in\A_{B\cap\ball_{R_n}(x)}^\st\).
For \(n\ge1\), one has \(D_n(t)=0\) once \(B\subset\ball_{R_{n-1}}(x)\), since both conditional expectations defining that shell then fix \(\tau_t^B(A_{n_{\max}})\).
Telescoping therefore gives the finite expansion \(\tau_t^B(A_{n_{\max}})=\sum_{n\ge0}D_n(t)\).

For \(n\ge1\), nesting and contractivity of the tracial expectations imply
\[
 \|D_n(t)\|
 \le\|(\operatorname{id}-\mathbb F^{\rm tr}_{\ball_{R_{n-1}}(x)})\tau_t^B(A_{n_{\max}})\|.
\]
Apply \eqref{eq:tail-integrated-locality} with \(k=w_f\) and \(R_{\rm in}=R_{n-1}\).
Since \(R_n=2R_{n-1}\) for \(n\ge2\), this yields
\[
 \int_0^\infty w_f(s)\max_{\epsilon=\pm1}
       \|D_n(\epsilon s)\|\,\dd s
 \le
 \begin{cases}
 C\E^{-mR_n^\theta},&\text{stretched-exponential case},\\
 C(1+R_n)^{-\nu},&\text{polynomial case},
 \end{cases}
\]
with a possibly smaller positive \(m\).
The same bounds hold for \(n=0,1\) after increasing the constants, because \(\|D_n(t)\|\le2\|A_{n_{\max}}\|\le2\) and \(w_f\in L^1([0,\infty))\).
Using the volume bound \(|\ball_{R_n}(x)|\le C_\Gamma(1+R_n)^{d_{\rm vol}}\) and summing the preceding estimate over \(n\) gives
\begin{equation}
 \sum_{n\ge0}|\ball_{R_n}(x)|
 \int_0^\infty w_f(s)\max_{\epsilon=\pm1}
 \|D_n(\epsilon s)\|\,\dd s
 \le C.
 \label{eq:tail-integrated-shell-moment}
\end{equation}
The polynomial sum \(\sum_n(1+R_n)^{d_{\rm vol}-\nu}\) is finite because \(\nu>d_{\rm vol}\); the stretched-exponential bound also gives a finite sum.

Let \(W\subset B\) be finite and \(D\in\A_W^\st\).
If \(Z\cap W=\varnothing\), evenness of \(\Psi(Z)\) gives \([\Psi^{(1)}(Z),D]=0\).
The remaining commutators form an absolutely norm-convergent series because \(\Psi\in\mathfrak J_1\).
Therefore
\[
 \|\delta_\Psi^{(1)}(D)\|
 \le2\|D\|\sum_{z\in W}\sum_{Z\ni z}\|\Psi(Z)\|
 \le2|W|\|D\|J_\Psi.
\]
If \(S\ne\varnothing\) and \(B\subset S^c\), then \eqref{eq:exterior-tail-intensity} gives the sharper bound
\[
 \|\delta_\Psi^{(1)}(D)\|
 \le2\|D\|\sum_{z\in W}t_{\Psi,S}(z)
 \le2|W|\|D\|\sup_{z\in B}t_{\Psi,S}(z).
\]
When \(B\subset S^c\), apply the exterior estimate to \(D_n(t)\), with \(W:=B\cap\ball_{R_n}(x)\), and sum the finite shell expansion:
\[
 \|\delta_\Psi^{(1)}(\tau_t^B(A_{n_{\max}}))\|
 \le2\Bigl(\sup_{z\in B}t_{\Psi,S}(z)\Bigr)
       \sum_{n\ge0}|\ball_{R_n}(x)|\,\|D_n(t)\|.
\]
Taking the maximum over \(t=\pm s\), integrating against \(w_f(s)\,\dd s\), and using \eqref{eq:tail-integrated-shell-moment} proves \eqref{eq:tail-integrated-commutator}.
For general \(B\), use the bound with \(J_\Psi\) in the same calculation to prove \eqref{eq:global-integrated-defect-commutator}.
\end{proof}

\paragraph{Filtered excitation densities.}

For a nonempty \(S\subset\Gamma\) and \(x\in\Gamma\), set \(D_x:=\dist(x,S)\).
Define the local supremum of the exterior tail strength by
\begin{equation}
 T_{\Psi,S}(x):=
 \begin{cases}
 \displaystyle\sup_{z\in\ball_{\lfloor D_x/3\rfloor}(x)}
                              t_{\Psi,S}(z),&x\in S^c,\\
 0,&x\in S.
 \end{cases}
 \label{eq:tail-local-intensity}
\end{equation}
For \(x\in S^c\), every site in this ball has distance at least \(2D_x/3\) to \(S\).

\begin{proposition}
\label{prop:tail-filtered-coercivity}
Assume Hypothesis~\ref{hyp:background-transport} in either locality regime.
In the polynomial case, use \(\nu\) from \eqref{eq:main-polynomial-range}.
There are constants \(C>0\) and, in the stretched-exponential case, \(m>0\), depending only on the model data, with the following property.
For every \(|\lambda|\le\lambda_0\), defect \(\Psi\in\mathfrak J_1\), nonempty \(S\subset\Gamma\), and unital representation \(\pi:\A\to\mathcal B(\mathcal H)\) in which a nonnegative self-adjoint \(L\) implements \(\tau^{\Phi_\lambda+\Psi}\), let \(\psi\in\1_{[0,3g/4]}(L)\mathcal H\) be a unit vector.
Define \(\rho(A):=\langle\psi,\pi(A)\psi\rangle\) for \(A\in\A\), and
\begin{equation}
 \rho_\st:=\rho\widehat\otimes\omega_{\rm aux},\qquad
 \rho_\at:=\rho_\st\circ\sigma_\lambda^{-1}.
 \label{eq:tail-coordinate-state}
\end{equation}
For every \(x\in\Gamma\) and \(j\in\mathscr I\),
\begin{equation}
 \rho_\at
  ((c_{f,x,j}^{(\lambda)})^*c_{f,x,j}^{(\lambda)})
 \le
 \begin{cases}
 C[\E^{-mD_x^\theta}+T_{\Psi,S}(x)^2],
                        &\text{stretched-exponential case},\\
 C[(1+D_x)^{-2\nu}+T_{\Psi,S}(x)^2],
                        &\text{polynomial case}.
 \end{cases}
 \label{eq:tail-filtered-coercivity}
\end{equation}
All constants are independent of \(\Psi,S,\pi,L,\psi\).
\end{proposition}

\begin{proof}
Use the stacked representation \(\pi_\st\) and positive Hamiltonian \(\mathbb L=L\otimes\1+\1\otimes L_{\rm aux}\) from Lemma~\ref{lem:stacked-positive-representation}.
The vector \(\psi\otimes\Omega_{\rm aux}\) lies in \(\1_{[0,3g/4]}(\mathbb L)(\mathcal H\otimes\mathcal H_{\rm aux})\).
For any finite \(B\Subset\Gamma\) and \(D\in\A_B^\st\), the implementing dynamics gives
\[
 \begin{aligned}
 [\mathbb L,\pi_\st(D)]
       -\pi_\st(\delta_{\Phi_\lambda^\st,B}(D))
 =\pi_\st\bigl(
       (\delta_{\Phi_\lambda^\st}-\delta_{\Phi_\lambda^\st,B})(D)
          +\delta_\Psi^{(1)}(D)\bigr).
 \end{aligned}
\]
The first difference contains only bulk interaction terms crossing the boundary of \(B\).
Taking norms and estimating the physical and auxiliary commutators separately, as in \eqref{eq:stacked-crossing-norms}, gives \eqref{eq:local-comparison-defect-discrepancy} with \(Q=\1\) and additional interaction \(\Psi^{(1)}\).
The domain conditions follow from the implementing dynamics as in the proof of Lemma~\ref{lem:positive-representation-locality}.

Fix \(x,j\), and first suppose that \(D_x=\dist(x,S)\ge48\).
Take \(R:=\lfloor D_x/3\rfloor\) and \(B:=\ball_R(x)\subset S^c\).
For the integral \(e_R\) in \eqref{eq:filtered-occupation-defect-error}, Lemma~\ref{lem:tail-integrated-commutator} gives
\[
 e_R\le C\sup_{z\in B}t_{\Psi,S}(z)=CT_{\Psi,S}(x).
\]
Since
\[
 \rho_\at\bigl((c_{f,x,j}^{(\lambda)})^*c_{f,x,j}^{(\lambda)}\bigr)
 =\|\pi_\st(\widetilde c_{f,x,j}^{\,(\lambda)})(\psi\otimes\Omega_{\rm aux})\|^2,
\]
Lemma~\ref{lem:common-filtered-occupation}, with \(R\ge D_x/4\), proves \eqref{eq:tail-filtered-coercivity}.
For \(D_x<48\), use \(\|c_{f,x,j}^{(\lambda)}\|\le\|f\|_{L^1(\R)}\) and increase \(C\).
\end{proof}

\subsection{From excitation bounds to local determination}

\begin{proof}[Proof of Theorem~\ref{thm:quasilocal-tails}]
Fix \(M\ge0\) in (i), or \(\xi>0\) in (ii).
Fix \(\lambda\in\mathcal J\), a defect \(\Psi\in\mathfrak J_1\), and a nonempty set \(S\subset\Gamma\) as in the corresponding part of the theorem.
Let \(\pi:\A\to\mathcal B(\mathcal H)\) be a unital representation in which a nonnegative self-adjoint \(L\) implements \(\tau^{\Phi_\lambda+\Psi}\), and set \(P:=\1_{[0,3g/4]}(L)\).
If \(P=0\), every compressed operator in the conclusion vanishes.
Assume henceforth that \(P\ne0\), fix a unit vector \(\psi\in P\mathcal H\), and use its states \(\rho,\rho_\st,\rho_\at\) from \eqref{eq:tail-coordinate-state}.
For \(x\in\Gamma\), write \(D_x:=\dist(x,S)\).
For every \(x\in\Gamma\), set
\[
 u_x:=\rho_\at(n_x),\qquad
 b_x:=2\rho_\at(n_{f,x}^{(\lambda)}).
\]
Write \(u:=(u_x)_{x\in\Gamma}\) and \(b:=(b_x)_{x\in\Gamma}\), viewed as bounded nonnegative functions on \(\Gamma\).
Proposition~\ref{prop:tail-filtered-coercivity}, summed over \(j\in\mathscr I\), bounds \(b_x\) by the sum of a bulk term and \(CT_{\Psi,S}(x)^2\).
Also \(b_x\le4r\|f\|_{L^1(\R)}^2\), since \(\|c_{f,x,j}^{(\lambda)}\|\le\|f\|_{L^1(\R)}\).
Using \(\min\{c,a+b\}\le a+\min\{c,b\}\) for nonnegative \(a,b,c\), we obtain \(b\le b^{\rm bulk}+b^{\rm tail}\) with suitable constants \(C>0\) and, in the stretched-exponential case, \(c>0\), depending only on the model data, where for each \(x\in\Gamma\) we define
\[
 b_x^{\rm bulk}:=
 \begin{cases}
 C\E^{-cD_x^\theta},&\text{stretched-exponential case},\\
 C(1+D_x)^{-2\nu},&\text{polynomial case},
 \end{cases}
 \qquad b_x^{\rm tail}:=C\min\{1,T_{\Psi,S}(x)^2\}.
\]
Both functions are bounded independently of the defect.
In the polynomial case, use the kernel \(\mathscr K_{\lambda,S}^{\rm loc}\) defined in \eqref{eq:localized-LED-kernel} and set
\[
 v^{\rm tail}:=(\1-\mathscr K_{\lambda,S}^{\rm loc})^{-1}b^{\rm tail}.
\]
Apply \eqref{eq:localized-density-with-tail} with \(\mathcal Q=\rho_\at\), using \(b\le b^{\rm bulk}+b^{\rm tail}\), to obtain
\[
 u_x\le C_0(1+D_x)^{-2\nu}+2v_x^{\rm tail},
\]
where \(C_0\) is independent of the defect and the tail parameters.
In the stretched-exponential case, put
\[
 v^{\rm bulk}:=(\1-\mathscr K_\lambda)^{-1}b^{\rm bulk},
 \qquad
 v^{\rm tail}:=(\1-\mathscr K_\lambda)^{-1}b^{\rm tail}.
\]
Equation~\eqref{eq:filtered-reconstruction-density} gives \(u\le v^{\rm bulk}+v^{\rm tail}\).
The identities
\[
 v^{\rm bulk}=b^{\rm bulk}+\mathscr K_\lambda v^{\rm bulk},
 \qquad
 v^{\rm tail}=b^{\rm tail}+\mathscr K_\lambda v^{\rm tail}
\]
allow us to apply Lemma~\ref{lem:absorption} with \(U=S^c\) and \(K=\mathscr K_\lambda\).

For \(x\notin S\), every \(z\in\ball_{\lfloor D_x/3\rfloor}(x)\) satisfies \(\dist(z,S)\ge D_x-d(x,z)\ge2D_x/3\).
For every \(x\in\Gamma\), the envelope definitions therefore give
\begin{align}
 T_{\Psi,S}(x)
 &\le (3/2)^M J_{\Psi,S}^{M}(1+D_x)^{-M},
 \label{eq:tail-local-polynomial}\\
 T_{\Psi,S}(x)
 &\le J_{\Psi,S}^{\xi,\theta}
                 \E^{-\xi(2/3)^\theta D_x^\theta}.
 \label{eq:tail-local-stretched}
\end{align}
For \(x\in S\), both follow from \(T_{\Psi,S}(x)=0\).
For \(M=0\), use \(T_{\Psi,S}(x)\le J_{\Psi,S}^{0}\) and the uniform row bound \(\vartheta_{\mathcal J}<1\).
The Neumann series gives
\[
 \|v^{\rm tail}\|_\infty
 \le \frac{C(J_{\Psi,S}^{0})^2}{1-\vartheta_{\mathcal J}}.
\]

For \(M>0\), use \(\sigma_M\) and \(\alpha_M\) from the theorem.
Since \(0<\alpha_M\le1\), the inequality \(\min\{1,z^2\}\le z^{2\alpha_M}\), for \(z\ge0\), and \eqref{eq:tail-local-polynomial} give
\[
 b_x^{\rm tail}
 \le C_M(J_{\Psi,S}^{M})^{2\alpha_M}
                         (1+D_x)^{-2\sigma_M}.
\]
The inequality \(b_x^{\rm tail}\le CT_{\Psi,S}(x)^2\) and \eqref{eq:tail-local-polynomial} also give
\[
 b_x^{\rm tail}
 \le C_M(J_{\Psi,S}^{M})^2(1+D_x)^{-2M}
 \le C_M(J_{\Psi,S}^{M})^2(1+D_x)^{-2\sigma_M},
\]
since \(\sigma_M\le M\).
Combining the two bounds yields, for every \(x\in\Gamma\),
\begin{equation}
 b_x^{\rm tail}\le C_M
 \min\{(J_{\Psi,S}^{M})^2,(J_{\Psi,S}^{M})^{2\alpha_M}\}
                         (1+D_x)^{-2\sigma_M}.
 \label{eq:tail-capped-polynomial}
\end{equation}
For \(M=0\), the same inequality follows from \(T_{\Psi,S}(x)\le J_{\Psi,S}^{0}\) and \(\alpha_0=1\).

For polynomial bulk locality, \(0\le 2\sigma_M\le2\nu\), so \eqref{eq:localized-kernel-profile} applies at order \(2\sigma_M\).
The Neumann series for \((\1-\mathscr K_{\lambda,S}^{\rm loc})^{-1}\) preserves pointwise inequalities.
Applying this inverse to \eqref{eq:tail-capped-polynomial} and using \eqref{eq:localized-kernel-profile} with \(s=2\sigma_M\) gives, including when \(M=0\),
\[
 v_x^{\rm tail}\le C_M
 \min\{(J_{\Psi,S}^{M})^2,(J_{\Psi,S}^{M})^{2\alpha_M}\}
                       (1+D_x)^{-2\sigma_M}.
\]
Substituting this bound on \(v_x^{\rm tail}\) into \eqref{eq:localized-density-with-tail}, with \(\mathcal Q=\rho_\at\), gives, for every \(x\in\Gamma\),
\begin{equation}
 u_x\le C_0(1+D_x)^{-2\nu}
       +C_M\min\{(J_{\Psi,S}^{M})^2,(J_{\Psi,S}^{M})^{2\alpha_M}\}
                                                   (1+D_x)^{-2\sigma_M}.
 \label{eq:tail-density-polynomial}
\end{equation}
The constant \(C_0\) and the order \(2\nu\) are independent of \(M\).

Under stretched-exponential bulk locality, \eqref{eq:LED-stretched-contraction} gives \(\mathfrak M_M(\mathscr K_\lambda)<\infty\) for every \(M>0\), so Lemma~\ref{lem:absorption}(ii) applies at every polynomial order.
Apply Lemma~\ref{lem:absorption}(i) to \(v^{\rm bulk}\).
For polynomial tails with \(M>0\) and \(J_{\Psi,S}^{M}>0\), equation~\eqref{eq:tail-local-polynomial} gives
\[
 \frac{b_x^{\rm tail}}{(J_{\Psi,S}^{M})^2}
 \le C_M(1+D_x)^{-2M}.
\]
Apply Lemma~\ref{lem:absorption}(ii) to \(v^{\rm tail}/(J_{\Psi,S}^{M})^2\) at order \(2\sigma_M=2M\), where \(\alpha_M=1\).
For \(M=0\), use the preceding bound on \(\|v^{\rm tail}\|_\infty\).
For stretched-exponential tails with \(J_{\Psi,S}^{\xi,\theta}>0\), equation~\eqref{eq:tail-local-stretched} gives
\[
 \frac{b_x^{\rm tail}}{(J_{\Psi,S}^{\xi,\theta})^2}
 \le C\E^{-2\xi(2/3)^\theta D_x^\theta}.
\]
Apply Lemma~\ref{lem:absorption}(i) to \(v^{\rm tail}/(J_{\Psi,S}^{\xi,\theta})^2\).
If \(J_{\Psi,S}^{M}=0\) or \(J_{\Psi,S}^{\xi,\theta}=0\), respectively, then \(b^{\rm tail}=v^{\rm tail}=0\).
For every \(x\in\Gamma\), the resulting estimates are, respectively,
\begin{align}
 u_x&\le C_0\E^{-c_0D_x^\theta}
       +C_M(J_{\Psi,S}^{M})^2(1+D_x)^{-2\sigma_M},
 \label{eq:tail-density-mixed}\\
 u_x&\le C_0\E^{-c_0D_x^\theta}
       +C_\xi (J_{\Psi,S}^{\xi,\theta})^2\E^{-c_\xi D_x^\theta},
 \label{eq:tail-density-stretched}
\end{align}
where \(C_0,c_0>0\) depend only on the model and LED data.
The other constants and rates depend additionally on the tail exponent \(M\) or \(\xi\).

Fix a nonempty finite \(Y\subset\Gamma\) and set \(D:=\dist(Y,S)\).
Define \(H(t)\), for \(t\ge0\), by replacing \(D_x\) with \(t\) on the right-hand side of the applicable estimate \eqref{eq:tail-density-polynomial}, \eqref{eq:tail-density-mixed}, or \eqref{eq:tail-density-stretched}.
Then \(H\) is bounded and nonincreasing.
The density estimate holds for every unit vector \(\psi\in P\mathcal H\), with the same function \(H\).
Apply Lemma~\ref{lem:positive-representation-cb} with the map \(\mathcal Q\) in \eqref{eq:tail-amplification-maps} and \(v_x:=H(D_x)\).
Use the difference map \(\mathcal D_Y\) on \(\A_Y\) defined after \eqref{eq:tail-amplification-maps}.
Fix \(k\in\mathbb N\) and \(A\in\A_Y\otimes M_k\).
The definition of \(\mathcal D_Y\) gives
\[
 \|(\mathcal D_Y\otimes\operatorname{id}_{M_k})(A)\|
 =\left\|(P\otimes\1)\left[(\pi\otimes\operatorname{id}_{M_k})(A)
       -\1\otimes(\omega_\lambda\otimes\operatorname{id}_{M_k})(A)\right]
       (P\otimes\1)\right\|.
\]
For \(D\ge4\), part (ii) gives the compression bound in \eqref{eq:positive-representation-cb}.
Use \(\alpha_{\rm rec}=d_{\rm vol}/(2p_\at)\) in the polynomial case and \(\alpha_{\rm rec}=q-1/2\) in the stretched-exponential case, as in the application preceding \eqref{eq:tail-amplification-maps}.
For polynomial bulk locality, subadditivity of the square root gives
\[
 \begin{aligned}
 \|(\mathcal D_Y\otimes\operatorname{id}_{M_k})(A)\|
 \le C\|A\|\Big[&
 (1+|Y|)^{1/2+\alpha_{\rm rec}}\bigl((1+D)^{-\nu}\\
 &+C_M\min\{J_{\Psi,S}^{M},(J_{\Psi,S}^{M})^{\alpha_M}\}
                           (1+D)^{-\sigma_M}\bigr)\\
 &+\min\{1,(1+|Y|)(1+D)^{-p_\at}\}\Big].
 \end{aligned}
\]
The fixed rescalings \(D/2,D/4\) change only the constants.
Since \(1/2+\alpha_{\rm rec}\le q\), the density terms have the support factor and decay powers of \eqref{eq:tail-short-polynomial}.
Estimate~\eqref{eq:support-tail-interpolation} bounds the last term by \((1+|Y|)^q(1+D)^{-\nu}\).
This bounds \(\|(\mathcal D_Y\otimes\operatorname{id}_{M_k})(A)\|\) by the right-hand side of \eqref{eq:tail-short-polynomial}, with the outer minimum replaced by its second argument.
For stretched-exponential bulk locality, the square-root inequality separates the bulk and tail density terms in \eqref{eq:tail-density-mixed} or \eqref{eq:tail-density-stretched}, each with support factor \((1+|Y|)^q\).
The term \(\min\{1,(1+|Y|)G_\at(D/4)\}\) in \eqref{eq:positive-representation-cb} is at most \((1+|Y|)^q\E^{-q\mu_\at D^\theta/4^\theta}\), where \(\mu_\at\) is defined in \eqref{eq:atomic-coordinate-profile}.
The bulk part of \(H(D/2)^{1/2}\) is bounded by \(C\E^{-c_0D^\theta/2^{1+\theta}}\).
Choose the common rate \(m\) in Theorem~\ref{thm:positive-energy-locality} no larger than \(c_0/2^{1+\theta}\) and \(q\mu_\at/4^\theta\).
Both depend only on the model and LED data and \(q\), so this choice is independent of \(M\) and \(\xi\).
In (i), taking the square root gives the tail exponent \(\sigma_M=M\).
In (ii), use \(m_\xi^{\rm tail}:=c_\xi/2^{1+\theta}\).
These choices give the right-hand sides of \eqref{eq:tail-short-polynomial} and \eqref{eq:tail-short-stretched}, with each outer minimum replaced by its second argument.

To obtain the minima in the statements, distinguish whether \(\min\{J_{\Psi,S}^{M},(J_{\Psi,S}^{M})^{\alpha_M}\} (1+D)^{-\sigma_M}\), respectively \(J_{\Psi,S}^{\xi,\theta}\E^{-m_\xi^{\rm tail}D^\theta}\), is at most one.
In that case the estimate just proved applies.
Otherwise use \(\|(\mathcal D_Y\otimes\operatorname{id}_{M_k})(A)\|\le2\|A\|\), since both \(A\mapsto V^*\pi(A)V\) and \(A\mapsto\omega_\lambda(A)\1_{\mathcal C}\) are unital and completely positive, and increase the tail constant to at least two.
For \(D<4\), the same norm bound is absorbed by the bulk term after increasing its constant.
Increase the common \(C\) from Theorem~\ref{thm:positive-energy-locality}, if necessary, to dominate the bulk prefactor established here.
This choice and, in the stretched-exponential case, the restriction on \(m\) depend only on the model and LED data and \(q\), uniformly in \(M\) and \(\xi\).
The bulk term is therefore \(CG(D)\), and the defect enters only through the tail envelopes.

These bounds hold for every matrix size \(k\) with the same constants.
Further compression gives the bounds on every smaller spectral window.
\end{proof}

\begin{proof}[Proof of Corollary~\ref{cor:uniformly-weak-defects}]
Fix \(\lambda\in\mathcal J\), a defect \(\Psi\in\mathfrak J_1\), and a unital representation \(\pi:\A\to\mathcal B(\mathcal H)\) in which a nonnegative self-adjoint \(L\) implements \(\tau^\Xi\), with \(\Xi:=\Phi_\lambda+\Psi\).
Set \(P:=\1_{[0,3g/4]}(L)\).
If \(P=0\), the compressed operator is zero.
Assume \(P\ne0\), and fix a unit vector \(\psi\in P\mathcal H\).
For \(A\in\A\), define
\[
 \rho(A):=\langle\psi,\pi(A)\psi\rangle,\qquad
 \rho_\at:=(\rho\widehat\otimes\omega_{\rm aux})
                              \circ\sigma_\lambda^{-1}.
\]

Use the stacked representation \(\pi_\st\) and nonnegative Hamiltonian \(\mathbb L=L\otimes\1+\1\otimes L_{\rm aux}\) from Lemma~\ref{lem:stacked-positive-representation}.
For every \(x,j\) and \(R\ge16\), we use the commutator calculation in the proof of Proposition~\ref{prop:tail-filtered-coercivity} to obtain the bound \eqref{eq:local-comparison-defect-discrepancy} required in Lemma~\ref{lem:common-filtered-occupation}, with \(Q=\1\), \(L=\mathbb L\), and additional interaction \(\Psi^{(1)}\).
Estimate~\eqref{eq:global-integrated-defect-commutator} bounds its integral \(e_R\) by \(CJ_\Psi\), uniformly in \(R,x,j\).
Apply \eqref{eq:common-filtered-occupation} to \(\psi\otimes\Omega_{\rm aux}\), which lies in the window \([0,3g/4]\) of \(\mathbb L\), and let \(R\to\infty\).
The bulk boundary contribution tends to zero, giving
\[
 \rho_\at\bigl((c_{f,x,j}^{(\lambda)})^*c_{f,x,j}^{(\lambda)}\bigr)
 =\|\pi_\st(\widetilde c_{f,x,j}^{\,(\lambda)})
                    (\psi\otimes\Omega_{\rm aux})\|^2
 \le CJ_\Psi^2.
\]
Summing over \(j\in\mathscr I\) yields, for every \(x\in\Gamma\),
\[
 b_x:=2\rho_\at(n_{f,x}^{(\lambda)})\le CJ_\Psi^2.
\]
Set \(u_x:=\rho_\at(n_x)\) for \(x\in\Gamma\), and write \(u:=(u_x)_{x\in\Gamma}\), \(b:=(b_x)_{x\in\Gamma}\).
The detectability inequality gives \(u\le b+\mathscr K_\lambda u\).
Since \(0\le u_x\le2r\) for every \(x\), taking suprema yields
\[
 \|u\|_\infty\le \frac{CJ_\Psi^2}{1-\vartheta_{\mathcal J}}.
\]

The density bound holds for every unit vector in \(P\mathcal H\), so apply Lemma~\ref{lem:positive-representation-cb}(i) with the map \(\mathcal Q\) in \eqref{eq:tail-amplification-maps} and \(v_x:=CJ_\Psi^2\).
Use the difference map \(\mathcal D_Y\) on \(\A_Y\) defined after \eqref{eq:tail-amplification-maps}.
For every nonempty finite \(Y\Subset\Gamma\), every \(k\in\mathbb N\), and every \(A\in\A_Y\otimes M_k\), it gives
\[
 \|(\mathcal D_Y\otimes\operatorname{id}_{M_k})(A)\|
 \le\min\{2,C(1+|Y|)^{1/2+\alpha_{\rm rec}}J_\Psi\}\|A\|.
\]
Here \(\alpha_{\rm rec}=q-1/2\) in the stretched-exponential case and \(\alpha_{\rm rec}=d_{\rm vol}/(2p_\at)\) in the polynomial case.
The density bound holds at every site, so part~(i) of the lemma sums all shells of the observable.
In the distance-dependent bound, the omitted shells give the additional support exponent \(\nu/p_\at\) in \eqref{eq:support-tail-interpolation}.
The amplified norm is the compression norm by \(P\otimes\1\), as in the preceding proof, which proves \eqref{eq:uniformly-weak-defect-comparison}.
\end{proof}

\section{Correlation bounds}
\label{sec:correlation-proofs}

In Subsection~\ref{sec:bulk-clustering-proof}, we prove clustering of the bulk ground state \(\omega_\lambda\) from locality of \(\sigma_\lambda\) and derive the extensive variance bound.
Under LED, the GNS gap allows us to apply the exponential-clustering argument in the metric \(d^{\theta_{\rm in}}\), giving the stronger stretched-exponential bound.
Subsection~\ref{sec:spatial-correlation-proofs} uses the excitation-density bounds to prove one-sided covariance estimates for low-energy states in positive-energy representations and for local ground states, together with an operator estimate on low-energy subspaces and their finite-sample specializations.

\subsection{Clustering of the bulk state}
\label{sec:bulk-clustering-proof}

\begin{proof}[Proof of Proposition~\ref{prop:bulk-clustering}]
Fix \(|\lambda|\le\lambda_0\), nonempty finite \(X,Y\), and \(A\in\A_X\), \(B\in\A_Y\).
Write \(d_{XY}:=\dist(X,Y)\); for \(d_{XY}<3\), the norm bound \(|\operatorname{Cov}_{\omega_\lambda}(A,B)|\le2\|A\|\|B\|\) suffices.
For \(d_{XY}\ge3\), put \(R=d_{XY}/3\) and set
\[
 A_\at:=\sigma_\lambda(A\widehat\otimes\1),\qquad
 A_R:=\mathbb F^{\rm tr}_{X^{(R)}}(A_\at),
\]
with the corresponding definitions for \(B\).
Equation~\eqref{eq:exact-atomic-state} identifies the bulk covariance with that of \(A_\at,B_\at\) in \(\omega_\at\).
The supports of \(A_R,B_R\) are disjoint, and the even product state factorizes on them, including for observables of arbitrary parity.
Bilinearity of covariance and contractivity therefore give
\begin{equation}
 |\operatorname{Cov}_{\omega_\at}(A_\at,B_\at)|
 \le2\|A_\at-A_R\|\|B\|+2\|A\|\|B_\at-B_R\|.
 \label{eq:transported-covariance-comparison}
\end{equation}
Lemma~\ref{lem:coordinate-locality} now proves \eqref{eq:bulk-correlation-polynomial}.

To obtain the boundary factors, use the path \((\Xi_u)_{u\in I}\) generating \(\sigma_\lambda\) from the proof of Lemma~\ref{lem:coordinate-locality}.
Its integrated first support moment is uniformly bounded with weight \(w(s)=(1+s)^{p_\at}\) in the polynomial case or \(w(s)=\E^{\mu_1s^\theta}\) for some \(\mu_1>0\) in the stretched-exponential case.
Apply \eqref{eq:sitewise-transport-restriction} to this path and use that \(\mathbb F^{\rm tr}_\Lambda\) fixes the observable evolved along the restricted path.
For \(X\subseteq\Lambda\Subset\Gamma\) this gives
\[
 \|A_\at-\mathbb F^{\rm tr}_\Lambda(A_\at)\|
 \le C\|A\|\sum_{x\in X}w(\dist(x,\Lambda^c))^{-1}.
\]

For each \(x\in X\), choose \(u_x\in X^c\) nearest to \(x\), and \(z_x\in X\) nearest to \(u_x\).
Then \(z_x\in\partial X\) and \(d(x,z_x)\le2\dist(x,X^c)\).
Volume growth, with a sum over balls and dyadic annuli about \(z_x\), gives, for \(p>d_{\rm vol}\), \(c>0\), and \(0<\alpha\le1\),
\begin{align}
 \sum_{x\in X}(1+\max\{R,\dist(x,X^c)\})^{-p}
 &\le C_p|\partial X|(1+R)^{d_{\rm vol}-p},\notag\\
 \sum_{x\in X}\E^{-c\max\{R,\dist(x,X^c)\}^\alpha}
 &\le C_{c,\alpha}|\partial X|\E^{-cR^\alpha/2}.
 \label{eq:clustering-boundary-sums}
\end{align}
For the polynomial sum, the annuli contribute the convergent series \(\sum_{n\ge0}2^{n(d_{\rm vol}-p)}\); for the second sum, split the exponent between \(R\) and \(d(x,z_x)/2\).
Take \(\Lambda=X^{(R)}\), for which \(\dist(x,\Lambda^c)\ge\max\{R,\dist(x,X^c)\}\).
The approximation estimate and \eqref{eq:clustering-boundary-sums} give
\[
 \|A_\at-A_R\|\le C|\partial X|\|A\|
 \begin{cases}
 (1+R)^{-(p_\at-d_{\rm vol})},&\text{polynomial case},\\
 \E^{-\mu_1R^\theta/2},&\text{stretched-exponential case}.
 \end{cases}
\]
Insert this bound and its counterpart for \(B\) into \eqref{eq:transported-covariance-comparison} to obtain \eqref{eq:bulk-correlation-polynomial-boundary}.
In the stretched-exponential case, combine the resulting estimate with the norm bound and \(\min\{2,u\}\le\sqrt{2u}\) to obtain \eqref{eq:bulk-correlation-transport-stretched}.
Taking this square root halves the stretched-exponential rate while preserving the order \(\theta\), whereas in \eqref{eq:bulk-correlation-polynomial-boundary} it would halve the decay power \(p_\at-d_{\rm vol}\).

For \eqref{eq:bulk-correlation-exponential}, suppose additionally that LED holds on \(\mathcal J\), and take \(\lambda\in\mathcal J\).
The interaction \(\Phi_\lambda\) has a uniformly bounded exponential norm in the metric \(d^{\theta_{\rm in}}\).
For \(b>0\) small enough, the matrix estimate of \cite[Lemma~3.1]{FermionicLR} bounds the off-diagonal propagation entries by
\[
 0\le(\E^{2|t|J_\lambda}-\1)(x,y)
 \le(\E^{v|t|}-1)\E^{-3b d(x,y)^{\theta_{\rm in}}},
 \qquad J_\lambda(x,y):=\sum_{Z\ni x,y}\|\Phi_\lambda(Z)\|.
\]
Since \(d(x,y)\ge d_{XY}\) and \(d(x,y)\ge\dist(x,X^c)\) for \(x\in X,y\in Y\), volume growth and \eqref{eq:clustering-boundary-sums} with \(R=0\) imply
\[
 \sum_{x\in X,y\in Y}\E^{-3b d(x,y)^{\theta_{\rm in}}}
 \le C_b\E^{-b d_{XY}^{\theta_{\rm in}}}
       \sum_{x\in X}\E^{-b\dist(x,X^c)^{\theta_{\rm in}}}
 \le C_b'|\partial X|\E^{-b d_{XY}^{\theta_{\rm in}}}.
\]
Interchanging \(X,Y\) gives the minimum of the two boundary sizes.
For odd \(A,B\), adjoin a decoupled CAR mode at a site of \(Y\) and extend \(\tau^{\Phi_\lambda}\) by the identity on this mode, as in \cite[Lemma~4.3]{FermionicLR}.
If \(c\) is a self-adjoint Majorana unitary of this mode, \(Bc\) is even and supported in \(Y\), and \(\|Bc\|=\|B\|\).
The interaction matrix \(J_\lambda\) is unchanged, so the commutator bound applies to \([\tau_t^{\Phi_\lambda}(A),Bc]=\{\tau_t^{\Phi_\lambda}(A),B\}c\) with the same constants.
For homogeneous \(A,B\) this gives
\begin{equation}
 \|[\tau_t^{\Phi_\lambda}(A),B]_{\rm gr}\|
 \le C\|A\|\|B\|\min\{|\partial X|,|\partial Y|\}
       (\E^{v|t|}-1)\E^{-a d_{XY}^{\theta_{\rm in}}}.
 \label{eq:clustering-graded-LR}
\end{equation}

Corollary~\ref{cor:ground-state-uniqueness} gives a simple ground vector and a GNS gap of at least \(3g/4\).
We use the Gaussian integral of Nachtergaele and Sims \cite[proof of Theorem~2]{NachtergaeleSims06} in the original ground-state representation, taking the anticommutator for odd pairs as in Hastings and Koma \cite[Section~3, proof of Theorem~2.6]{HastingsKoma06}.
Centering \(A,B\) in the even state \(\omega_\lambda\) preserves parity and removes the ground-vector contribution.
Write \(r_{XY}=d_{XY}^{\theta_{\rm in}}\), \(K=\min\{|\partial X|,|\partial Y|\}\), and \(\gamma=3g/4\).
Split the time integral at \(|t|=T:=ar_{XY}/(2v)\) and use the Gaussian \(\E^{-\alpha t^2}\) with \(\alpha=\gamma/(2T)\).
By \eqref{eq:clustering-graded-LR}, the part with \(|t|\le T\) is bounded by
\[
 C\|A\|\|B\|K\E^{-ar_{XY}}\int_0^T\frac{\E^{vt}-1}{t}\,\dd t
 \le C\|A\|\|B\|K\E^{-ar_{XY}/2}.
\]
For \(|t|>T\), the norm bound on the graded commutator leaves the Gaussian tail \(\int_T^\infty\E^{-\alpha t^2}\,\dd t/t\).
The spectral error for energies at least \(\gamma\) is bounded by \(C\|A\|\|B\|\E^{-\gamma^2/(4\alpha)}\).
Since \(\alpha T^2=\gamma^2/(4\alpha)=\gamma T/2\) and \(d_{XY}\ge3\), both errors are at most \(C\|A\|\|B\|\E^{-\gamma T/2}\).
Adding the three bounds gives, for some \(\mu_0>0\) depending only on the model and LED data,
\[
 |\operatorname{Cov}_{\omega_\lambda}(A,B)|
 \le C(1+\min\{|\partial X|,|\partial Y|\})\|A\|\|B\|
       \E^{-\mu_0d_{XY}^{\theta_{\rm in}}}.
\]
Combining this with the norm bound and \(\min\{2,u\}\le\sqrt{2u}\) proves \eqref{eq:bulk-correlation-exponential}, with \(\mu=\mu_0/2\).
Decomposing arbitrary observables into their even and odd parts completes the proof.
\end{proof}

\begin{proof}[Proof of \eqref{eq:extensive-variance}]
Assume Hypothesis~\ref{hyp:background-transport}, and fix \(s>d_{\rm vol}\), a self-adjoint family \((A_x)_{x\in\Gamma}\) with \(J_s<\infty\), and \(|\lambda|\le\lambda_0\).
For each \(x\), use the dyadic shells \(A_{x,n}:=\Delta_{\{x\},n}^{\rm tr}A_x\), supported in \(\ball_{R_n}(x)\).
The localization norm and the ratio of successive radii give
\[
 \|A_{x,n}\|\le C_sJ_s(1+R_n)^{-s},
 \qquad
 \sum_{n\ge0}(1+R_n)^{d_{\rm vol}}\|A_{x,n}\|
       \le C_s'J_s.
\]
The last series converges because \(s>d_{\rm vol}\).
Let \(F_{\rm cl}(R):=\E^{-\mu_{\rm tr}R^\theta}\) in the stretched-exponential case and \(F_{\rm cl}(R):=(1+R)^{-p_\at}\) in the polynomial case, using \eqref{eq:bulk-correlation-transport-stretched} and \eqref{eq:bulk-correlation-polynomial}, respectively.
For \(x,y\in\Gamma\) with \(D:=d(x,y)\ge8\), let \(n_{\max}\) be the largest dyadic index with \(R_{n_{\max}}\le D/4\), and put \(A_x^{[n_{\max}]}:=\sum_{n=0}^{n_{\max}} A_{x,n}\), and similarly for \(y\).
For \(n,k\le n_{\max}\), the supports of \(A_{x,n}\) and \(A_{y,k}\) are separated by at least \(D/2\).
Bound the support factors in Proposition~\ref{prop:bulk-clustering} by \(C[(1+R_n)^{d_{\rm vol}}+(1+R_k)^{d_{\rm vol}}]\), using volume growth.
Bilinearity of covariance and the shell bound then imply
\[
 \begin{aligned}
 |\operatorname{Cov}_{\omega_\lambda}(A_x^{[n_{\max}]},A_y^{[n_{\max}]})|
 &\le C F_{\rm cl}(D/2)
     \sum_{n,k=0}^{n_{\max}}
       \bigl[(1+R_n)^{d_{\rm vol}}+(1+R_k)^{d_{\rm vol}}\bigr]
        \|A_{x,n}\|\|A_{y,k}\|\\
 &\le C_sJ_s^2 F_{\rm cl}(D/2).
 \end{aligned}
\]
Since \(A_x^{[n_{\max}]}=\mathbb F^{\rm tr}_{\ball_{R_{n_{\max}}}(x)}(A_x)\) and \(R_{n_{\max}}>D/8\), the omitted part has norm at most \(C_sJ_s(1+D)^{-s}\), and the same holds for \(y\).
Using \(|\operatorname{Cov}_\rho(E,F)|\le2\|E\|\|F\|\) for the two remainder terms yields
\[
 |\operatorname{Cov}_{\omega_\lambda}(A_x,A_y)|
 \le C_sJ_s^2\bigl[F_{\rm cl}(d(x,y)/2)+(1+d(x,y))^{-s}\bigr].
\]
After increasing \(C_s\), this bound also holds for \(d(x,y)<8\).
Volume growth, \(s>d_{\rm vol}\), and \(p_\at>d_{\rm vol}\) in the polynomial case make the right-hand side summable over \(y\), uniformly in \(x\).
Thus expansion of the square gives
\[
 \operatorname{Var}_{\omega_\lambda}
       \left(\sum_{x\in\Lambda}A_x\right)
 =\sum_{x,y\in\Lambda}\operatorname{Cov}_{\omega_\lambda}(A_x,A_y)
 \le\sum_{x,y\in\Lambda}
       |\operatorname{Cov}_{\omega_\lambda}(A_x,A_y)|
 \le C_sJ_s^2|\Lambda|.\qedhere
\]
\end{proof}

\subsection{Correlations in local ground states and low-energy states}
\label{sec:spatial-correlation-proofs}

For a state \(\varrho\) and a finite region \(Z\), smallness of \(\sum_{x\in Z}\varrho(n_x)\) implies both local agreement with \(\omega_\at\) on \(Z\) and small correlations with observables supported outside \(Z\).

\begin{lemma}
\label{lem:atomic-decoupling}
Let \(\omega_{\rm p}\), \(P_x\), and \(Q_x\) be as in Lemma~\ref{lem:coordinate-reconstruction}, and put \(P_Z:=\prod_{z\in Z}P_z\) for finite \(Z\Subset\Gamma\).
For every nonempty finite \(Z\Subset\Gamma\), state \(\varrho\) on \(\A^\st\), \(C\in\A_Z^\st\), and \(D\in\A_{Z^c}^\st\), one has
\begin{equation}
 |\operatorname{Cov}_\varrho(C,D)|
 \le6\|C\|\|D\|\sqrt{\varrho(\1-P_Z)}
 \le6\|C\|\|D\|
          \left(\sum_{z\in Z}\varrho(Q_z)\right)^{1/2}.
 \label{eq:atomic-decoupling}
\end{equation}
\end{lemma}

\begin{proof}
The projection \(P_Z\) is even and has rank one in the local algebra, so
\[
 P_ZCP_Z=\omega_{\rm p}(C)P_Z,\qquad \1-P_Z\le\sum_{z\in Z}Q_z.
\]
Since \(P_Z\) is even, \(P_ZD=DP_Z\).

Suppose first that \(\varrho(P_Z)>0\), and define the state \(\varrho^{P_Z}(T):=\varrho(P_ZTP_Z)/\varrho(P_Z)\) on \(\A^\st\).
Estimate~\eqref{eq:vacuum-projection-state-distance} gives \(\|\varrho-\varrho^{P_Z}\|\le2\sqrt{1-\varrho(P_Z)}\).
Moreover,
\[
 \varrho^{P_Z}(CD)=\frac{\varrho(P_ZCP_ZD)}{\varrho(P_Z)}
           =\omega_{\rm p}(C)\varrho^{P_Z}(D),\qquad
 \varrho^{P_Z}(C)=\omega_{\rm p}(C).
\]
Thus \(\operatorname{Cov}_{\varrho^{P_Z}}(C,D)=0\).
Expanding the difference of the two covariances, with \(\|\varrho\|=\|\varrho^{P_Z}\|=1\), yields
\[
 |\operatorname{Cov}_\varrho(C,D)|
 \le3\|C\|\|D\|\|\varrho-\varrho^{P_Z}\|
 \le6\|C\|\|D\|\sqrt{\varrho(\1-P_Z)}.
\]
If \(\varrho(P_Z)=0\), the same bound follows from \(|\operatorname{Cov}_\varrho(C,D)|\le2\|C\|\|D\|\).
The inequality \(\1-P_Z\le\sum_{z\in Z}Q_z\) gives the second inequality in \eqref{eq:atomic-decoupling}.
\end{proof}

We apply this lemma with \(\omega_{\rm p}=\omega_\at\), \(Q_x=n_x\), and \(P_Z=P_Z^\vac\).

\begin{proof}[Proof of Theorem~\ref{thm:spatial-correlations}]
Fix \(\lambda\in\mathcal J\).
We apply Lemma~\ref{lem:atomic-decoupling} to the transported observables.
For a state \(\rho\) on \(\A\) and a set \(U\subset\Gamma\), put
\[
 \varrho:=(\rho\widehat\otimes\omega_{\rm aux})
                   \circ\sigma_\lambda^{-1},\qquad
 A_\at:=\sigma_\lambda(A\widehat\otimes\1),\qquad
 B_\at:=\sigma_\lambda(B\widehat\otimes\1).
\]
By construction, \(\operatorname{Cov}_\rho(A,B) =\operatorname{Cov}_\varrho(A_\at,B_\at)\).
For the moment suppose that \(U\ne\Gamma\) and that a bounded nonincreasing function \(H:[0,\infty)\to[0,\infty)\) satisfies, for every \(z\in\Gamma\),
\[
 \varrho(n_z)\le H(\dist(z,U^c)).
\]
Set \(d_{XY}:=\dist(X,Y)\) and \(D_X:=\dist(X,U^c)\), and first assume \(d_{XY},D_X\ge8\).
Define
\[
 B_{d_{XY}/4}:=\mathbb F^{\rm tr}_{Y^{(d_{XY}/4)}}(B_\at),\qquad
 T:=\min\{d_{XY}/4,D_X/2\},
\]
and let \(n_{\max}\) be the largest dyadic index with \(R_{n_{\max}}\le T\).
Thus \(T/2<R_{n_{\max}}\le T\).
Use the dyadic tracial shells \(A_{\at,n}:=\Delta_{X,n}^{\rm tr}A_\at\).
Use \(\alpha_{\rm rec}=d_{\rm vol}/(2p_\at)\) in the polynomial case and \(\alpha_{\rm rec}=q-1/2\) in the stretched-exponential case.
The shell estimate \eqref{eq:coordinate-weighted-shells}, established in the proof of Lemma~\ref{lem:coordinate-reconstruction}, and the approximation bound \eqref{eq:gamma-locality-explicit} for \(\sigma_\lambda\) give
\begin{align}
 \sum_{n\ge0}(1+R_n)^{d_{\rm vol}/2}\|A_{\at,n}\|
 &\le C(1+|X|)^{\alpha_{\rm rec}}\|A\|,
 \label{eq:correlation-weighted-shells}\\
 \left\|A_\at-\sum_{n=0}^{n_{\max}} A_{\at,n}\right\|
 &\le C\|A\|\min\{1,(1+|X|)G_\at(R_{n_{\max}})\},
 \label{eq:correlation-shell-remainder}\\
 \|B_\at-B_{d_{XY}/4}\|
 &\le C(1+|Y|)\|B\|G_\at(d_{XY}/4).
 \label{eq:correlation-second-approximation}
\end{align}

For \(n\le n_{\max}\), the sets \(X^{(R_n)}\) and \(Y^{(d_{XY}/4)}\) are disjoint.
Every site in \(X^{(R_n)}\) has distance at least \(D_X/2\) to \(U^c\), and
\[
 |X^{(R_n)}|\le C_\Gamma|X|(1+R_n)^{d_{\rm vol}}.
\]
Lemma~\ref{lem:atomic-decoupling}, applied on \(X^{(R_n)}\), therefore gives
\[
 \begin{aligned}
 \sum_{n=0}^{n_{\max}}
 |\operatorname{Cov}_\varrho(A_{\at,n},B_{d_{XY}/4})|
 &\le C|X|^{1/2}\|B\|H(D_X/2)^{1/2}
       \sum_{n=0}^{n_{\max}}(1+R_n)^{d_{\rm vol}/2}\|A_{\at,n}\|\\
 &\le C(1+|X|)^{1/2+\alpha_{\rm rec}}\|A\|\|B\|H(D_X/2)^{1/2}.
 \end{aligned}
\]
Here \(\|B_{d_{XY}/4}\|\le\|B\|\).
Bilinearity and the norm bound for covariance control the two remaining errors by \eqref{eq:correlation-shell-remainder} and \eqref{eq:correlation-second-approximation}.
Since \(R_{n_{\max}}>\min\{d_{XY}/8,D_X/4\}\), monotonicity bounds \(G_\at(R_{n_{\max}})\) by the larger of \(G_\at(d_{XY}/8)\) and \(G_\at(D_X/4)\).
Taking the minimum with one preserves this comparison, and bounding the larger term by their sum gives
\begin{equation}
 \begin{split}
 |\operatorname{Cov}_\rho(A,B)|
 \le C\|A\|\|B\|\bigl[
 &(1+|X|)^{1/2+\alpha_{\rm rec}}H(D_X/2)^{1/2}\\
 &+\min\{1,(1+|X|)G_\at(D_X/4)\}\\
 &+(1+|X|+|Y|)G_\at(d_{XY}/8)\bigr].
 \end{split}
 \label{eq:one-sided-density-comparison}
\end{equation}

Suppose that the function \(H\) bounding \(\varrho(n_z)\) has the form
\[
 H(D)=
 \begin{cases}
 C\E^{-cD^\theta},&\text{stretched-exponential locality},\\
 C(1+D)^{-2\nu},&\text{polynomial locality},
 \end{cases}
 \qquad H(\infty)=0.
\]
In the polynomial case, the definition of \(q\) bounds \((1+|X|)^{1/2+\alpha_{\rm rec}}H(D_X/2)^{1/2}\) in \eqref{eq:one-sided-density-comparison} by \(C(1+|X|)^q(1+D_X)^{-\nu}\).
Estimate~\eqref{eq:support-tail-interpolation} bounds \(\min\{1,(1+|X|)G_\at(D_X/4)\}\) by the same expression.
The separation term has power \(p_\at\).
In the stretched-exponential case, \(1/2+\alpha_{\rm rec}=q\), and \(\min\{1,(1+|X|)G_\at(D_X/4)\}\le(1+|X|)^qG_\at(D_X/4)^q\).
Choose \(m>0\) no larger than the decay rates of \(H(D_X/2)^{1/2}\), \(G_\at(D_X/4)^q\), and \(G_\at(d_{XY}/8)\), and no larger than \(\mu_{\rm tr}\) in Proposition~\ref{prop:bulk-clustering}.
Then \eqref{eq:one-sided-density-comparison} gives the right-hand side of \eqref{eq:positive-energy-clustering}.
If \(d_{XY}<8\) or \(D_X<8\), the norm bound for covariance proves the same estimate after increasing \(C\).
If \(U=\Gamma\), zero excitation density implies \(\varrho=\omega_\at\) by Lemma~\ref{lem:exact-atomic-reconstruction}, hence \(\rho=\omega_\lambda\); Proposition~\ref{prop:bulk-clustering} gives the remaining separation bound.

\emph{States in positive-energy windows.}
Fix \(U,\Xi,\pi,L\) satisfying the hypotheses of part (i), and set \(P:=\1_{[0,3g/4]}(L)\).
If \(P=0\), the compressed operator is zero and no state satisfies \(\varphi(P)=1\), so assume \(P\ne0\).
Let \(\mathcal C:=P\mathcal H\), let \(V:\mathcal C\to\mathcal H\) be the inclusion, and use the unital completely positive map \(\mathcal Q:\A^\st\to\mathcal B(\mathcal C)\) from \eqref{eq:tail-amplification-maps}.
Lemma~\ref{lem:positive-representation-locality} bounds the excitation densities uniformly over all unit vectors in \(\mathcal C\), so
\[
 0\le\mathcal Q(n_x)\le H(\dist(x,U^c))\1_{\mathcal C}
 \qquad(x\in\Gamma),
\]
with the envelope in \eqref{eq:low-energy-density}, including zero when \(U=\Gamma\).
For any state \(\varphi\) on \(\mathcal B(\mathcal H)\) with \(\varphi(P)=1\), Cauchy--Schwarz gives \(\varphi(T)=\varphi(PTP)\) for every \(T\in\mathcal B(\mathcal H)\).
Define
\[
 \zeta(C):=\varphi(VCV^*)\quad(C\in\mathcal B(\mathcal C)),
 \qquad \rho:=\varphi\circ\pi.
\]
These states satisfy
\[
 \zeta\circ\mathcal Q
   =(\rho\widehat\otimes\omega_{\rm aux})\circ\sigma_\lambda^{-1}.
\]
This identity follows on elementary graded tensors from the definition of the stacked representation and then extends by norm continuity.
Applying \(\zeta\) to the operator density bound gives the hypothesis of \eqref{eq:one-sided-density-comparison}.
The preceding argument proves \eqref{eq:positive-energy-clustering}, without any normality or stationarity assumption on \(\varphi\).
For every Borel \(I\subset[0,3g/4]\), a state supported in \(P_I:=\1_I(L)\) is also supported in \(P\), so the same state estimate holds on that subwindow.

\emph{Operator estimate.}
Let \(R\) be either \(P\) or any \(P_I\) as above, and suppose \(R\ne0\).
Regard \(A_R:=R\pi(A)R\) and \(B_R:=R\pi(B)R\) as operators on \(R\mathcal H\), and set
\[
 T_R:=R\pi(A)(\1-R)\pi(B)R
     =R\pi(AB)R-A_RB_R.
\]
For a unit vector \(\psi\in R\mathcal H\), let \(\rho_\psi(A):=\langle\psi,\pi(A)\psi\rangle\).
Then
\[
 \langle\psi,T_R\psi\rangle
 =\operatorname{Cov}_{\rho_\psi}(A,B)
  -\bigl(\langle\psi,A_RB_R\psi\rangle
       -\langle\psi,A_R\psi\rangle\langle\psi,B_R\psi\rangle\bigr).
\]
Subtracting \(\omega_\lambda(A)R\) from \(A_R\) leaves the expression in parentheses unchanged and bounds its absolute value by
\[
 2\|A_R-\omega_\lambda(A)R\|\,\|B_R\|
 \le 2C(1+|X|)^q\|A\|\|B\|G(\dist(X,U^c)),
\]
where Theorem~\ref{thm:positive-energy-locality} applies by further compression to \(R\).
Together with the state bound, this controls the numerical radius of \(T_R\) by the right-hand side of \eqref{eq:operator-one-sided-clustering}, up to a constant depending only on the model and LED data and \(q\).
The elementary inequality
\[
 \|T_R\|\le2\sup_{\substack{\psi\in R\mathcal H\\\|\psi\|=1}}
               |\langle\psi,T_R\psi\rangle|
\]
proves \eqref{eq:operator-one-sided-clustering} after increasing \(C\).
This argument applies to arbitrary \(A,B\), without self-adjointness or parity restrictions.
The same proof applies to the subwindow projection \(P_I\), with middle factor \(\1-P_I\).
If \(R=0\), the estimate is immediate.

\emph{Local ground states.}
Let \(\rho\in\LG_{\Phi_\lambda}(U)\).
Set \(\varrho:=(\rho\widehat\otimes\omega_{\rm aux})\circ\sigma_\lambda^{-1}\).
Estimates \eqref{eq:density-finite-polynomial} and \eqref{eq:density-stretched} give \(\varrho(n_x)\le H(\dist(x,U^c))\), with \(H(D)=C(1+D)^{-2\nu}\) in the polynomial case and \(H(D)=C\E^{-\mu_{\rm den}D^\theta}\) in the stretched-exponential case, uniformly over \(\rho\in\LG_{\Phi_\lambda}(U)\).
For \(U\ne\Gamma\), substituting this bound into \eqref{eq:one-sided-density-comparison} proves \eqref{eq:one-sided-clustering}.
For \(U=\Gamma\), the zero-density argument above applies.

All constants used above depend only on the model and LED data and the chosen exponents.
In the stretched-exponential case, make the preceding choice of \(m\) for both the local-ground-state density estimate \eqref{eq:density-stretched} and the low-energy density estimate \eqref{eq:low-energy-density}.
This single choice is positive and preserves Theorems~\ref{thm:positive-energy-locality} and~\ref{thm:bulk-uniqueness} and their corollaries, since decreasing \(m\) only enlarges \(G\) at finite distance and leaves \(G(\infty)=0\).
Finally choose one constant \(C\), depending only on the model and LED data and \(q\), large enough that \eqref{eq:positive-energy-clustering}, \eqref{eq:operator-one-sided-clustering}, and \eqref{eq:one-sided-clustering} hold with this same \(C\).
\end{proof}

\begin{proof}[Proof of Corollary~\ref{cor:finite-sample-correlations}(i)]
Fix \(\lambda,\Lambda,w,K\) as in Corollary~\ref{cor:finite-sample-bulk}, and abbreviate \(H:=H_{\Lambda,K}^{(\lambda)}\), \(E:=E_{\Lambda,K}^{(\lambda)}\), and \(U:=\Lambda\setminus C_{\Lambda,w}=S_{\Lambda,w}^c\).
The cut interaction \(\Xi\) from \eqref{eq:cut-comparison-interaction} satisfies \eqref{eq:comparison-interaction-criterion}: terms supported in \(U\) agree with the bulk, and terms crossing its boundary are unchanged if their supports lie in \(\Lambda\) and zero otherwise.
The boundary operator \(K\) is supported in \(U^c\).
As in the proof of Corollary~\ref{cor:finite-sample-bulk}, its dynamics is \(\tau_t^\Xi=\operatorname{Ad}(\E^{\I tH})\); this also covers an empty collar, whose boundary term is scalar.

Choose a faithful unital representation \(\pi\) of \(\A\).
Then \(L:=\pi(H-E\1)\ge0\) implements \(\tau^\Xi\).
For every Borel \(I\subset[0,3g/4]\), set \(Q_I:=\1_{E+I}(H)\).
Since \(H\) has finite spectrum,
\[
 \1_I(L)=\pi(Q_I).
\]
Apply the spectral-subwindow form of \eqref{eq:operator-one-sided-clustering} in Theorem~\ref{thm:spatial-correlations}.
Faithfulness gives
\[
 \|\pi(Q_I)\pi(A)(\1-\pi(Q_I))\pi(B)\pi(Q_I)\|
 =\|Q_IA(\1-Q_I)BQ_I\|,
\]
so the operator bound transfers to the sample.

For the covariance bound, let \(\rho\) be any sample state with \(\rho(Q_I)=1\), choose any state extension \(\rho^\Gamma\) to \(\A\), and use its GNS representation \((\pi_\rho,\mathcal H_\rho,\Omega_\rho)\).
The operator \(L_\rho:=\pi_\rho(H-E\1)\ge0\) implements the same dynamics, and
\[
 \1_I(L_\rho)\Omega_\rho
 =\pi_\rho(Q_I)\Omega_\rho=\Omega_\rho.
\]
The vector-state case of \eqref{eq:positive-energy-clustering} therefore applies.
For \(A,B\in\A_\Lambda\), its covariance equals \(\operatorname{Cov}_\rho(A,B)\).
Taking \(I=[0,3g/4]\) gives the full sample window, and all constants are independent of \(\Lambda,w,K,I\), and the chosen state.
\end{proof}

\begin{proof}[Proof of Corollary~\ref{cor:finite-sample-correlations}(ii)]
Use the cut interaction and faithful representation from the proof of part (i).
Functional calculus gives
\[
 \E^{-t\pi(H-E\1)}=\pi(\E^{-t(H-E\1)}),
 \qquad
 (\pi(H-E\1)+z)^{-1}=\pi((H-E\1+z\1)^{-1}),
\]
and identifies the spectral projections with their sample counterparts.
Faithfulness therefore transfers estimates \eqref{eq:low-energy-spectral-weight}, \eqref{eq:mixed-imaginary-time-correlation}, and \eqref{eq:elementary-inverse-energy} to the sample algebra \(\A_\Lambda\).
For every sample state \(\rho\) with \(\rho(P_{\Lambda,K}^{(\lambda)})=1\), apply \eqref{eq:centered-low-energy-vector} and \eqref{eq:imaginary-time-correlation} with \(\varphi=\rho\), \(A_\rho=A-\rho(A)\1\), and \(B_\rho=B-\rho(B)\1\).
Integrating \eqref{eq:imaginary-time-correlation} against \(\E^{-zt}\) gives the bound on \(|\rho(A_\rho^*(L+z)^{-1}B_\rho)|\).
\end{proof}

\appendix

\section{Infinite-volume dynamics}
\label{app:first-moment-dynamics}

\begin{proof}[Proof of Proposition~\ref{prop:first-moment-dynamics}]
Define \(J(x,y):=\sum_{Z\ni x,y}\|\Xi(Z)\|\).
This nonnegative symmetric matrix satisfies \(\sup_x\sum_yJ(x,y)=\|\Xi\|_1\).
For finite \(X\subset\Lambda\), \(Y\Subset\Gamma\), \(A\in\A_X\), and \(B\in\A_Y\) with at least one observable even, \cite[Lemma~3.1]{FermionicLR}, applied to the constant interaction path with weight \(w=1\), gives
\begin{equation}
 \|[\tau_t^{\Xi,\Lambda}(A),B]\|
 \le2\|A\|\|B\|\sum_{x\in X,y\in Y}\E^{2|t|J}(x,y).
 \label{eq:first-moment-LR}
\end{equation}
Apply this estimate to \(B=\Xi(Z)\) and sum over crossing terms, counting a site \(y\in Z\setminus\Lambda\), to obtain
\begin{equation}
 \sum_{Z\text{ crossing }\Lambda}
 \|[\Xi(Z),\tau_t^{\Xi,\Lambda}(A)]\|
 \le2\|A\|\sum_{x\in X,y\notin\Lambda}
             (\E^{2|t|J}J)(x,y).
 \label{eq:first-moment-crossing}
\end{equation}
All sums converge absolutely because the matrix on the right has row sums at most \(\|\Xi\|_1\E^{2|t|\|\Xi\|_1}\).
For \(X\subset\Lambda\subset\Lambda'\), Duhamel's formula therefore gives
\begin{equation}
 \sup_{|t|\le T}\|\tau_t^{\Xi,\Lambda'}(A)-\tau_t^{\Xi,\Lambda}(A)\|
 \le2T\|A\|\sum_{x\in X,y\notin\Lambda}(\E^{2TJ}J)(x,y).
 \label{eq:finite-volume-dynamics-Cauchy}
\end{equation}
The right side tends to zero along every finite exhaustion.
The finite-volume dynamics thus converge on the local algebra, uniformly on compact time intervals, and isometry and density extend this convergence to \(\A\).
Their algebraic relations pass to the limit and give a strongly continuous parity-preserving automorphism group \(\tau^\Xi\); comparing two exhaustions through their unions shows that the limit is independent of the exhaustion.

Let \(\mathcal G\) be its generator.
For local \(A\), absolute convergence of the defining commutator sum gives \(\delta_{\Xi,\Lambda}(A)\to\delta_\Xi(A)\) in norm.
Passing to the limit in the finite-volume integral equation gives
\[
 \tau_t^\Xi(A)-A
 =\I\int_0^t\tau_s^\Xi(\delta_\Xi(A))\,\dd s.
\]
Hence \(\mathcal GA=\I\delta_\Xi(A)\), and \(\delta_\Xi\) is closable.
To show that its closure gives the entire generator, we use the resolvent construction of \cite[Lemmas~4.4--4.7 and Theorem~4.8]{BruPedra17}, with \eqref{eq:first-moment-crossing} providing the required estimate under the first support moment alone.
Fix \(z>2\|\Xi\|_1\) and, for local \(A\in\A_X\) and finite \(\Lambda\supset X\), set
\[
 R_{z,\Lambda}A:=\int_0^\infty\E^{-zt}\tau_t^{\Xi,\Lambda}(A)\,\dd t
 \in\A_\Lambda.
\]
Integrating \eqref{eq:first-moment-crossing} gives
\begin{equation}
 \|(\delta_\Xi-\delta_{\Xi,\Lambda})R_{z,\Lambda}A\|
 \le2\|A\|\sum_{x\in X,y\notin\Lambda}((z-2J)^{-1}J)(x,y)
 \longrightarrow0,
 \label{eq:resolvent-boundary-range}
\end{equation}
since the nonnegative matrix \((z-2J)^{-1}J\) has summable rows.
The finite-volume resolvent identity now implies
\[
 (z-\I\delta_\Xi)R_{z,\Lambda}A
 =A-\I(\delta_\Xi-\delta_{\Xi,\Lambda})R_{z,\Lambda}A\longrightarrow A.
\]
Thus \((z-\I\delta_\Xi)(\A_{\loc})\) is dense in \(\A\).
For \(B\in\operatorname{Dom}\mathcal G\), choose local \(B_n\) with \((z-\I\delta_\Xi)B_n\to(z-\mathcal G)B\).
Applying the bounded inverse \((z-\mathcal G)^{-1}\) gives \(B_n\to B\) and \(\I\delta_\Xi(B_n)\to\mathcal GB\).
Thus \(\A_{\loc}\) is a core for \(\mathcal G=\I\overline\delta_\Xi\).
\end{proof}

\begin{proof}[Proof of Lemma~\ref{lem:local-ground-characterization}]
Only the converse requires proof.
Inserting \(\1+zB\) into \eqref{eq:local-characterization-ground-state} and letting \(z\to0\) along the real and imaginary axes gives \(\rho(\delta_\Phi(B))=0\) for local \(B\).
Applying this identity to \(B^*B\) shows that \(\rho(B^*\delta_\Phi(B))\) is real, and hence nonnegative by assumption.
Proposition~\ref{prop:first-moment-dynamics} extends this inequality by graph-norm continuity to \(\operatorname{Dom}\overline\delta_\Phi\), giving the ground-state condition of \cite[Definition~5.3.18 and Proposition~5.3.19]{BR97}.
\end{proof}

\section{Proof of the linear-field comparison}
\label{app:odd-locality-proofs}

\begin{proof}[Proof of Lemma~\ref{lem:dynamics-comparison}]
By \eqref{eq:input-metric-reduction}, it suffices to prove the estimate for \(\theta_{\rm in}=1\): applying it in \(d'\) replaces the coefficient weight by \(\E^{\mu d(x,x_u)^{\theta_{\rm in}}}\) and the volume exponent by \(d_{\rm vol}/\theta_{\rm in}\).
Fix \(\mu,x,v\) as in the statement and \(\lambda\in\R\).
Regard \(v\) as a row vector.
For every \(t\in\R\), \eqref{eq:Majorana-free-dynamics} gives
\[
 \tau_t^{\Phi_h}(A_v)=\sum_u v_u(t)\gamma_u,
 \qquad v(t):=v\E^{-\I th}.
\]
Submultiplicativity of the Schur norm and \eqref{eq:h-locality} give \(\|\E^{-\I th}\|_{\mu,1}^{\rm Sch}\le\E^{C_h|t|}\).
The triangle inequality for the distance, followed by summation over the \(2r\) Majorana labels at each site, therefore gives, for every \(t\in\R\),
\[
 \sum_u\E^{\mu d(x,x_u)}|v_u(t)|
 \le C_r\E^{C_h|t|}\sum_u\E^{\mu d(x,x_u)}|v_u|.
\]
Unitarity also gives \(\|v(t)\|_2=\|v\|_2\).
Define \(R_t:=1+2C_h|t|/\mu\).
Cauchy--Schwarz and the volume bound yield
\[
 \sum_{u:\,d(x,x_u)\le R_t}|v_u(t)|
 \le\bigl(2r|\ball_{R_t}(x)|\bigr)^{1/2}\|v(t)\|_2
 \le C_\mu(1+|t|)^{d_{\rm vol}/2}\|v\|_2.
\]
For the remaining coefficients, the weighted estimate gives
\[
 \sum_{u:\,d(x,x_u)>R_t}|v_u(t)|
 \le\E^{-\mu R_t}\sum_u\E^{\mu d(x,x_u)}|v_u(t)|
 \le C_r\sum_u\E^{\mu d(x,x_u)}|v_u|,
\]
since \(-\mu R_t+C_h|t|\le0\).
Combining these bounds and using \(\|v\|_2\le\sum_u\E^{\mu d(x,x_u)}|v_u|\), we obtain
\[
 \sum_u|v_u(t)|
 \le C_\mu(1+|t|)^{d_{\rm vol}/2}
               \sum_u\E^{\mu d(x,x_u)}|v_u|.
\]

We next justify applying the generators to these fields.
For a coefficient vector \(w\in\ell^1(\Gamma\times\{1,\ldots,2r\})\), write \(A_w:=\sum_u w_u\gamma_u\); this map satisfies \(\|A_w\|\le\|w\|_{\ell^1(\Gamma\times\{1,\ldots,2r\})}\).
For any \(\Xi\in\mathfrak J_1\) and any finitely supported coefficient vector \(w\), evenness of \(\Xi(Z)\) implies \([\Xi(Z),\gamma_u]=0\) when \(x_u\notin Z\).
Thus
\[
 \|\delta_\Xi(A_w)\|
 \le2\sum_u|w_u|\sum_{Z\ni x_u}\|\Xi(Z)\|
 \le2\|\Xi\|_1\|w\|_{\ell^1(\Gamma\times\{1,\ldots,2r\})}.
\]
For \(w\in\ell^1\), take finitely supported vectors converging to \(w\) in \(\ell^1\).
Their fields converge to \(A_w\) in norm, and the displayed estimate makes their images under \(\delta_\Xi\) a Cauchy sequence.
Proposition~\ref{prop:first-moment-dynamics} therefore gives \(A_w\in\operatorname{Dom}(\overline\delta_\Xi)\) and
\[
 \|\overline\delta_\Xi(A_w)\|\le2\|\Xi\|_1\|w\|_{\ell^1(\Gamma\times\{1,\ldots,2r\})}.
\]
Applying this approximation with \(\Xi=\Phi_h,\mathcal V,\Phi_\lambda\) also gives
\[
 \overline\delta_{\Phi_\lambda}(A_w)
 =\overline\delta_{\Phi_h}(A_w)
       +\lambda\overline\delta_{\mathcal V}(A_w).
\]

The Schur bound makes right multiplication by \(h\) bounded on coefficient \(\ell^1\), so \(s\mapsto v(s)\) is continuously differentiable there.
The preceding estimates show that \(s\mapsto A_{v(s)}\) is norm-differentiable and that \(s\mapsto\overline\delta_{\Phi_\lambda}(A_{v(s)})\) is norm-continuous.
The product rule for the dynamics therefore gives, for fixed \(t\in\R\),
\[
 \frac{\dd}{\dd s}
 \tau_{t-s}^{\Phi_\lambda}(\tau_s^{\Phi_h}(A_v))
 =-\I\lambda\tau_{t-s}^{\Phi_\lambda}
       \bigl(\overline\delta_{\mathcal V}
                      (\tau_s^{\Phi_h}(A_v))\bigr).
\]
Integrating from \(0\) to \(t\) yields
\[
 \tau_t^{\Phi_\lambda}(A_v)-\tau_t^{\Phi_h}(A_v)
 =\I\lambda\int_0^t\tau_{t-s}^{\Phi_\lambda}
       \bigl(\overline\delta_{\mathcal V}
                      (\tau_s^{\Phi_h}(A_v))\bigr)\,\dd s.
\]
For negative \(t\), the integral is oriented from \(0\) to \(t\).
Isometry of the interacting dynamics and the estimate for \(\overline\delta_{\mathcal V}\), together with the bound on \(\sum_u|v_u(s)|\), give
\begin{align*}
 \|\tau_t^{\Phi_\lambda}(A_v)-\tau_t^{\Phi_h}(A_v)\|
 &\le C_\mu|\lambda|
       \left(\int_0^{|t|}(1+s)^{d_{\rm vol}/2}\,\dd s\right)
       \sum_u\E^{\mu d(x,x_u)}|v_u|\\
 &\le C_\mu|\lambda|\,|t|(1+|t|)^{d_{\rm vol}/2}
       \sum_u\E^{\mu d(x,x_u)}|v_u|.\qedhere
\end{align*}
\end{proof}

\section*{Acknowledgments}

The author thanks Lennart Becker, Marius Lemm, Harman Preet Singh, Marius Wesle, and Tom Wessel for helpful remarks and discussions.
He also thanks Domenico Monaco and Gianluca Panati at Sapienza University of Rome and Massimo Moscolari at Politecnico di Milano for their hospitality during his two-month sabbatical visits to their respective institutions.
This work was supported by the Deutsche Forschungsgemeinschaft (DFG, German Research Foundation) -- Project numbers 470903074 and 465199066.

\section*{Declaration on the use of generative AI}

During the preparation of this manuscript, the author used OpenAI's ChatGPT Pro (GPT-5.6 Sol) to assist in exploring and critically examining mathematical arguments and literature, as well as for creating the TikZ figures.
The final version was checked for consistency and polished using GPT-6 Astra.
The author reviewed and revised all AI-assisted material.
The author takes full responsibility for the content of the manuscript.

\end{document}